\documentclass[12pt,prc,showpacs,superscriptaddress]{revtex4}
\usepackage{graphicx}

\begin{document}
\title{UrQMD v2.3 - Changes and Comparisons}
\author{Hannah Petersen}
\affiliation{Frankfurt Institute for Advanced Studies (FIAS),
Ruth-Moufang-Str.~1, D-60438 Frankfurt am Main,
Germany}
\affiliation{Institut f\"ur Theoretische Physik, Johann Wolfgang Goethe-Universit\"at, Max-von-Laue-Str.~1, 
D-60438 Frankfurt am Main, Germany} 
\author{Marcus Bleicher}
\affiliation{Institut f\"ur Theoretische Physik, Johann Wolfgang Goethe-Universit\"at, Max-von-Laue-Str.~1,
 D-60438 Frankfurt am Main, Germany}
\author{Steffen A. Bass}
\affiliation{Department of Physics, Duke University, Durham, NC 27708, USA}
\author{Horst St\"ocker}
\affiliation{Frankfurt Institute for Advanced Studies (FIAS),
Ruth-Moufang-Str.~1, D-60438 Frankfurt am Main,
Germany}
\affiliation{Institut f\"ur Theoretische Physik, Johann Wolfgang Goethe-Universit\"at, Max-von-Laue-Str.~1, 
D-60438 Frankfurt am Main, Germany} 
\affiliation{Gesellschaft f\"ur Schwerionenforschung (GSI), Planckstr.~1, D-64291 Darmstadt, Germany}

\begin{abstract}
The new version of the Ultra-relativistic Quantum Molecular Dynamics model (UrQMD-2.3) is presented. The Ultra-relativistic Quantum Molecular Dynamics model (UrQMD) is a microscopic many body approach to pp, pA and AA interactions at relativistic energies. The major updates and changes are explained and a comparison to the previous version (UrQMD-1.3p1) in the context of the available data is performed. The plots and numerical data tables for hadron (i.e. $\pi$, K, p, $\bar{p}$, $\Lambda$, $\bar{\Lambda}$, $\Xi$, $\bar{\Xi}$, $\Omega$, $\bar{\Omega}$) multiplicities, and mean transverse mass momenta, $\langle m_T \rangle -m_0$ excitation functions, transverse mass spectra and rapidity distributions in pp and central Au+Au/Pb+Pb reactions from $E_{\rm lab}=2A~$GeV to $\sqrt{s}_{\rm NN}=200$ GeV are provided in this paper. The source code of UrQMD-2.3 is available at www.th.physik.uni-frankfurt.de/\textasciitilde urqmd.  
\end{abstract}

\maketitle
\section{Introduction}
The Ultra-relativistic Quantum Molecular Dynamics model (UrQMD) \cite{Bleicher:1999xi,Bass:1998ca} is a microscopic many body approach and can be applied 
to study hadron-hadron, hadron-nucleus and heavy ion reactions from $E_{\rm lab}= 100A$~MeV to $\sqrt{s_{NN}}=200$ GeV. 
This microscopic transport approach is based on the covariant propagation of color strings, 
constituent quarks and diquarks (as string ends) accompanied by mesonic and baryonic 
degrees of freedom. It simulates multiple interactions of 
in-going and newly produced particles, the excitation
and fragmentation of colour strings and the formation and decay of
hadronic resonances. 
Towards higher energies, the treatment of sub-hadronic degrees of freedom is
of major importance.
In the present model, these degrees of freedom enter via
the introduction of a formation time for hadrons produced in the 
fragmentation of strings \cite{Andersson:1986gw,NilssonAlmqvist:1986rx,Sjostrand:1993yb} and by hard (pQCD) scatterings via the PYTHIA model.
A phase transition to a quark-gluon state is 
not incorporated explicitly into the model dynamics. Let us shortly review the major physics questions and topics in which UrQMD has been used in the past: 

\begin{itemize}
\item
The thermal properties of the UrQMD model have been investigated. It was shown that a detailed analysis of the model in equilibrium yields an effective equation of state of 
Hagedorn type \cite{Belkacem:1998gy,Bravina:1999dh}. Further studies involve the exploration of the systems evolution in the QCD phase diagram and the equilibation time scales of QCD matter. This includes also studies on the active degrees of freedom and the relation between pressure and energy density (equation of state) \cite{nucl-th/9711032,nucl-th/9804008,nucl-th/9804058,nucl-th/9808021,Bravina:1999kk,hep-ph/9906548,hep-ph/0010172,nucl-th/0011011,nucl-th/0601062}. 

\item 
The UrQMD transport model has been successfully used to predict and interprete experimental data at various energies and for a multitude of observables and reaction systems, e.g. hadron yields, transverse and longitudinal spectra \cite{nucl-th/9907090,hep-ph/9911420,nucl-th/0402026}, 

\item
strangeness production, multi-strange baryons and antiprotons \cite{hep-ph/9811459,nucl-th/9907026,hep-ph/0004045,hep-ph/0007215,nucl-th/0402026}, 
\item
hadron resonance production e.g. $K^*$,$\rho$,$\Lambda^*$,$\Delta$ \cite{hep-ph/0201123,hep-ph/0212378,hep-ph/0312278,nucl-th/0509105,Vogel:2006ss,arXiv:0710.1158}, 
\item
radial, directed and elliptic flow \cite{hep-ph/9803346,nucl-th/9903061,hep-ph/0006147,nucl-th/0205069,nucl-th/0509081,Zhu:2006yv,nucl-th/0601049,nucl-th/0602009,hep-ph/0608189,nucl-th/0703031}, 
\item
event-by-event fluctuations \cite{Bleicher:1998wd,hep-ph/9803345,hep-ph/0006201,nucl-th/0006047,nucl-th/0103084,nucl-th/0506025,hep-ph/0507189,nucl-th/0511083,nucl-th/0608021,arXiv:0707.1788}, 
\item
particle correlations and HBT \cite{nucl-th/9904080,Li:2006gp,nucl-th/0612030,arXiv:0706.2091,arXiv:0709.1409,arXiv:0802.3618}, 
\item
real photon and dilepton production \cite{hep-ph/9709487,nucl-th/9712069,hep-ph/0008119,nucl-th/0608041,arXiv:0710.4463},
\item
Drell-Yan, charm, D-mesons and J/$\Psi$ production and dynamics \cite{hep-ph/9706525,Spieles:1998wz,hep-ph/9809441,hep-ph/9810486,Spieles:1999pm,hep-ph/9902337,nucl-th/0301067,hep-ph/0604178} and 
\item
studies at low beam energies to explore potential effects and isospin asymmetries \cite{nucl-th/0506030,nucl-th/0507068,nucl-th/0509070,nucl-th/0601047}.
\end{itemize}

Furthermore, the UrQMD model has been used within various hybrid model studies ranging from air shower simulations \cite{astro-ph/0305429,astro-ph/0307453} to hybrid models for relativistic heavy ion reactions. Most noteworthy are the pioneering studies related to a coupling between UrQMD and hydrodynamics, see e.g. \cite{nucl-th/9901046,nucl-th/9902055,nucl-th/9902062,nucl-th/0001033,nucl-th/0012085,nucl-th/0109055,hep-ph/0111187,nucl-th/0312015,nucl-th/0510038,nucl-th/0607018,arXiv:0710.0332}. 

The aim of this article is to describe the major changes between the last publically available version UrQMD-1.3p1 and the present state-of-the-art version UrQMD-2.3 of the model and to compare the results of both versions to each other and to the available data. This paper is organized as follows: In Section \ref{pythia} the inclusion of the Pythia model for (initial) hard scatterings is described. The new treatment of high mass resonance states is explained in Section \ref{highmassresonances}. In Section \ref{changes} various small changes of lesser importance are described. Section \ref{data} consists of the comparison of UrQMD results (from the two different versions) to the available experimental data. First, the multiplicities and $\langle p_{T} \rangle$ excitation functions in elementary p-p collisions are shown for different particle species. Then, excitation functions of multiplicities and mean transverse momentum and transverse mass and rapidity spectra for Pb+Pb/Au+Au collisions in the whole energy range are investigated. All the figures are contained in the Appendix \ref{appfigs}. The numerical data for all figures are provided in Appendix \ref{appnumdata}. Section \ref{summary} summarizes the paper.

For further details concerning the implementation and the usage of the new version the reader is referred to the User Guide which is available on the UrQMD homepage.

\section{Inclusion of Pythia}
\label{pythia}

To employ the UrQMD transport approach at higher energies (above $\sqrt{s}_{\rm NN}\cong 10~$GeV) it is important to treat the initial hard collisions carefully. Therefore, we have implemented the latest version (6.409) of Pythia \cite{Sjostrand:2006za} to perform those hard collisions instead of the normal UrQMD string excitation and fragmentation routine. Note that Pythia 6.4 is technically not anymore the latest version since there is a new C++ implementation (current version 8.1). However, Pythia 6.4 is the latest stable and full-featured Fortran implementation, which is still considered to be the benchmark for the physics processes. 

The minimal center of mass energy in the individual two particle reactions for a Pythia call is $\sqrt{s}_{\rm min} = 10 $ GeV (applicability limit of Pythia). Hard collisions are defined as collisions with momentum transfer $Q > 1.5$~GeV. The transition between 
the low energy string routine and Pythia is smooth and given by the probability distribution for hard scatterings determined from Pythia. The standard low energy string routine is called to perform the string excitation and fragmentation calculation for soft collisions only. 

Leading particles produced by Pythia strings are treated in analogy to the leading particles created in the standard UrQMD string fragmentation procedure. Leading particles are the particles that contain the quarks or diquarks of the original hadrons. Those leading particles are allowed to interact with a fraction of one third, two third (for diquarks) or a half (for mesons) of their normal cross section during their formation time of $\sim 1 $fm/c, while all the other newly produced particles do not interact until they are fully formed. To account for coherence effects the cross sections for leading particles from Pythia are additionally suppressed by a factor 0.4.

\section{Treatment of high mass resonances}
\label{highmassresonances}

In the previous version UrQMD-1.3p1 the resonances with masses up to $2.2$ GeV are included with all their vacuum properties and decay dynamics. For processes at higher energies string excitation and fragmentation dominates the interaction in UrQMD-1-3p1. Since the angular distributions of the particles produced by strings are forward-backward peaked the resulting mean transverse momenta were found to be too low compared to experimental data. To reproduce the experimentally measured high $\langle p_T \rangle$ values a modified treatment of high mass resonances similar to RQMD is introduced. This modified treatment of meson-baryon interactions in the intermediate energy regime is described in the following. 

A continuous spectrum of high mass resonance states is included in the energy regime between $\sqrt{s}_{coll}=1.67$ GeV and $\sqrt{s}_{coll}=3$~GeV for meson-baryon reactions. These particle excitations are treated as pseudo-resonances instead of strings. Below $\sqrt{s}_{coll}=1.67$~GeV normal resonance excitation takes place. Above $\sqrt{s}_{coll}=3$~GeV the standard UrQMD string fragmentation is called. The properties for the unknown resonances are extrapolated from the in mass closest known resonance of same type.

To fix the strangeness production in the decay process of these new resonance states which was reduced because of the new production of high mass 
resonances instead of strings, the branching ratios of high lying resonances are changed to the 
corresponding branching ratios obtained from string decays of the same mass.
Further adjustments are made to keep the particle properties in line with the Particle Data Book 2006. All branching ratios and other resonance properties are within the limits of the Particle Data Book 2006. 

\section{Other important changes}
\label{changes}

The following list contains the most important smaller changes that have been implemented:
\begin{itemize}
\item
New Regge-parametrisations for total and elastic cross-sections at high energies are implemented for all the elementary reactions for which they are available.

\item
The mass distribution of the nucleon resonances $N^{*}$ has been fixed via inclusion of the $\Delta$ resonances.

\item  
Adjustment of the $\Xi$ and $\Omega$ production rates in p-p-collisions to newly available data 
via a change of the double strange diquark suppression factor. 

\item 
The single strange diquark suppression factor is set to 0.5 to reproduce the measured $\bar{\Lambda}$ production in p-p collisions.

\item
A new subroutine which provides a faster initialization that is needed for cosmic air shower simulations is introduced.

\item
Changes that need to be made to run UrQMD at LHC energies have been studied. The necessary adjustments are described in detail in the User Guide, but are not implemented in the default version UrQMD-2.3.
\end{itemize}

\section{Comparison to data}
\label{data}

In the following subsections the results of the new version UrQMD-2.3 are compared to the calculations using UrQMD-1.3p1 in the context of the available experimental data. We have concentrated on bulk observables like multiplicities and particle spectra to demonstrate the major differences. Here we refrain from explaining all the details of the shown figures, but concentrate on the effects of the changes that are described above. The full/dotted lines refer always to the UrQMD-2.3 (shown as full lines) and UrQMD-1.3p1 (shown as dotted lines) results, while experimental data are depicted as symbols. All the figures are contained in the appendix \ref{appfigs} with an explanatory title, key and caption. Numerical data is given in Appendix \ref{appnumdata}.  

\subsection{p-p collisions}
\label{ppcoll}

\begin{figure}
\centering
\includegraphics[width=15cm]{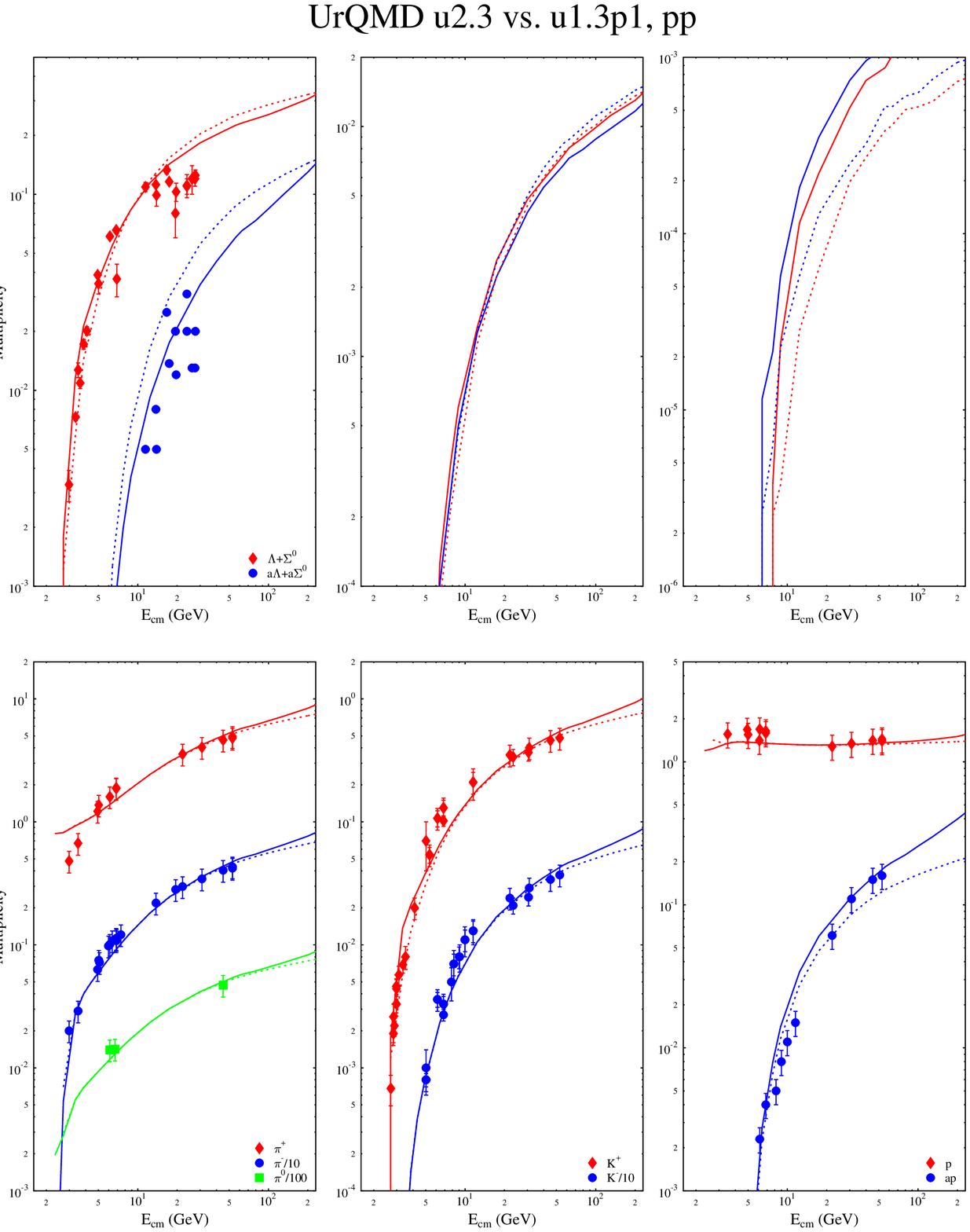}
\caption{(Color online) Excitation function of particle multiplicities ($4\pi$) in inelastic pp collisions from $E_{\rm lab}=2~A$GeV to $\sqrt{s_{NN}}=200$ GeV. UrQMD-2.3 calculations are depicted with full lines, while UrQMD-1.3p1 calculations are depicted with dotted lines. The corresponding data from different experiments \cite{Antinucci:1972ib,Rossi:1974if}are depicted with symbols.}
\label{figmulallypp}
\end{figure}

First we show the excitation functions of the total multiplicities and the mean transverse momentum for elementary p-p collisions. In Fig. \ref{figmulallypp} an enhanced production of pions, kaons and antiprotons due the implementation of Pythia is visible for energies above $\sqrt{s}_{\rm NN}\gtrsim 50~$GeV. The $\bar{\Lambda}$ yield is reduced in UrQMD-2.3 and has been adjusted by the single strange diquark suppression factor. The $\Omega$ yield is increased in UrQMD-2.3 and was adjusted by a change of the double strange diquark suppression factor. Preliminary NA49 data for $\Xi$ and $\Omega$ production in p-p was used to adjust the multiplicities. 

\begin{figure}
\centering
\includegraphics[width=15cm]{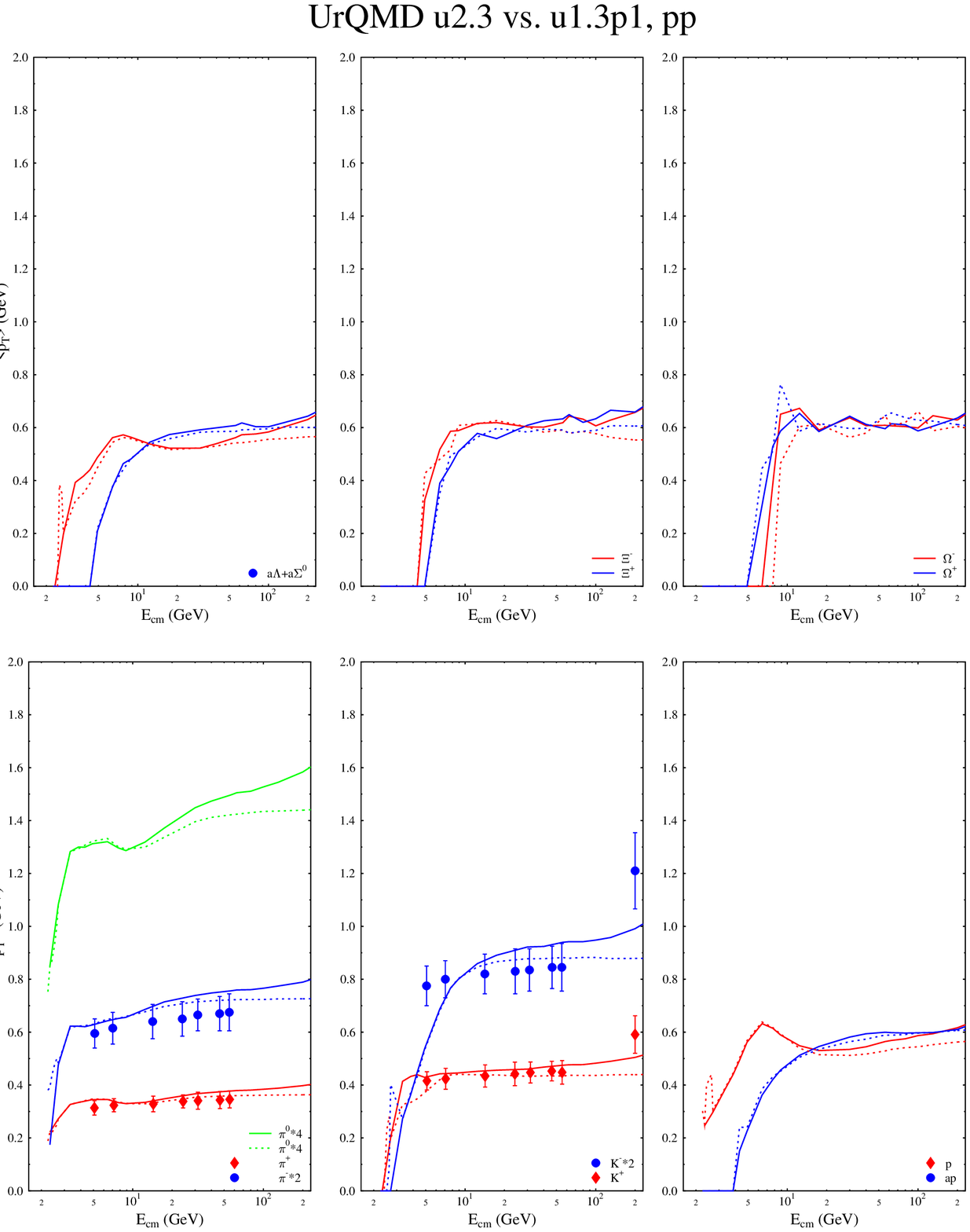}
\caption{(Color online) Excitation function of $\langle p_T \rangle$ values for different particle species at midrapidity ($|y|<0.5$) in inelastic pp collisions from $E_{\rm lab}=2~A$GeV to $\sqrt{s_{NN}}=200$ GeV. UrQMD-2.3 calculations are depicted with full lines, while UrQMD-1.3p1 calculations are depicted with dotted lines. The corresponding data from different experiments \cite{Rossi:1974if} are depicted with symbols.}
\label{figmptpp}
\end{figure}

Fig. \ref{figmptpp} shows the $\langle p_T \rangle$ of produced particles in p-p collisions. The inclusion of pQCD hard scatterings in UrQMD-2.3 leads to a slight increase of the $\langle p_T \rangle$ at higher energies compared to UrQMD-1.3p1.  

\subsection{A-A Collisions}
\label{aacoll}

In the following subsections the explanation of the changes on the results for heavy ion collisions (Au+Au/Pb+Pb) are given. All the calculations have been performed for central collisions ($b< 3.4$ fm).

\subsubsection{Multiplicities and excitation functions}
\label{mulsexc}
\begin{figure}
\centering
\includegraphics[width=15cm]{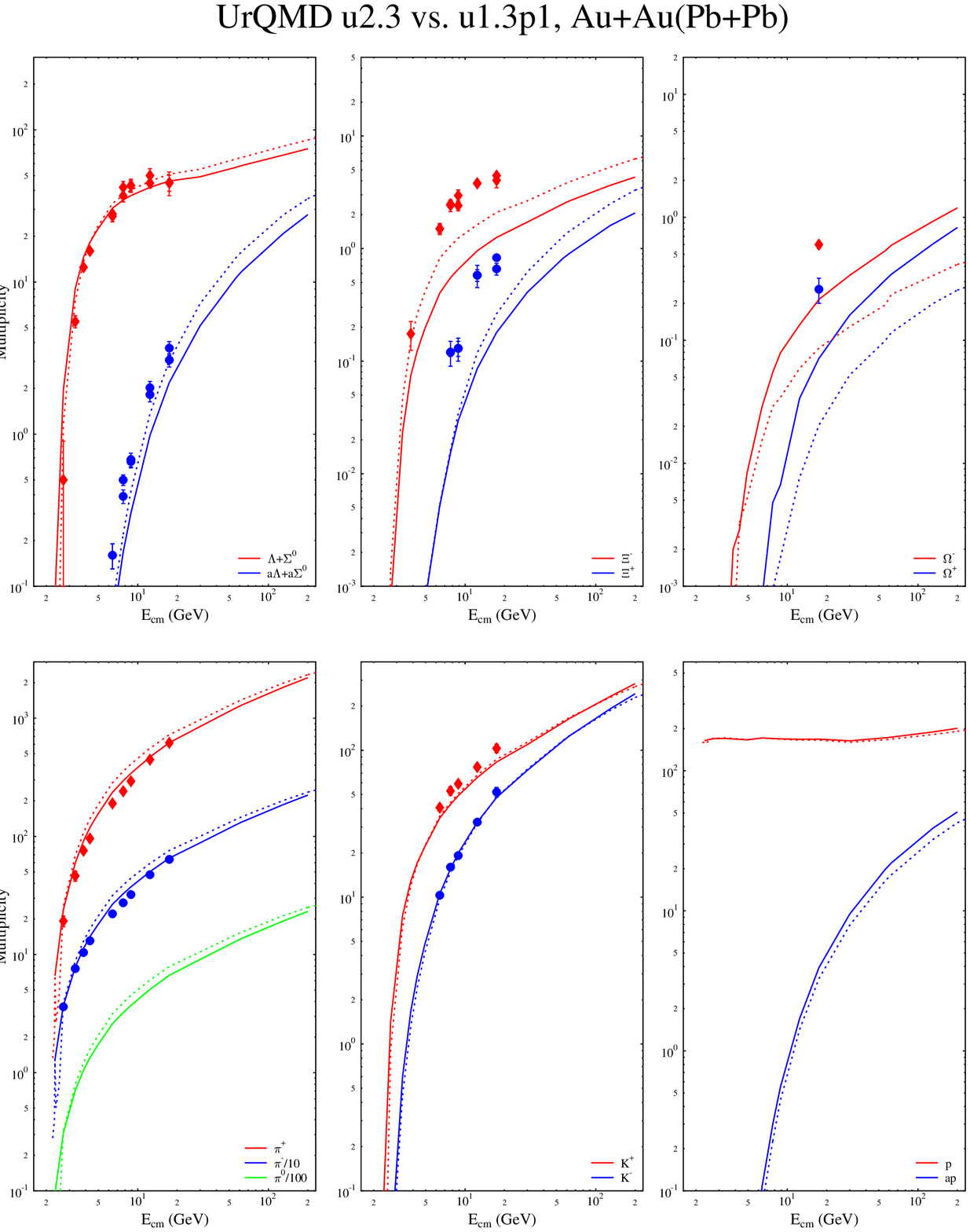}
\caption{(Color online) Excitation function of particle multiplicities ($4\pi$) in Au+Au/Pb+Pb collisions from $E_{\rm lab}=2~A$GeV to $\sqrt{s_{NN}}=200$ GeV. UrQMD-2.3 calculations are depicted with full lines, while UrQMD-1.3p1 calculations are depicted with dotted lines. The corresponding data from different experiments \cite{Klay:2003zf,Pinkenburg:2001fj,Chung:2003zr,:2007fe,Afanasiev:2002mx,Anticic:2003ux,Richard:2005rx,Mitrovski:2006js,arXiv:0804.3770,Blume:2004ci,Afanasiev:2002he,Alt:2004kq} are depicted with symbols.}
\label{figmulallyaa}
\end{figure}

\begin{figure}
\centering
\includegraphics[width=15cm]{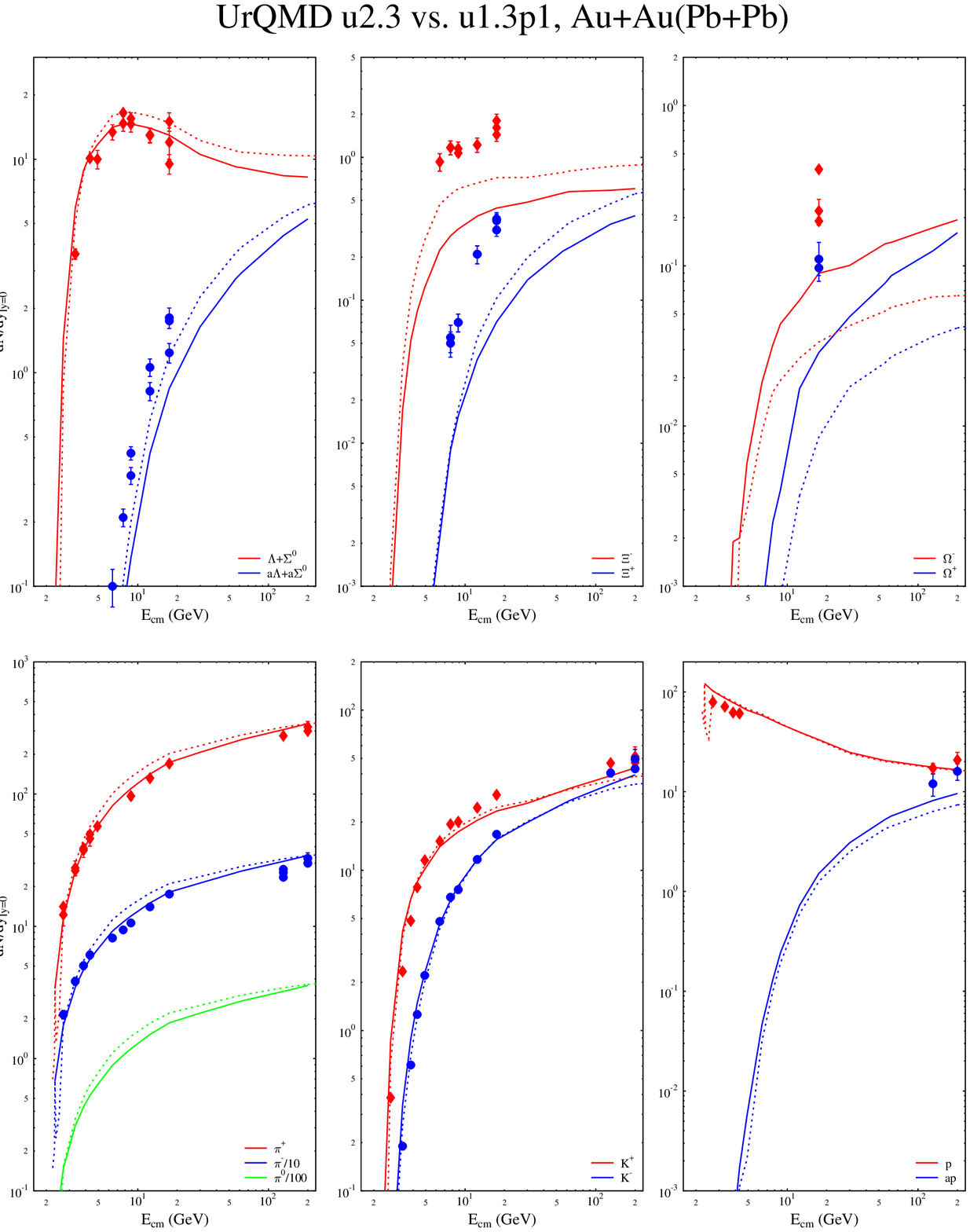}
\caption{(Color online) Excitation function of particle yields at midrapidity ($|y|<0.5$) in Au+Au/Pb+Pb collisions from $E_{\rm lab}=2~A$GeV to $\sqrt{s_{NN}}=200$ GeV. UrQMD-2.3 calculations are depicted with full lines, while UrQMD-1.3p1 calculations are depicted with dotted lines. The corresponding data from different experiments \cite{Ahle:1999uy,Klay:2003zf,Afanasiev:2002mx,Adcox:2003nr,Lee:2004bx,Ahle:2000wq,Antinori:1999hy,Ouerdane:2002gm,Ahmad:1991nv,Mitrovski:2006js,arXiv:0804.3770,Alt:2004kq,Mischke:2002wt,Adams:2003xp} are depicted with symbols.}
\label{figmulmidyaa}
\end{figure}

Fig. \ref{figmulallyaa} shows the excitation function of $4\pi$ multiplicities for different particle species. The yields of the (multi-)strange baryons has changed because of the new treatment of high mass resonances and the adjustments of the strangeness suppression factors. The same effect is also visible for the yields at midrapidity that are shown in Fig. \ref{figmulmidyaa}.

\begin{figure}
\centering
\includegraphics[width=15cm]{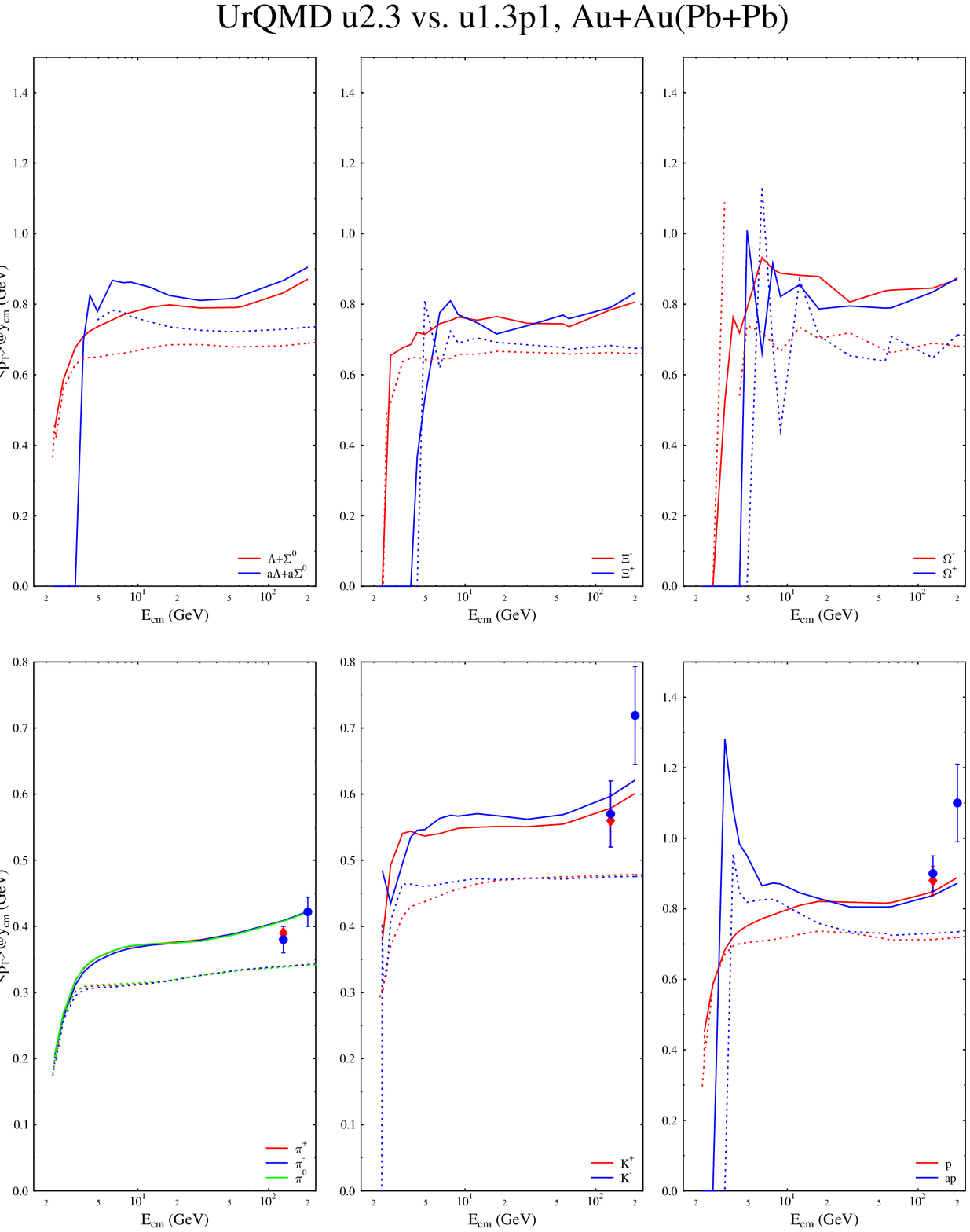}
\caption{(Color online) Excitation function of $\langle p_T \rangle$ values for different particle species at midrapidity ($|y|<0.5$) in Au+Au/Pb+Pb collisions from $E_{\rm lab}=2~A$GeV to $\sqrt{s_{NN}}=200$ GeV. UrQMD-2.3 calculations are depicted with full lines, while UrQMD-1.3p1 calculations are depicted with dotted lines. The corresponding data from different experiments \cite{Adcox:2003nr,Adams:2003xp} are depicted with symbols.}
\label{figmptaa}
\end{figure}

\begin{figure}
\centering
\includegraphics[width=15cm]{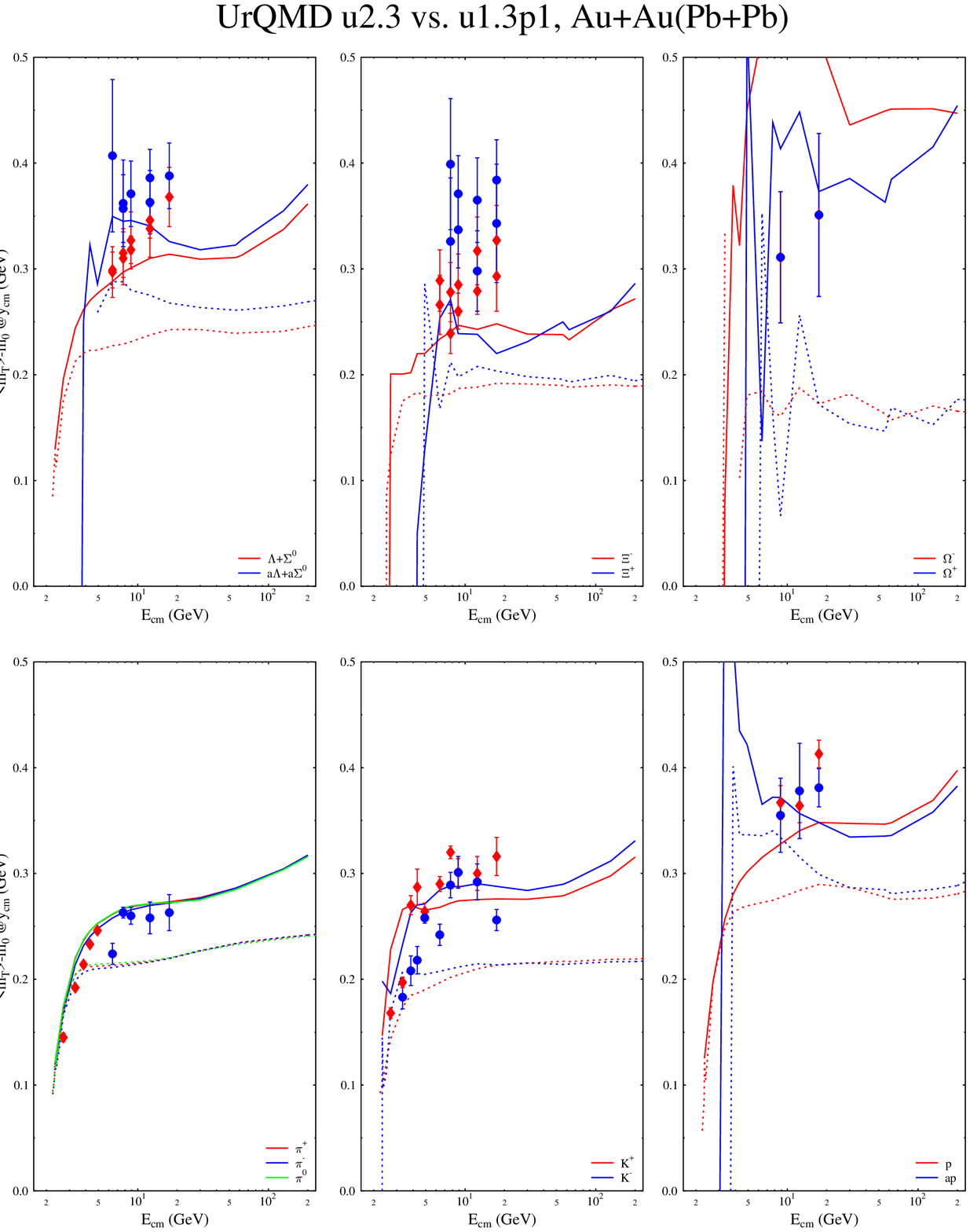}
\caption{(Color online) Excitation function of $\langle m_T \rangle -m_0$ values for different particle species at midrapidity ($|y|<0.5$) in Au+Au/Pb+Pb collisions from $E_{\rm lab}=2~A$GeV to $\sqrt{s_{NN}}=200$ GeV. UrQMD-2.3 calculations are depicted with full lines, while UrQMD-1.3p1 calculations are depicted with dotted lines. The corresponding data from different experiments \cite{Ahle:1999uy,Ahle:2000wq,Afanasiev:2002mx,Anticic:2004yj,Richard:2005rx,:2007fe,Mitrovski:2006js,arXiv:0804.3770,Afanasiev:2002fk,Veres:1999ip,Anticic:2003ux,Afanasiev:2002he,Alt:2004kq} are depicted with symbols.}
\label{figmmtaa}
\end{figure}

In Figs. \ref{figmptaa} and \ref{figmmtaa} the excitation functions of the mean transverse momentum and transverse mass for different particle species are shown. The new calculations with UrQMD-2.3 generally result in an increase in transverse momentum, resulting in a better description of the experimental data. This increase is due to the modified treatment of the high mass resonances that decay isotropically. 

\subsubsection{Transverse mass spectra}
\label{mtspectra}

\begin{figure}
\centering
\includegraphics[width=15cm]{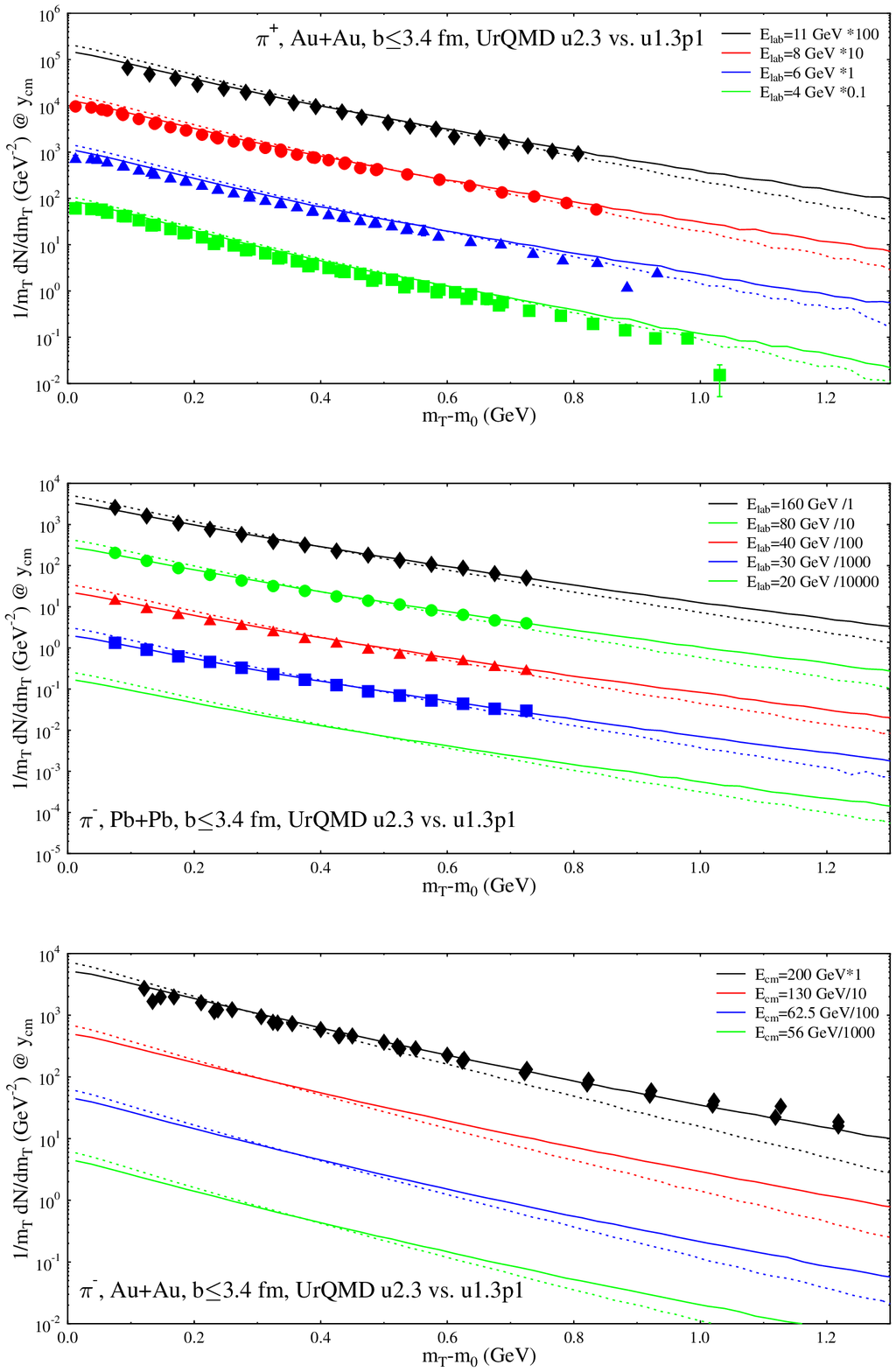}
\caption{(Color online) Transverse mass spectra of $\pi^-$ ($\pi^+$ for AGS energies) at midrapidity ($|y|<0.5$) for central ($b<3.4$ fm) Au+Au/Pb+Pb collisions from $E_{\rm lab}=2~A$GeV to $\sqrt{s_{NN}}=200$ GeV. UrQMD-2.3 calculations are depicted with full lines, while UrQMD-1.3p1 calculations are depicted with dotted lines. The corresponding data from different experiments \cite{Klay:2003zf,Afanasiev:2002mx,:2007fe,Adams:2003xp,Adler:2003cb,Arsene:2005mr} are depicted with symbols.}
\label{figdndmtpi}
\end{figure}

\begin{figure}
\centering
\includegraphics[width=15cm]{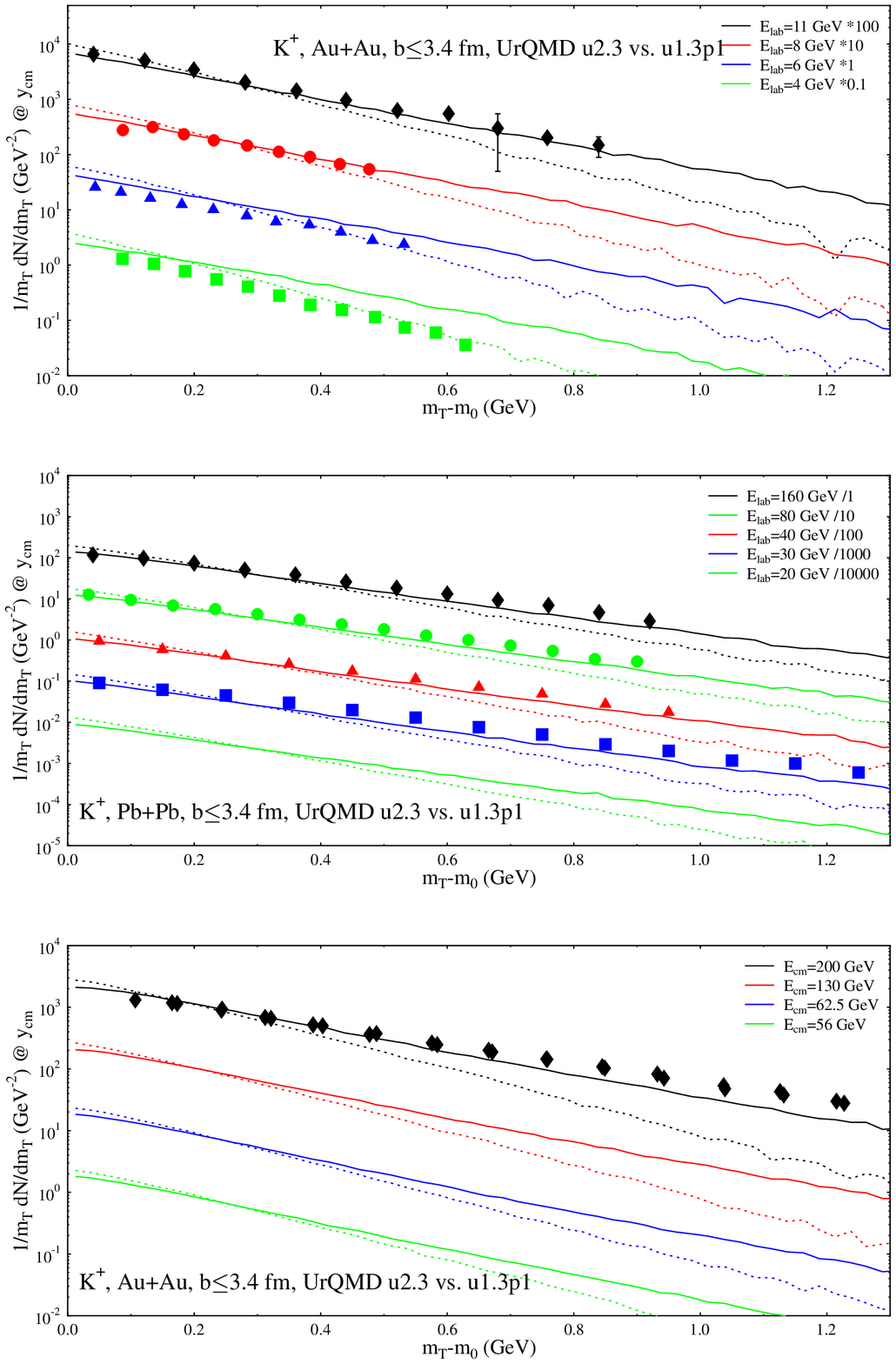}
\caption{(Color online) Transverse mass spectra of $K^+$ at midrapidity ($|y|<0.5$) for central ($b<3.4$ fm) Au+Au/Pb+Pb collisions from $E_{\rm lab}=2~A$GeV to $\sqrt{s_{NN}}=200$ GeV. UrQMD-2.3 calculations are depicted with full lines, while UrQMD-1.3p1 calculations are depicted with dotted lines. The corresponding data from different experiments \cite{Ahle:2000wq,Afanasiev:2002mx,:2007fe,Adams:2003xp,Adler:2003cb} are depicted with symbols.}
\label{figdndmtkplus}
\end{figure}

\begin{figure}
\centering
\includegraphics[width=15cm]{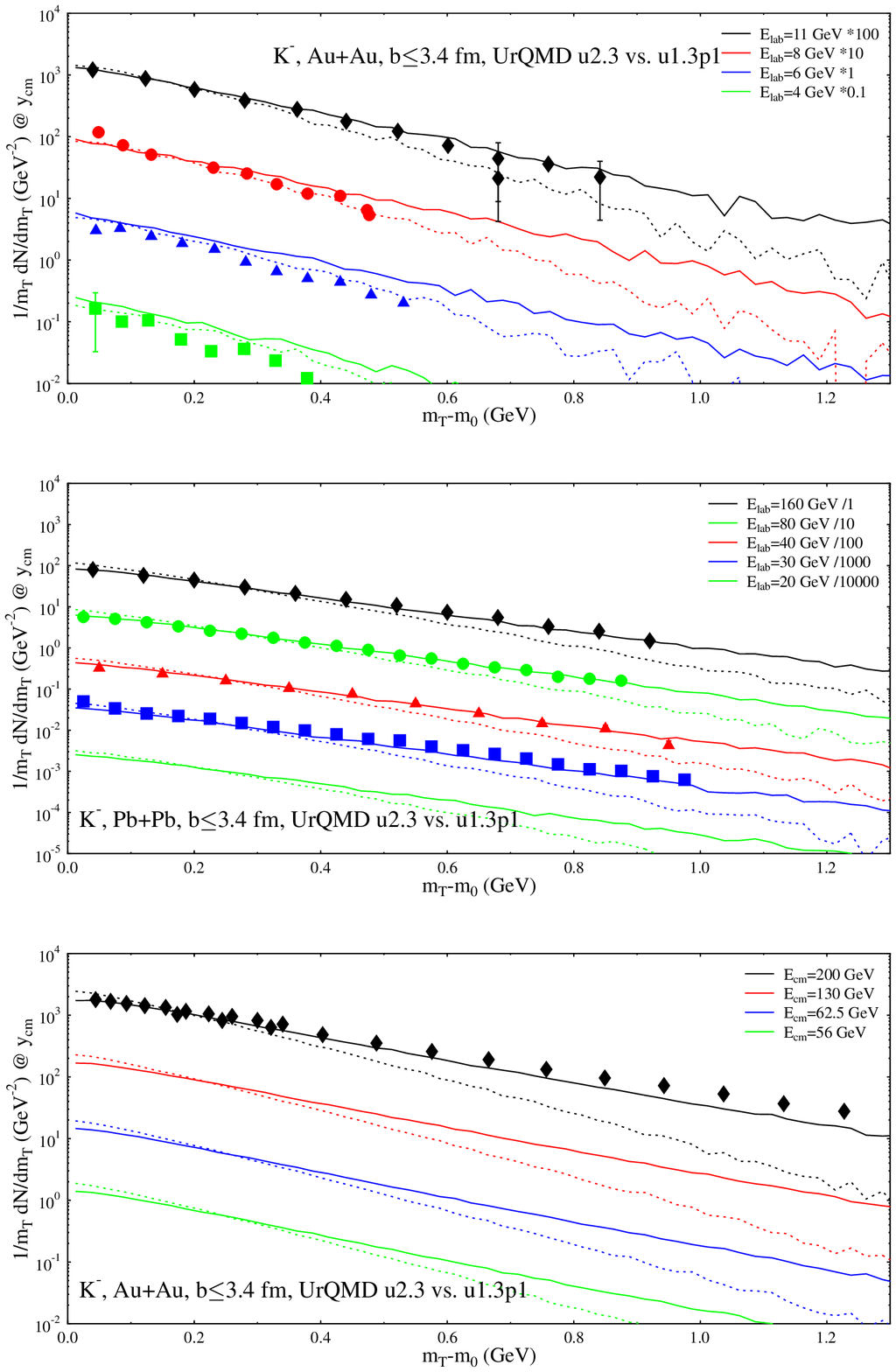}
\caption{(Color online) Transverse mass spectra of $K^-$ at midrapidity ($|y|<0.5$) for central ($b<3.4$ fm) Au+Au/Pb+Pb collisions from $E_{\rm lab}=2~A$GeV to $\sqrt{s_{NN}}=200$ GeV. UrQMD-2.3 calculations are depicted with full lines, while UrQMD-1.3p1 calculations are depicted with dotted lines. The corresponding data from different experiments \cite{Ahle:2000wq,Afanasiev:2002mx,:2007fe,Adams:2003xp,Adler:2003cb} are depicted with symbols.}
\label{figdndmtkminus}
\end{figure}

\begin{figure}
\centering
\includegraphics[width=15cm]{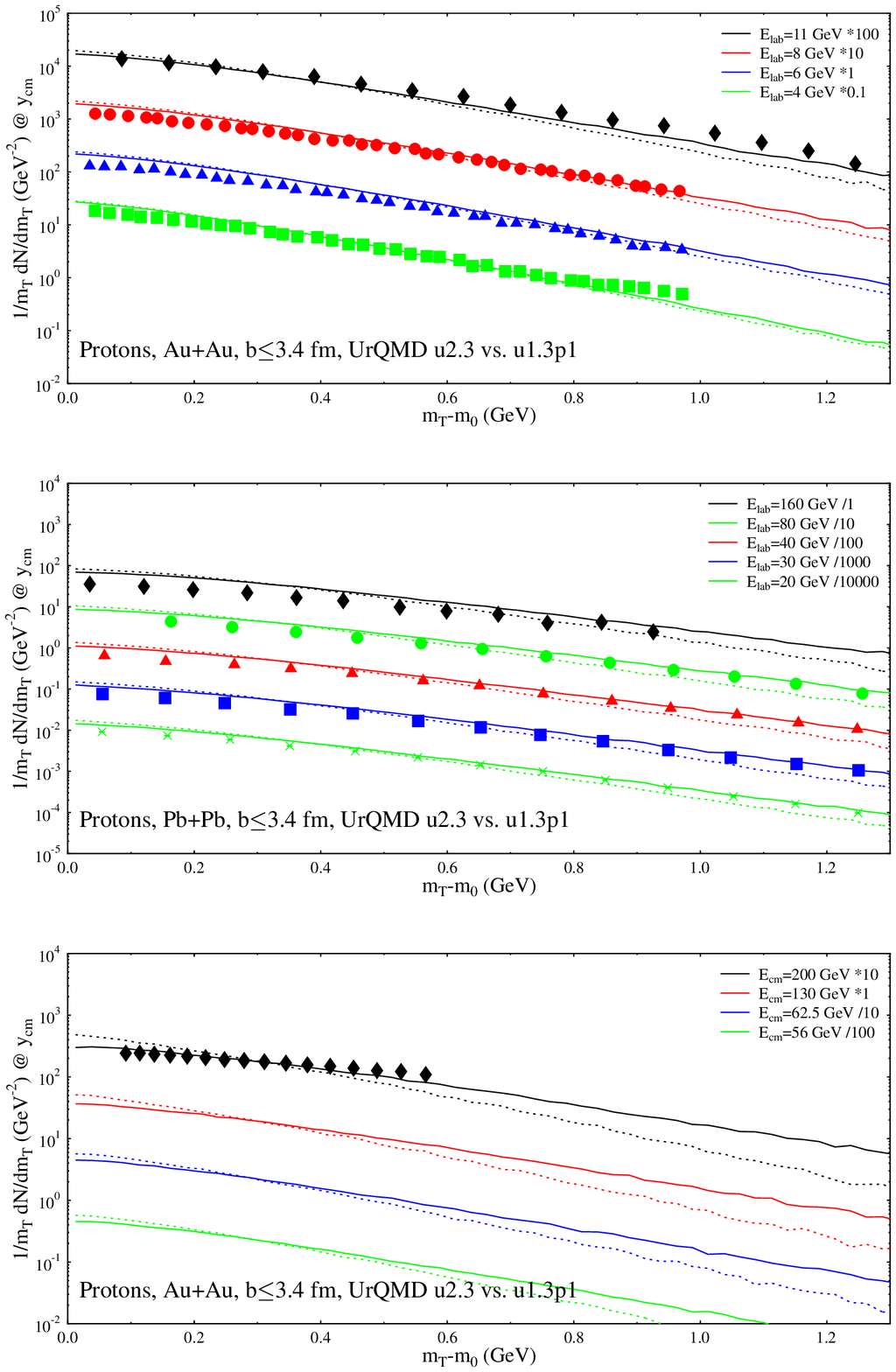}
\caption{(Color online) Transverse mass spectra of protons at midrapidity ($|y|<0.5$) for central ($b<3.4$ fm) Au+Au/Pb+Pb collisions from $E_{\rm lab}=2~A$GeV to $\sqrt{s_{NN}}=200$ GeV. UrQMD-2.3 calculations are depicted with full lines, while UrQMD-1.3p1 calculations are depicted with dotted lines. The corresponding data from different experiments \cite{Klay:2001tf,Akiba:1996xf,Alt:2006dk,Adams:2003xp} are depicted with symbols.}
\label{figdndmtp}
\end{figure}

Figs. \ref{figdndmtpi},\ref{figdndmtkplus},\ref{figdndmtkminus} and \ref{figdndmtp} show differential transverse mass spectra for pions, kaons and protons in the whole energy regime from lower AGS to the highest RHIC energy. Overall, a flatening of the slopes of the spectra due to the higher mean transverse momenta is observed. This is again due to the new treatment of the high mass resonances.

\subsubsection{Rapidity spectra}
\label{rapspectra}

In this subsection rapidity spectra for $\pi$, K, p, $\Lambda$, $\Xi$, $\Omega$ and their respective antiparticles are shown for beam energies from $E_{\rm lab}=2~A$GeV to $\sqrt{s_{NN}}=200$GeV.  

\begin{figure}
\centering
\includegraphics[width=15cm]{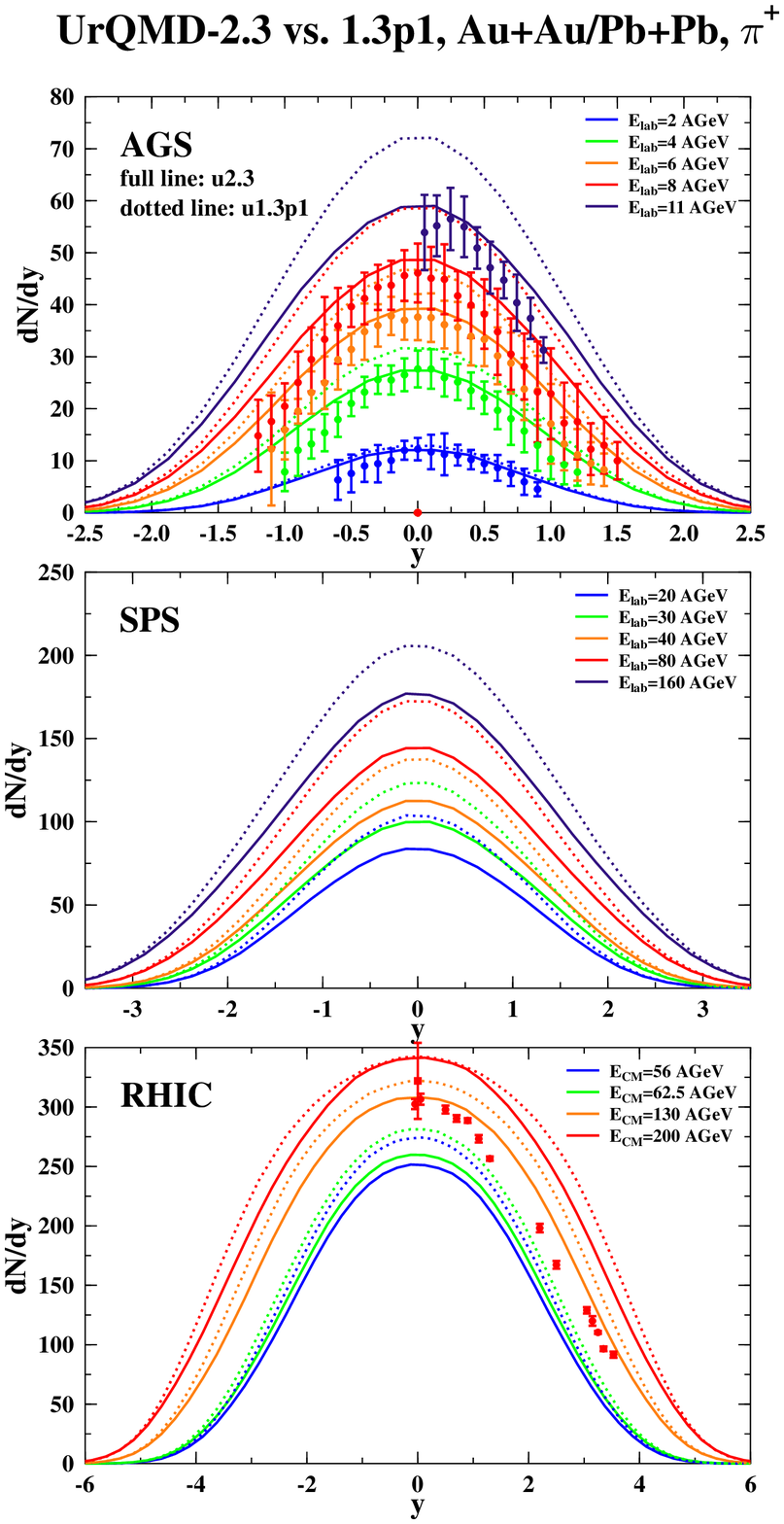}
\caption{(Color online) Rapidity spectra of $\pi^+$ for central ($b<3.4$ fm) Au+Au/Pb+Pb collisions from $E_{\rm lab}=2~A$GeV to $\sqrt{s_{NN}}=200$ GeV. UrQMD-2.3 calculations are depicted with full lines, while UrQMD-1.3p1 calculations are depicted with dotted lines. The corresponding data from different experiments \cite{Klay:2003zf,Akiba:1996xf,Bearden:2004yx} are depicted with symbols.}
\label{figdndypiplus}
\end{figure}

\begin{figure}
\centering
\includegraphics[width=15cm]{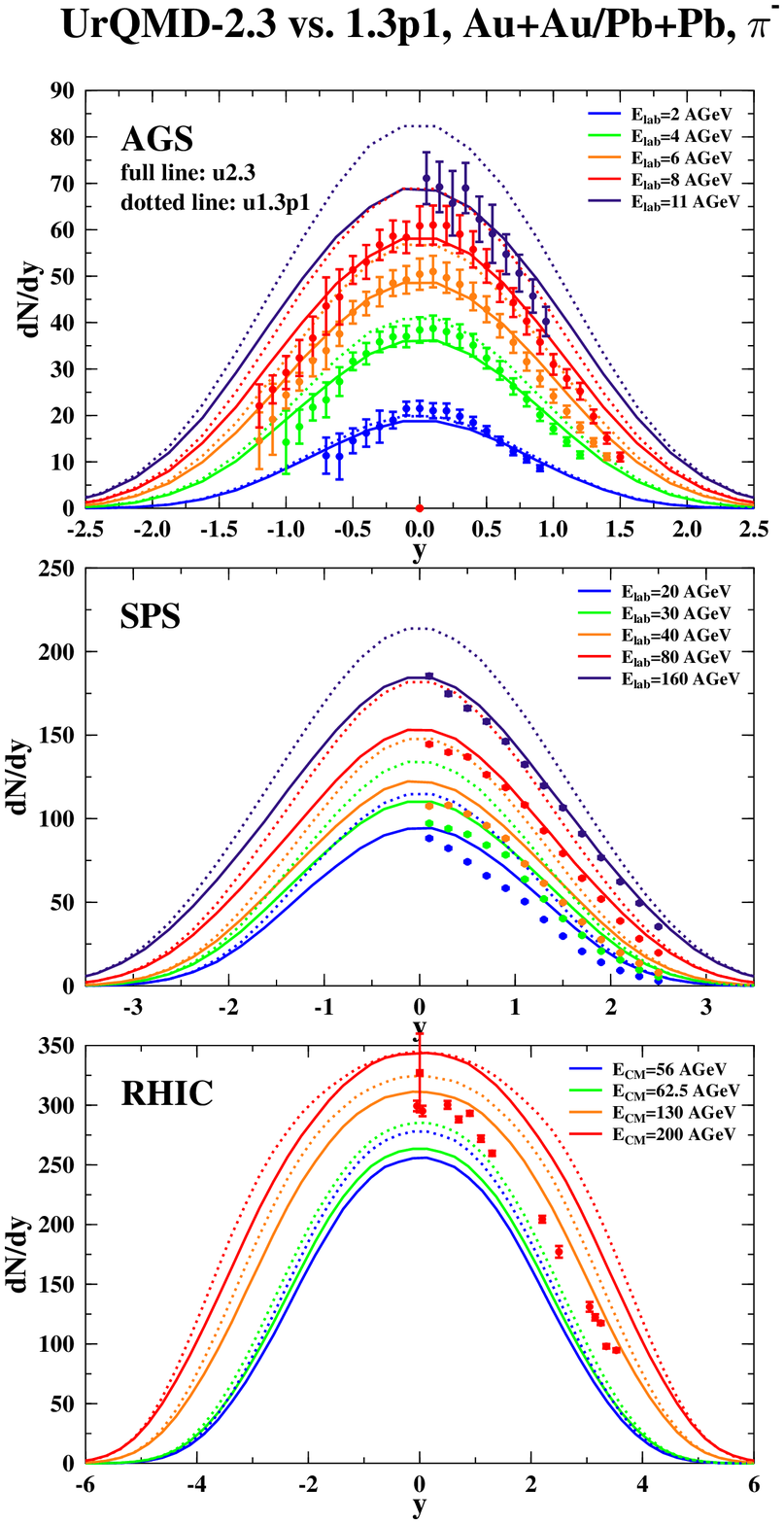}
\caption{(Color online) Rapidity spectra of $\pi^-$ for central ($b<3.4$ fm) Au+Au/Pb+Pb collisions from $E_{\rm lab}=2~A$GeV to $\sqrt{s_{NN}}=200$ GeV. UrQMD-2.3 calculations are depicted with full lines, while UrQMD-1.3p1 calculations are depicted with dotted lines. The corresponding data from different experiments \cite{Klay:2003zf,Akiba:1996xf,:2007fe,Afanasiev:2002mx,Bearden:2004yx} are depicted with symbols.}
\label{figdndypiminus}
\end{figure}

\begin{figure}
\centering
\includegraphics[width=15cm]{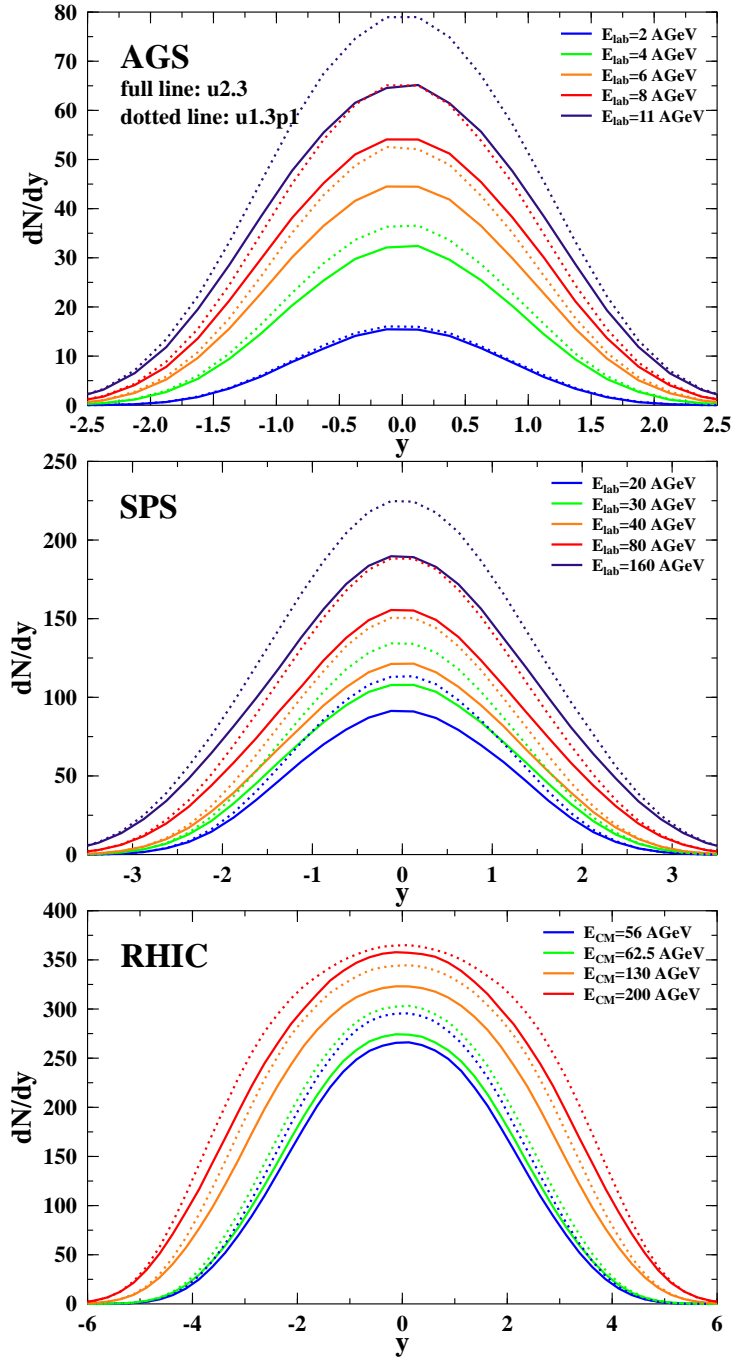}
\caption{(Color online) Rapidity spectra of $\pi_0$ for central ($b<3.4$ fm) Au+Au/Pb+Pb collisions from $E_{\rm lab}=2~A$GeV to $\sqrt{s_{NN}}=200$ GeV. UrQMD-2.3 calculations are depicted with full lines, while UrQMD-1.3p1 calculations are depicted with dotted lines.}
\label{figdndypinull}
\end{figure}

Fig. \ref{figdndypiplus}, \ref{figdndypiminus} and \ref{figdndypinull} show the results for pions. The UrQMD-2.3 calculations lead to a lower pion yield which is in better agreement with the experimental data over the whole energy range. The shape of the distribution looks very similar to the observed one. Please note that at RHIC energies the pion yield at midrapidity differs between the different experiments. 

\begin{figure}
\centering
\includegraphics[width=15cm]{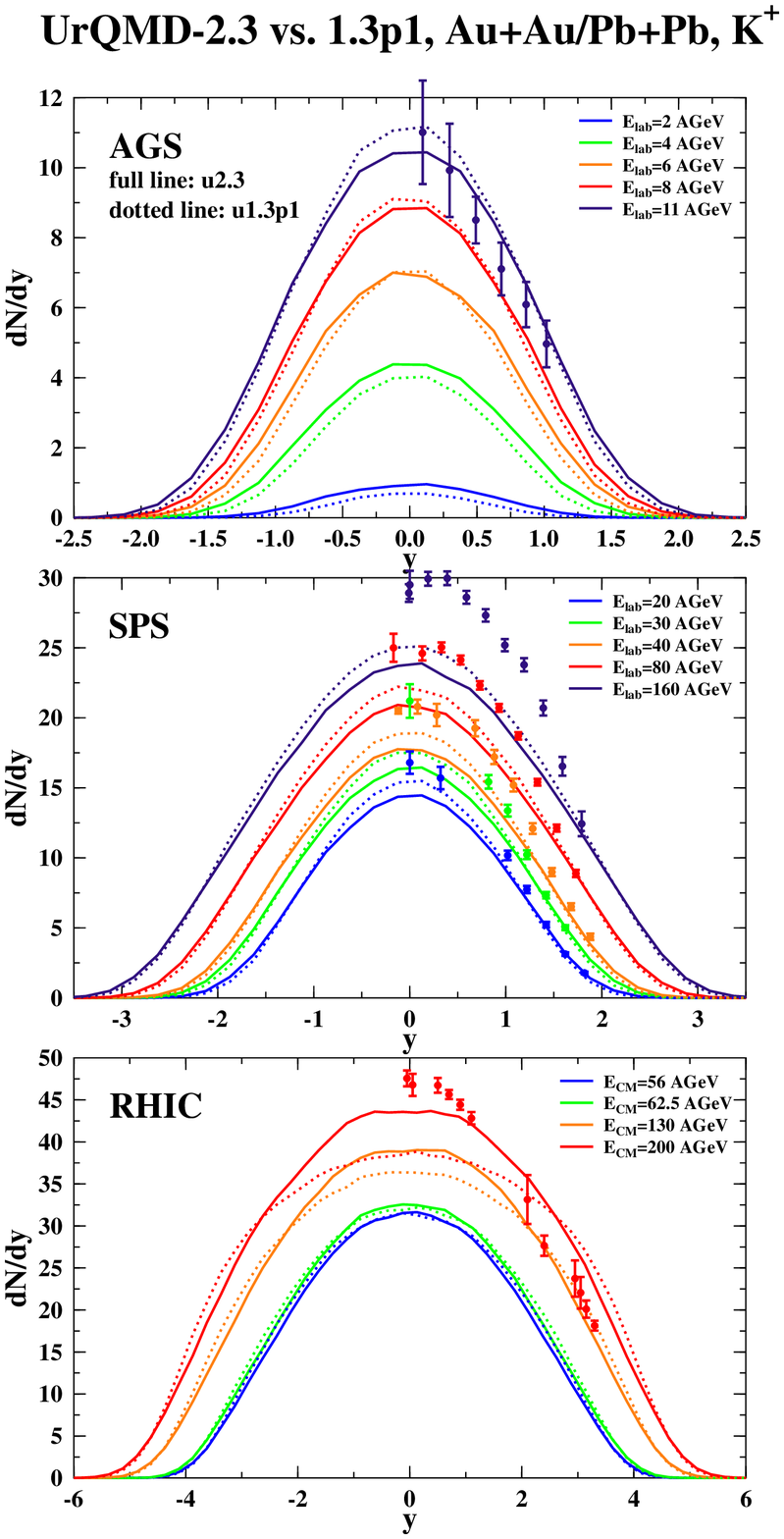}
\caption{(Color online) Rapidity spectra of $K^+$ for central ($b<3.4$ fm) Au+Au/Pb+Pb collisions from $E_{\rm lab}=2~A$GeV to $\sqrt{s_{NN}}=200$ GeV. UrQMD-2.3 calculations are depicted with full lines, while UrQMD-1.3p1 calculations are depicted with dotted lines. The corresponding data from different experiments \cite{Akiba:1996xf,:2007fe,Afanasiev:2002mx,Bearden:2004yx} are depicted with symbols.}
\label{figdndykaplus}
\end{figure}

\begin{figure}
\includegraphics[width=15cm]{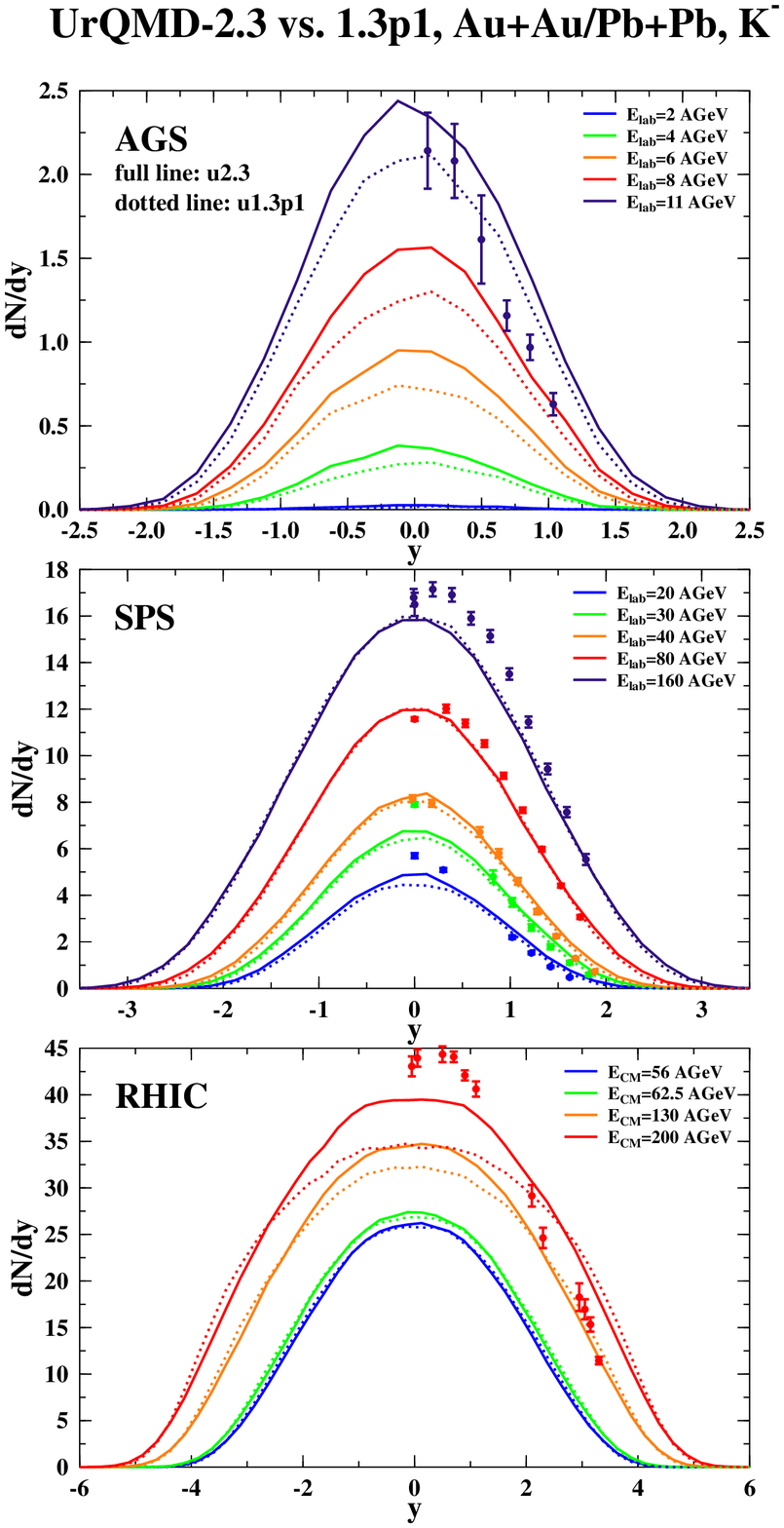}
\caption{(Color online) Rapidity spectra of $K^-$ for central ($b<3.4$ fm) Au+Au/Pb+Pb collisions from $E_{\rm lab}=2~A$GeV to $\sqrt{s_{NN}}=200$ GeV. UrQMD-2.3 calculations are depicted with full lines, while UrQMD-1.3p1 calculations are depicted with dotted lines. The corresponding data from different experiments \cite{Akiba:1996xf,:2007fe,Afanasiev:2002mx,Bearden:2004yx} are depicted with symbols.}
\label{figdndykaminus}
\end{figure}

\begin{figure}
\centering
\includegraphics[width=15cm]{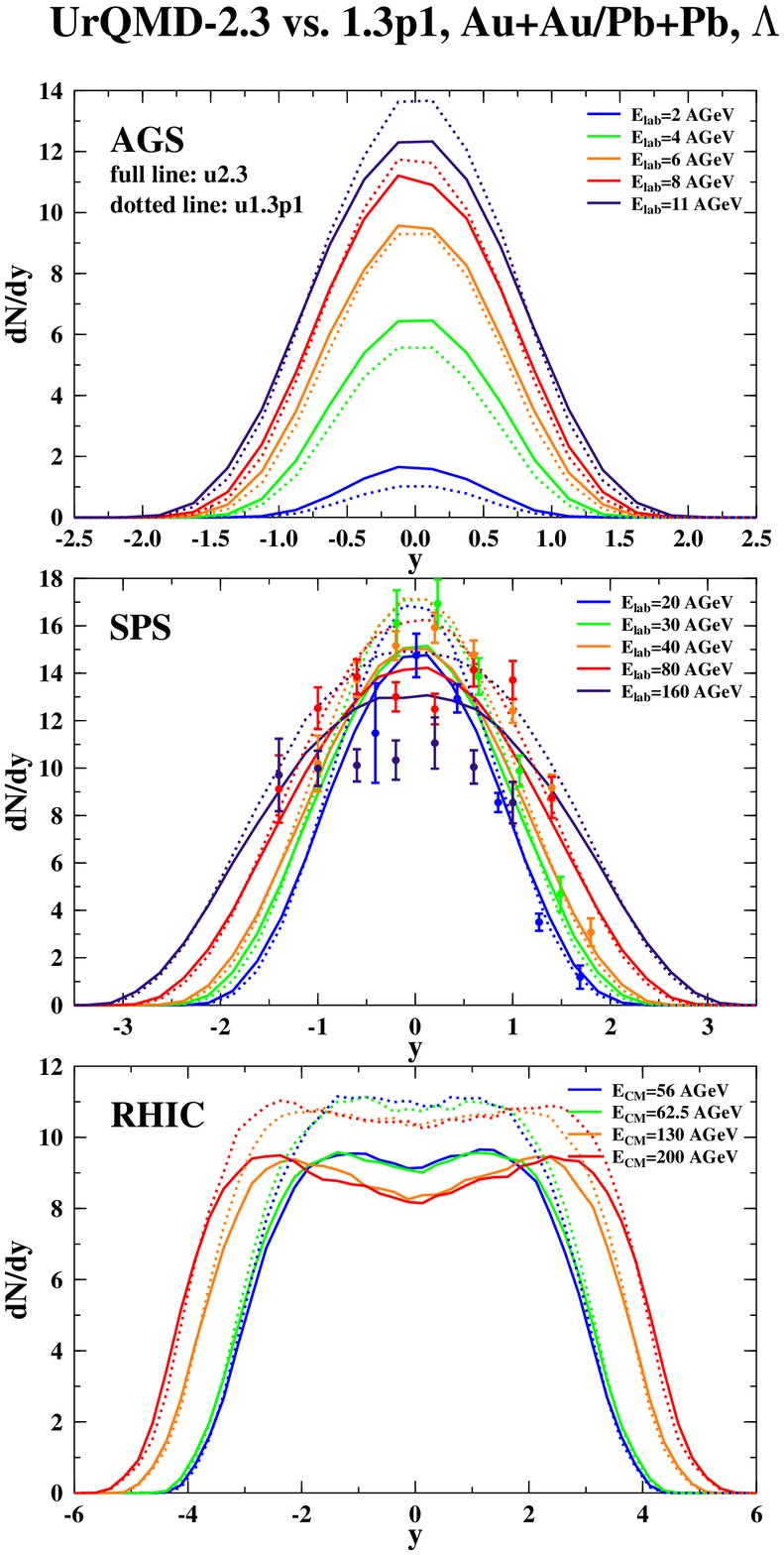}
\caption{(Color online) Rapidity spectra of $\Lambda+\Sigma_0$ for central ($b<3.4$ fm) Au+Au/Pb+Pb collisions from $E_{\rm lab}=2~A$GeV to $\sqrt{s_{NN}}=200$ GeV. UrQMD-2.3 calculations are depicted with full lines, while UrQMD-1.3p1 calculations are depicted with dotted lines. The corresponding data from NA49 \cite{Richard:2005rx,Mitrovski:2006js,Anticic:2003ux} are depicted with symbols.}
\label{figdndylambda}
\end{figure}

\begin{figure}
\centering
\includegraphics[width=15cm]{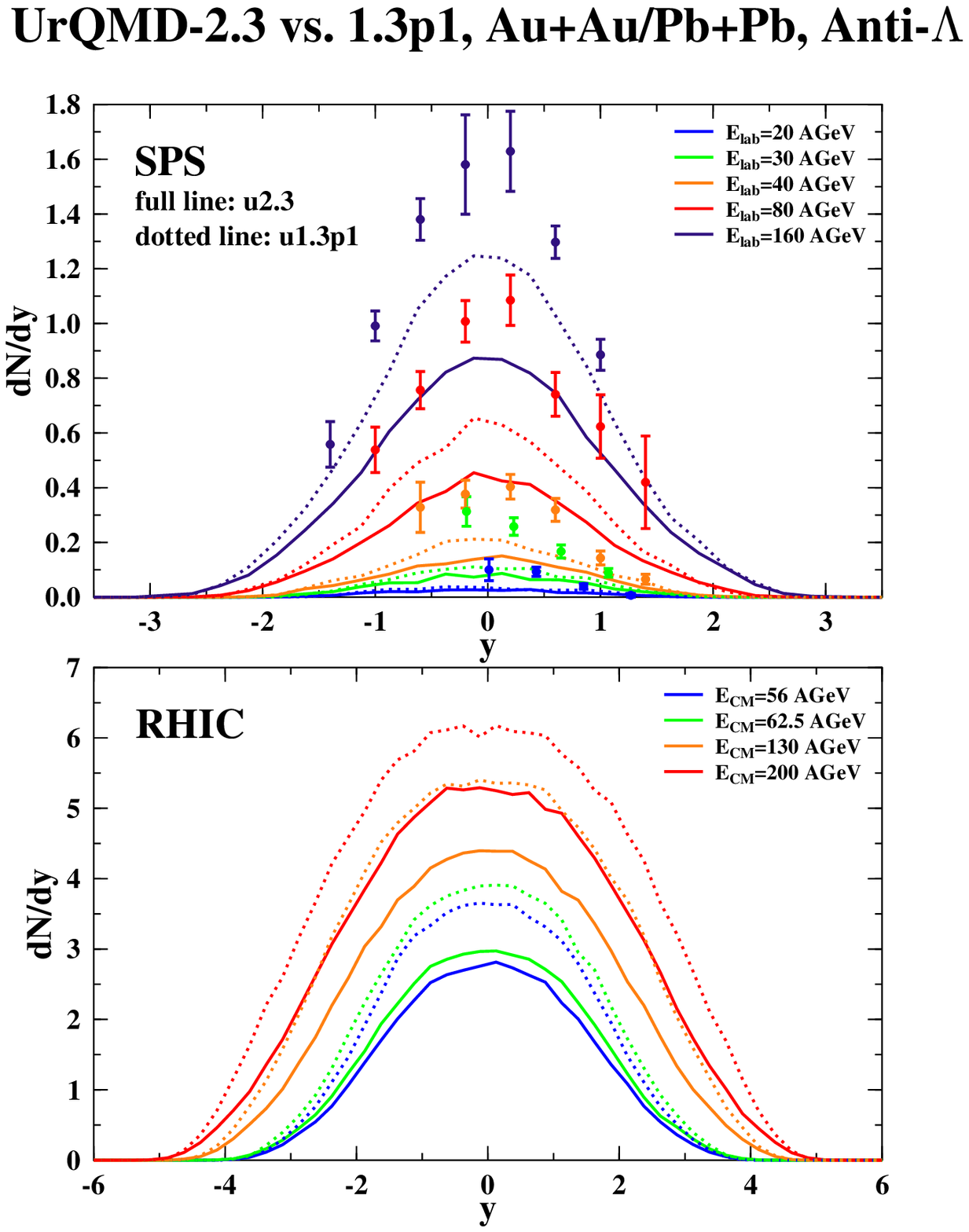}
\caption{(Color online) Rapidity spectra of $\bar{\Lambda}+\bar{\Sigma_0}$ for central ($b<3.4$ fm) Au+Au/Pb+Pb collisions from $E_{\rm lab}=20~A$GeV to $\sqrt{s_{NN}}=200$ GeV. UrQMD-2.3 calculations are depicted with full lines, while UrQMD-1.3p1 calculations are depicted with dotted lines. The corresponding data from NA49 \cite{Richard:2005rx,Mitrovski:2006js,Anticic:2003ux} are depicted with symbols.}
\label{figdndyalambda}
\end{figure}

\begin{figure}
\centering
\includegraphics[width=15cm]{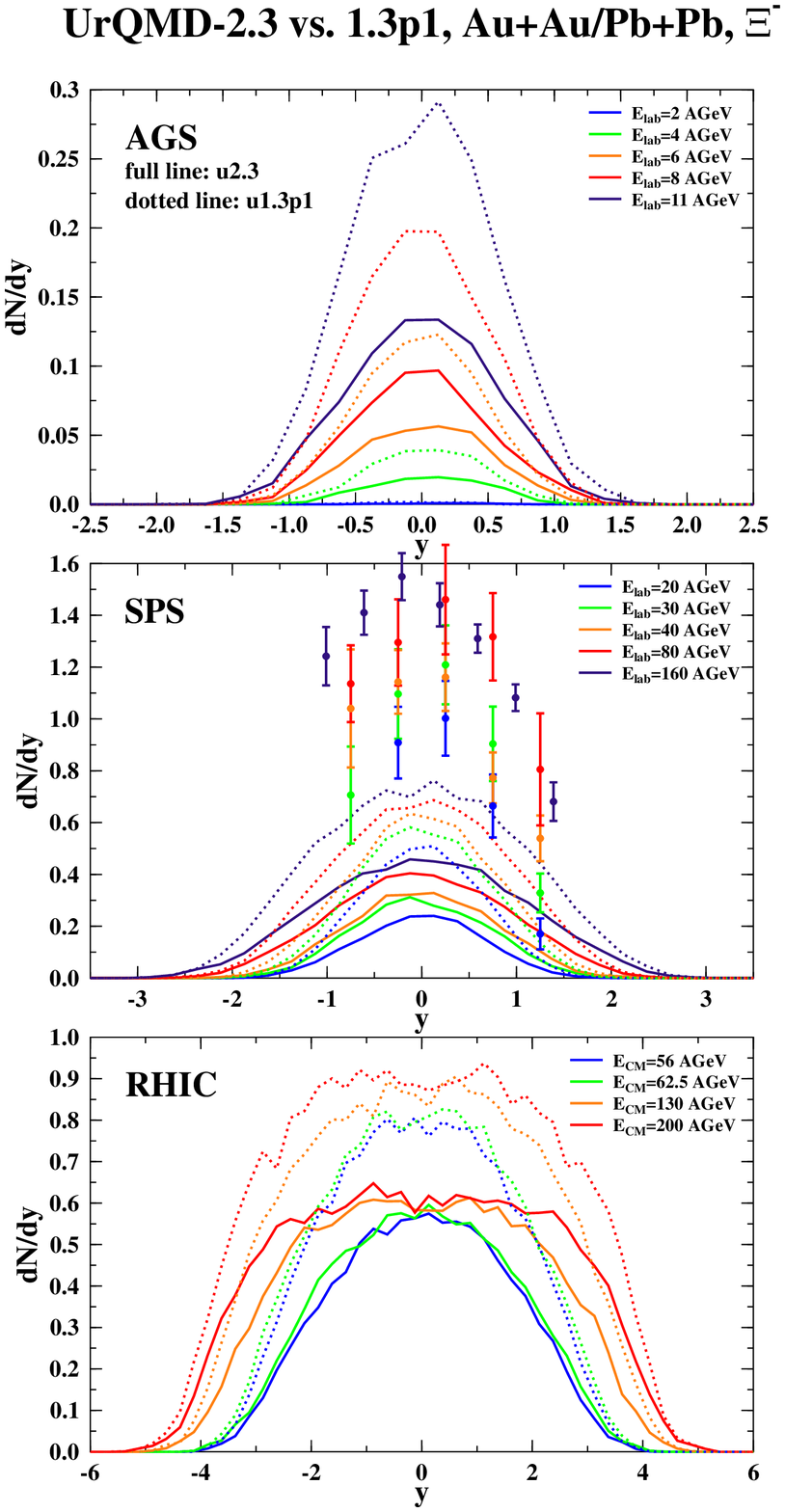}
\caption{(Color online) Rapidity spectra of $\Xi^-$ for central ($b<3.4$ fm) Au+Au/Pb+Pb collisions from $E_{\rm lab}=2~A$GeV to $\sqrt{s_{NN}}=200$ GeV. UrQMD-2.3 calculations are depicted with full lines, while UrQMD-1.3p1 calculations are depicted with dotted lines. The corresponding data from NA49 \cite{Mitrovski:2006js,Afanasiev:2002he} are depicted with symbols.}
\label{figdndyxi}
\end{figure}

\begin{figure}
\centering
\includegraphics[width=15cm]{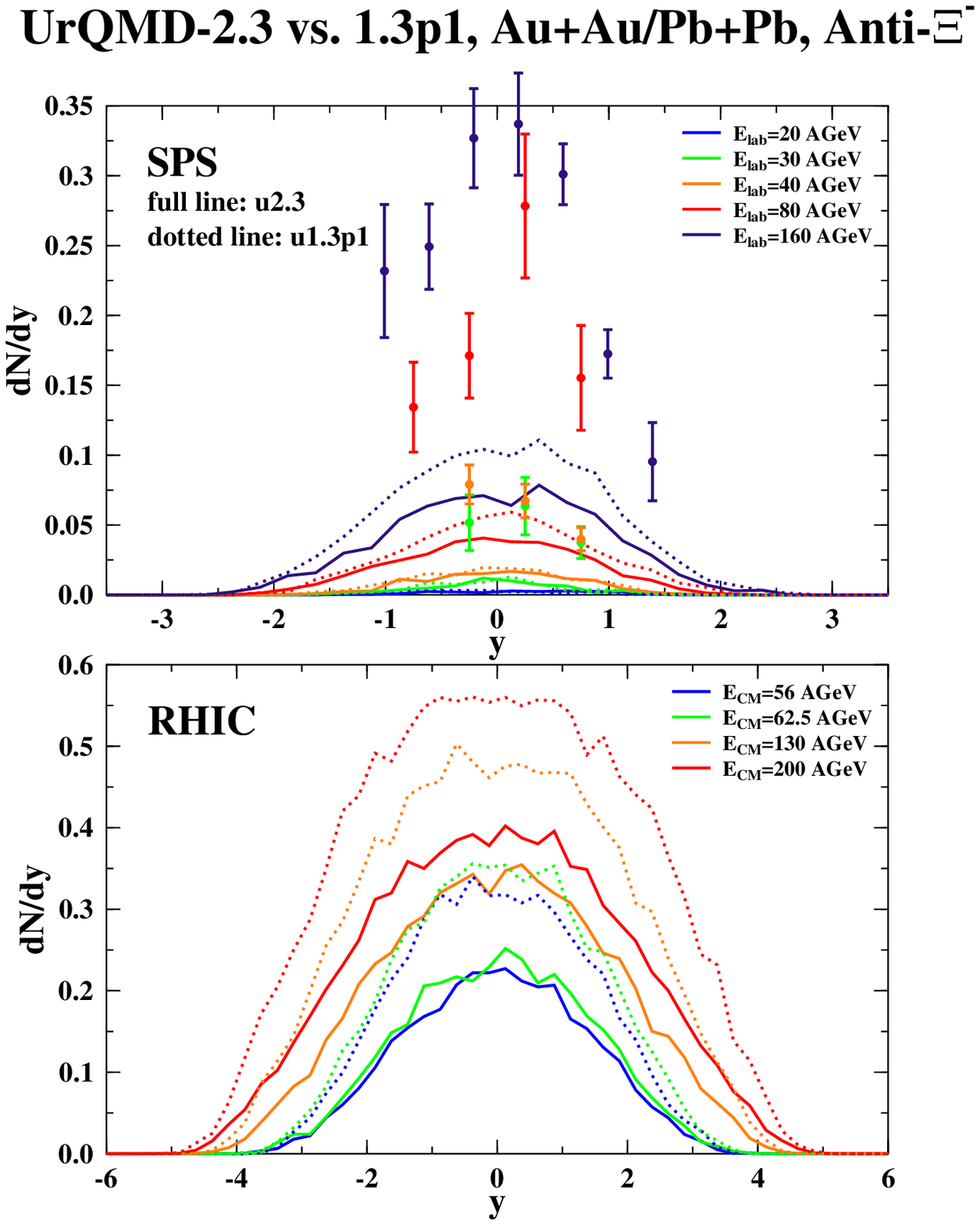}
\caption{(Color online) Rapidity spectra of $\bar{\Xi^+}$ for central ($b<3.4$ fm) Au+Au/Pb+Pb collisions from $E_{\rm lab}=20~A$GeV to $\sqrt{s_{NN}}=200$ GeV. UrQMD-2.3 calculations are depicted with full lines, while UrQMD-1.3p1 calculations are depicted with dotted lines. The corresponding data from  NA49 \cite{Mitrovski:2006js,Afanasiev:2002he} are depicted with symbols.}
\label{figdndyaxi}
\end{figure}

\begin{figure}
\centering
\includegraphics[width=15cm]{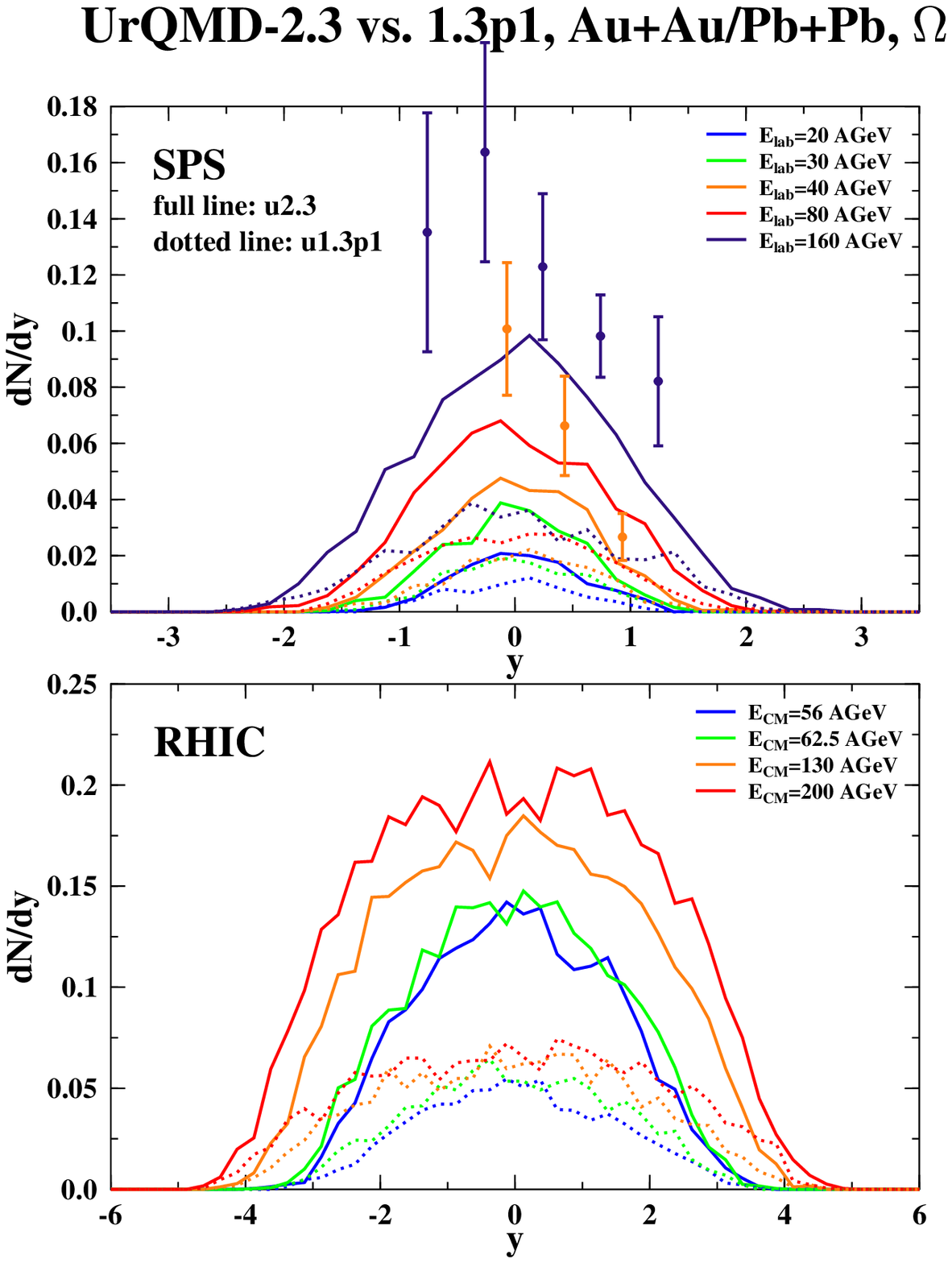}
\caption{(Color online) Rapidity spectra of $\Omega$ for central ($b<3.4$ fm) Au+Au/Pb+Pb collisions from $E_{\rm lab}=20~A$GeV to $\sqrt{s_{NN}}=200$ GeV. UrQMD-2.3 calculations are depicted with full lines, while UrQMD-1.3p1 calculations are depicted with dotted lines. The corresponding data from NA49 \cite{Alt:2004kq} are depicted with symbols.}
\label{figdndyomega}
\end{figure}

\begin{figure}
\centering
\includegraphics[width=15cm]{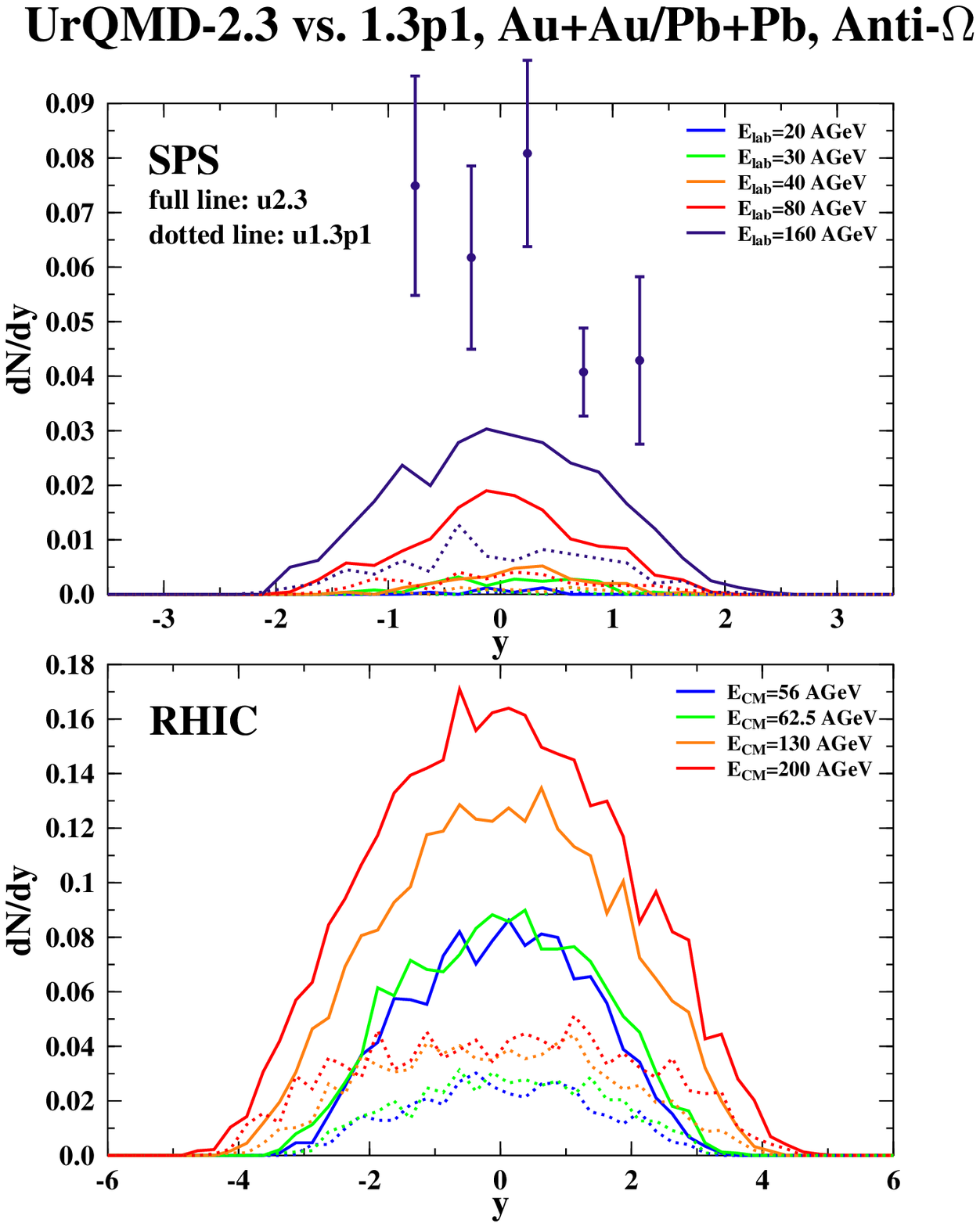}
\caption{(Color online) Rapidity spectra of $\bar{\Omega}$ for central ($b<3.4$ fm) Au+Au/Pb+Pb collisions from $E_{\rm lab}=20~A$GeV to $\sqrt{s_{NN}}=200$ GeV. UrQMD-2.3 calculations are depicted with full lines, while UrQMD-1.3p1 calculations are depicted with dotted lines. The corresponding data from NA49 \cite{Alt:2004kq} are depicted with symbols.}
\label{figdndyaomega}
\end{figure}

The $K^+$ distribution stays within 20\% the same in both UrQMD versions (Fig. \ref{figdndykaplus}). Because of the replacement of string excitation with high mass resonances in the intermediate energy regime a slightly lower yield of the strange particles is observed in UrQMD-2.3. The effect of the further adjustments of the strangeness production, e.g. changes in the resonance cross sections and the strangeness suppression factors for the string dynamics, can be seen in Figs. \ref{figdndykaminus},\ref{figdndylambda},\ref{figdndyalambda},\ref{figdndyxi},\ref{figdndyaxi},\ref{figdndyomega} and \ref{figdndyaomega}. 

\begin{figure}
\centering
\includegraphics[width=15cm]{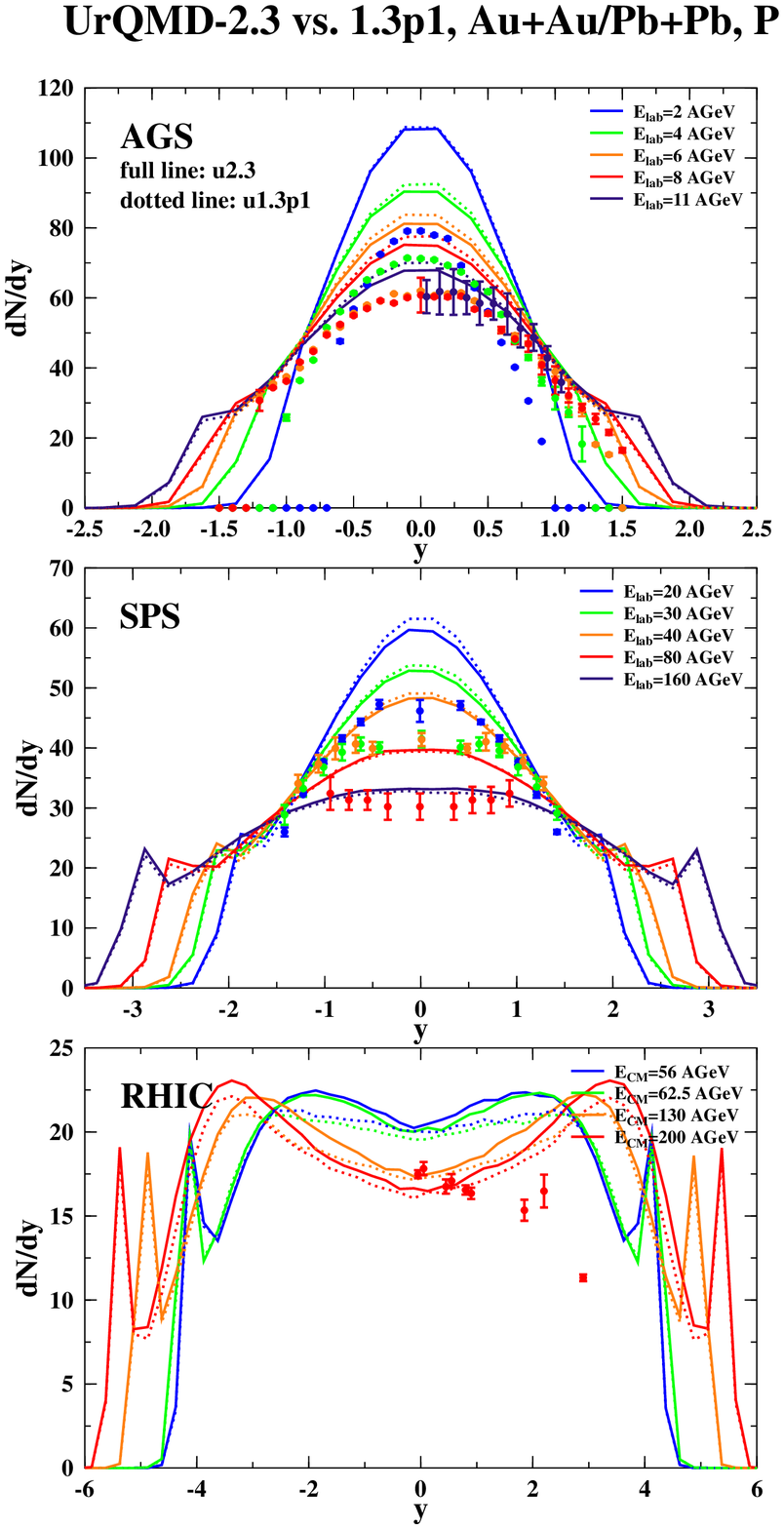}
\caption{(Color online) Rapidity spectra of protons for central ($b<3.4$ fm) Au+Au/Pb+Pb collisions from $E_{\rm lab}=2~A$GeV to $\sqrt{s_{NN}}=200$ GeV. UrQMD-2.3 calculations are depicted with full lines, while UrQMD-1.3p1 calculations are depicted with dotted lines. The corresponding data from different experiments \cite{Klay:2001tf,Akiba:1996xf,Blume:2007kw,Bearden:2003hx} are depicted with symbols.}
\label{figdndyproton}
\end{figure}

\begin{figure}
\centering
\includegraphics[width=15cm]{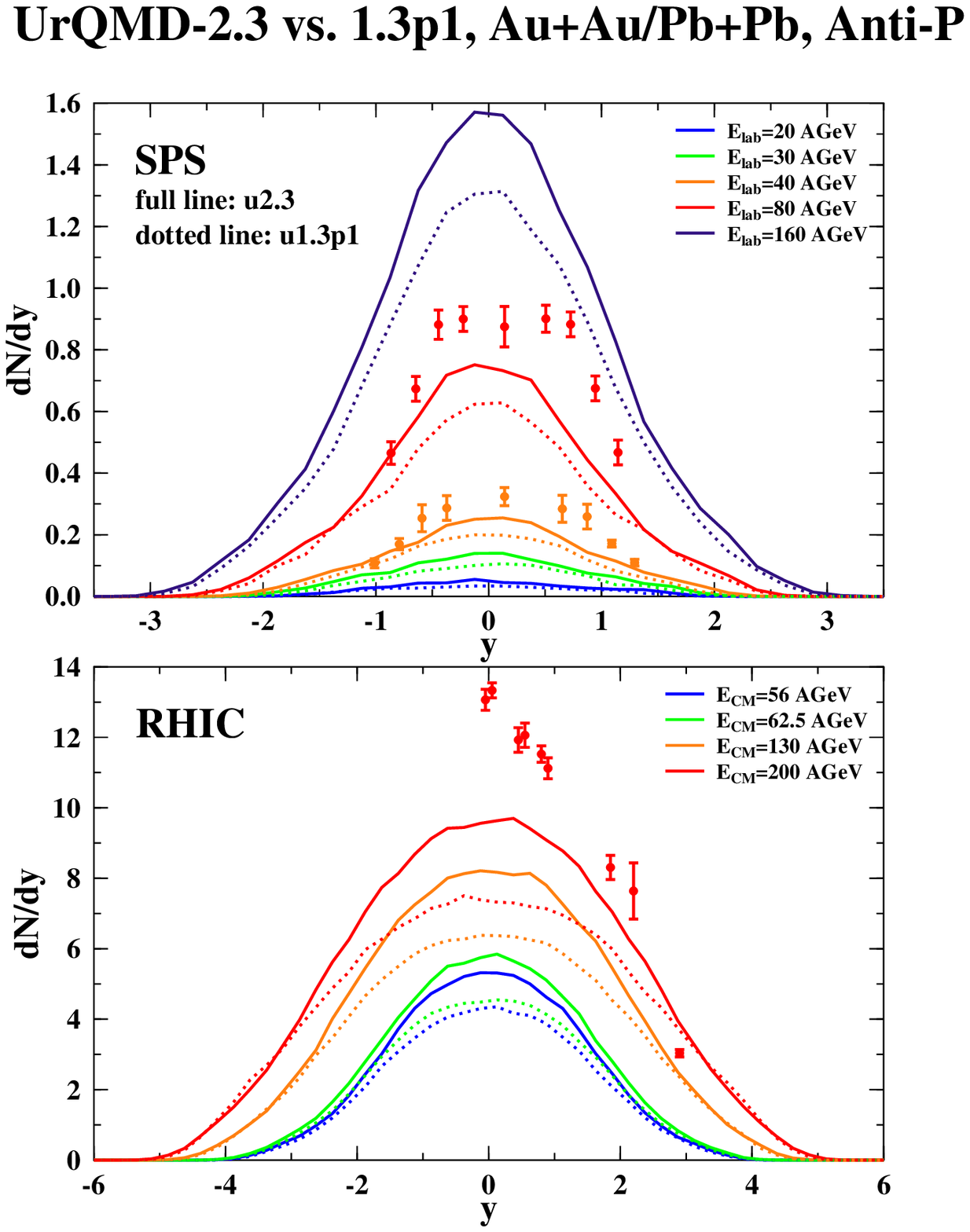}
\caption{(Color online) Rapidity spectra of antiprotons for central ($b<3.4$ fm) Au+Au/Pb+Pb collisions from $E_{\rm lab}=20~A$GeV to $\sqrt{s_{NN}}=200$ GeV. UrQMD-2.3 calculations are depicted with full lines, while UrQMD-1.3p1 calculations are depicted with dotted lines. The corresponding data from different experiments \cite{Blume:2007kw,Bearden:2003hx} are depicted with symbols.}
\label{figdndyaproton}
\end{figure}

The rapidity distributions of protons (see Fig. \ref{figdndyproton}) shows peaks at high rapidities in both UrQMD calculations from the spectators which usually are not measured by the experiments, but end up in a veto calorimeter for the centrality selection. The protons are only weakly affected by the updates and changes in the new version. The shape of the distributions looks different in the experimental data compared to both UrQMD calculations. There are more antiprotons produced because of the inclusion of Pythia (see Fig. \ref{figdndyaproton}).
 
\section{Summary}
\label{summary}
The new version UrQMD-2.3 of the Ultra-relativistic Quantum Molecular Dynamics model was presented. The most important changes and updates with respect to the previous publically available version have been explained. To illustrate the results, a comparison of the particle spectra between the two versions has been shown in the context of the available data. We encourage all users to submit potential problems and bug reports to the following email address: bleicher@th.physik.uni-frankfurt.de\,. 

\begin{acknowledgments}

We would like to thank everybody who has been sending suggestions, bug reports and ideas how to fix them. Especially, Drs. Hajo Drescher, Dieter Heck and Tanguy Pierog (CORSIKA), Vladimir Uzhinsky, the HADES collaboration. 

We are grateful to the Center for the Scientific Computing (CSC) at Frankfurt for the computing resources. H. Petersen gratefully acknowledges financial support by the Deutsche Telekom Stiftung and support from the Helmholtz Research School on Quark Matter Studies. This work was supported by GSI and BMBF. 
\end{acknowledgments}

\begin{appendix}
\section{Figures}
\label{appfigs}

Note that statistically insignificant (first low multiplicity) points have been removed from the plots, however are still present in the tables for completeness. 

\clearpage

\clearpage

\clearpage

\clearpage

\clearpage 

\clearpage

\clearpage

\clearpage

\clearpage

\clearpage

\clearpage

\clearpage

\clearpage

\clearpage

\clearpage

\clearpage

\clearpage

\clearpage

\clearpage

\clearpage

\clearpage 

\clearpage

\clearpage

\section{Numerical Data}
\label{appnumdata}
The following Section contains all the numerical data for the UrQMD-2.3 results that have been shown in the Figs in Section \ref{appfigs}. The following abbreviations for the particle types are used:

\begin{tabular}{lcl}
pi+&:&$\pi^+$\\
pi-&:&$\pi^-$\\
pi0&:&$\pi^0$\\
K+&:&$K^+$\\
K-&:&$K^-$\\
P&:&Protons\\
aP&:&Antiprotons\\
L+S0&:&$\Lambda + \Sigma_0$\\
a(L+S0)&:&$\bar{\Lambda}+\bar{\Sigma_0}$\\
Xi-&:&$\Xi^-$\\
aXi-&:&$\Xi^+$\\
Om&:&$\Omega$\\
aOm&:&$\bar{\Omega}$\\
\end{tabular}

\newpage
\subsection*{Total multiplicities in p-p collisions}
The results for the energy dependence (ecm is $\sqrt{s}_{\rm NN}$) are given.  

{\tiny
\begin{verbatim}
!ecm pi+ pi- pi0 K+ K- P aP L+S0 a(L+S0) Xi- aXi- Om aOm
2.325E+00 8.034E-01 5.202E-04 1.951E-01 0.000E+00 0.000E+00 1.197E+00 0.000E+00 0.000E+00 0.000E+00 0.000E+00 0.000E+00 0.000E+00 0.000E+00 
2.695E+00 8.170E-01 5.357E-02 2.809E-01 1.778E-03 0.000E+00 1.235E+00 0.000E+00 1.773E-03 0.000E+00 0.000E+00 0.000E+00 0.000E+00 0.000E+00 
3.325E+00 9.272E-01 2.727E-01 5.477E-01 1.366E-02 1.405E-04 1.327E+00 0.000E+00 1.151E-02 0.000E+00 0.000E+00 0.000E+00 0.000E+00 0.000E+00 
3.845E+00 1.002E+00 3.975E-01 6.892E-01 2.114E-02 1.421E-03 1.368E+00 0.000E+00 2.124E-02 0.000E+00 0.000E+00 0.000E+00 0.000E+00 0.000E+00 
4.305E+00 1.070E+00 4.790E-01 7.869E-01 2.727E-02 3.815E-03 1.377E+00 1.358E-06 2.683E-02 0.000E+00 0.000E+00 0.000E+00 0.000E+00 0.000E+00 
4.915E+00 1.170E+00 5.813E-01 9.183E-01 3.711E-02 7.917E-03 1.372E+00 1.232E-04 3.539E-02 1.325E-05 1.325E-06 0.000E+00 0.000E+00 0.000E+00 
6.405E+00 1.434E+00 8.425E-01 1.232E+00 6.595E-02 2.326E-02 1.353E+00 2.745E-03 5.510E-02 6.112E-04 1.256E-04 1.025E-04 0.000E+00 1.153E-05 
7.735E+00 1.671E+00 1.074E+00 1.508E+00 9.364E-02 4.069E-02 1.339E+00 7.919E-03 7.141E-02 1.980E-03 3.377E-04 2.520E-04 3.780E-06 2.142E-05 
8.865E+00 1.866E+00 1.265E+00 1.727E+00 1.162E-01 5.536E-02 1.332E+00 1.399E-02 8.357E-02 3.622E-03 5.996E-04 4.960E-04 2.374E-05 5.747E-05 
1.240E+01 2.431E+00 1.815E+00 2.342E+00 1.828E-01 1.058E-01 1.316E+00 3.398E-02 1.144E-01 9.201E-03 1.339E-03 1.270E-03 1.154E-04 1.830E-04 
1.743E+01 3.079E+00 2.450E+00 3.025E+00 2.603E-01 1.694E-01 1.311E+00 6.033E-02 1.428E-01 1.753E-02 2.609E-03 2.221E-03 2.202E-04 3.528E-04 
3.000E+01 4.172E+00 3.524E+00 4.160E+00 3.924E-01 2.859E-01 1.321E+00 1.100E-01 1.830E-01 3.460E-02 4.818E-03 4.230E-03 5.172E-04 7.395E-04 
4.001E+01 4.767E+00 4.110E+00 4.762E+00 4.650E-01 3.537E-01 1.337E+00 1.411E-01 2.016E-01 4.543E-02 5.895E-03 5.450E-03 7.397E-04 9.573E-04 
5.601E+01 5.535E+00 4.862E+00 5.528E+00 5.591E-01 4.428E-01 1.360E+00 1.841E-01 2.249E-01 6.023E-02 7.435E-03 6.743E-03 8.734E-04 1.136E-03 
6.251E+01 5.762E+00 5.086E+00 5.753E+00 5.867E-01 4.686E-01 1.368E+00 1.970E-01 2.310E-01 6.510E-02 8.079E-03 7.285E-03 1.006E-03 1.191E-03 
8.000E+01 6.162E+00 5.473E+00 6.136E+00 6.382E-01 5.180E-01 1.381E+00 2.237E-01 2.433E-01 7.356E-02 8.952E-03 7.959E-03 1.092E-03 1.408E-03 
1.000E+02 6.631E+00 5.932E+00 6.594E+00 6.997E-01 5.778E-01 1.403E+00 2.571E-01 2.550E-01 8.435E-02 9.900E-03 8.822E-03 1.124E-03 1.365E-03 
1.300E+02 7.219E+00 6.508E+00 7.162E+00 7.763E-01 6.527E-01 1.433E+00 3.004E-01 2.729E-01 9.962E-02 1.117E-02 9.798E-03 1.182E-03 1.382E-03 
2.000E+02 8.391E+00 7.653E+00 8.279E+00 9.343E-01 8.084E-01 1.503E+00 3.976E-01 3.062E-01 1.298E-01 1.299E-02 1.167E-02 1.080E-03 1.371E-03 
5.460E+02 1.327E+01 1.248E+01 1.306E+01 1.599E+00 1.469E+00 1.869E+00 8.142E-01 4.411E-01 2.618E-01 2.211E-02 2.033E-02 7.696E-04 8.871E-04 
9.000E+02 1.528E+01 1.447E+01 1.503E+01 1.868E+00 1.738E+00 2.029E+00 9.856E-01 4.963E-01 3.176E-01 2.628E-02 2.430E-02 7.125E-04 7.823E-04 
1.800E+03 1.995E+01 1.914E+01 1.974E+01 2.474E+00 2.343E+00 2.394E+00 1.355E+00 6.187E-01 4.373E-01 3.544E-02 3.312E-02 8.820E-04 9.213E-04 
5.500E+03 2.919E+01 2.838E+01 2.905E+01 3.671E+00 3.542E+00 3.121E+00 2.081E+00 8.565E-01 6.750E-01 5.353E-02 5.112E-02 1.519E-03 1.432E-03 
1.400E+04 3.872E+01 3.791E+01 3.867E+01 4.905E+00 4.775E+00 3.871E+00 2.832E+00 1.104E+00 9.208E-01 7.280E-02 7.042E-02 2.071E-03 2.130E-03 
\end{verbatim}
}

\subsection*{$\langle p_T \rangle$ excitation function in p-p collisions}
All $\langle p_T \rangle$ values are given at midrapidity ($|y|<0.5$).

{\tiny
\begin{verbatim}
!ecm pi+ pi- pi0 K+ K- P aP L+S0 a(L+S0) Xi- aXi- Om aOm
2.325E+00 2.221E-01 8.720E-02 2.124E-01 0.000E+00 0.000E+00 2.459E-01 0.000E+00 0.000E+00 0.000E+00 0.000E+00 0.000E+00 0.000E+00 0.000E+00 
2.695E+00 2.724E-01 2.382E-01 2.707E-01 2.013E-01 0.000E+00 2.976E-01 0.000E+00 1.959E-01 0.000E+00 0.000E+00 0.000E+00 0.000E+00 0.000E+00 
3.325E+00 3.266E-01 3.112E-01 3.207E-01 4.140E-01 1.364E-01 3.826E-01 0.000E+00 3.927E-01 0.000E+00 0.000E+00 0.000E+00 0.000E+00 0.000E+00 
3.845E+00 3.342E-01 3.112E-01 3.250E-01 4.327E-01 1.825E-01 4.397E-01 0.000E+00 4.167E-01 0.000E+00 0.000E+00 0.000E+00 0.000E+00 0.000E+00 
4.305E+00 3.390E-01 3.103E-01 3.248E-01 4.391E-01 2.240E-01 4.933E-01 1.505E-01 4.405E-01 0.000E+00 0.000E+00 0.000E+00 0.000E+00 0.000E+00 
4.915E+00 3.450E-01 3.136E-01 3.281E-01 4.290E-01 2.694E-01 5.617E-01 2.253E-01 4.875E-01 2.096E-01 3.271E-01 0.000E+00 0.000E+00 0.000E+00 
6.405E+00 3.454E-01 3.214E-01 3.300E-01 4.424E-01 3.436E-01 6.321E-01 3.618E-01 5.619E-01 3.752E-01 5.173E-01 3.899E-01 0.000E+00 3.058E-01 
7.735E+00 3.338E-01 3.258E-01 3.242E-01 4.443E-01 3.832E-01 6.147E-01 4.206E-01 5.726E-01 4.638E-01 5.862E-01 4.556E-01 3.854E-01 5.285E-01 
8.865E+00 3.297E-01 3.280E-01 3.216E-01 4.458E-01 3.997E-01 5.896E-01 4.564E-01 5.637E-01 4.812E-01 5.884E-01 5.087E-01 6.506E-01 5.871E-01 
1.240E+01 3.345E-01 3.429E-01 3.296E-01 4.519E-01 4.295E-01 5.486E-01 5.124E-01 5.380E-01 5.452E-01 6.159E-01 5.782E-01 6.729E-01 6.536E-01 
1.743E+01 3.479E-01 3.572E-01 3.429E-01 4.561E-01 4.454E-01 5.303E-01 5.456E-01 5.222E-01 5.737E-01 6.188E-01 5.583E-01 5.908E-01 5.852E-01 
3.000E+01 3.668E-01 3.693E-01 3.621E-01 4.602E-01 4.610E-01 5.348E-01 5.813E-01 5.224E-01 5.917E-01 6.027E-01 6.088E-01 6.372E-01 6.429E-01 
4.001E+01 3.719E-01 3.740E-01 3.684E-01 4.679E-01 4.620E-01 5.446E-01 5.939E-01 5.393E-01 5.992E-01 6.017E-01 6.256E-01 6.074E-01 6.122E-01 
5.601E+01 3.774E-01 3.780E-01 3.741E-01 4.748E-01 4.699E-01 5.644E-01 5.996E-01 5.621E-01 6.080E-01 6.185E-01 6.330E-01 6.087E-01 5.959E-01 
6.251E+01 3.792E-01 3.797E-01 3.763E-01 4.766E-01 4.710E-01 5.686E-01 5.980E-01 5.730E-01 6.176E-01 6.431E-01 6.490E-01 6.067E-01 6.146E-01 
8.000E+01 3.808E-01 3.803E-01 3.777E-01 4.772E-01 4.710E-01 5.749E-01 5.965E-01 5.760E-01 6.035E-01 6.318E-01 6.200E-01 6.039E-01 6.103E-01 
1.000E+02 3.841E-01 3.830E-01 3.818E-01 4.826E-01 4.740E-01 5.869E-01 5.974E-01 5.834E-01 6.032E-01 6.067E-01 6.334E-01 5.987E-01 5.875E-01 
1.300E+02 3.881E-01 3.872E-01 3.862E-01 4.898E-01 4.789E-01 5.940E-01 5.985E-01 6.018E-01 6.181E-01 6.285E-01 6.657E-01 6.450E-01 6.051E-01 
2.000E+02 3.976E-01 3.945E-01 3.959E-01 5.045E-01 4.956E-01 6.165E-01 6.101E-01 6.306E-01 6.433E-01 6.575E-01 6.588E-01 6.285E-01 6.361E-01 
5.460E+02 4.307E-01 4.291E-01 4.304E-01 5.668E-01 5.596E-01 6.997E-01 6.894E-01 7.501E-01 7.508E-01 7.746E-01 8.045E-01 7.790E-01 7.672E-01 
9.000E+02 4.430E-01 4.415E-01 4.423E-01 5.897E-01 5.824E-01 7.328E-01 7.229E-01 7.909E-01 7.825E-01 8.463E-01 8.399E-01 8.533E-01 8.260E-01 
1.800E+03 4.749E-01 4.743E-01 4.753E-01 6.458E-01 6.414E-01 8.222E-01 8.152E-01 9.026E-01 8.978E-01 9.592E-01 1.002E+00 1.034E+00 1.059E+00 
5.500E+03 5.420E-01 5.414E-01 5.421E-01 7.582E-01 7.597E-01 1.006E+00 1.000E+00 1.107E+00 1.113E+00 1.228E+00 1.223E+00 1.515E+00 1.225E+00 
1.400E+04 6.110E-01 6.106E-01 6.107E-01 8.792E-01 8.780E-01 1.187E+00 1.180E+00 1.314E+00 1.315E+00 1.481E+00 1.455E+00 1.592E+00 1.605E+00 
\end{verbatim}
}
\newpage
\subsection*{Multiplicities in A-A collisions}
The total multiplicites ($4\pi$) are displayed as a function of $\sqrt{s}_{\rm NN}$. All results are for central Au+Au/Pb+Pb collisions $b \le 3.4$ fm. 

{\tiny
\begin{verbatim}
!ecm pi+ pi- pi0 K+ K- P aP L+S0 a(L+S0) Xi- aXi- Om aOm
2.325E+00 6.601E+00 1.294E+01 9.499E+00 5.090E-02 1.000E-04 1.643E+02 0.000E+00 8.370E-02 0.000E+00 0.000E+00 0.000E+00 0.000E+00 0.000E+00 
2.695E+00 2.387E+01 3.685E+01 3.076E+01 1.425E+00 3.620E-02 1.698E+02 0.000E+00 1.937E+00 0.000E+00 7.000E-04 0.000E+00 0.000E+00 0.000E+00 
3.325E+00 5.968E+01 7.818E+01 7.059E+01 7.516E+00 5.988E-01 1.703E+02 4.000E-04 9.084E+00 1.000E-04 2.340E-02 0.000E+00 1.000E-04 0.000E+00 
3.845E+00 9.146E+01 1.126E+02 1.048E+02 1.288E+01 1.677E+00 1.689E+02 1.200E-03 1.457E+01 3.000E-04 7.450E-02 0.000E+00 2.000E-03 0.000E+00 
4.305E+00 1.188E+02 1.416E+02 1.339E+02 1.713E+01 2.902E+00 1.678E+02 3.600E-03 1.825E+01 1.400E-03 1.231E-01 2.000E-04 2.900E-03 1.000E-04 
4.915E+00 1.522E+02 1.765E+02 1.696E+02 2.199E+01 4.757E+00 1.664E+02 1.330E-02 2.209E+01 5.600E-03 1.937E-01 7.000E-04 8.300E-03 1.000E-04 
6.405E+00 2.362E+02 2.662E+02 2.598E+02 3.461E+01 1.056E+01 1.716E+02 1.074E-01 3.055E+01 5.600E-02 4.028E-01 5.200E-03 2.890E-02 8.000E-04 
7.735E+00 2.953E+02 3.266E+02 3.220E+02 4.276E+01 1.555E+01 1.702E+02 2.990E-01 3.465E+01 1.673E-01 5.516E-01 1.530E-02 5.490E-02 4.800E-03 
8.865E+00 3.417E+02 3.736E+02 3.710E+02 4.907E+01 1.991E+01 1.691E+02 5.544E-01 3.702E+01 3.029E-01 6.546E-01 2.960E-02 7.930E-02 6.700E-03 
1.240E+01 4.695E+02 5.027E+02 5.064E+02 6.523E+01 3.235E+01 1.680E+02 1.695E+00 4.192E+01 9.819E-01 9.569E-01 8.582E-02 1.334E-01 3.369E-02 
1.743E+01 6.210E+02 6.547E+02 6.672E+02 8.326E+01 4.773E+01 1.684E+02 3.923E+00 4.615E+01 2.186E+00 1.256E+00 1.799E-01 2.149E-01 7.114E-02 
3.000E+01 8.510E+02 8.813E+02 9.085E+02 1.095E+02 7.315E+01 1.638E+02 9.480E+00 4.919E+01 5.161E+00 1.720E+00 4.098E-01 3.376E-01 1.604E-01 
5.601E+01 1.218E+03 1.248E+03 1.290E+03 1.542E+02 1.155E+02 1.719E+02 1.973E+01 5.667E+01 1.057E+01 2.488E+00 8.172E-01 5.339E-01 3.084E-01 
6.251E+01 1.293E+03 1.323E+03 1.369E+03 1.640E+02 1.248E+02 1.739E+02 2.199E+01 5.831E+01 1.176E+01 2.645E+00 9.015E-01 5.933E-01 3.445E-01 
1.300E+02 1.828E+03 1.858E+03 1.924E+03 2.335E+02 1.923E+02 1.895E+02 3.892E+01 6.834E+01 2.086E+01 3.659E+00 1.611E+00 9.277E-01 6.075E-01 
2.000E+02 2.203E+03 2.234E+03 2.315E+03 2.842E+02 2.424E+02 2.011E+02 5.085E+01 7.518E+01 2.771E+01 4.314E+00 2.072E+00 1.193E+00 8.260E-01 
\end{verbatim}
}

\subsection*{Midrapidity yields in A-A collisions}

The particle yields at midrapidity ($|y|<0.5$) are displayed as a function of beam energy (ecm is $\sqrt{s}_{\rm NN}$). All results are for central Au+Au/Pb+Pb collisions $b \le 3.4$ fm. 

{\tiny
\begin{verbatim}
!ecm pi+ pi- pi0 K+ K- P aP L+S0 a(L+S0) Xi- aXi- Om aOm
2.325E+00 3.432E+00 6.784E+00 4.954E+00 3.650E-02 1.000E-04 1.213E+02 0.000E+00 7.340E-02 0.000E+00 0.000E+00 0.000E+00 0.000E+00 0.000E+00 
2.695E+00 1.157E+01 1.785E+01 1.476E+01 8.693E-01 2.250E-02 1.021E+02 0.000E+00 1.442E+00 0.000E+00 6.000E-04 0.000E+00 0.000E+00 0.000E+00 
3.325E+00 2.632E+01 3.463E+01 3.097E+01 4.161E+00 3.419E-01 8.664E+01 2.000E-04 5.920E+00 0.000E+00 1.710E-02 0.000E+00 1.000E-04 0.000E+00 
3.845E+00 3.791E+01 4.691E+01 4.310E+01 6.644E+00 8.896E-01 7.814E+01 4.000E-04 8.855E+00 3.000E-04 5.210E-02 0.000E+00 1.900E-03 0.000E+00 
4.305E+00 4.713E+01 5.634E+01 5.260E+01 8.479E+00 1.485E+00 7.238E+01 1.700E-03 1.043E+01 9.000E-04 8.370E-02 1.000E-04 2.000E-03 0.000E+00 
4.915E+00 5.732E+01 6.673E+01 6.317E+01 1.016E+01 2.291E+00 6.567E+01 5.700E-03 1.170E+01 3.000E-03 1.230E-01 3.000E-04 5.900E-03 1.000E-04 
6.405E+00 8.175E+01 9.208E+01 8.905E+01 1.403E+01 4.649E+00 5.816E+01 4.670E-02 1.412E+01 2.650E-02 2.247E-01 2.200E-03 1.880E-02 7.000E-04 
7.735E+00 9.768E+01 1.078E+02 1.057E+02 1.596E+01 6.493E+00 5.185E+01 1.299E-01 1.459E+01 7.710E-02 2.821E-01 8.900E-03 3.200E-02 2.500E-03 
8.865E+00 1.100E+02 1.196E+02 1.186E+02 1.740E+01 8.001E+00 4.758E+01 2.437E-01 1.463E+01 1.355E-01 3.149E-01 1.550E-02 4.350E-02 4.000E-03 
1.240E+01 1.416E+02 1.502E+02 1.522E+02 2.054E+01 1.172E+01 3.957E+01 7.259E-01 1.398E+01 4.193E-01 3.876E-01 3.855E-02 6.097E-02 1.712E-02 
1.743E+01 1.737E+02 1.815E+02 1.863E+02 2.344E+01 1.556E+01 3.314E+01 1.518E+00 1.296E+01 8.455E-01 4.413E-01 7.062E-02 8.983E-02 2.877E-02 
3.000E+01 2.071E+02 2.125E+02 2.208E+02 2.624E+01 1.992E+01 2.477E+01 3.079E+00 1.053E+01 1.642E+00 4.857E-01 1.386E-01 1.006E-01 4.812E-02 
5.601E+01 2.501E+02 2.542E+02 2.644E+02 3.136E+01 2.597E+01 2.050E+01 5.271E+00 9.218E+00 2.750E+00 5.622E-01 2.207E-01 1.372E-01 7.802E-02 
6.251E+01 2.585E+02 2.622E+02 2.730E+02 3.238E+01 2.718E+01 2.012E+01 5.708E+00 9.124E+00 2.945E+00 5.756E-01 2.330E-01 1.401E-01 8.678E-02 
1.300E+02 3.073E+02 3.101E+02 3.222E+02 3.894E+01 3.451E+01 1.757E+01 8.152E+00 8.370E+00 4.385E+00 5.879E-01 3.408E-01 1.726E-01 1.240E-01 
2.000E+02 3.403E+02 3.429E+02 3.567E+02 4.355E+01 3.942E+01 1.662E+01 9.583E+00 8.247E+00 5.250E+00 6.036E-01 3.900E-01 1.935E-01 1.608E-01 
\end{verbatim}
}
\newpage
\subsection*{$\langle p_T \rangle$ excitation function in A-A collisions}
All $\langle p_T \rangle$ values are given at midrapidity ($|y|<0.5$).

{\tiny
\begin{verbatim}
!ecm pi+ pi- pi0 K+ K- P aP L+S0 a(L+S0) Xi- aXi- Om aOm
2.325E+00 2.086E-01 2.064E-01 2.087E-01 3.772E-01 4.848E-01 4.552E-01 0.000E+00 4.537E-01 0.000E+00 0.000E+00 0.000E+00 0.000E+00 0.000E+00 
2.695E+00 2.674E-01 2.621E-01 2.689E-01 4.927E-01 4.346E-01 5.862E-01 0.000E+00 5.857E-01 0.000E+00 6.548E-01 0.000E+00 0.000E+00 0.000E+00 
3.325E+00 3.182E-01 3.113E-01 3.171E-01 5.405E-01 4.955E-01 6.836E-01 1.281E+00 6.770E-01 0.000E+00 6.762E-01 0.000E+00 5.248E-01 0.000E+00 
3.845E+00 3.359E-01 3.305E-01 3.371E-01 5.439E-01 5.351E-01 7.201E-01 1.082E+00 7.091E-01 6.964E-01 6.858E-01 0.000E+00 7.630E-01 0.000E+00 
4.305E+00 3.455E-01 3.397E-01 3.458E-01 5.398E-01 5.453E-01 7.378E-01 9.830E-01 7.241E-01 8.251E-01 7.204E-01 3.674E-01 7.185E-01 0.000E+00 
4.915E+00 3.531E-01 3.478E-01 3.535E-01 5.366E-01 5.464E-01 7.515E-01 9.501E-01 7.358E-01 7.794E-01 7.144E-01 5.301E-01 7.886E-01 1.009E+00 
6.405E+00 3.626E-01 3.587E-01 3.630E-01 5.403E-01 5.634E-01 7.719E-01 8.650E-01 7.559E-01 8.678E-01 7.448E-01 7.764E-01 9.320E-01 6.626E-01 
7.735E+00 3.683E-01 3.642E-01 3.684E-01 5.452E-01 5.680E-01 7.828E-01 8.730E-01 7.701E-01 8.610E-01 7.539E-01 8.097E-01 8.996E-01 9.155E-01 
8.865E+00 3.701E-01 3.669E-01 3.708E-01 5.483E-01 5.667E-01 7.907E-01 8.706E-01 7.767E-01 8.625E-01 7.636E-01 7.704E-01 8.876E-01 8.217E-01 
1.240E+01 3.732E-01 3.715E-01 3.734E-01 5.499E-01 5.704E-01 8.095E-01 8.454E-01 7.913E-01 8.482E-01 7.546E-01 7.472E-01 8.820E-01 8.553E-01 
1.743E+01 3.752E-01 3.740E-01 3.744E-01 5.510E-01 5.674E-01 8.209E-01 8.293E-01 7.983E-01 8.248E-01 7.651E-01 7.157E-01 8.785E-01 7.865E-01 
3.000E+01 3.796E-01 3.788E-01 3.774E-01 5.508E-01 5.620E-01 8.186E-01 8.052E-01 7.894E-01 8.106E-01 7.458E-01 7.391E-01 8.064E-01 7.950E-01 
5.601E+01 3.892E-01 3.892E-01 3.877E-01 5.546E-01 5.690E-01 8.159E-01 8.052E-01 7.906E-01 8.169E-01 7.445E-01 7.691E-01 8.381E-01 7.889E-01 
6.251E+01 3.915E-01 3.917E-01 3.903E-01 5.575E-01 5.720E-01 8.175E-01 8.057E-01 7.929E-01 8.242E-01 7.360E-01 7.590E-01 8.401E-01 7.897E-01 
1.300E+02 4.086E-01 4.086E-01 4.078E-01 5.787E-01 5.970E-01 8.478E-01 8.375E-01 8.322E-01 8.673E-01 7.842E-01 7.907E-01 8.458E-01 8.346E-01 
2.000E+02 4.220E-01 4.225E-01 4.210E-01 6.012E-01 6.212E-01 8.887E-01 8.724E-01 8.713E-01 9.051E-01 8.061E-01 8.324E-01 8.706E-01 8.743E-01 
\end{verbatim}
}

\subsection*{$\langle m_T \rangle-m_0$ excitation function in A-A collisions}
All $\langle m_T \rangle-m_0$ values are given at midrapidity ($|y|<0.5$).
{\tiny
\begin{verbatim}
!ecm pi+ pi- pi0 K+ K- P aP L+S0 a(L+S0) Xi- aXi- Om aOm
2.325E+00 1.211E-01 1.191E-01 1.209E-01 1.467E-01 1.981E-01 1.261E-01 0.000E+00 1.293E-01 0.000E+00 0.000E+00 0.000E+00 0.000E+00 0.000E+00 
2.695E+00 1.735E-01 1.687E-01 1.747E-01 2.282E-01 1.864E-01 1.973E-01 0.000E+00 1.959E-01 0.000E+00 0.000E+00 0.000E+00 0.000E+00 0.000E+00
3.325E+00 2.202E-01 2.138E-01 2.191E-01 2.659E-01 2.328E-01 2.568E-01 6.535E-01 2.441E-01 0.000E+00 2.006E-01 0.000E+00 8.043E-02 0.000E+00 
3.845E+00 2.366E-01 2.317E-01 2.376E-01 2.700E-01 2.629E-01 2.803E-01 5.221E-01 2.615E-01 2.504E-01 2.019E-01 0.000E+00 3.791E-01 0.000E+00 
4.305E+00 2.456E-01 2.403E-01 2.457E-01 2.674E-01 2.703E-01 2.920E-01 4.347E-01 2.701E-01 3.226E-01 2.198E-01 5.037E-02 3.220E-01 0.000E+00 
4.915E+00 2.526E-01 2.477E-01 2.529E-01 2.652E-01 2.712E-01 3.014E-01 4.220E-01 2.769E-01 2.858E-01 2.200E-01 1.295E-01 4.502E-01 5.358E-01 
6.405E+00 2.614E-01 2.579E-01 2.618E-01 2.680E-01 2.844E-01 3.151E-01 3.653E-01 2.882E-01 3.496E-01 2.338E-01 2.533E-01 5.225E-01 1.374E-01 
7.735E+00 2.668E-01 2.631E-01 2.669E-01 2.720E-01 2.883E-01 3.226E-01 3.721E-01 2.970E-01 3.450E-01 2.405E-01 2.708E-01 5.700E-01 4.385E-01 
8.865E+00 2.685E-01 2.656E-01 2.692E-01 2.741E-01 2.872E-01 3.278E-01 3.718E-01 3.009E-01 3.458E-01 2.467E-01 2.388E-01 5.183E-01 4.136E-01 
1.240E+01 2.713E-01 2.698E-01 2.715E-01 2.753E-01 2.903E-01 3.404E-01 3.566E-01 3.100E-01 3.406E-01 2.429E-01 2.382E-01 5.073E-01 4.482E-01 
1.743E+01 2.731E-01 2.720E-01 2.724E-01 2.759E-01 2.878E-01 3.481E-01 3.481E-01 3.139E-01 3.259E-01 2.481E-01 2.200E-01 5.188E-01 3.731E-01 
3.000E+01 2.772E-01 2.764E-01 2.751E-01 2.756E-01 2.839E-01 3.473E-01 3.344E-01 3.092E-01 3.181E-01 2.384E-01 2.312E-01 4.360E-01 3.855E-01 
5.601E+01 2.861E-01 2.861E-01 2.847E-01 2.788E-01 2.897E-01 3.466E-01 3.353E-01 3.107E-01 3.225E-01 2.379E-01 2.499E-01 4.493E-01 3.631E-01 
6.251E+01 2.882E-01 2.884E-01 2.871E-01 2.810E-01 2.921E-01 3.479E-01 3.359E-01 3.129E-01 3.275E-01 2.330E-01 2.425E-01 4.511E-01 3.847E-01 
1.300E+02 3.042E-01 3.042E-01 3.035E-01 2.977E-01 3.117E-01 3.690E-01 3.580E-01 3.372E-01 3.548E-01 2.610E-01 2.600E-01 4.513E-01 4.152E-01 
2.000E+02 3.168E-01 3.173E-01 3.159E-01 3.154E-01 3.308E-01 3.973E-01 3.826E-01 3.615E-01 3.798E-01 2.717E-01 2.862E-01 4.471E-01 4.543E-01 
\end{verbatim}
}
\newpage
\subsection*{Transverse mass spectra}

All results are calculated at midrapidity ($|y|<0.5$). For each beam energy there is a table with the spectra for all particle species.

Au+Au \@ $E_{\rm lab}=2A~$GeV:
{\tiny
\begin{verbatim}
! m_t-m0 , 1/mt dN/dmt(pi+ pi- pi0 K+ K- P aP L+S0 a(L+S0) Xi- aXi- Om aOm) 
1.250E-02 3.316E+02 5.224E+02 4.139E+02 5.749E+00 1.816E-01 4.360E+02 0.000E+00 6.497E+00 0.000E+00 6.026E-03 0.000E+00 0.000E+00 0.000E+00 
3.750E-02 2.881E+02 4.461E+02 3.564E+02 5.381E+00 1.656E-01 3.991E+02 0.000E+00 5.642E+00 0.000E+00 0.000E+00 0.000E+00 0.000E+00 0.000E+00 
6.250E-02 2.372E+02 3.739E+02 3.010E+02 4.636E+00 1.438E-01 3.618E+02 0.000E+00 4.925E+00 0.000E+00 2.904E-03 0.000E+00 0.000E+00 0.000E+00 
8.750E-02 1.897E+02 3.033E+02 2.451E+02 4.175E+00 1.238E-01 3.246E+02 0.000E+00 4.101E+00 0.000E+00 0.000E+00 0.000E+00 0.000E+00 0.000E+00 
1.125E-01 1.526E+02 2.458E+02 1.960E+02 3.495E+00 9.233E-02 2.879E+02 0.000E+00 3.712E+00 0.000E+00 0.000E+00 0.000E+00 0.000E+00 0.000E+00 
1.375E-01 1.224E+02 1.920E+02 1.564E+02 3.420E+00 1.520E-01 2.536E+02 0.000E+00 3.130E+00 0.000E+00 0.000E+00 0.000E+00 0.000E+00 0.000E+00 
1.625E-01 9.617E+01 1.505E+02 1.236E+02 3.260E+00 9.749E-02 2.223E+02 0.000E+00 2.785E+00 0.000E+00 0.000E+00 0.000E+00 0.000E+00 0.000E+00 
1.875E-01 7.623E+01 1.163E+02 9.955E+01 2.876E+00 7.043E-02 1.936E+02 0.000E+00 2.286E+00 0.000E+00 0.000E+00 0.000E+00 0.000E+00 0.000E+00 
2.125E-01 6.018E+01 9.172E+01 7.782E+01 2.327E+00 5.096E-02 1.701E+02 0.000E+00 1.963E+00 0.000E+00 0.000E+00 0.000E+00 0.000E+00 0.000E+00 
2.375E-01 4.794E+01 7.213E+01 6.200E+01 2.122E+00 6.015E-02 1.462E+02 0.000E+00 1.631E+00 0.000E+00 0.000E+00 0.000E+00 0.000E+00 0.000E+00 
2.625E-01 3.860E+01 5.816E+01 4.985E+01 1.787E+00 3.173E-02 1.271E+02 0.000E+00 1.332E+00 0.000E+00 2.536E-03 0.000E+00 0.000E+00 0.000E+00 
2.875E-01 3.067E+01 4.607E+01 3.974E+01 1.587E+00 4.095E-02 1.094E+02 0.000E+00 1.120E+00 0.000E+00 0.000E+00 0.000E+00 0.000E+00 0.000E+00 
3.125E-01 2.523E+01 3.655E+01 3.213E+01 1.384E+00 9.919E-03 9.410E+01 0.000E+00 1.022E+00 0.000E+00 0.000E+00 0.000E+00 0.000E+00 0.000E+00 
3.375E-01 1.992E+01 2.983E+01 2.608E+01 1.400E+00 1.924E-02 8.042E+01 0.000E+00 8.063E-01 0.000E+00 0.000E+00 0.000E+00 0.000E+00 0.000E+00 
3.625E-01 1.602E+01 2.475E+01 2.019E+01 9.761E-01 3.736E-02 6.867E+01 0.000E+00 6.547E-01 0.000E+00 0.000E+00 0.000E+00 0.000E+00 0.000E+00 
3.875E-01 1.406E+01 1.930E+01 1.740E+01 8.803E-01 3.176E-02 5.927E+01 0.000E+00 6.784E-01 0.000E+00 0.000E+00 0.000E+00 0.000E+00 0.000E+00 
4.125E-01 1.109E+01 1.622E+01 1.431E+01 7.590E-01 1.765E-02 5.028E+01 0.000E+00 5.155E-01 0.000E+00 2.315E-03 0.000E+00 0.000E+00 0.000E+00 
4.375E-01 8.778E+00 1.314E+01 1.177E+01 7.042E-01 1.288E-02 4.307E+01 0.000E+00 4.660E-01 0.000E+00 2.282E-03 0.000E+00 0.000E+00 0.000E+00 
4.625E-01 7.507E+00 1.101E+01 9.825E+00 6.315E-01 8.364E-03 3.633E+01 0.000E+00 3.497E-01 0.000E+00 0.000E+00 0.000E+00 0.000E+00 0.000E+00 
4.875E-01 6.401E+00 8.556E+00 8.083E+00 5.135E-01 8.151E-03 3.147E+01 0.000E+00 2.993E-01 0.000E+00 0.000E+00 0.000E+00 0.000E+00 0.000E+00 
5.125E-01 4.956E+00 7.071E+00 6.204E+00 4.014E-01 0.000E+00 2.695E+01 0.000E+00 2.530E-01 0.000E+00 0.000E+00 0.000E+00 0.000E+00 0.000E+00 
5.375E-01 4.252E+00 6.147E+00 5.531E+00 4.615E-01 0.000E+00 2.260E+01 0.000E+00 2.250E-01 0.000E+00 0.000E+00 0.000E+00 0.000E+00 0.000E+00 
5.625E-01 3.603E+00 4.825E+00 4.208E+00 3.597E-01 0.000E+00 1.968E+01 0.000E+00 1.478E-01 0.000E+00 0.000E+00 0.000E+00 0.000E+00 0.000E+00 
5.875E-01 2.939E+00 3.904E+00 3.600E+00 2.404E-01 7.397E-03 1.641E+01 0.000E+00 1.080E-01 0.000E+00 0.000E+00 0.000E+00 0.000E+00 0.000E+00 
6.125E-01 2.462E+00 3.411E+00 3.241E+00 3.145E-01 3.615E-03 1.411E+01 0.000E+00 1.597E-01 0.000E+00 0.000E+00 0.000E+00 0.000E+00 0.000E+00 
6.375E-01 1.945E+00 2.713E+00 2.460E+00 2.262E-01 3.535E-03 1.163E+01 0.000E+00 1.118E-01 0.000E+00 0.000E+00 0.000E+00 0.000E+00 0.000E+00 
6.625E-01 1.664E+00 2.229E+00 1.974E+00 2.006E-01 3.459E-03 1.014E+01 0.000E+00 9.446E-02 0.000E+00 0.000E+00 0.000E+00 0.000E+00 0.000E+00 
6.875E-01 1.439E+00 2.103E+00 1.822E+00 1.490E-01 0.000E+00 8.635E+00 0.000E+00 6.432E-02 0.000E+00 0.000E+00 0.000E+00 0.000E+00 0.000E+00 
7.125E-01 1.096E+00 1.590E+00 1.307E+00 9.615E-02 0.000E+00 7.397E+00 0.000E+00 6.125E-02 0.000E+00 0.000E+00 0.000E+00 0.000E+00 0.000E+00 
7.375E-01 8.818E-01 1.243E+00 1.307E+00 9.419E-02 3.248E-03 6.138E+00 0.000E+00 3.021E-02 0.000E+00 0.000E+00 0.000E+00 0.000E+00 0.000E+00 
7.625E-01 8.440E-01 1.110E+00 9.506E-01 7.959E-02 6.367E-03 5.318E+00 0.000E+00 5.323E-02 0.000E+00 0.000E+00 0.000E+00 0.000E+00 0.000E+00 
7.875E-01 5.619E-01 9.595E-01 8.082E-01 6.867E-02 0.000E+00 4.632E+00 0.000E+00 4.413E-02 0.000E+00 0.000E+00 0.000E+00 0.000E+00 0.000E+00 
8.125E-01 5.471E-01 7.533E-01 6.523E-01 5.205E-02 0.000E+00 3.805E+00 0.000E+00 3.319E-02 0.000E+00 0.000E+00 0.000E+00 0.000E+00 0.000E+00 
8.375E-01 3.854E-01 6.520E-01 5.618E-01 6.008E-02 0.000E+00 3.226E+00 0.000E+00 2.662E-02 0.000E+00 0.000E+00 0.000E+00 0.000E+00 0.000E+00 
8.625E-01 3.518E-01 5.117E-01 4.718E-01 2.359E-02 0.000E+00 2.701E+00 0.000E+00 2.628E-02 0.000E+00 0.000E+00 0.000E+00 0.000E+00 0.000E+00 
8.875E-01 2.886E-01 4.135E-01 4.135E-01 2.606E-02 0.000E+00 2.277E+00 0.000E+00 1.198E-02 0.000E+00 0.000E+00 0.000E+00 0.000E+00 0.000E+00 
9.125E-01 2.208E-01 3.922E-01 3.122E-01 4.550E-02 0.000E+00 1.967E+00 0.000E+00 9.860E-03 0.000E+00 0.000E+00 0.000E+00 0.000E+00 0.000E+00 
9.375E-01 2.380E-01 2.901E-01 2.975E-01 1.677E-02 0.000E+00 1.734E+00 0.000E+00 9.739E-03 0.000E+00 0.000E+00 0.000E+00 0.000E+00 0.000E+00 
9.625E-01 1.745E-01 2.290E-01 2.290E-01 1.648E-02 0.000E+00 1.391E+00 0.000E+00 3.849E-03 0.000E+00 0.000E+00 0.000E+00 0.000E+00 0.000E+00 
9.875E-01 1.208E-01 2.132E-01 2.203E-01 2.160E-02 0.000E+00 1.222E+00 0.000E+00 1.141E-02 0.000E+00 0.000E+00 0.000E+00 0.000E+00 0.000E+00 
1.012E+00 1.495E-01 2.051E-01 1.738E-01 2.390E-02 0.000E+00 9.618E-01 0.000E+00 1.879E-03 0.000E+00 0.000E+00 0.000E+00 0.000E+00 0.000E+00 
1.038E+00 1.021E-01 1.497E-01 1.531E-01 1.567E-02 2.612E-03 8.464E-01 0.000E+00 7.430E-03 0.000E+00 0.000E+00 0.000E+00 0.000E+00 0.000E+00 
1.062E+00 9.663E-02 1.066E-01 9.996E-02 2.570E-03 0.000E+00 7.398E-01 0.000E+00 1.102E-02 0.000E+00 0.000E+00 0.000E+00 0.000E+00 0.000E+00 
1.087E+00 5.222E-02 1.142E-01 7.507E-02 1.012E-02 2.529E-03 5.529E-01 0.000E+00 1.815E-03 0.000E+00 0.000E+00 0.000E+00 0.000E+00 0.000E+00 
1.113E+00 5.758E-02 7.357E-02 8.317E-02 7.470E-03 0.000E+00 4.662E-01 0.000E+00 5.385E-03 0.000E+00 0.000E+00 0.000E+00 0.000E+00 0.000E+00 
1.137E+00 4.077E-02 7.526E-02 9.408E-02 1.716E-02 0.000E+00 3.874E-01 0.000E+00 3.550E-03 0.000E+00 0.000E+00 0.000E+00 0.000E+00 0.000E+00 
1.163E+00 5.844E-02 5.536E-02 4.306E-02 7.244E-03 0.000E+00 3.561E-01 0.000E+00 1.756E-03 0.000E+00 0.000E+00 0.000E+00 0.000E+00 0.000E+00 
1.188E+00 5.432E-02 4.828E-02 6.035E-02 0.000E+00 0.000E+00 3.218E-01 0.000E+00 1.736E-03 0.000E+00 0.000E+00 0.000E+00 0.000E+00 0.000E+00 
1.213E+00 2.073E-02 3.554E-02 3.554E-02 7.032E-03 0.000E+00 2.344E-01 0.000E+00 0.000E+00 0.000E+00 0.000E+00 0.000E+00 0.000E+00 0.000E+00 
1.238E+00 1.745E-02 3.199E-02 2.908E-02 2.310E-03 0.000E+00 2.023E-01 0.000E+00 1.700E-03 0.000E+00 0.000E+00 0.000E+00 0.000E+00 0.000E+00 
1.262E+00 2.856E-02 3.427E-02 1.714E-02 4.555E-03 0.000E+00 1.927E-01 0.000E+00 0.000E+00 0.000E+00 0.000E+00 0.000E+00 0.000E+00 0.000E+00 
1.288E+00 8.418E-03 2.525E-02 2.525E-02 2.245E-03 0.000E+00 1.761E-01 0.000E+00 1.664E-03 0.000E+00 0.000E+00 0.000E+00 0.000E+00 0.000E+00 
1.312E+00 0.000E+00 3.033E-02 2.206E-02 2.214E-03 0.000E+00 1.369E-01 0.000E+00 1.647E-03 0.000E+00 0.000E+00 0.000E+00 0.000E+00 0.000E+00 
\end{verbatim}
}

\newpage
Au+Au \@ $E_{\rm lab}=4A~$GeV:
{\tiny
\begin{verbatim}
! m_t-m0 , 1/mt dN/dmt(pi+ pi- pi0 K+ K- P aP L+S0 a(L+S0) Xi- aXi- Om aOm) 
1.250E-02 6.294E+02 8.475E+02 7.276E+02 2.458E+01 2.472E+00 2.706E+02 0.000E+00 2.013E+01 0.000E+00 5.424E-02 0.000E+00 0.000E+00 0.000E+00 
3.750E-02 5.323E+02 7.207E+02 6.281E+02 2.237E+01 2.115E+00 2.565E+02 0.000E+00 1.832E+01 0.000E+00 6.506E-02 0.000E+00 0.000E+00 0.000E+00 
6.250E-02 4.458E+02 6.064E+02 5.295E+02 2.023E+01 1.941E+00 2.396E+02 0.000E+00 1.628E+01 0.000E+00 5.227E-02 0.000E+00 0.000E+00 0.000E+00 
8.750E-02 3.722E+02 5.024E+02 4.397E+02 1.784E+01 1.761E+00 2.236E+02 0.000E+00 1.482E+01 0.000E+00 3.137E-02 0.000E+00 2.273E-03 0.000E+00 
1.125E-01 3.044E+02 4.118E+02 3.630E+02 1.680E+01 1.477E+00 2.070E+02 0.000E+00 1.339E+01 0.000E+00 3.082E-02 0.000E+00 0.000E+00 0.000E+00 
1.375E-01 2.530E+02 3.376E+02 2.983E+02 1.508E+01 1.311E+00 1.906E+02 0.000E+00 1.164E+01 0.000E+00 3.029E-02 0.000E+00 0.000E+00 0.000E+00 
1.625E-01 2.067E+02 2.731E+02 2.439E+02 1.322E+01 1.152E+00 1.732E+02 0.000E+00 1.052E+01 0.000E+00 2.707E-02 0.000E+00 0.000E+00 0.000E+00 
1.875E-01 1.698E+02 2.255E+02 2.029E+02 1.187E+01 9.567E-01 1.568E+02 0.000E+00 9.252E+00 0.000E+00 1.597E-02 0.000E+00 0.000E+00 0.000E+00 
2.125E-01 1.402E+02 1.827E+02 1.665E+02 1.079E+01 9.738E-01 1.418E+02 0.000E+00 8.316E+00 0.000E+00 2.881E-02 0.000E+00 0.000E+00 0.000E+00 
2.375E-01 1.142E+02 1.517E+02 1.356E+02 9.465E+00 7.820E-01 1.282E+02 0.000E+00 7.613E+00 0.000E+00 2.061E-02 0.000E+00 0.000E+00 0.000E+00 
2.625E-01 9.591E+01 1.232E+02 1.104E+02 8.423E+00 6.292E-01 1.146E+02 0.000E+00 6.421E+00 0.000E+00 1.014E-02 0.000E+00 0.000E+00 0.000E+00 
2.875E-01 7.794E+01 1.008E+02 9.210E+01 7.816E+00 5.118E-01 1.024E+02 0.000E+00 5.615E+00 0.000E+00 1.997E-02 0.000E+00 0.000E+00 0.000E+00 
3.125E-01 6.584E+01 8.479E+01 7.676E+01 6.864E+00 5.307E-01 9.085E+01 0.000E+00 5.063E+00 0.000E+00 9.831E-03 0.000E+00 0.000E+00 0.000E+00 
3.375E-01 5.483E+01 6.986E+01 6.519E+01 6.374E+00 5.244E-01 8.176E+01 0.000E+00 4.233E+00 0.000E+00 4.841E-03 0.000E+00 0.000E+00 0.000E+00 
3.625E-01 4.638E+01 5.906E+01 5.323E+01 5.231E+00 4.250E-01 7.290E+01 0.000E+00 3.688E+00 0.000E+00 9.538E-03 0.000E+00 0.000E+00 0.000E+00 
3.875E-01 3.984E+01 4.912E+01 4.564E+01 4.560E+00 3.539E-01 6.438E+01 0.000E+00 3.379E+00 0.000E+00 7.048E-03 0.000E+00 0.000E+00 0.000E+00 
4.125E-01 3.332E+01 4.111E+01 4.023E+01 4.426E+00 3.089E-01 5.588E+01 0.000E+00 2.850E+00 0.000E+00 2.315E-03 0.000E+00 0.000E+00 0.000E+00 
4.375E-01 2.738E+01 3.566E+01 3.341E+01 3.706E+00 2.362E-01 5.009E+01 0.000E+00 2.480E+00 0.000E+00 6.847E-03 0.000E+00 0.000E+00 0.000E+00 
4.625E-01 2.359E+01 2.925E+01 2.805E+01 3.174E+00 2.049E-01 4.513E+01 0.000E+00 2.187E+00 0.000E+00 0.000E+00 0.000E+00 0.000E+00 0.000E+00 
4.875E-01 2.030E+01 2.543E+01 2.390E+01 2.849E+00 1.549E-01 3.970E+01 0.000E+00 1.871E+00 0.000E+00 2.219E-03 0.000E+00 0.000E+00 0.000E+00 
5.125E-01 1.744E+01 2.216E+01 2.078E+01 2.603E+00 1.947E-01 3.452E+01 2.758E-03 1.759E+00 0.000E+00 0.000E+00 0.000E+00 0.000E+00 0.000E+00 
5.375E-01 1.572E+01 1.890E+01 1.752E+01 2.261E+00 1.629E-01 3.029E+01 0.000E+00 1.401E+00 0.000E+00 0.000E+00 0.000E+00 0.000E+00 0.000E+00 
5.625E-01 1.299E+01 1.651E+01 1.484E+01 1.829E+00 1.249E-01 2.668E+01 0.000E+00 1.225E+00 0.000E+00 2.130E-03 0.000E+00 0.000E+00 0.000E+00 
5.875E-01 1.089E+01 1.357E+01 1.238E+01 1.646E+00 8.137E-02 2.388E+01 0.000E+00 1.040E+00 0.000E+00 2.102E-03 0.000E+00 0.000E+00 0.000E+00 
6.125E-01 9.604E+00 1.206E+01 1.106E+01 1.547E+00 1.048E-01 2.059E+01 0.000E+00 9.095E-01 0.000E+00 4.150E-03 0.000E+00 0.000E+00 0.000E+00 
6.375E-01 8.258E+00 1.032E+01 9.516E+00 1.382E+00 8.131E-02 1.846E+01 0.000E+00 7.687E-01 0.000E+00 0.000E+00 0.000E+00 0.000E+00 0.000E+00 
6.625E-01 6.941E+00 8.819E+00 8.235E+00 1.266E+00 9.339E-02 1.590E+01 0.000E+00 7.332E-01 0.000E+00 4.046E-03 0.000E+00 0.000E+00 0.000E+00 
6.875E-01 5.984E+00 7.724E+00 6.852E+00 9.479E-01 7.787E-02 1.408E+01 0.000E+00 6.365E-01 0.000E+00 3.995E-03 0.000E+00 0.000E+00 0.000E+00 
7.125E-01 5.432E+00 6.486E+00 6.180E+00 8.554E-01 4.310E-02 1.261E+01 0.000E+00 5.425E-01 0.000E+00 1.973E-03 0.000E+00 0.000E+00 0.000E+00 
7.375E-01 4.377E+00 5.629E+00 5.158E+00 7.795E-01 4.872E-02 1.069E+01 0.000E+00 4.791E-01 0.000E+00 1.949E-03 0.000E+00 0.000E+00 0.000E+00 
7.625E-01 4.091E+00 4.797E+00 4.615E+00 7.449E-01 3.502E-02 9.110E+00 0.000E+00 3.897E-01 0.000E+00 0.000E+00 0.000E+00 0.000E+00 0.000E+00 
7.875E-01 3.363E+00 4.218E+00 3.873E+00 6.211E-01 3.433E-02 8.232E+00 0.000E+00 3.761E-01 0.000E+00 1.902E-03 0.000E+00 0.000E+00 0.000E+00 
8.125E-01 2.887E+00 3.653E+00 3.560E+00 5.603E-01 2.755E-02 7.401E+00 2.285E-03 3.028E-01 0.000E+00 1.880E-03 0.000E+00 0.000E+00 0.000E+00 
8.375E-01 2.600E+00 3.026E+00 2.911E+00 4.536E-01 2.403E-02 6.400E+00 0.000E+00 2.682E-01 0.000E+00 0.000E+00 0.000E+00 0.000E+00 0.000E+00 
8.625E-01 2.255E+00 2.595E+00 2.527E+00 4.128E-01 5.898E-03 5.503E+00 0.000E+00 2.406E-01 0.000E+00 1.837E-03 0.000E+00 0.000E+00 0.000E+00 
8.875E-01 2.059E+00 2.473E+00 2.188E+00 3.735E-01 2.316E-02 4.884E+00 0.000E+00 1.917E-01 0.000E+00 0.000E+00 0.000E+00 0.000E+00 0.000E+00 
9.125E-01 1.519E+00 1.991E+00 1.877E+00 2.901E-01 1.706E-02 4.243E+00 0.000E+00 1.755E-01 0.000E+00 0.000E+00 0.000E+00 0.000E+00 0.000E+00 
9.375E-01 1.447E+00 1.581E+00 1.659E+00 2.683E-01 1.677E-02 3.856E+00 0.000E+00 1.909E-01 0.000E+00 0.000E+00 0.000E+00 0.000E+00 0.000E+00 
9.625E-01 1.269E+00 1.508E+00 1.621E+00 2.609E-01 1.648E-02 3.315E+00 0.000E+00 1.251E-01 0.000E+00 0.000E+00 0.000E+00 0.000E+00 0.000E+00 
9.875E-01 1.116E+00 1.279E+00 1.276E+00 1.836E-01 2.430E-02 2.773E+00 0.000E+00 1.483E-01 0.000E+00 0.000E+00 0.000E+00 0.000E+00 0.000E+00 
1.012E+00 8.970E-01 1.123E+00 1.095E+00 1.726E-01 7.965E-03 2.451E+00 0.000E+00 9.396E-02 0.000E+00 0.000E+00 0.000E+00 0.000E+00 0.000E+00 
1.038E+00 8.575E-01 1.041E+00 9.460E-01 1.254E-01 5.224E-03 2.124E+00 0.000E+00 8.730E-02 0.000E+00 0.000E+00 0.000E+00 0.000E+00 0.000E+00 
1.062E+00 7.497E-01 8.330E-01 8.730E-01 1.388E-01 1.285E-02 1.942E+00 0.000E+00 6.610E-02 0.000E+00 0.000E+00 0.000E+00 0.000E+00 0.000E+00 
1.087E+00 6.495E-01 8.160E-01 7.997E-01 1.138E-01 5.058E-03 1.742E+00 0.000E+00 6.717E-02 0.000E+00 1.665E-03 0.000E+00 0.000E+00 0.000E+00 
1.113E+00 5.790E-01 6.333E-01 6.269E-01 9.213E-02 4.980E-03 1.463E+00 0.000E+00 5.205E-02 0.000E+00 0.000E+00 0.000E+00 0.000E+00 0.000E+00 
1.137E+00 5.174E-01 6.335E-01 5.363E-01 1.030E-01 2.452E-03 1.260E+00 0.000E+00 8.165E-02 0.000E+00 0.000E+00 0.000E+00 0.000E+00 0.000E+00 
1.163E+00 4.737E-01 5.106E-01 4.767E-01 7.486E-02 1.207E-02 1.084E+00 0.000E+00 3.511E-02 0.000E+00 0.000E+00 0.000E+00 0.000E+00 0.000E+00 
1.188E+00 3.742E-01 4.768E-01 4.647E-01 6.423E-02 2.379E-03 9.748E-01 0.000E+00 2.952E-02 0.000E+00 1.598E-03 0.000E+00 0.000E+00 0.000E+00 
1.213E+00 3.228E-01 3.939E-01 2.873E-01 7.735E-02 4.688E-03 8.352E-01 0.000E+00 3.092E-02 0.000E+00 0.000E+00 0.000E+00 0.000E+00 0.000E+00 
1.238E+00 2.356E-01 3.402E-01 2.850E-01 5.082E-02 6.930E-03 7.060E-01 0.000E+00 1.700E-02 0.000E+00 0.000E+00 0.000E+00 0.000E+00 0.000E+00 
1.262E+00 2.456E-01 2.742E-01 2.999E-01 4.782E-02 4.555E-03 6.071E-01 0.000E+00 2.018E-02 0.000E+00 0.000E+00 0.000E+00 0.000E+00 0.000E+00 
1.288E+00 1.992E-01 2.441E-01 1.964E-01 4.491E-02 2.245E-03 6.003E-01 0.000E+00 2.663E-02 0.000E+00 0.000E+00 0.000E+00 0.000E+00 0.000E+00 
1.312E+00 1.848E-01 1.986E-01 1.848E-01 3.543E-02 4.428E-03 4.817E-01 0.000E+00 1.482E-02 0.000E+00 0.000E+00 0.000E+00 0.000E+00 0.000E+00 
\end{verbatim}
}
\newpage
Au+Au \@ $E_{\rm lab}=6A~$GeV:
{\tiny
\begin{verbatim}
! m_t-m0 , 1/mt dN/dmt(pi+ pi- pi0 K+ K- P aP L+S0 a(L+S0) Xi- aXi- Om aOm) 
1.250E-02 8.349E+02 1.081E+03 9.348E+02 4.155E+01 5.797E+00 2.185E+02 0.000E+00 2.700E+01 0.000E+00 1.627E-01 0.000E+00 9.498E-03 0.000E+00 
3.750E-02 7.243E+02 9.195E+02 8.170E+02 3.694E+01 4.832E+00 2.092E+02 0.000E+00 2.524E+01 0.000E+00 1.597E-01 0.000E+00 2.340E-03 0.000E+00 
6.250E-02 6.105E+02 7.737E+02 6.917E+02 3.324E+01 4.471E+00 1.984E+02 0.000E+00 2.299E+01 0.000E+00 1.510E-01 0.000E+00 4.612E-03 0.000E+00 
8.750E-02 5.096E+02 6.413E+02 5.772E+02 2.949E+01 3.955E+00 1.862E+02 0.000E+00 2.142E+01 3.324E-03 1.312E-01 0.000E+00 2.273E-03 0.000E+00 
1.125E-01 4.220E+02 5.271E+02 4.829E+02 2.635E+01 3.561E+00 1.745E+02 3.808E-03 1.936E+01 3.256E-03 8.967E-02 0.000E+00 2.242E-03 0.000E+00 
1.375E-01 3.512E+02 4.363E+02 4.007E+02 2.297E+01 3.313E+00 1.621E+02 0.000E+00 1.684E+01 0.000E+00 8.812E-02 0.000E+00 0.000E+00 0.000E+00 
1.625E-01 2.916E+02 3.638E+02 3.323E+02 2.098E+01 2.821E+00 1.493E+02 0.000E+00 1.551E+01 0.000E+00 6.497E-02 0.000E+00 0.000E+00 0.000E+00 
1.875E-01 2.393E+02 2.986E+02 2.764E+02 1.830E+01 2.547E+00 1.372E+02 0.000E+00 1.363E+01 0.000E+00 6.123E-02 0.000E+00 2.151E-03 0.000E+00 
2.125E-01 1.988E+02 2.462E+02 2.261E+02 1.672E+01 2.259E+00 1.249E+02 0.000E+00 1.234E+01 0.000E+00 5.499E-02 0.000E+00 6.368E-03 0.000E+00 
2.375E-01 1.672E+02 2.039E+02 1.912E+02 1.479E+01 1.985E+00 1.140E+02 0.000E+00 1.067E+01 0.000E+00 7.472E-02 0.000E+00 4.190E-03 0.000E+00 
2.625E-01 1.384E+02 1.687E+02 1.582E+02 1.294E+01 1.708E+00 1.036E+02 0.000E+00 9.465E+00 0.000E+00 4.311E-02 0.000E+00 0.000E+00 0.000E+00 
2.875E-01 1.195E+02 1.426E+02 1.335E+02 1.161E+01 1.510E+00 9.317E+01 0.000E+00 8.735E+00 0.000E+00 3.495E-02 0.000E+00 0.000E+00 0.000E+00 
3.125E-01 9.680E+01 1.184E+02 1.119E+02 1.031E+01 1.364E+00 8.504E+01 0.000E+00 7.474E+00 0.000E+00 3.687E-02 0.000E+00 2.016E-03 0.000E+00 
3.375E-01 8.229E+01 9.972E+01 9.471E+01 9.414E+00 1.280E+00 7.700E+01 0.000E+00 6.795E+00 0.000E+00 3.873E-02 0.000E+00 0.000E+00 0.000E+00 
3.625E-01 6.979E+01 8.408E+01 7.758E+01 7.944E+00 1.149E+00 6.809E+01 0.000E+00 5.971E+00 0.000E+00 2.861E-02 0.000E+00 0.000E+00 0.000E+00 
3.875E-01 5.981E+01 6.988E+01 6.842E+01 7.342E+00 1.085E+00 6.124E+01 0.000E+00 5.249E+00 0.000E+00 2.349E-02 0.000E+00 0.000E+00 0.000E+00 
4.125E-01 5.010E+01 6.183E+01 5.798E+01 6.610E+00 8.869E-01 5.399E+01 0.000E+00 4.622E+00 0.000E+00 1.389E-02 0.000E+00 0.000E+00 0.000E+00 
4.375E-01 4.443E+01 5.188E+01 4.819E+01 5.445E+00 7.128E-01 4.922E+01 0.000E+00 3.973E+00 0.000E+00 2.967E-02 0.000E+00 1.896E-03 0.000E+00 
4.625E-01 3.732E+01 4.391E+01 4.266E+01 5.110E+00 6.858E-01 4.348E+01 2.856E-03 3.837E+00 0.000E+00 1.350E-02 0.000E+00 1.874E-03 0.000E+00 
4.875E-01 3.185E+01 3.808E+01 3.639E+01 4.687E+00 6.032E-01 3.859E+01 0.000E+00 3.196E+00 2.495E-03 1.553E-02 0.000E+00 0.000E+00 0.000E+00 
5.125E-01 2.695E+01 3.269E+01 3.174E+01 3.923E+00 5.166E-01 3.481E+01 0.000E+00 2.793E+00 0.000E+00 1.094E-02 0.000E+00 0.000E+00 0.000E+00 
5.375E-01 2.396E+01 2.877E+01 2.737E+01 3.571E+00 4.382E-01 3.083E+01 0.000E+00 2.535E+00 0.000E+00 4.318E-03 0.000E+00 1.810E-03 0.000E+00 
5.625E-01 2.105E+01 2.515E+01 2.317E+01 3.264E+00 4.203E-01 2.743E+01 0.000E+00 2.121E+00 0.000E+00 4.261E-03 0.000E+00 0.000E+00 0.000E+00 
5.875E-01 1.773E+01 2.087E+01 2.044E+01 2.741E+00 3.847E-01 2.475E+01 2.622E-03 1.883E+00 0.000E+00 1.682E-02 0.000E+00 0.000E+00 0.000E+00 
6.125E-01 1.548E+01 1.846E+01 1.768E+01 2.433E+00 2.675E-01 2.170E+01 0.000E+00 1.636E+00 0.000E+00 2.075E-03 0.000E+00 0.000E+00 0.000E+00 
6.375E-01 1.326E+01 1.588E+01 1.499E+01 2.209E+00 3.323E-01 1.885E+01 0.000E+00 1.419E+00 0.000E+00 2.049E-03 0.000E+00 0.000E+00 0.000E+00 
6.625E-01 1.185E+01 1.375E+01 1.346E+01 2.020E+00 2.110E-01 1.702E+01 0.000E+00 1.293E+00 0.000E+00 8.091E-03 0.000E+00 0.000E+00 0.000E+00 
6.875E-01 1.035E+01 1.230E+01 1.201E+01 1.723E+00 2.234E-01 1.495E+01 0.000E+00 1.045E+00 0.000E+00 1.998E-03 0.000E+00 0.000E+00 0.000E+00 
7.125E-01 8.908E+00 1.049E+01 1.054E+01 1.525E+00 1.989E-01 1.326E+01 0.000E+00 1.002E+00 0.000E+00 7.891E-03 0.000E+00 0.000E+00 0.000E+00 
7.375E-01 7.776E+00 9.485E+00 8.690E+00 1.218E+00 1.364E-01 1.199E+01 0.000E+00 9.193E-01 0.000E+00 1.949E-03 0.000E+00 0.000E+00 0.000E+00 
7.625E-01 7.134E+00 8.093E+00 7.689E+00 1.245E+00 1.464E-01 1.038E+01 0.000E+00 7.964E-01 0.000E+00 0.000E+00 0.000E+00 0.000E+00 0.000E+00 
7.875E-01 5.995E+00 6.941E+00 6.531E+00 1.058E+00 1.092E-01 9.303E+00 0.000E+00 7.208E-01 0.000E+00 1.902E-03 0.000E+00 0.000E+00 0.000E+00 
8.125E-01 5.201E+00 6.144E+00 5.908E+00 8.909E-01 9.797E-02 8.226E+00 0.000E+00 6.181E-01 0.000E+00 1.880E-03 0.000E+00 0.000E+00 0.000E+00 
8.375E-01 4.470E+00 5.511E+00 5.019E+00 7.630E-01 9.012E-02 7.383E+00 0.000E+00 4.730E-01 0.000E+00 5.575E-03 0.000E+00 0.000E+00 0.000E+00 
8.625E-01 4.058E+00 4.622E+00 4.670E+00 7.018E-01 9.141E-02 6.387E+00 0.000E+00 4.953E-01 0.000E+00 1.837E-03 0.000E+00 0.000E+00 0.000E+00 
8.875E-01 3.370E+00 4.181E+00 3.858E+00 6.225E-01 6.370E-02 5.513E+00 0.000E+00 3.434E-01 0.000E+00 1.816E-03 0.000E+00 0.000E+00 0.000E+00 
9.125E-01 3.020E+00 3.743E+00 3.358E+00 6.228E-01 5.688E-02 4.883E+00 0.000E+00 3.707E-01 0.000E+00 0.000E+00 0.000E+00 0.000E+00 0.000E+00 
9.375E-01 2.693E+00 3.459E+00 3.109E+00 5.086E-01 6.706E-02 4.558E+00 2.133E-03 3.136E-01 0.000E+00 0.000E+00 0.000E+00 0.000E+00 0.000E+00 
9.625E-01 2.264E+00 2.741E+00 2.672E+00 4.174E-01 5.218E-02 4.020E+00 0.000E+00 2.867E-01 0.000E+00 0.000E+00 0.000E+00 0.000E+00 0.000E+00 
9.875E-01 2.072E+00 2.470E+00 2.434E+00 4.374E-01 5.670E-02 3.442E+00 0.000E+00 2.358E-01 0.000E+00 0.000E+00 0.000E+00 0.000E+00 0.000E+00 
1.012E+00 1.714E+00 2.166E+00 2.187E+00 3.903E-01 4.514E-02 2.986E+00 0.000E+00 2.067E-01 0.000E+00 0.000E+00 0.000E+00 0.000E+00 0.000E+00 
1.038E+00 1.504E+00 1.797E+00 1.797E+00 2.037E-01 2.612E-02 2.594E+00 0.000E+00 1.839E-01 0.000E+00 0.000E+00 0.000E+00 0.000E+00 0.000E+00 
1.062E+00 1.429E+00 1.696E+00 1.596E+00 2.518E-01 3.598E-02 2.403E+00 0.000E+00 1.340E-01 0.000E+00 0.000E+00 0.000E+00 0.000E+00 0.000E+00 
1.087E+00 1.149E+00 1.397E+00 1.446E+00 2.251E-01 2.782E-02 2.056E+00 0.000E+00 1.234E-01 0.000E+00 0.000E+00 0.000E+00 0.000E+00 0.000E+00 
1.113E+00 1.100E+00 1.228E+00 1.257E+00 2.017E-01 2.490E-02 1.834E+00 0.000E+00 9.872E-02 0.000E+00 0.000E+00 0.000E+00 0.000E+00 0.000E+00 
1.137E+00 8.781E-01 1.123E+00 1.120E+00 1.790E-01 1.961E-02 1.652E+00 0.000E+00 1.083E-01 0.000E+00 0.000E+00 0.000E+00 0.000E+00 0.000E+00 
1.163E+00 7.782E-01 9.812E-01 1.006E+00 1.449E-01 2.898E-02 1.341E+00 0.000E+00 1.141E-01 0.000E+00 0.000E+00 0.000E+00 0.000E+00 0.000E+00 
1.188E+00 7.605E-01 9.114E-01 8.118E-01 1.118E-01 1.665E-02 1.214E+00 0.000E+00 1.007E-01 0.000E+00 0.000E+00 0.000E+00 0.000E+00 0.000E+00 
1.213E+00 6.161E-01 7.582E-01 6.990E-01 1.594E-01 2.110E-02 1.129E+00 0.000E+00 6.528E-02 0.000E+00 0.000E+00 0.000E+00 0.000E+00 0.000E+00 
1.238E+00 5.903E-01 7.037E-01 7.008E-01 1.063E-01 1.848E-02 1.013E+00 0.000E+00 5.099E-02 0.000E+00 0.000E+00 0.000E+00 0.000E+00 0.000E+00 
1.262E+00 4.598E-01 5.798E-01 5.741E-01 1.025E-01 1.139E-02 9.143E-01 0.000E+00 7.231E-02 0.000E+00 0.000E+00 0.000E+00 0.000E+00 0.000E+00 
1.288E+00 4.069E-01 5.893E-01 5.107E-01 7.185E-02 1.347E-02 7.800E-01 0.000E+00 5.658E-02 0.000E+00 0.000E+00 0.000E+00 0.000E+00 0.000E+00 
1.312E+00 4.109E-01 5.267E-01 4.964E-01 6.643E-02 1.329E-02 6.825E-01 0.000E+00 2.471E-02 0.000E+00 0.000E+00 0.000E+00 0.000E+00 0.000E+00 
\end{verbatim}
}

\newpage
Au+Au \@ $E_{\rm lab}=8A~$GeV:
{\tiny
\begin{verbatim}
! m_t-m0 , 1/mt dN/dmt(pi+ pi- pi0 K+ K- P aP L+S0 a(L+S0) Xi- aXi- Om aOm) 
1.250E-02 1.011E+03 1.243E+03 1.102E+03 5.380E+01 9.082E+00 1.937E+02 0.000E+00 3.109E+01 0.000E+00 2.139E-01 0.000E+00 9.498E-03 0.000E+00 
3.750E-02 8.735E+02 1.070E+03 9.673E+02 4.773E+01 7.767E+00 1.855E+02 0.000E+00 2.861E+01 0.000E+00 2.396E-01 0.000E+00 2.340E-03 0.000E+00 
6.250E-02 7.345E+02 8.972E+02 8.144E+02 4.325E+01 7.468E+00 1.762E+02 0.000E+00 2.624E+01 3.394E-03 2.178E-01 2.904E-03 6.918E-03 0.000E+00 
8.750E-02 6.141E+02 7.537E+02 6.866E+02 3.977E+01 6.459E+00 1.660E+02 0.000E+00 2.396E+01 0.000E+00 1.968E-01 0.000E+00 0.000E+00 0.000E+00 
1.125E-01 5.107E+02 6.192E+02 5.816E+02 3.437E+01 5.553E+00 1.567E+02 0.000E+00 2.194E+01 0.000E+00 1.849E-01 0.000E+00 0.000E+00 0.000E+00 
1.375E-01 4.248E+02 5.180E+02 4.833E+02 3.071E+01 5.251E+00 1.451E+02 3.719E-03 1.997E+01 6.382E-03 1.294E-01 0.000E+00 4.421E-03 0.000E+00 
1.625E-01 3.569E+02 4.302E+02 3.981E+02 2.674E+01 4.972E+00 1.364E+02 3.635E-03 1.799E+01 3.129E-03 1.029E-01 0.000E+00 4.361E-03 0.000E+00 
1.875E-01 2.962E+02 3.572E+02 3.354E+02 2.358E+01 4.067E+00 1.240E+02 0.000E+00 1.605E+01 0.000E+00 1.038E-01 0.000E+00 0.000E+00 0.000E+00 
2.125E-01 2.495E+02 2.981E+02 2.759E+02 2.101E+01 3.890E+00 1.144E+02 3.477E-03 1.420E+01 3.011E-03 8.380E-02 0.000E+00 0.000E+00 0.000E+00 
2.375E-01 2.064E+02 2.452E+02 2.315E+02 1.878E+01 3.254E+00 1.043E+02 3.403E-03 1.257E+01 0.000E+00 9.018E-02 0.000E+00 2.095E-03 0.000E+00 
2.625E-01 1.710E+02 2.029E+02 1.936E+02 1.682E+01 3.167E+00 9.454E+01 0.000E+00 1.145E+01 0.000E+00 1.166E-01 0.000E+00 4.135E-03 0.000E+00 
2.875E-01 1.480E+02 1.713E+02 1.639E+02 1.452E+01 2.800E+00 8.658E+01 3.264E-03 9.984E+00 2.850E-03 6.989E-02 0.000E+00 2.041E-03 0.000E+00 
3.125E-01 1.240E+02 1.460E+02 1.366E+02 1.290E+01 2.410E+00 7.817E+01 0.000E+00 9.089E+00 0.000E+00 7.373E-02 0.000E+00 4.031E-03 0.000E+00 
3.375E-01 1.055E+02 1.233E+02 1.168E+02 1.153E+01 2.242E+00 7.186E+01 0.000E+00 8.162E+00 0.000E+00 2.905E-02 0.000E+00 0.000E+00 0.000E+00 
3.625E-01 9.004E+01 1.029E+02 9.988E+01 9.989E+00 1.798E+00 6.460E+01 6.151E-03 7.313E+00 0.000E+00 5.484E-02 0.000E+00 1.966E-03 0.000E+00 
3.875E-01 7.657E+01 8.875E+01 8.594E+01 8.576E+00 1.593E+00 5.808E+01 3.018E-03 6.348E+00 0.000E+00 3.524E-02 0.000E+00 0.000E+00 0.000E+00 
4.125E-01 6.506E+01 7.580E+01 7.308E+01 7.771E+00 1.465E+00 5.145E+01 0.000E+00 5.768E+00 0.000E+00 3.242E-02 0.000E+00 0.000E+00 0.000E+00 
4.375E-01 5.563E+01 6.597E+01 6.272E+01 7.081E+00 1.168E+00 4.627E+01 5.816E-03 4.918E+00 0.000E+00 2.967E-02 0.000E+00 0.000E+00 0.000E+00 
4.625E-01 4.819E+01 5.529E+01 5.405E+01 6.256E+00 1.154E+00 4.116E+01 2.856E-03 4.399E+00 2.534E-03 2.025E-02 0.000E+00 0.000E+00 0.000E+00 
4.875E-01 4.200E+01 4.867E+01 4.644E+01 5.123E+00 9.414E-01 3.719E+01 2.806E-03 3.842E+00 0.000E+00 1.997E-02 0.000E+00 0.000E+00 0.000E+00 
5.125E-01 3.606E+01 4.120E+01 4.030E+01 4.948E+00 9.379E-01 3.348E+01 0.000E+00 3.399E+00 0.000E+00 2.408E-02 0.000E+00 0.000E+00 0.000E+00 
5.375E-01 3.095E+01 3.542E+01 3.403E+01 4.390E+00 7.601E-01 3.005E+01 0.000E+00 2.978E+00 2.419E-03 1.296E-02 0.000E+00 0.000E+00 0.000E+00 
5.625E-01 2.631E+01 3.069E+01 3.000E+01 3.850E+00 6.550E-01 2.665E+01 2.666E-03 2.517E+00 0.000E+00 1.704E-02 0.000E+00 1.790E-03 0.000E+00 
5.875E-01 2.324E+01 2.598E+01 2.563E+01 3.499E+00 6.583E-01 2.386E+01 2.622E-03 2.374E+00 0.000E+00 8.410E-03 0.000E+00 0.000E+00 0.000E+00 
6.125E-01 2.047E+01 2.362E+01 2.282E+01 2.906E+00 5.676E-01 2.169E+01 2.580E-03 2.085E+00 2.314E-03 1.660E-02 0.000E+00 0.000E+00 0.000E+00 
6.375E-01 1.763E+01 2.075E+01 1.946E+01 2.549E+00 4.631E-01 1.908E+01 0.000E+00 1.814E+00 0.000E+00 8.195E-03 0.000E+00 0.000E+00 0.000E+00 
6.625E-01 1.536E+01 1.745E+01 1.746E+01 2.435E+00 4.773E-01 1.738E+01 0.000E+00 1.626E+00 0.000E+00 1.011E-02 0.000E+00 0.000E+00 0.000E+00 
6.875E-01 1.342E+01 1.551E+01 1.534E+01 2.116E+00 3.995E-01 1.522E+01 0.000E+00 1.411E+00 0.000E+00 5.993E-03 0.000E+00 0.000E+00 0.000E+00 
7.125E-01 1.194E+01 1.332E+01 1.317E+01 1.989E+00 3.216E-01 1.355E+01 0.000E+00 1.295E+00 0.000E+00 1.578E-02 0.000E+00 0.000E+00 0.000E+00 
7.375E-01 1.021E+01 1.223E+01 1.140E+01 1.757E+00 2.631E-01 1.178E+01 0.000E+00 1.127E+00 0.000E+00 5.847E-03 0.000E+00 0.000E+00 0.000E+00 
7.625E-01 9.182E+00 1.047E+01 1.035E+01 1.493E+00 2.388E-01 1.047E+01 0.000E+00 9.241E-01 0.000E+00 5.776E-03 0.000E+00 0.000E+00 0.000E+00 
7.875E-01 8.121E+00 9.098E+00 8.493E+00 1.464E+00 2.653E-01 9.231E+00 2.318E-03 9.393E-01 0.000E+00 5.707E-03 0.000E+00 0.000E+00 0.000E+00 
8.125E-01 7.230E+00 7.819E+00 7.512E+00 1.166E+00 2.174E-01 8.551E+00 0.000E+00 8.400E-01 0.000E+00 7.521E-03 0.000E+00 0.000E+00 0.000E+00 
8.375E-01 5.942E+00 7.106E+00 7.114E+00 1.036E+00 1.983E-01 7.466E+00 0.000E+00 6.757E-01 0.000E+00 5.575E-03 0.000E+00 0.000E+00 0.000E+00 
8.625E-01 5.177E+00 5.929E+00 6.041E+00 9.200E-01 1.356E-01 6.691E+00 0.000E+00 6.247E-01 0.000E+00 0.000E+00 0.000E+00 0.000E+00 0.000E+00 
8.875E-01 4.825E+00 5.539E+00 5.207E+00 7.875E-01 9.844E-02 5.842E+00 2.191E-03 5.530E-01 0.000E+00 1.816E-03 0.000E+00 0.000E+00 0.000E+00 
9.125E-01 3.990E+00 4.691E+00 4.664E+00 6.769E-01 1.422E-01 5.248E+00 0.000E+00 4.752E-01 0.000E+00 1.796E-03 0.000E+00 0.000E+00 0.000E+00 
9.375E-01 3.719E+00 4.392E+00 3.991E+00 5.980E-01 8.942E-02 4.519E+00 0.000E+00 3.740E-01 0.000E+00 0.000E+00 0.000E+00 0.000E+00 0.000E+00 
9.625E-01 3.202E+00 3.606E+00 3.762E+00 5.136E-01 8.788E-02 4.195E+00 0.000E+00 3.887E-01 0.000E+00 1.756E-03 0.000E+00 0.000E+00 0.000E+00 
9.875E-01 2.854E+00 3.287E+00 3.064E+00 5.616E-01 9.720E-02 3.442E+00 0.000E+00 3.347E-01 0.000E+00 5.212E-03 0.000E+00 0.000E+00 0.000E+00 
1.012E+00 2.729E+00 2.858E+00 2.830E+00 4.700E-01 7.965E-02 3.129E+00 0.000E+00 3.251E-01 0.000E+00 0.000E+00 0.000E+00 0.000E+00 0.000E+00 
1.038E+00 2.191E+00 2.603E+00 2.590E+00 3.761E-01 5.746E-02 2.843E+00 0.000E+00 2.285E-01 0.000E+00 0.000E+00 0.000E+00 0.000E+00 0.000E+00 
1.062E+00 1.913E+00 2.102E+00 2.076E+00 3.289E-01 6.682E-02 2.557E+00 0.000E+00 2.056E-01 0.000E+00 1.682E-03 0.000E+00 0.000E+00 0.000E+00 
1.087E+00 1.763E+00 2.148E+00 1.860E+00 2.985E-01 4.047E-02 2.180E+00 0.000E+00 2.088E-01 0.000E+00 0.000E+00 0.000E+00 0.000E+00 0.000E+00 
1.113E+00 1.603E+00 1.759E+00 1.705E+00 2.589E-01 4.482E-02 1.941E+00 0.000E+00 1.562E-01 0.000E+00 3.296E-03 0.000E+00 0.000E+00 0.000E+00 
1.137E+00 1.455E+00 1.646E+00 1.530E+00 2.329E-01 3.187E-02 1.819E+00 0.000E+00 1.615E-01 0.000E+00 1.631E-03 0.000E+00 0.000E+00 0.000E+00 
1.163E+00 1.212E+00 1.436E+00 1.418E+00 2.463E-01 2.898E-02 1.582E+00 0.000E+00 1.387E-01 0.000E+00 0.000E+00 0.000E+00 0.000E+00 0.000E+00 
1.188E+00 1.098E+00 1.201E+00 1.295E+00 2.046E-01 3.092E-02 1.295E+00 0.000E+00 1.042E-01 0.000E+00 0.000E+00 0.000E+00 0.000E+00 0.000E+00 
1.213E+00 9.448E-01 1.117E+00 1.143E+00 1.547E-01 2.813E-02 1.213E+00 0.000E+00 1.065E-01 0.000E+00 0.000E+00 0.000E+00 0.000E+00 0.000E+00 
1.238E+00 8.811E-01 1.018E+00 9.655E-01 1.455E-01 2.079E-02 1.114E+00 0.000E+00 9.688E-02 0.000E+00 0.000E+00 0.000E+00 0.000E+00 0.000E+00 
1.262E+00 7.340E-01 8.511E-01 8.711E-01 1.344E-01 1.139E-02 8.634E-01 0.000E+00 8.409E-02 0.000E+00 3.104E-03 0.000E+00 0.000E+00 0.000E+00 
1.288E+00 7.240E-01 7.913E-01 8.053E-01 1.123E-01 1.347E-02 8.771E-01 0.000E+00 6.823E-02 0.000E+00 0.000E+00 0.000E+00 0.000E+00 0.000E+00 
1.312E+00 5.708E-01 6.508E-01 6.591E-01 9.300E-02 1.107E-02 7.341E-01 0.000E+00 6.588E-02 0.000E+00 0.000E+00 0.000E+00 0.000E+00 0.000E+00 
\end{verbatim}
}

\newpage
Au+Au \@ $E_{\rm lab}=11A~$GeV:
{\tiny
\begin{verbatim}
 ! m_t-m0 , 1/mt dN/dmt(pi+ pi- pi0 K+ K- P aP L+S0 a(L+S0) Xi- aXi- Om aOm) 
1.250E-02 1.192E+03 1.420E+03 1.288E+03 6.600E+01 1.318E+01 1.696E+02 4.208E-03 3.387E+01 3.545E-03 3.887E-01 0.000E+00 2.137E-02 0.000E+00 
3.750E-02 1.035E+03 1.236E+03 1.132E+03 5.834E+01 1.254E+01 1.629E+02 8.201E-03 3.106E+01 0.000E+00 3.579E-01 5.915E-03 1.404E-02 0.000E+00 
6.250E-02 8.724E+02 1.040E+03 9.530E+02 5.201E+01 1.119E+01 1.564E+02 3.998E-03 2.876E+01 6.788E-03 3.107E-01 0.000E+00 1.384E-02 0.000E+00 
8.750E-02 7.370E+02 8.681E+02 8.112E+02 4.748E+01 1.021E+01 1.472E+02 0.000E+00 2.664E+01 6.647E-03 2.881E-01 0.000E+00 1.364E-02 0.000E+00 
1.125E-01 6.102E+02 7.261E+02 6.782E+02 4.131E+01 9.029E+00 1.372E+02 7.615E-03 2.414E+01 3.256E-03 2.298E-01 0.000E+00 6.725E-03 0.000E+00 
1.375E-01 5.146E+02 5.991E+02 5.697E+02 3.686E+01 8.399E+00 1.293E+02 7.438E-03 2.192E+01 1.915E-02 2.286E-01 0.000E+00 6.632E-03 0.000E+00 
1.625E-01 4.277E+02 5.034E+02 4.738E+02 3.256E+01 7.561E+00 1.198E+02 3.635E-03 2.017E+01 0.000E+00 1.733E-01 0.000E+00 2.180E-03 0.000E+00 
1.875E-01 3.606E+02 4.186E+02 3.965E+02 2.912E+01 6.415E+00 1.110E+02 0.000E+00 1.768E+01 9.206E-03 1.225E-01 0.000E+00 6.453E-03 0.000E+00 
2.125E-01 2.959E+02 3.477E+02 3.309E+02 2.496E+01 5.905E+00 1.028E+02 1.043E-02 1.577E+01 1.204E-02 1.126E-01 0.000E+00 6.368E-03 0.000E+00 
2.375E-01 2.504E+02 2.896E+02 2.789E+02 2.224E+01 5.157E+00 9.452E+01 1.021E-02 1.381E+01 2.955E-03 1.262E-01 0.000E+00 4.190E-03 1.817E-03 
2.625E-01 2.111E+02 2.458E+02 2.349E+02 1.992E+01 4.711E+00 8.592E+01 3.332E-03 1.301E+01 0.000E+00 1.040E-01 0.000E+00 2.068E-03 0.000E+00 
2.875E-01 1.780E+02 2.072E+02 1.993E+02 1.740E+01 4.105E+00 7.921E+01 3.264E-03 1.149E+01 0.000E+00 1.098E-01 0.000E+00 4.083E-03 0.000E+00 
3.125E-01 1.509E+02 1.741E+02 1.667E+02 1.523E+01 3.660E+00 7.185E+01 1.279E-02 1.031E+01 2.800E-03 8.111E-02 2.458E-03 2.016E-03 0.000E+00 
3.375E-01 1.291E+02 1.461E+02 1.423E+02 1.312E+01 3.108E+00 6.405E+01 1.568E-02 9.335E+00 0.000E+00 8.230E-02 0.000E+00 0.000E+00 0.000E+00 
3.625E-01 1.099E+02 1.246E+02 1.237E+02 1.253E+01 2.816E+00 5.900E+01 3.076E-03 8.095E+00 0.000E+00 8.346E-02 0.000E+00 3.932E-03 0.000E+00 
3.875E-01 9.351E+01 1.066E+02 1.044E+02 1.058E+01 2.691E+00 5.263E+01 1.207E-02 6.915E+00 2.660E-03 4.934E-02 0.000E+00 5.827E-03 0.000E+00 
4.125E-01 8.146E+01 9.261E+01 8.912E+01 9.244E+00 2.268E+00 4.748E+01 5.924E-03 6.490E+00 0.000E+00 5.094E-02 0.000E+00 1.919E-03 0.000E+00 
4.375E-01 7.000E+01 8.002E+01 7.753E+01 8.305E+00 2.005E+00 4.309E+01 2.908E-03 5.711E+00 2.575E-03 5.250E-02 0.000E+00 3.792E-03 0.000E+00 
4.625E-01 6.062E+01 6.824E+01 6.666E+01 7.101E+00 1.761E+00 3.856E+01 2.856E-03 4.853E+00 0.000E+00 3.376E-02 0.000E+00 0.000E+00 0.000E+00 
4.875E-01 5.316E+01 5.978E+01 5.698E+01 6.590E+00 1.447E+00 3.446E+01 1.403E-02 4.662E+00 2.495E-03 3.551E-02 0.000E+00 1.852E-03 0.000E+00 
5.125E-01 4.544E+01 5.168E+01 4.964E+01 5.520E+00 1.415E+00 3.139E+01 2.758E-03 3.800E+00 2.456E-03 1.970E-02 0.000E+00 1.831E-03 0.000E+00 
5.375E-01 3.906E+01 4.467E+01 4.274E+01 5.123E+00 1.136E+00 2.775E+01 5.422E-03 3.619E+00 4.838E-03 2.591E-02 0.000E+00 0.000E+00 0.000E+00 
5.625E-01 3.506E+01 3.832E+01 3.737E+01 4.365E+00 1.083E+00 2.511E+01 5.332E-03 3.150E+00 2.383E-03 2.130E-02 0.000E+00 1.790E-03 0.000E+00 
5.875E-01 2.947E+01 3.362E+01 3.266E+01 3.891E+00 1.017E+00 2.233E+01 2.622E-03 2.783E+00 2.348E-03 1.892E-02 0.000E+00 0.000E+00 0.000E+00 
6.125E-01 2.556E+01 2.958E+01 2.923E+01 3.470E+00 9.254E-01 1.963E+01 2.580E-03 2.492E+00 0.000E+00 1.453E-02 0.000E+00 0.000E+00 0.000E+00 
6.375E-01 2.289E+01 2.535E+01 2.514E+01 3.044E+00 6.717E-01 1.815E+01 2.539E-03 2.265E+00 0.000E+00 1.844E-02 0.000E+00 0.000E+00 0.000E+00 
6.625E-01 1.998E+01 2.249E+01 2.179E+01 2.888E+00 6.641E-01 1.629E+01 0.000E+00 2.033E+00 0.000E+00 1.820E-02 0.000E+00 0.000E+00 0.000E+00 
6.875E-01 1.751E+01 1.970E+01 1.881E+01 2.394E+00 5.451E-01 1.412E+01 2.461E-03 1.792E+00 2.218E-03 1.598E-02 0.000E+00 0.000E+00 0.000E+00 
7.125E-01 1.508E+01 1.658E+01 1.657E+01 2.261E+00 4.542E-01 1.278E+01 0.000E+00 1.547E+00 0.000E+00 1.578E-02 0.000E+00 0.000E+00 0.000E+00 
7.375E-01 1.300E+01 1.509E+01 1.449E+01 2.024E+00 4.352E-01 1.131E+01 0.000E+00 1.301E+00 0.000E+00 9.744E-03 0.000E+00 0.000E+00 0.000E+00 
7.625E-01 1.157E+01 1.331E+01 1.294E+01 1.713E+00 3.756E-01 1.012E+01 2.352E-03 1.229E+00 0.000E+00 7.702E-03 0.000E+00 0.000E+00 0.000E+00 
7.875E-01 1.026E+01 1.176E+01 1.134E+01 1.548E+00 3.028E-01 9.073E+00 4.636E-03 1.063E+00 0.000E+00 5.707E-03 0.000E+00 1.626E-03 0.000E+00 
8.125E-01 8.875E+00 1.030E+01 9.995E+00 1.399E+00 3.184E-01 7.998E+00 0.000E+00 9.417E-01 0.000E+00 5.640E-03 0.000E+00 0.000E+00 0.000E+00 
8.375E-01 7.914E+00 8.541E+00 8.689E+00 1.199E+00 3.094E-01 7.453E+00 2.253E-03 7.965E-01 0.000E+00 5.575E-03 0.000E+00 1.594E-03 0.000E+00 
8.625E-01 7.072E+00 7.908E+00 7.900E+00 9.701E-01 2.418E-01 6.576E+00 4.443E-03 7.076E-01 0.000E+00 0.000E+00 0.000E+00 0.000E+00 0.000E+00 
8.875E-01 6.206E+00 6.674E+00 7.017E+00 1.005E+00 2.258E-01 5.732E+00 0.000E+00 6.608E-01 0.000E+00 1.816E-03 0.000E+00 0.000E+00 0.000E+00 
9.125E-01 5.517E+00 6.062E+00 5.906E+00 8.617E-01 1.593E-01 5.036E+00 0.000E+00 4.989E-01 0.000E+00 3.591E-03 0.000E+00 0.000E+00 0.000E+00 
9.375E-01 4.828E+00 5.404E+00 5.229E+00 8.159E-01 1.453E-01 4.705E+00 0.000E+00 5.026E-01 0.000E+00 3.552E-03 0.000E+00 0.000E+00 0.000E+00 
9.625E-01 4.249E+00 4.605E+00 4.365E+00 6.619E-01 1.318E-01 4.180E+00 0.000E+00 4.061E-01 0.000E+00 0.000E+00 0.000E+00 0.000E+00 0.000E+00 
9.875E-01 3.806E+00 4.197E+00 4.016E+00 5.589E-01 1.107E-01 3.827E+00 0.000E+00 3.651E-01 0.000E+00 1.737E-03 0.000E+00 0.000E+00 0.000E+00 
1.012E+00 3.411E+00 3.518E+00 3.793E+00 5.390E-01 1.142E-01 3.216E+00 0.000E+00 2.875E-01 0.000E+00 0.000E+00 0.000E+00 0.000E+00 0.000E+00 
1.038E+00 2.967E+00 3.318E+00 3.318E+00 4.858E-01 5.224E-02 2.823E+00 0.000E+00 3.065E-01 0.000E+00 1.700E-03 0.000E+00 0.000E+00 0.000E+00 
1.062E+00 2.626E+00 2.829E+00 2.862E+00 4.549E-01 1.079E-01 2.409E+00 2.000E-03 2.791E-01 0.000E+00 1.682E-03 0.000E+00 0.000E+00 0.000E+00 
1.087E+00 2.386E+00 2.565E+00 2.575E+00 3.617E-01 8.852E-02 2.216E+00 0.000E+00 2.251E-01 0.000E+00 0.000E+00 0.000E+00 0.000E+00 0.000E+00 
1.113E+00 2.121E+00 2.415E+00 2.194E+00 3.336E-01 4.731E-02 1.988E+00 0.000E+00 2.136E-01 0.000E+00 1.648E-03 0.000E+00 0.000E+00 0.000E+00 
1.137E+00 1.982E+00 2.051E+00 2.020E+00 2.501E-01 4.903E-02 1.873E+00 1.927E-03 1.935E-01 0.000E+00 0.000E+00 0.000E+00 0.000E+00 0.000E+00 
1.163E+00 1.547E+00 1.741E+00 1.935E+00 2.656E-01 4.347E-02 1.603E+00 0.000E+00 1.984E-01 0.000E+00 3.229E-03 0.000E+00 0.000E+00 0.000E+00 
1.188E+00 1.452E+00 1.675E+00 1.551E+00 2.189E-01 5.471E-02 1.509E+00 0.000E+00 1.493E-01 0.000E+00 1.598E-03 0.000E+00 0.000E+00 0.000E+00 
1.213E+00 1.345E+00 1.422E+00 1.437E+00 2.063E-01 4.454E-02 1.362E+00 0.000E+00 1.323E-01 0.000E+00 3.165E-03 0.000E+00 0.000E+00 0.000E+00 
1.238E+00 1.221E+00 1.224E+00 1.245E+00 1.779E-01 3.927E-02 1.125E+00 0.000E+00 1.394E-01 0.000E+00 3.134E-03 0.000E+00 0.000E+00 0.000E+00 
1.262E+00 1.054E+00 1.080E+00 1.237E+00 1.344E-01 4.099E-02 9.943E-01 0.000E+00 1.043E-01 0.000E+00 3.104E-03 0.000E+00 0.000E+00 0.000E+00 
1.288E+00 8.558E-01 1.072E+00 9.737E-01 1.280E-01 4.491E-02 8.555E-01 0.000E+00 8.488E-02 0.000E+00 3.074E-03 0.000E+00 0.000E+00 0.000E+00 
1.312E+00 7.915E-01 8.825E-01 1.001E+00 1.129E-01 3.321E-02 8.780E-01 0.000E+00 6.094E-02 0.000E+00 0.000E+00 0.000E+00 0.000E+00 0.000E+00
\end{verbatim}
}

\newpage
Pb+Pb \@ $E_{\rm lab}=20A~$GeV:
{\tiny
\begin{verbatim}
! m_t-m0 , 1/mt dN/dmt(pi+ pi- pi0 K+ K- P aP L+S0 a(L+S0) Xi- aXi- Om aOm) 
1.250E-02 1.641E+03 1.885E+03 1.758E+03 8.756E+01 2.541E+01 1.424E+02 2.104E-02 3.813E+01 4.253E-02 6.237E-01 6.026E-03 2.850E-02 2.375E-03 
3.750E-02 1.418E+03 1.628E+03 1.551E+03 8.075E+01 2.332E+01 1.382E+02 4.100E-02 3.639E+01 2.774E-02 5.412E-01 0.000E+00 3.276E-02 0.000E+00 
6.250E-02 1.206E+03 1.388E+03 1.308E+03 7.250E+01 2.214E+01 1.301E+02 6.797E-02 3.375E+01 3.055E-02 4.762E-01 2.904E-03 2.767E-02 0.000E+00 
8.750E-02 1.017E+03 1.160E+03 1.121E+03 6.515E+01 1.959E+01 1.238E+02 6.241E-02 3.000E+01 4.321E-02 5.219E-01 1.141E-02 4.092E-02 0.000E+00 
1.125E-01 8.476E+02 9.684E+02 9.264E+02 5.802E+01 1.813E+01 1.191E+02 7.615E-02 2.806E+01 3.907E-02 4.567E-01 2.802E-03 2.017E-02 4.483E-03 
1.375E-01 7.195E+02 8.117E+02 7.841E+02 4.998E+01 1.669E+01 1.106E+02 9.670E-02 2.575E+01 7.020E-02 3.828E-01 5.508E-03 2.211E-02 4.421E-03 
1.625E-01 6.000E+02 6.805E+02 6.544E+02 4.545E+01 1.478E+01 1.022E+02 8.723E-02 2.367E+01 2.816E-02 3.113E-01 8.122E-03 2.617E-02 0.000E+00 
1.875E-01 5.089E+02 5.715E+02 5.532E+02 3.980E+01 1.368E+01 9.664E+01 8.174E-02 2.137E+01 4.603E-02 3.062E-01 2.662E-03 2.366E-02 0.000E+00 
2.125E-01 4.221E+02 4.783E+02 4.694E+02 3.514E+01 1.153E+01 8.859E+01 1.008E-01 1.889E+01 2.108E-02 2.566E-01 0.000E+00 1.486E-02 2.123E-03 
2.375E-01 3.564E+02 4.027E+02 3.904E+02 3.056E+01 1.044E+01 8.212E+01 9.528E-02 1.718E+01 1.773E-02 2.499E-01 2.576E-03 2.514E-02 0.000E+00 
2.625E-01 3.074E+02 3.429E+02 3.303E+02 2.683E+01 9.375E+00 7.561E+01 5.331E-02 1.549E+01 2.321E-02 2.054E-01 2.536E-03 1.241E-02 2.068E-03 
2.875E-01 2.575E+02 2.890E+02 2.822E+02 2.396E+01 8.374E+00 6.923E+01 6.854E-02 1.438E+01 2.280E-02 1.872E-01 0.000E+00 1.021E-02 0.000E+00 
3.125E-01 2.192E+02 2.445E+02 2.371E+02 2.105E+01 7.092E+00 6.269E+01 3.199E-02 1.250E+01 4.760E-02 1.843E-01 0.000E+00 6.047E-03 0.000E+00 
3.375E-01 1.886E+02 2.099E+02 2.044E+02 1.878E+01 6.749E+00 5.799E+01 7.526E-02 1.115E+01 3.302E-02 1.307E-01 2.421E-03 5.972E-03 0.000E+00 
3.625E-01 1.620E+02 1.767E+02 1.757E+02 1.634E+01 6.118E+00 5.253E+01 3.691E-02 1.003E+01 4.329E-02 1.431E-01 0.000E+00 1.376E-02 0.000E+00 
3.875E-01 1.397E+02 1.533E+02 1.495E+02 1.418E+01 5.296E+00 4.757E+01 2.414E-02 8.918E+00 2.927E-02 1.081E-01 0.000E+00 1.360E-02 0.000E+00 
4.125E-01 1.197E+02 1.323E+02 1.312E+02 1.304E+01 4.713E+00 4.366E+01 4.739E-02 8.139E+00 2.355E-02 1.019E-01 0.000E+00 7.676E-03 0.000E+00 
4.375E-01 1.039E+02 1.150E+02 1.122E+02 1.177E+01 4.221E+00 3.959E+01 5.816E-02 7.163E+00 1.802E-02 7.760E-02 2.282E-03 1.138E-02 0.000E+00 
4.625E-01 9.000E+01 9.879E+01 9.675E+01 1.002E+01 3.350E+00 3.570E+01 4.284E-02 6.378E+00 2.281E-02 7.876E-02 0.000E+00 9.370E-03 0.000E+00 
4.875E-01 7.690E+01 8.478E+01 8.324E+01 9.353E+00 3.179E+00 3.211E+01 3.087E-02 5.777E+00 1.497E-02 5.992E-02 0.000E+00 9.261E-03 0.000E+00 
5.125E-01 6.650E+01 7.370E+01 7.338E+01 7.642E+00 2.770E+00 2.867E+01 2.758E-02 5.077E+00 2.211E-02 4.596E-02 0.000E+00 5.493E-03 0.000E+00 
5.375E-01 5.726E+01 6.406E+01 6.321E+01 7.042E+00 2.734E+00 2.574E+01 3.795E-02 4.560E+00 4.838E-03 6.046E-02 0.000E+00 3.621E-03 0.000E+00 
5.625E-01 5.166E+01 5.682E+01 5.549E+01 6.406E+00 2.283E+00 2.295E+01 2.399E-02 4.037E+00 7.149E-03 5.752E-02 0.000E+00 1.790E-03 0.000E+00 
5.875E-01 4.500E+01 4.848E+01 4.840E+01 5.407E+00 2.049E+00 2.082E+01 1.573E-02 3.579E+00 9.392E-03 4.415E-02 2.102E-03 7.081E-03 0.000E+00 
6.125E-01 3.892E+01 4.239E+01 4.255E+01 5.068E+00 1.959E+00 1.929E+01 1.290E-02 2.955E+00 4.628E-03 2.698E-02 0.000E+00 0.000E+00 0.000E+00 
6.375E-01 3.366E+01 3.811E+01 3.699E+01 4.366E+00 1.580E+00 1.687E+01 2.285E-02 2.867E+00 4.562E-03 3.688E-02 0.000E+00 0.000E+00 0.000E+00 
6.625E-01 2.983E+01 3.267E+01 3.263E+01 3.808E+00 1.377E+00 1.528E+01 1.500E-02 2.649E+00 4.498E-03 1.618E-02 0.000E+00 5.140E-03 0.000E+00 
6.875E-01 2.574E+01 2.816E+01 2.774E+01 3.345E+00 1.249E+00 1.360E+01 1.969E-02 2.335E+00 2.218E-03 1.598E-02 1.998E-03 1.695E-03 0.000E+00 
7.125E-01 2.292E+01 2.568E+01 2.535E+01 3.047E+00 1.058E+00 1.251E+01 9.694E-03 1.925E+00 1.313E-02 1.381E-02 0.000E+00 0.000E+00 0.000E+00 
7.375E-01 2.041E+01 2.227E+01 2.203E+01 2.758E+00 8.380E-01 1.126E+01 1.194E-02 1.692E+00 4.316E-03 2.533E-02 1.949E-03 1.660E-03 0.000E+00 
7.625E-01 1.775E+01 1.926E+01 1.941E+01 2.321E+00 9.423E-01 9.830E+00 9.409E-03 1.525E+00 4.259E-03 2.118E-02 1.925E-03 1.643E-03 0.000E+00 
7.875E-01 1.558E+01 1.781E+01 1.734E+01 2.119E+00 7.897E-01 8.832E+00 6.955E-03 1.395E+00 2.101E-03 1.712E-02 0.000E+00 0.000E+00 0.000E+00 
8.125E-01 1.357E+01 1.528E+01 1.505E+01 1.855E+00 6.858E-01 7.975E+00 9.140E-03 1.178E+00 0.000E+00 1.316E-02 0.000E+00 0.000E+00 0.000E+00 
8.375E-01 1.225E+01 1.358E+01 1.306E+01 1.920E+00 6.459E-01 7.103E+00 2.253E-03 1.134E+00 0.000E+00 1.487E-02 0.000E+00 0.000E+00 0.000E+00 
8.625E-01 1.088E+01 1.164E+01 1.184E+01 1.389E+00 5.485E-01 6.274E+00 1.555E-02 9.725E-01 2.022E-03 1.286E-02 0.000E+00 0.000E+00 0.000E+00 
8.875E-01 9.981E+00 1.005E+01 1.041E+01 1.315E+00 5.299E-01 5.857E+00 4.382E-03 8.685E-01 1.997E-03 1.635E-02 0.000E+00 0.000E+00 0.000E+00 
9.125E-01 8.537E+00 9.462E+00 8.933E+00 1.240E+00 3.925E-01 5.266E+00 2.162E-03 7.927E-01 1.972E-03 8.979E-03 0.000E+00 1.548E-03 0.000E+00 
9.375E-01 7.241E+00 8.145E+00 8.264E+00 1.053E+00 4.108E-01 4.366E+00 4.266E-03 7.246E-01 5.844E-03 1.776E-03 0.000E+00 0.000E+00 0.000E+00 
9.625E-01 6.990E+00 7.397E+00 7.506E+00 9.228E-01 3.405E-01 4.003E+00 0.000E+00 6.505E-01 7.698E-03 3.513E-03 0.000E+00 0.000E+00 0.000E+00 
9.875E-01 5.900E+00 6.532E+00 6.546E+00 8.532E-01 3.078E-01 3.731E+00 0.000E+00 5.534E-01 0.000E+00 5.212E-03 0.000E+00 3.008E-03 0.000E+00 
1.012E+00 5.226E+00 5.552E+00 6.032E+00 6.930E-01 2.602E-01 3.173E+00 8.203E-03 4.698E-01 0.000E+00 5.156E-03 0.000E+00 0.000E+00 0.000E+00 
1.038E+00 4.505E+00 5.288E+00 5.033E+00 6.347E-01 2.063E-01 2.780E+00 4.050E-03 4.291E-01 0.000E+00 5.101E-03 0.000E+00 0.000E+00 0.000E+00 
1.062E+00 4.385E+00 4.775E+00 4.625E+00 5.834E-01 2.082E-01 2.551E+00 3.999E-03 4.003E-01 0.000E+00 5.047E-03 0.000E+00 0.000E+00 0.000E+00 
1.087E+00 3.698E+00 4.047E+00 3.976E+00 5.413E-01 2.276E-01 2.232E+00 1.975E-03 3.340E-01 0.000E+00 4.995E-03 0.000E+00 0.000E+00 0.000E+00 
1.113E+00 3.141E+00 3.483E+00 3.554E+00 4.407E-01 1.867E-01 2.066E+00 1.951E-03 3.267E-01 0.000E+00 1.648E-03 0.000E+00 0.000E+00 0.000E+00 
1.137E+00 2.898E+00 3.127E+00 3.227E+00 3.996E-01 1.373E-01 1.866E+00 0.000E+00 2.858E-01 0.000E+00 3.262E-03 0.000E+00 0.000E+00 0.000E+00 
1.163E+00 2.553E+00 2.811E+00 2.922E+00 3.550E-01 1.207E-01 1.615E+00 1.904E-03 2.598E-01 1.756E-03 0.000E+00 0.000E+00 0.000E+00 0.000E+00 
1.188E+00 2.287E+00 2.601E+00 2.659E+00 3.092E-01 1.189E-01 1.402E+00 1.882E-03 1.980E-01 0.000E+00 1.598E-03 0.000E+00 0.000E+00 0.000E+00 
1.213E+00 2.106E+00 2.372E+00 2.248E+00 3.328E-01 1.149E-01 1.401E+00 1.860E-03 1.993E-01 0.000E+00 0.000E+00 0.000E+00 0.000E+00 0.000E+00 
1.238E+00 1.881E+00 1.925E+00 1.951E+00 2.957E-01 1.016E-01 1.210E+00 0.000E+00 1.836E-01 1.700E-03 0.000E+00 0.000E+00 0.000E+00 0.000E+00 
1.262E+00 1.805E+00 1.951E+00 1.862E+00 2.414E-01 7.743E-02 1.056E+00 0.000E+00 1.564E-01 0.000E+00 0.000E+00 0.000E+00 1.363E-03 0.000E+00 
1.288E+00 1.510E+00 1.647E+00 1.653E+00 1.931E-01 8.083E-02 9.562E-01 0.000E+00 1.331E-01 1.664E-03 3.074E-03 0.000E+00 0.000E+00 0.000E+00 
1.312E+00 1.349E+00 1.365E+00 1.464E+00 1.860E-01 5.978E-02 7.963E-01 0.000E+00 1.235E-01 0.000E+00 1.522E-03 0.000E+00 0.000E+00 0.000E+00 
\end{verbatim}
}
\newpage
Pb+Pb \@ $E_{\rm lab}=30A~$GeV:
{\tiny
\begin{verbatim}
! m_t-m0 , 1/mt dN/dmt(pi+ pi- pi0 K+ K- P aP L+S0 a(L+S0) Xi- aXi- Om aOm) 
1.250E-02 1.913E+03 2.172E+03 2.062E+03 9.946E+01 3.492E+01 1.260E+02 4.629E-02 3.844E+01 1.063E-01 7.473E-01 2.109E-02 5.165E-02 4.749E-03 
3.750E-02 1.675E+03 1.874E+03 1.794E+03 8.972E+01 3.269E+01 1.185E+02 1.640E-01 3.640E+01 1.248E-01 7.009E-01 2.070E-02 6.428E-02 4.680E-03 
6.250E-02 1.407E+03 1.571E+03 1.538E+03 8.150E+01 3.028E+01 1.139E+02 1.519E-01 3.383E+01 1.561E-01 6.853E-01 1.162E-02 5.045E-02 2.306E-03 
8.750E-02 1.186E+03 1.334E+03 1.293E+03 7.313E+01 2.808E+01 1.076E+02 2.184E-01 3.091E+01 1.329E-01 6.103E-01 1.426E-02 3.694E-02 2.273E-03 
1.125E-01 9.980E+02 1.114E+03 1.086E+03 6.393E+01 2.550E+01 1.023E+02 2.285E-01 2.841E+01 1.237E-01 5.632E-01 2.802E-03 3.652E-02 0.000E+00 
1.375E-01 8.514E+02 9.369E+02 9.144E+02 5.686E+01 2.279E+01 9.711E+01 2.752E-01 2.573E+01 1.085E-01 5.067E-01 2.478E-02 2.347E-02 8.842E-03 
1.625E-01 7.109E+02 7.931E+02 7.669E+02 5.170E+01 2.069E+01 8.961E+01 1.926E-01 2.337E+01 1.126E-01 4.115E-01 1.624E-02 5.177E-02 0.000E+00 
1.875E-01 6.020E+02 6.657E+02 6.491E+02 4.570E+01 1.814E+01 8.395E+01 2.168E-01 2.170E+01 1.197E-01 3.275E-01 1.331E-02 3.178E-02 0.000E+00 
2.125E-01 5.113E+02 5.613E+02 5.454E+02 3.935E+01 1.697E+01 7.813E+01 2.538E-01 1.968E+01 1.204E-01 2.959E-01 1.833E-02 2.095E-02 2.123E-03 
2.375E-01 4.294E+02 4.705E+02 4.651E+02 3.535E+01 1.463E+01 7.330E+01 2.178E-01 1.760E+01 8.570E-02 2.989E-01 1.288E-02 1.382E-02 4.190E-03 
2.625E-01 3.656E+02 4.006E+02 3.919E+02 3.098E+01 1.298E+01 6.620E+01 1.733E-01 1.600E+01 1.074E-01 2.307E-01 1.014E-02 1.538E-02 6.203E-03 
2.875E-01 3.102E+02 3.403E+02 3.356E+02 2.713E+01 1.162E+01 6.181E+01 1.795E-01 1.444E+01 9.975E-02 2.197E-01 1.248E-02 2.536E-02 4.083E-03 
3.125E-01 2.653E+02 2.904E+02 2.853E+02 2.414E+01 1.019E+01 5.649E+01 1.311E-01 1.287E+01 9.240E-02 2.114E-01 0.000E+00 1.338E-02 2.016E-03 
3.375E-01 2.254E+02 2.493E+02 2.464E+02 2.157E+01 9.087E+00 5.103E+01 1.505E-01 1.174E+01 3.302E-02 1.815E-01 2.421E-03 1.159E-02 0.000E+00 
3.625E-01 1.950E+02 2.134E+02 2.139E+02 1.919E+01 8.061E+00 4.682E+01 1.230E-01 1.049E+01 5.681E-02 1.621E-01 0.000E+00 9.833E-03 1.966E-03 
3.875E-01 1.659E+02 1.829E+02 1.798E+02 1.734E+01 6.834E+00 4.284E+01 1.237E-01 9.493E+00 5.055E-02 1.551E-01 2.349E-03 8.111E-03 1.942E-03 
4.125E-01 1.462E+02 1.555E+02 1.589E+02 1.443E+01 6.482E+00 3.931E+01 1.451E-01 8.521E+00 4.972E-02 1.227E-01 2.315E-03 1.285E-02 1.919E-03 
4.375E-01 1.244E+02 1.370E+02 1.367E+02 1.299E+01 5.780E+00 3.560E+01 9.015E-02 7.547E+00 4.635E-02 1.369E-01 4.565E-03 1.272E-02 1.896E-03 
4.625E-01 1.082E+02 1.181E+02 1.185E+02 1.182E+01 5.470E+00 3.237E+01 1.257E-01 6.781E+00 6.082E-02 1.170E-01 9.001E-03 9.446E-03 0.000E+00 
4.875E-01 9.406E+01 1.033E+02 1.024E+02 1.023E+01 4.495E+00 2.908E+01 7.857E-02 6.084E+00 3.492E-02 6.657E-02 4.438E-03 3.118E-03 0.000E+00 
5.125E-01 8.309E+01 8.932E+01 8.791E+01 8.974E+00 3.923E+00 2.646E+01 9.652E-02 5.514E+00 4.176E-02 5.910E-02 0.000E+00 3.088E-03 0.000E+00 
5.375E-01 7.124E+01 7.725E+01 7.811E+01 8.027E+00 3.575E+00 2.374E+01 1.003E-01 4.657E+00 4.838E-02 6.910E-02 0.000E+00 1.223E-02 0.000E+00 
5.625E-01 6.192E+01 6.752E+01 6.831E+01 6.985E+00 3.313E+00 2.166E+01 6.931E-02 4.416E+00 3.813E-02 3.835E-02 4.261E-03 7.573E-03 0.000E+00 
5.875E-01 5.538E+01 6.016E+01 5.882E+01 6.325E+00 2.811E+00 1.938E+01 6.555E-02 3.776E+00 3.053E-02 3.995E-02 2.102E-03 3.001E-03 0.000E+00 
6.125E-01 4.786E+01 5.134E+01 5.127E+01 5.686E+00 2.437E+00 1.744E+01 6.192E-02 3.497E+00 1.851E-02 4.773E-02 2.075E-03 1.487E-03 0.000E+00 
6.375E-01 4.212E+01 4.535E+01 4.520E+01 5.038E+00 2.277E+00 1.569E+01 5.078E-02 3.139E+00 2.281E-02 3.892E-02 4.097E-03 2.946E-03 1.732E-03 
6.625E-01 3.667E+01 3.917E+01 4.022E+01 4.164E+00 1.965E+00 1.423E+01 3.999E-02 2.748E+00 2.249E-02 3.641E-02 0.000E+00 1.459E-03 1.713E-03 
6.875E-01 3.193E+01 3.471E+01 3.521E+01 4.140E+00 1.784E+00 1.250E+01 3.445E-02 2.449E+00 2.218E-02 3.196E-02 0.000E+00 2.892E-03 0.000E+00 
7.125E-01 2.905E+01 2.982E+01 3.075E+01 3.441E+00 1.634E+00 1.147E+01 3.635E-02 2.181E+00 1.313E-02 3.157E-02 3.946E-03 0.000E+00 0.000E+00 
7.375E-01 2.506E+01 2.681E+01 2.728E+01 2.952E+00 1.393E+00 1.006E+01 2.865E-02 1.957E+00 1.079E-02 2.339E-02 1.949E-03 2.841E-03 0.000E+00 
7.625E-01 2.209E+01 2.388E+01 2.389E+01 2.900E+00 1.210E+00 9.315E+00 2.117E-02 1.721E+00 2.129E-03 3.081E-02 0.000E+00 1.408E-03 0.000E+00 
7.875E-01 2.005E+01 2.091E+01 2.148E+01 2.438E+00 1.067E+00 8.355E+00 2.782E-02 1.507E+00 1.051E-02 1.902E-02 1.902E-03 0.000E+00 0.000E+00 
8.125E-01 1.718E+01 1.908E+01 1.894E+01 2.235E+00 9.981E-01 7.161E+00 2.514E-02 1.435E+00 1.659E-02 2.068E-02 0.000E+00 0.000E+00 0.000E+00 
8.375E-01 1.529E+01 1.650E+01 1.688E+01 2.043E+00 8.922E-01 6.502E+00 2.929E-02 1.227E+00 8.190E-03 1.858E-02 0.000E+00 1.372E-03 0.000E+00 
8.625E-01 1.348E+01 1.442E+01 1.464E+01 1.737E+00 7.696E-01 5.734E+00 2.444E-02 1.118E+00 8.087E-03 1.653E-02 0.000E+00 5.441E-03 0.000E+00 
8.875E-01 1.205E+01 1.270E+01 1.267E+01 1.598E+00 7.586E-01 5.568E+00 1.753E-02 1.006E+00 5.990E-03 7.264E-03 0.000E+00 1.349E-03 0.000E+00 
9.125E-01 1.032E+01 1.185E+01 1.135E+01 1.436E+00 6.683E-01 4.950E+00 1.081E-02 9.051E-01 1.775E-02 1.796E-02 0.000E+00 0.000E+00 0.000E+00 
9.375E-01 9.755E+00 9.915E+00 1.013E+01 1.271E+00 5.812E-01 4.121E+00 6.398E-03 7.869E-01 3.896E-03 8.879E-03 0.000E+00 0.000E+00 0.000E+00 
9.625E-01 8.342E+00 8.585E+00 8.803E+00 1.096E+00 5.218E-01 3.927E+00 4.209E-03 6.659E-01 7.698E-03 8.782E-03 1.756E-03 1.315E-03 0.000E+00 
9.875E-01 7.428E+00 7.684E+00 7.908E+00 8.667E-01 4.536E-01 3.467E+00 1.246E-02 6.865E-01 0.000E+00 5.212E-03 0.000E+00 0.000E+00 0.000E+00 
1.012E+00 6.634E+00 7.054E+00 7.367E+00 8.019E-01 3.160E-01 2.910E+00 2.051E-03 5.300E-01 1.879E-03 5.156E-03 0.000E+00 1.294E-03 0.000E+00 
1.038E+00 5.928E+00 6.472E+00 6.428E+00 7.600E-01 3.004E-01 2.825E+00 4.050E-03 4.532E-01 3.715E-03 5.101E-03 0.000E+00 1.284E-03 0.000E+00 
1.062E+00 5.155E+00 5.664E+00 5.644E+00 6.502E-01 2.955E-01 2.449E+00 1.400E-02 4.811E-01 1.836E-03 1.514E-02 0.000E+00 0.000E+00 0.000E+00 
1.087E+00 4.586E+00 4.876E+00 5.219E+00 6.525E-01 3.060E-01 2.247E+00 3.950E-03 4.629E-01 3.631E-03 3.330E-03 0.000E+00 0.000E+00 0.000E+00 
1.113E+00 4.098E+00 4.488E+00 4.501E+00 5.826E-01 2.614E-01 1.923E+00 5.852E-03 3.303E-01 0.000E+00 1.648E-03 1.648E-03 0.000E+00 0.000E+00 
1.137E+00 3.666E+00 3.813E+00 4.045E+00 5.345E-01 2.648E-01 1.754E+00 1.927E-03 3.302E-01 0.000E+00 6.524E-03 0.000E+00 0.000E+00 0.000E+00 
1.163E+00 3.337E+00 3.411E+00 3.426E+00 4.854E-01 1.811E-01 1.632E+00 1.904E-03 2.563E-01 3.511E-03 3.229E-03 0.000E+00 0.000E+00 0.000E+00 
1.188E+00 2.972E+00 3.241E+00 3.259E+00 3.711E-01 1.832E-01 1.451E+00 1.882E-03 2.448E-01 3.473E-03 3.197E-03 0.000E+00 0.000E+00 0.000E+00 
1.213E+00 2.760E+00 2.757E+00 2.737E+00 3.703E-01 1.805E-01 1.259E+00 3.720E-03 2.336E-01 1.718E-03 1.583E-03 0.000E+00 1.215E-03 0.000E+00 
1.238E+00 2.428E+00 2.469E+00 2.475E+00 3.373E-01 1.571E-01 1.160E+00 1.839E-03 1.955E-01 3.399E-03 1.567E-03 0.000E+00 0.000E+00 0.000E+00 
1.262E+00 2.171E+00 2.248E+00 2.322E+00 3.052E-01 1.457E-01 1.011E+00 0.000E+00 1.766E-01 1.682E-03 1.552E-03 1.552E-03 1.197E-03 0.000E+00 
1.288E+00 1.936E+00 2.015E+00 2.144E+00 2.694E-01 1.145E-01 9.472E-01 1.797E-03 1.564E-01 0.000E+00 1.537E-03 0.000E+00 0.000E+00 0.000E+00 
1.312E+00 1.644E+00 1.826E+00 1.848E+00 2.170E-01 1.063E-01 7.732E-01 1.777E-03 1.697E-01 0.000E+00 4.567E-03 0.000E+00 0.000E+00 0.000E+00 
\end{verbatim}
}
\newpage
Pb+Pb \@ $E_{\rm lab}=40A~$GeV:
{\tiny
\begin{verbatim}
! m_t-m0 , 1/mt dN/dmt(pi+ pi- pi0 K+ K- P aP L+S0 a(L+S0) Xi- aXi- Om aOm) 
1.250E-02 2.156E+03 2.380E+03 2.278E+03 1.058E+02 4.357E+01 1.106E+02 2.609E-01 3.793E+01 2.056E-01 8.618E-01 2.712E-02 1.116E-01 1.425E-02 
3.750E-02 1.864E+03 2.059E+03 2.026E+03 9.592E+01 4.078E+01 1.073E+02 2.624E-01 3.578E+01 2.046E-01 7.956E-01 3.845E-02 7.020E-02 0.000E+00 
6.250E-02 1.577E+03 1.740E+03 1.713E+03 8.712E+01 3.738E+01 1.027E+02 3.238E-01 3.326E+01 2.274E-01 7.521E-01 2.613E-02 9.455E-02 9.225E-03 
8.750E-02 1.331E+03 1.469E+03 1.434E+03 8.004E+01 3.354E+01 9.829E+01 3.745E-01 3.032E+01 2.592E-01 5.733E-01 3.422E-02 6.365E-02 9.093E-03 
1.125E-01 1.121E+03 1.233E+03 1.210E+03 7.087E+01 3.164E+01 9.315E+01 3.655E-01 2.798E+01 2.149E-01 5.660E-01 2.802E-02 6.949E-02 8.966E-03 
1.375E-01 9.445E+02 1.032E+03 1.018E+03 6.235E+01 2.855E+01 8.763E+01 3.794E-01 2.570E+01 2.010E-01 5.205E-01 1.652E-02 4.421E-02 8.842E-03 
1.625E-01 8.000E+02 8.687E+02 8.573E+02 5.512E+01 2.458E+01 8.169E+01 3.925E-01 2.372E+01 1.596E-01 5.008E-01 1.624E-02 6.323E-02 4.361E-03 
1.875E-01 6.753E+02 7.340E+02 7.257E+02 4.948E+01 2.267E+01 7.609E+01 3.696E-01 2.148E+01 2.332E-01 4.978E-01 3.727E-02 4.302E-02 2.151E-03 
2.125E-01 5.657E+02 6.189E+02 6.154E+02 4.419E+01 2.044E+01 7.197E+01 3.755E-01 1.965E+01 1.897E-01 3.195E-01 2.357E-02 4.033E-02 0.000E+00 
2.375E-01 4.821E+02 5.230E+02 5.190E+02 3.928E+01 1.819E+01 6.602E+01 3.777E-01 1.783E+01 1.950E-01 2.963E-01 2.061E-02 5.237E-02 2.095E-03 
2.625E-01 4.119E+02 4.427E+02 4.443E+02 3.471E+01 1.607E+01 6.080E+01 3.365E-01 1.632E+01 1.828E-01 3.043E-01 2.789E-02 2.068E-02 4.135E-03 
2.875E-01 3.533E+02 3.825E+02 3.819E+02 2.960E+01 1.426E+01 5.670E+01 3.884E-01 1.475E+01 1.425E-01 2.147E-01 2.995E-02 2.041E-02 2.041E-03 
3.125E-01 2.991E+02 3.229E+02 3.197E+02 2.660E+01 1.248E+01 5.243E+01 3.455E-01 1.319E+01 1.428E-01 2.482E-01 2.458E-03 4.837E-02 2.016E-03 
3.375E-01 2.564E+02 2.762E+02 2.767E+02 2.380E+01 1.155E+01 4.796E+01 3.575E-01 1.152E+01 1.459E-01 1.936E-01 1.694E-02 1.991E-02 3.981E-03 
3.625E-01 2.206E+02 2.364E+02 2.397E+02 2.091E+01 1.008E+01 4.352E+01 3.260E-01 1.067E+01 1.028E-01 1.979E-01 2.385E-03 1.180E-02 3.932E-03 
3.875E-01 1.909E+02 2.062E+02 2.061E+02 1.801E+01 8.935E+00 3.957E+01 2.444E-01 9.314E+00 1.038E-01 1.316E-01 2.349E-03 7.769E-03 0.000E+00 
4.125E-01 1.667E+02 1.782E+02 1.765E+02 1.576E+01 8.115E+00 3.602E+01 2.606E-01 8.670E+00 9.421E-02 9.957E-02 6.946E-03 1.151E-02 1.919E-03 
4.375E-01 1.416E+02 1.525E+02 1.521E+02 1.404E+01 7.115E+00 3.313E+01 1.948E-01 7.629E+00 7.982E-02 1.255E-01 6.847E-03 1.517E-02 0.000E+00 
4.625E-01 1.219E+02 1.327E+02 1.322E+02 1.275E+01 6.411E+00 2.993E+01 1.685E-01 6.804E+00 8.869E-02 1.193E-01 0.000E+00 1.687E-02 1.874E-03 
4.875E-01 1.076E+02 1.139E+02 1.158E+02 1.121E+01 5.188E+00 2.711E+01 1.964E-01 6.089E+00 7.733E-02 1.021E-01 1.110E-02 9.261E-03 0.000E+00 
5.125E-01 9.361E+01 9.970E+01 1.002E+02 9.784E+00 5.071E+00 2.434E+01 1.020E-01 5.703E+00 5.895E-02 1.029E-01 8.755E-03 7.324E-03 1.831E-03 
5.375E-01 8.354E+01 8.725E+01 8.916E+01 9.047E+00 4.553E+00 2.200E+01 1.355E-01 4.916E+00 6.774E-02 8.205E-02 0.000E+00 1.448E-02 1.810E-03 
5.625E-01 7.141E+01 7.619E+01 7.748E+01 8.102E+00 3.983E+00 1.987E+01 1.440E-01 4.592E+00 6.911E-02 6.178E-02 4.261E-03 3.580E-03 0.000E+00 
5.875E-01 6.156E+01 6.662E+01 6.771E+01 6.831E+00 3.547E+00 1.824E+01 1.023E-01 4.015E+00 3.992E-02 5.256E-02 2.102E-03 7.081E-03 0.000E+00 
6.125E-01 5.494E+01 5.768E+01 5.922E+01 5.997E+00 3.109E+00 1.607E+01 1.032E-01 3.598E+00 4.165E-02 6.226E-02 2.075E-03 1.051E-02 1.751E-03 
6.375E-01 4.773E+01 5.047E+01 5.105E+01 5.289E+00 2.849E+00 1.446E+01 8.378E-02 3.159E+00 4.334E-02 5.531E-02 0.000E+00 8.660E-03 0.000E+00 
6.625E-01 4.163E+01 4.482E+01 4.510E+01 4.825E+00 2.504E+00 1.303E+01 8.997E-02 2.820E+00 4.048E-02 2.630E-02 0.000E+00 1.713E-03 1.713E-03 
6.875E-01 3.640E+01 3.989E+01 4.021E+01 4.185E+00 1.960E+00 1.196E+01 5.414E-02 2.464E+00 2.883E-02 4.594E-02 0.000E+00 6.781E-03 0.000E+00 
7.125E-01 3.203E+01 3.481E+01 3.525E+01 3.773E+00 1.959E+00 1.080E+01 5.332E-02 2.166E+00 1.969E-02 3.748E-02 1.973E-03 1.678E-03 0.000E+00 
7.375E-01 2.884E+01 3.019E+01 3.131E+01 3.391E+00 1.721E+00 9.841E+00 6.685E-02 1.858E+00 1.942E-02 1.949E-02 1.949E-03 0.000E+00 0.000E+00 
7.625E-01 2.564E+01 2.732E+01 2.742E+01 3.142E+00 1.423E+00 8.539E+00 5.881E-02 1.744E+00 1.491E-02 2.696E-02 1.925E-03 4.929E-03 0.000E+00 
7.875E-01 2.174E+01 2.361E+01 2.444E+01 2.716E+00 1.323E+00 7.495E+00 4.173E-02 1.662E+00 1.681E-02 2.663E-02 0.000E+00 3.253E-03 0.000E+00 
8.125E-01 1.923E+01 2.047E+01 2.101E+01 2.400E+00 1.228E+00 6.887E+00 4.342E-02 1.342E+00 1.659E-02 1.692E-02 1.880E-03 3.220E-03 0.000E+00 
8.375E-01 1.734E+01 1.854E+01 1.844E+01 2.106E+00 1.121E+00 6.171E+00 3.830E-02 1.274E+00 8.190E-03 2.416E-02 0.000E+00 1.594E-03 0.000E+00 
8.625E-01 1.563E+01 1.625E+01 1.648E+01 1.937E+00 9.259E-01 5.601E+00 4.221E-02 1.262E+00 1.820E-02 2.388E-02 0.000E+00 3.156E-03 0.000E+00 
8.875E-01 1.391E+01 1.428E+01 1.511E+01 1.749E+00 8.426E-01 5.048E+00 4.163E-02 1.066E+00 5.990E-03 1.998E-02 0.000E+00 0.000E+00 0.000E+00 
9.125E-01 1.220E+01 1.326E+01 1.287E+01 1.504E+00 7.565E-01 4.459E+00 2.810E-02 8.952E-01 1.183E-02 8.979E-03 0.000E+00 3.095E-03 0.000E+00 
9.375E-01 1.061E+01 1.140E+01 1.202E+01 1.436E+00 6.119E-01 3.997E+00 1.706E-02 8.279E-01 5.844E-03 5.327E-03 0.000E+00 1.533E-03 0.000E+00 
9.625E-01 9.705E+00 9.876E+00 1.025E+01 1.211E+00 6.591E-01 3.681E+00 1.473E-02 7.698E-01 1.924E-02 1.405E-02 0.000E+00 0.000E+00 0.000E+00 
9.875E-01 8.711E+00 9.038E+00 9.272E+00 1.112E+00 5.535E-01 3.413E+00 1.662E-02 7.074E-01 5.705E-03 1.042E-02 0.000E+00 0.000E+00 0.000E+00 
1.012E+00 7.934E+00 8.087E+00 8.268E+00 1.065E+00 5.124E-01 2.820E+00 1.846E-02 6.089E-01 5.638E-03 2.062E-02 5.156E-03 1.490E-03 0.000E+00 
1.038E+00 7.030E+00 7.095E+00 7.132E+00 9.220E-01 4.727E-01 2.669E+00 1.417E-02 4.811E-01 7.430E-03 6.801E-03 0.000E+00 0.000E+00 0.000E+00 
1.062E+00 5.901E+00 6.534E+00 6.567E+00 7.992E-01 3.598E-01 2.465E+00 5.999E-03 5.013E-01 9.181E-03 5.047E-03 0.000E+00 1.463E-03 0.000E+00 
1.087E+00 5.431E+00 5.617E+00 5.852E+00 7.259E-01 3.718E-01 2.133E+00 5.924E-03 4.157E-01 3.631E-03 3.330E-03 0.000E+00 1.450E-03 0.000E+00 
1.113E+00 4.629E+00 5.128E+00 5.185E+00 6.573E-01 3.262E-01 1.923E+00 1.951E-03 3.590E-01 5.385E-03 1.648E-03 0.000E+00 0.000E+00 0.000E+00 
1.137E+00 4.171E+00 4.478E+00 4.588E+00 5.835E-01 2.819E-01 1.688E+00 1.156E-02 3.532E-01 1.775E-03 4.893E-03 0.000E+00 0.000E+00 0.000E+00 
1.163E+00 3.854E+00 4.060E+00 4.208E+00 4.998E-01 2.801E-01 1.594E+00 9.522E-03 3.020E-01 7.022E-03 3.229E-03 0.000E+00 1.411E-03 0.000E+00 
1.188E+00 3.353E+00 3.479E+00 3.769E+00 4.900E-01 2.189E-01 1.344E+00 9.410E-03 2.744E-01 1.736E-03 1.598E-03 0.000E+00 0.000E+00 0.000E+00 
1.213E+00 2.906E+00 3.178E+00 3.225E+00 4.899E-01 1.828E-01 1.215E+00 5.580E-03 2.216E-01 0.000E+00 6.330E-03 0.000E+00 0.000E+00 0.000E+00 
1.238E+00 2.611E+00 2.754E+00 3.115E+00 3.558E-01 1.964E-01 1.105E+00 5.516E-03 1.887E-01 1.700E-03 6.268E-03 0.000E+00 0.000E+00 0.000E+00 
1.262E+00 2.479E+00 2.568E+00 2.656E+00 3.234E-01 1.731E-01 1.036E+00 9.089E-03 1.850E-01 1.682E-03 1.552E-03 0.000E+00 0.000E+00 0.000E+00 
1.288E+00 2.088E+00 2.295E+00 2.441E+00 2.447E-01 1.437E-01 8.609E-01 3.595E-03 1.781E-01 1.664E-03 3.074E-03 0.000E+00 0.000E+00 0.000E+00 
1.312E+00 1.961E+00 2.093E+00 2.082E+00 2.502E-01 1.019E-01 7.643E-01 3.555E-03 1.334E-01 1.647E-03 0.000E+00 0.000E+00 0.000E+00 0.000E+00 
\end{verbatim}
}
\newpage
Pb+Pb \@ $E_{\rm lab}=80A~$GeV:
{\tiny
\begin{verbatim}
! m_t-m0 , 1/mt dN/dmt(pi+ pi- pi0 K+ K- P aP L+S0 a(L+S0) Xi- aXi- Om aOm)
1.250E-02 2.714E+03 2.907E+03 2.881E+03 1.224E+02 6.285E+01 8.670E+01 8.971E-01 3.496E+01 7.243E-01 1.125E+00 1.265E-01 1.364E-01 4.721E-02 
3.750E-02 2.380E+03 2.536E+03 2.561E+03 1.143E+02 5.800E+01 8.375E+01 1.055E+00 3.312E+01 7.354E-01 9.767E-01 9.799E-02 1.421E-01 4.652E-02 
6.250E-02 2.010E+03 2.149E+03 2.174E+03 1.027E+02 5.504E+01 8.200E+01 1.086E+00 3.080E+01 7.797E-01 9.269E-01 5.132E-02 1.146E-01 2.802E-02 
8.750E-02 1.687E+03 1.806E+03 1.834E+03 9.406E+01 4.951E+01 7.790E+01 1.142E+00 2.898E+01 8.260E-01 8.852E-01 1.008E-01 8.035E-02 3.013E-02 
1.125E-01 1.426E+03 1.522E+03 1.540E+03 8.282E+01 4.678E+01 7.471E+01 1.333E+00 2.657E+01 6.689E-01 6.654E-01 7.118E-02 8.170E-02 2.723E-02 
1.375E-01 1.200E+03 1.282E+03 1.305E+03 7.381E+01 4.050E+01 7.001E+01 1.105E+00 2.432E+01 7.225E-01 6.722E-01 5.779E-02 8.545E-02 1.465E-02 
1.625E-01 1.013E+03 1.079E+03 1.096E+03 6.576E+01 3.776E+01 6.660E+01 1.236E+00 2.205E+01 5.771E-01 6.309E-01 6.578E-02 8.429E-02 1.445E-02 
1.875E-01 8.586E+02 9.197E+02 9.235E+02 5.733E+01 3.351E+01 6.348E+01 1.150E+00 2.044E+01 6.236E-01 4.587E-01 4.705E-02 6.652E-02 1.426E-02 
2.125E-01 7.349E+02 7.765E+02 7.833E+02 5.190E+01 2.949E+01 5.862E+01 1.252E+00 1.801E+01 5.687E-01 4.107E-01 3.760E-02 7.033E-02 2.110E-02 
2.375E-01 6.236E+02 6.623E+02 6.715E+02 4.648E+01 2.633E+01 5.415E+01 1.154E+00 1.665E+01 5.255E-01 3.500E-01 3.984E-02 4.396E-02 1.620E-02 
2.625E-01 5.353E+02 5.605E+02 5.692E+02 3.926E+01 2.403E+01 5.101E+01 1.130E+00 1.489E+01 4.871E-01 3.641E-01 2.521E-02 6.623E-02 1.142E-02 
2.875E-01 4.588E+02 4.824E+02 4.860E+02 3.591E+01 2.100E+01 4.640E+01 9.770E-01 1.405E+01 4.407E-01 2.647E-01 3.033E-02 2.706E-02 1.353E-02 
3.125E-01 3.925E+02 4.158E+02 4.168E+02 3.139E+01 1.842E+01 4.347E+01 9.044E-01 1.270E+01 4.021E-01 2.226E-01 3.800E-02 3.117E-02 0.000E+00 
3.375E-01 3.379E+02 3.538E+02 3.634E+02 2.766E+01 1.622E+01 3.990E+01 8.140E-01 1.123E+01 3.982E-01 2.486E-01 2.941E-02 3.518E-02 2.199E-03 
3.625E-01 2.870E+02 3.037E+02 3.099E+02 2.465E+01 1.464E+01 3.685E+01 7.813E-01 1.012E+01 3.675E-01 2.054E-01 1.580E-02 1.737E-02 8.686E-03 
3.875E-01 2.502E+02 2.641E+02 2.687E+02 2.192E+01 1.301E+01 3.357E+01 8.333E-01 9.391E+00 3.467E-01 1.297E-01 2.854E-02 1.287E-02 2.145E-03 
4.125E-01 2.159E+02 2.283E+02 2.318E+02 1.912E+01 1.179E+01 3.052E+01 6.575E-01 8.278E+00 2.804E-01 1.586E-01 1.534E-02 1.484E-02 8.478E-03 
4.375E-01 1.868E+02 1.982E+02 1.996E+02 1.715E+01 9.856E+00 2.819E+01 6.616E-01 7.587E+00 2.730E-01 1.815E-01 1.765E-02 2.304E-02 4.189E-03 
4.625E-01 1.623E+02 1.709E+02 1.742E+02 1.495E+01 9.178E+00 2.574E+01 4.763E-01 6.860E+00 2.155E-01 1.392E-01 9.942E-03 1.656E-02 1.242E-02 
4.875E-01 1.408E+02 1.482E+02 1.512E+02 1.356E+01 8.165E+00 2.275E+01 5.176E-01 5.855E+00 1.901E-01 1.201E-01 2.206E-02 1.227E-02 2.046E-03 
5.125E-01 1.225E+02 1.280E+02 1.310E+02 1.200E+01 7.515E+00 2.104E+01 4.447E-01 5.217E+00 1.845E-01 1.209E-01 9.670E-03 1.213E-02 6.067E-03 
5.375E-01 1.061E+02 1.124E+02 1.149E+02 1.070E+01 6.442E+00 1.894E+01 3.952E-01 5.066E+00 1.897E-01 1.049E-01 4.770E-03 9.998E-03 2.000E-03 
5.625E-01 9.371E+01 9.895E+01 9.985E+01 9.488E+00 5.578E+00 1.746E+01 3.533E-01 4.382E+00 1.790E-01 8.471E-02 9.412E-03 1.186E-02 5.931E-03 
5.875E-01 8.183E+01 8.516E+01 8.649E+01 8.387E+00 5.147E+00 1.565E+01 3.215E-01 3.945E+00 1.400E-01 5.805E-02 4.644E-03 1.173E-02 0.000E+00 
6.125E-01 7.197E+01 7.447E+01 7.713E+01 7.203E+00 4.428E+00 1.373E+01 2.422E-01 3.553E+00 1.227E-01 6.647E-02 4.584E-03 1.160E-02 1.934E-03 
6.375E-01 6.281E+01 6.561E+01 6.625E+01 6.478E+00 4.111E+00 1.288E+01 2.944E-01 3.112E+00 9.574E-02 6.562E-02 4.525E-03 1.722E-02 1.913E-03 
6.625E-01 5.507E+01 5.813E+01 5.912E+01 5.795E+00 3.392E+00 1.131E+01 2.429E-01 2.700E+00 1.043E-01 5.362E-02 6.702E-03 3.785E-03 1.892E-03 
6.875E-01 4.823E+01 5.055E+01 5.143E+01 4.969E+00 3.279E+00 1.068E+01 2.473E-01 2.589E+00 7.594E-02 7.501E-02 4.412E-03 5.617E-03 5.617E-03 
7.125E-01 4.250E+01 4.431E+01 4.525E+01 4.526E+00 2.922E+00 9.430E+00 1.633E-01 2.356E+00 8.457E-02 4.358E-02 8.716E-03 9.264E-03 3.706E-03 
7.375E-01 3.742E+01 3.849E+01 3.984E+01 3.978E+00 2.583E+00 8.411E+00 1.292E-01 2.129E+00 6.912E-02 3.659E-02 2.152E-03 7.334E-03 1.834E-03 
7.625E-01 3.261E+01 3.427E+01 3.595E+01 3.484E+00 2.416E+00 7.488E+00 1.689E-01 1.696E+00 6.350E-02 3.190E-02 2.127E-03 1.815E-03 3.629E-03 
7.875E-01 2.895E+01 3.070E+01 3.123E+01 3.120E+00 2.165E+00 7.067E+00 8.961E-02 1.724E+00 6.267E-02 3.782E-02 4.203E-03 5.389E-03 0.000E+00 
8.125E-01 2.544E+01 2.651E+01 2.817E+01 2.851E+00 1.823E+00 6.297E+00 1.186E-01 1.400E+00 5.040E-02 2.907E-02 0.000E+00 1.067E-02 0.000E+00 
8.375E-01 2.281E+01 2.421E+01 2.441E+01 2.552E+00 1.699E+00 5.546E+00 8.460E-02 1.368E+00 7.689E-02 1.847E-02 0.000E+00 5.281E-03 3.521E-03 
8.625E-01 2.005E+01 2.113E+01 2.144E+01 2.404E+00 1.466E+00 4.949E+00 9.324E-02 1.130E+00 4.466E-02 1.826E-02 2.029E-03 1.743E-03 0.000E+00 
8.875E-01 1.790E+01 1.880E+01 1.920E+01 2.079E+00 1.388E+00 4.501E+00 7.986E-02 1.032E+00 5.292E-02 2.006E-02 0.000E+00 0.000E+00 0.000E+00 
9.125E-01 1.612E+01 1.730E+01 1.679E+01 1.778E+00 1.131E+00 4.114E+00 5.730E-02 1.035E+00 3.049E-02 5.950E-03 1.983E-03 0.000E+00 0.000E+00 
9.375E-01 1.392E+01 1.451E+01 1.524E+01 1.531E+00 1.065E+00 3.522E+00 5.182E-02 8.842E-01 2.367E-02 5.884E-03 5.884E-03 0.000E+00 0.000E+00 
9.625E-01 1.218E+01 1.327E+01 1.369E+01 1.356E+00 8.766E-01 3.313E+00 6.276E-02 7.567E-01 1.488E-02 5.819E-03 0.000E+00 0.000E+00 0.000E+00 
9.875E-01 1.129E+01 1.151E+01 1.189E+01 1.357E+00 8.290E-01 2.893E+00 5.966E-02 7.015E-01 2.940E-02 1.727E-02 0.000E+00 0.000E+00 0.000E+00 
1.012E+00 9.719E+00 1.067E+01 1.062E+01 1.167E+00 7.947E-01 2.646E+00 4.757E-02 5.915E-01 8.302E-03 9.491E-03 0.000E+00 0.000E+00 0.000E+00 
1.038E+00 8.625E+00 9.103E+00 9.670E+00 1.007E+00 6.721E-01 2.520E+00 4.025E-02 5.826E-01 6.155E-03 1.315E-02 0.000E+00 0.000E+00 0.000E+00 
1.062E+00 7.938E+00 8.240E+00 8.545E+00 8.941E-01 5.478E-01 2.105E+00 3.092E-02 4.928E-01 2.434E-02 7.433E-03 0.000E+00 1.616E-03 0.000E+00 
1.087E+00 7.051E+00 7.315E+00 7.437E+00 8.017E-01 5.587E-01 2.002E+00 2.617E-02 4.571E-01 8.020E-03 5.517E-03 1.839E-03 1.601E-03 1.601E-03 
1.113E+00 6.126E+00 6.529E+00 6.716E+00 6.683E-01 4.538E-01 1.782E+00 3.447E-02 4.282E-01 1.388E-02 1.092E-02 0.000E+00 1.587E-03 0.000E+00 
1.137E+00 5.542E+00 5.867E+00 6.041E+00 7.041E-01 3.926E-01 1.594E+00 2.341E-02 3.411E-01 1.764E-02 1.441E-02 0.000E+00 0.000E+00 0.000E+00 
1.163E+00 5.164E+00 5.119E+00 5.279E+00 5.841E-01 3.841E-01 1.521E+00 2.314E-02 3.025E-01 1.357E-02 8.916E-03 0.000E+00 0.000E+00 0.000E+00 
1.188E+00 4.616E+00 4.586E+00 4.873E+00 5.701E-01 3.153E-01 1.297E+00 1.455E-02 2.302E-01 9.590E-03 8.827E-03 0.000E+00 0.000E+00 0.000E+00 
1.213E+00 3.998E+00 3.998E+00 4.377E+00 4.298E-01 2.925E-01 1.169E+00 1.233E-02 2.504E-01 9.487E-03 5.244E-03 0.000E+00 0.000E+00 0.000E+00 
1.238E+00 3.446E+00 3.655E+00 3.774E+00 4.414E-01 2.603E-01 9.809E-01 1.015E-02 2.478E-01 3.754E-03 6.923E-03 0.000E+00 1.518E-03 0.000E+00 
1.262E+00 3.095E+00 3.312E+00 3.539E+00 3.974E-01 2.163E-01 9.215E-01 1.205E-02 2.210E-01 3.715E-03 5.142E-03 0.000E+00 0.000E+00 0.000E+00 
1.288E+00 2.954E+00 3.037E+00 3.133E+00 3.224E-01 2.058E-01 8.338E-01 1.588E-02 2.224E-01 1.103E-02 0.000E+00 0.000E+00 0.000E+00 0.000E+00 
1.312E+00 2.534E+00 2.668E+00 2.778E+00 3.033E-01 1.834E-01 7.852E-01 1.178E-02 1.492E-01 3.638E-03 0.000E+00 0.000E+00 0.000E+00 0.000E+00
\end{verbatim}
}

\newpage
Pb+Pb \@ $E_{\rm lab}=160A~$GeV:
{\tiny
\begin{verbatim}
! m_t-m0 , 1/mt dN/dmt(pi+ pi- pi0 K+ K- P aP L+S0 a(L+S0) Xi- aXi- Om aOm) 
1.250E-02 3.289E+03 3.447E+03 3.491E+03 1.374E+02 8.257E+01 6.951E+01 2.229E+00 3.118E+01 1.590E+00 1.227E+00 1.971E-01 2.047E-01 6.165E-02 
3.750E-02 2.868E+03 3.022E+03 3.105E+03 1.289E+02 7.809E+01 6.770E+01 2.444E+00 2.994E+01 1.559E+00 1.078E+00 2.519E-01 1.701E-01 7.290E-02 
6.250E-02 2.425E+03 2.574E+03 2.655E+03 1.162E+02 7.451E+01 6.624E+01 2.674E+00 2.806E+01 1.611E+00 9.741E-01 1.538E-01 2.036E-01 6.466E-02 
8.750E-02 2.067E+03 2.154E+03 2.239E+03 1.059E+02 6.799E+01 6.401E+01 2.625E+00 2.642E+01 1.705E+00 8.649E-01 1.836E-01 1.558E-01 4.958E-02 
1.125E-01 1.739E+03 1.822E+03 1.874E+03 9.440E+01 6.080E+01 6.046E+01 2.855E+00 2.431E+01 1.549E+00 8.294E-01 1.019E-01 1.304E-01 6.053E-02 
1.375E-01 1.466E+03 1.544E+03 1.574E+03 8.391E+01 5.599E+01 5.859E+01 2.773E+00 2.284E+01 1.515E+00 7.350E-01 1.230E-01 1.286E-01 2.984E-02 
1.625E-01 1.247E+03 1.299E+03 1.340E+03 7.553E+01 4.884E+01 5.537E+01 2.673E+00 2.076E+01 1.436E+00 6.748E-01 1.012E-01 9.964E-02 3.850E-02 
1.875E-01 1.062E+03 1.104E+03 1.137E+03 6.753E+01 4.413E+01 5.214E+01 2.532E+00 1.862E+01 1.237E+00 5.751E-01 9.124E-02 9.159E-02 4.915E-02 
2.125E-01 9.013E+02 9.410E+02 9.694E+02 5.904E+01 3.935E+01 4.899E+01 2.491E+00 1.700E+01 1.166E+00 5.412E-01 7.887E-02 8.377E-02 3.527E-02 
2.375E-01 7.674E+02 8.041E+02 8.216E+02 5.356E+01 3.507E+01 4.576E+01 2.276E+00 1.575E+01 1.142E+00 4.469E-01 5.352E-02 8.484E-02 1.523E-02 
2.625E-01 6.530E+02 6.890E+02 7.017E+02 4.653E+01 3.134E+01 4.236E+01 2.111E+00 1.434E+01 9.583E-01 3.818E-01 8.163E-02 4.939E-02 2.792E-02 
2.875E-01 5.643E+02 5.847E+02 6.000E+02 4.096E+01 2.816E+01 3.948E+01 2.017E+00 1.290E+01 8.879E-01 3.655E-01 4.666E-02 6.148E-02 1.484E-02 
3.125E-01 4.833E+02 5.035E+02 5.122E+02 3.568E+01 2.447E+01 3.581E+01 1.794E+00 1.169E+01 8.375E-01 2.706E-01 5.615E-02 4.815E-02 1.465E-02 
3.375E-01 4.202E+02 4.337E+02 4.428E+02 3.164E+01 2.154E+01 3.321E+01 1.736E+00 1.025E+01 6.059E-01 2.288E-01 4.274E-02 5.168E-02 1.034E-02 
3.625E-01 3.597E+02 3.705E+02 3.792E+02 2.873E+01 1.991E+01 3.136E+01 1.613E+00 9.530E+00 6.547E-01 2.427E-01 3.962E-02 3.675E-02 1.225E-02 
3.875E-01 3.102E+02 3.234E+02 3.322E+02 2.521E+01 1.737E+01 2.853E+01 1.244E+00 8.493E+00 6.079E-01 2.440E-01 2.196E-02 4.438E-02 1.412E-02 
4.125E-01 2.696E+02 2.753E+02 2.860E+02 2.229E+01 1.537E+01 2.558E+01 1.357E+00 7.917E+00 6.414E-01 2.501E-01 4.569E-02 2.591E-02 7.971E-03 
4.375E-01 2.324E+02 2.409E+02 2.481E+02 1.966E+01 1.339E+01 2.349E+01 1.172E+00 6.827E+00 6.043E-01 1.849E-01 2.607E-02 2.363E-02 7.877E-03 
4.625E-01 2.001E+02 2.094E+02 2.169E+02 1.716E+01 1.213E+01 2.149E+01 1.154E+00 6.379E+00 4.342E-01 1.496E-01 2.571E-02 3.309E-02 1.946E-03 
4.875E-01 1.734E+02 1.829E+02 1.870E+02 1.528E+01 1.081E+01 1.948E+01 8.888E-01 5.743E+00 4.041E-01 1.314E-01 2.074E-02 1.731E-02 3.847E-03 
5.125E-01 1.531E+02 1.595E+02 1.624E+02 1.390E+01 9.291E+00 1.791E+01 8.792E-01 5.000E+00 2.678E-01 1.227E-01 9.092E-03 2.852E-02 0.000E+00 
5.375E-01 1.345E+02 1.385E+02 1.419E+02 1.164E+01 8.340E+00 1.628E+01 7.489E-01 4.407E+00 2.889E-01 9.867E-02 1.121E-02 1.692E-02 1.880E-03 
5.625E-01 1.161E+02 1.203E+02 1.228E+02 1.030E+01 7.573E+00 1.432E+01 6.921E-01 4.153E+00 2.772E-01 8.850E-02 1.106E-02 1.859E-02 0.000E+00 
5.875E-01 1.010E+02 1.049E+02 1.069E+02 9.403E+00 6.799E+00 1.347E+01 5.964E-01 3.894E+00 2.414E-01 7.642E-02 4.367E-03 7.354E-03 1.839E-03 
6.125E-01 8.970E+01 9.156E+01 9.512E+01 8.301E+00 5.875E+00 1.234E+01 6.591E-01 3.353E+00 2.259E-01 8.836E-02 2.155E-03 1.818E-02 1.818E-03 
6.375E-01 7.767E+01 8.007E+01 8.209E+01 7.420E+00 5.342E+00 1.103E+01 4.377E-01 2.938E+00 1.729E-01 8.723E-02 6.383E-03 1.079E-02 3.597E-03 
6.625E-01 6.740E+01 7.123E+01 7.260E+01 6.487E+00 4.953E+00 1.029E+01 4.309E-01 2.602E+00 2.055E-01 5.042E-02 1.050E-02 8.897E-03 3.559E-03 
6.875E-01 5.963E+01 6.214E+01 6.323E+01 5.840E+00 4.272E+00 8.939E+00 4.702E-01 2.402E+00 1.797E-01 7.053E-02 6.223E-03 8.803E-03 3.521E-03 
7.125E-01 5.276E+01 5.512E+01 5.638E+01 5.065E+00 3.478E+00 8.203E+00 3.725E-01 2.220E+00 1.522E-01 6.761E-02 1.024E-02 5.226E-03 1.742E-03 
7.375E-01 4.680E+01 4.793E+01 4.862E+01 4.399E+00 3.424E+00 7.237E+00 3.669E-01 1.842E+00 1.659E-01 4.857E-02 6.072E-03 5.172E-03 0.000E+00 
7.625E-01 4.075E+01 4.191E+01 4.304E+01 4.238E+00 2.797E+00 6.720E+00 2.516E-01 1.667E+00 1.194E-01 3.999E-02 5.999E-03 8.532E-03 0.000E+00 
7.875E-01 3.609E+01 3.760E+01 3.795E+01 3.728E+00 2.729E+00 6.038E+00 2.696E-01 1.615E+00 9.821E-02 2.964E-02 7.903E-03 3.378E-03 0.000E+00 
8.125E-01 3.182E+01 3.331E+01 3.439E+01 3.285E+00 2.362E+00 5.302E+00 2.444E-01 1.381E+00 8.616E-02 2.538E-02 5.858E-03 0.000E+00 0.000E+00 
8.375E-01 2.786E+01 2.920E+01 3.020E+01 2.780E+00 2.175E+00 4.670E+00 1.731E-01 1.278E+00 8.719E-02 2.702E-02 1.158E-02 1.655E-03 0.000E+00 
8.625E-01 2.431E+01 2.528E+01 2.689E+01 2.533E+00 1.859E+00 4.478E+00 1.846E-01 1.119E+00 8.608E-02 3.243E-02 5.723E-03 3.278E-03 3.278E-03 
8.875E-01 2.172E+01 2.291E+01 2.387E+01 2.252E+00 1.654E+00 4.080E+00 1.889E-01 1.080E+00 6.842E-02 9.430E-03 0.000E+00 1.623E-03 1.623E-03 
9.125E-01 1.961E+01 1.982E+01 2.044E+01 2.085E+00 1.568E+00 3.545E+00 1.437E-01 9.093E-01 5.120E-02 1.305E-02 1.865E-03 3.215E-03 0.000E+00 
9.375E-01 1.706E+01 1.798E+01 1.818E+01 1.950E+00 1.320E+00 3.307E+00 1.484E-01 7.566E-01 6.069E-02 1.107E-02 1.844E-03 0.000E+00 0.000E+00 
9.625E-01 1.541E+01 1.551E+01 1.636E+01 1.777E+00 1.244E+00 3.036E+00 1.246E-01 7.695E-01 4.597E-02 1.824E-02 0.000E+00 3.154E-03 0.000E+00 
9.875E-01 1.333E+01 1.340E+01 1.480E+01 1.511E+00 9.730E-01 2.561E+00 1.381E-01 6.853E-01 4.740E-02 1.263E-02 1.804E-03 1.562E-03 1.562E-03 
1.012E+00 1.175E+01 1.221E+01 1.307E+01 1.329E+00 9.789E-01 2.439E+00 1.044E-01 5.465E-01 5.074E-02 7.139E-03 0.000E+00 0.000E+00 0.000E+00 
1.038E+00 1.078E+01 1.109E+01 1.157E+01 1.134E+00 9.087E-01 2.145E+00 9.463E-02 4.514E-01 3.472E-02 8.829E-03 1.766E-03 4.600E-03 0.000E+00 
1.062E+00 9.578E+00 1.004E+01 1.061E+01 1.102E+00 7.153E-01 1.956E+00 7.060E-02 4.843E-01 3.432E-02 6.989E-03 0.000E+00 3.038E-03 0.000E+00 
1.087E+00 8.603E+00 9.417E+00 9.227E+00 9.955E-01 7.171E-01 1.698E+00 8.204E-02 4.562E-01 3.582E-02 3.458E-03 0.000E+00 0.000E+00 0.000E+00 
1.113E+00 7.687E+00 7.993E+00 8.315E+00 7.292E-01 6.258E-01 1.596E+00 5.065E-02 4.642E-01 2.610E-02 1.198E-02 1.711E-03 0.000E+00 1.492E-03 
1.137E+00 6.784E+00 6.979E+00 7.266E+00 6.900E-01 6.136E-01 1.335E+00 5.204E-02 3.576E-01 3.134E-02 5.081E-03 0.000E+00 1.479E-03 0.000E+00 
1.163E+00 6.206E+00 6.267E+00 6.580E+00 6.219E-01 4.890E-01 1.216E+00 5.735E-02 2.917E-01 1.641E-02 5.030E-03 0.000E+00 1.466E-03 0.000E+00 
1.188E+00 5.510E+00 5.720E+00 5.926E+00 6.201E-01 4.645E-01 1.079E+00 4.104E-02 2.994E-01 1.803E-02 1.660E-03 0.000E+00 0.000E+00 0.000E+00 
1.213E+00 4.759E+00 5.143E+00 5.122E+00 5.648E-01 3.432E-01 9.832E-01 4.443E-02 2.408E-01 1.784E-02 8.218E-03 0.000E+00 0.000E+00 0.000E+00 
1.238E+00 4.415E+00 4.279E+00 4.666E+00 4.870E-01 3.839E-01 9.127E-01 3.628E-02 2.189E-01 1.412E-02 1.627E-03 0.000E+00 0.000E+00 0.000E+00 
1.262E+00 3.856E+00 4.010E+00 4.253E+00 4.635E-01 3.074E-01 7.872E-01 3.964E-02 2.305E-01 1.397E-02 1.612E-03 0.000E+00 0.000E+00 0.000E+00 
1.288E+00 3.471E+00 3.733E+00 3.853E+00 3.894E-01 2.728E-01 8.176E-01 1.493E-02 1.832E-01 8.642E-03 3.192E-03 0.000E+00 0.000E+00 0.000E+00 
1.312E+00 3.165E+00 3.173E+00 3.365E+00 3.403E-01 2.782E-01 6.700E-01 2.030E-02 1.574E-01 3.421E-03 1.581E-03 1.581E-03 0.000E+00 0.000E+00 
\end{verbatim}
}
\newpage
Au+Au \@ $E_{\rm CM}=56A~$GeV:
{\tiny
\begin{verbatim}
! m_t-m0 , 1/mt dN/dmt(pi+ pi- pi0 K+ K- P aP L+S0 a(L+S0) Xi- aXi- Om aOm) 
1.250E-02 4.366E+03 4.445E+03 4.647E+03 1.805E+02 1.396E+02 4.524E+01 1.015E+01 2.375E+01 5.707E+00 1.656E+00 5.989E-01 3.690E-01 1.755E-01 
3.750E-02 3.860E+03 3.920E+03 4.122E+03 1.707E+02 1.343E+02 4.509E+01 1.055E+01 2.299E+01 5.653E+00 1.482E+00 5.628E-01 2.919E-01 1.352E-01 
6.250E-02 3.301E+03 3.365E+03 3.528E+03 1.548E+02 1.238E+02 4.356E+01 1.125E+01 2.119E+01 5.469E+00 1.421E+00 4.697E-01 3.048E-01 1.428E-01 
8.750E-02 2.828E+03 2.878E+03 3.015E+03 1.411E+02 1.129E+02 4.129E+01 1.033E+01 1.929E+01 5.387E+00 1.161E+00 4.432E-01 2.644E-01 8.229E-02 
1.125E-01 2.405E+03 2.457E+03 2.561E+03 1.254E+02 1.012E+02 3.846E+01 1.026E+01 1.773E+01 4.950E+00 1.182E+00 3.940E-01 1.848E-01 1.133E-01 
1.375E-01 2.061E+03 2.102E+03 2.188E+03 1.140E+02 9.143E+01 3.612E+01 9.389E+00 1.567E+01 4.487E+00 9.287E-01 3.756E-01 1.706E-01 7.358E-02 
1.625E-01 1.772E+03 1.793E+03 1.866E+03 1.020E+02 8.257E+01 3.441E+01 9.222E+00 1.409E+01 4.555E+00 8.214E-01 3.091E-01 1.452E-01 6.242E-02 
1.875E-01 1.509E+03 1.539E+03 1.597E+03 9.014E+01 7.241E+01 3.252E+01 8.533E+00 1.346E+01 4.341E+00 7.064E-01 2.674E-01 1.501E-01 5.492E-02 
2.125E-01 1.306E+03 1.322E+03 1.365E+03 7.946E+01 6.433E+01 3.011E+01 7.994E+00 1.221E+01 3.842E+00 6.367E-01 2.907E-01 1.436E-01 4.587E-02 
2.375E-01 1.125E+03 1.137E+03 1.176E+03 7.089E+01 5.883E+01 2.770E+01 7.537E+00 1.082E+01 3.493E+00 4.957E-01 2.560E-01 1.019E-01 5.215E-02 
2.625E-01 9.581E+02 9.780E+02 1.013E+03 6.224E+01 5.269E+01 2.595E+01 7.165E+00 9.662E+00 3.043E+00 4.611E-01 1.716E-01 8.525E-02 3.831E-02 
2.875E-01 8.297E+02 8.334E+02 8.761E+02 5.451E+01 4.663E+01 2.367E+01 6.511E+00 8.975E+00 2.862E+00 4.275E-01 1.583E-01 7.985E-02 3.299E-02 
3.125E-01 7.146E+02 7.282E+02 7.589E+02 4.844E+01 4.089E+01 2.162E+01 6.182E+00 7.924E+00 2.694E+00 3.560E-01 1.611E-01 8.311E-02 3.430E-02 
3.375E-01 6.142E+02 6.310E+02 6.569E+02 4.292E+01 3.631E+01 2.024E+01 5.460E+00 7.135E+00 2.429E+00 3.224E-01 1.638E-01 5.892E-02 4.368E-02 
3.625E-01 5.377E+02 5.448E+02 5.679E+02 3.875E+01 3.177E+01 1.824E+01 4.965E+00 6.271E+00 2.065E+00 2.748E-01 1.286E-01 6.651E-02 2.404E-02 
3.875E-01 4.659E+02 4.744E+02 4.902E+02 3.355E+01 2.940E+01 1.663E+01 4.498E+00 6.077E+00 1.873E+00 2.558E-01 1.143E-01 5.133E-02 2.222E-02 
4.125E-01 4.038E+02 4.087E+02 4.275E+02 2.864E+01 2.522E+01 1.554E+01 4.274E+00 5.224E+00 1.726E+00 2.644E-01 1.028E-01 4.869E-02 1.415E-02 
4.375E-01 3.508E+02 3.579E+02 3.701E+02 2.607E+01 2.249E+01 1.387E+01 3.702E+00 4.653E+00 1.631E+00 2.027E-01 8.204E-02 4.210E-02 1.558E-02 
4.625E-01 3.069E+02 3.068E+02 3.224E+02 2.355E+01 2.026E+01 1.274E+01 3.355E+00 4.239E+00 1.431E+00 1.856E-01 9.516E-02 3.566E-02 1.544E-02 
4.875E-01 2.650E+02 2.718E+02 2.813E+02 2.054E+01 1.755E+01 1.142E+01 3.183E+00 3.864E+00 1.287E+00 1.525E-01 7.507E-02 2.937E-02 1.377E-02 
5.125E-01 2.361E+02 2.338E+02 2.454E+02 1.768E+01 1.608E+01 1.024E+01 2.781E+00 3.506E+00 1.109E+00 1.435E-01 7.173E-02 3.097E-02 7.581E-03 
5.375E-01 2.036E+02 2.080E+02 2.127E+02 1.578E+01 1.394E+01 9.452E+00 2.439E+00 3.093E+00 1.000E-00 1.050E-01 4.794E-02 1.914E-02 4.508E-03 
5.625E-01 1.780E+02 1.805E+02 1.864E+02 1.415E+01 1.227E+01 8.545E+00 2.373E+00 2.789E+00 9.070E-01 1.216E-01 4.955E-02 2.839E-02 2.979E-03 
5.875E-01 1.582E+02 1.576E+02 1.625E+02 1.255E+01 1.123E+01 8.103E+00 2.198E+00 2.595E+00 7.844E-01 9.336E-02 3.779E-02 2.059E-02 2.953E-03 
6.125E-01 1.365E+02 1.380E+02 1.414E+02 1.133E+01 9.910E+00 7.105E+00 1.827E+00 2.251E+00 7.217E-01 1.053E-01 4.168E-02 1.111E-02 2.928E-03 
6.375E-01 1.200E+02 1.223E+02 1.251E+02 1.001E+01 8.768E+00 6.579E+00 1.739E+00 2.077E+00 6.753E-01 8.880E-02 3.032E-02 9.155E-03 1.451E-03 
6.625E-01 1.041E+02 1.062E+02 1.103E+02 8.794E+00 7.982E+00 5.940E+00 1.564E+00 1.986E+00 6.016E-01 6.201E-02 3.422E-02 7.246E-03 4.317E-03 
6.875E-01 9.221E+01 9.295E+01 9.649E+01 7.756E+00 6.661E+00 5.513E+00 1.423E+00 1.644E+00 5.815E-01 7.180E-02 1.901E-02 1.434E-02 8.560E-03 
7.125E-01 8.217E+01 8.295E+01 8.395E+01 7.052E+00 6.327E+00 4.909E+00 1.189E+00 1.418E+00 4.741E-01 7.717E-02 2.086E-02 8.867E-03 2.830E-03 
7.375E-01 7.251E+01 7.301E+01 7.485E+01 6.280E+00 5.460E+00 4.551E+00 1.164E+00 1.289E+00 4.175E-01 7.005E-02 2.266E-02 3.510E-03 2.806E-03 
7.625E-01 6.289E+01 6.472E+01 6.687E+01 5.607E+00 5.072E+00 4.073E+00 8.530E-01 1.245E+00 3.692E-01 3.867E-02 1.628E-02 1.737E-02 2.783E-03 
7.875E-01 5.474E+01 5.673E+01 5.842E+01 4.904E+00 4.267E+00 3.786E+00 9.215E-01 1.140E+00 3.332E-01 3.218E-02 1.207E-02 5.158E-03 6.901E-03 
8.125E-01 4.931E+01 5.146E+01 5.200E+01 4.405E+00 3.839E+00 3.442E+00 7.972E-01 9.738E-01 3.026E-01 4.572E-02 1.988E-02 5.106E-03 2.738E-03 
8.375E-01 4.375E+01 4.483E+01 4.653E+01 3.913E+00 3.424E+00 3.032E+00 6.431E-01 9.375E-01 2.489E-01 2.947E-02 1.965E-02 1.685E-03 1.358E-03 
8.625E-01 3.891E+01 4.020E+01 4.018E+01 3.476E+00 3.077E+00 2.692E+00 5.942E-01 7.888E-01 2.714E-01 3.301E-02 9.710E-03 1.668E-03 1.347E-03 
8.875E-01 3.477E+01 3.491E+01 3.621E+01 3.104E+00 2.825E+00 2.504E+00 5.838E-01 6.418E-01 2.786E-01 1.920E-02 3.840E-03 8.261E-03 2.673E-03 
9.125E-01 3.069E+01 3.024E+01 3.202E+01 2.736E+00 2.501E+00 2.208E+00 5.599E-01 6.609E-01 1.668E-01 1.898E-02 9.492E-03 1.636E-03 1.326E-03 
9.375E-01 2.727E+01 2.842E+01 2.831E+01 2.354E+00 2.316E+00 2.032E+00 4.352E-01 5.911E-01 1.956E-01 2.441E-02 5.632E-03 3.241E-03 1.316E-03 
9.625E-01 2.414E+01 2.475E+01 2.510E+01 2.256E+00 1.980E+00 1.827E+00 4.139E-01 5.331E-01 1.221E-01 2.414E-02 7.427E-03 4.815E-03 2.611E-03 
9.875E-01 2.142E+01 2.136E+01 2.281E+01 1.950E+00 1.727E+00 1.553E+00 3.360E-01 4.745E-01 1.086E-01 1.837E-02 5.510E-03 1.590E-03 1.296E-03 
1.012E+00 1.918E+01 1.959E+01 1.980E+01 1.659E+00 1.592E+00 1.576E+00 3.382E-01 4.272E-01 1.391E-01 2.180E-02 5.451E-03 3.151E-03 1.286E-03 
1.038E+00 1.767E+01 1.729E+01 1.800E+01 1.590E+00 1.513E+00 1.409E+00 2.740E-01 3.849E-01 9.818E-02 1.258E-02 3.595E-03 1.561E-03 0.000E+00 
1.062E+00 1.563E+01 1.554E+01 1.621E+01 1.377E+00 1.348E+00 1.201E+00 2.515E-01 3.728E-01 9.123E-02 3.557E-03 3.557E-03 1.546E-03 0.000E+00 
1.087E+00 1.326E+01 1.406E+01 1.418E+01 1.209E+00 1.075E+00 1.098E+00 2.317E-01 2.975E-01 1.113E-01 1.408E-02 7.041E-03 4.597E-03 1.257E-03 
1.113E+00 1.194E+01 1.228E+01 1.266E+01 1.069E+00 1.003E+00 9.755E-01 2.227E-01 3.018E-01 1.082E-01 6.968E-03 1.742E-03 4.556E-03 1.248E-03 
1.137E+00 1.077E+01 1.128E+01 1.142E+01 9.927E-01 9.538E-01 9.800E-01 1.610E-01 2.534E-01 6.756E-02 1.035E-02 0.000E+00 3.010E-03 0.000E+00 
1.163E+00 9.931E+00 9.713E+00 1.022E+01 9.497E-01 8.297E-01 8.617E-01 1.792E-01 2.302E-01 7.053E-02 6.827E-03 1.707E-03 0.000E+00 1.230E-03 
1.188E+00 9.061E+00 8.607E+00 8.994E+00 8.123E-01 7.117E-01 7.381E-01 1.293E-01 2.534E-01 5.507E-02 1.014E-02 3.380E-03 0.000E+00 1.221E-03 
1.213E+00 7.872E+00 7.978E+00 8.091E+00 7.013E-01 7.335E-01 6.509E-01 1.298E-01 1.889E-01 5.448E-02 5.019E-03 1.673E-03 1.466E-03 1.212E-03 
1.238E+00 6.880E+00 7.363E+00 7.440E+00 6.643E-01 6.814E-01 6.162E-01 1.205E-01 1.420E-01 4.672E-02 9.940E-03 3.313E-03 0.000E+00 0.000E+00 
1.262E+00 6.181E+00 6.444E+00 6.712E+00 6.067E-01 5.658E-01 5.592E-01 1.153E-01 1.743E-01 3.911E-02 1.641E-03 1.641E-03 0.000E+00 0.000E+00 
1.288E+00 5.286E+00 5.716E+00 5.764E+00 5.911E-01 4.510E-01 4.427E-01 1.026E-01 1.513E-01 2.111E-02 6.500E-03 3.250E-03 0.000E+00 1.187E-03 
1.312E+00 5.003E+00 5.102E+00 5.361E+00 4.448E-01 4.822E-01 4.622E-01 7.328E-02 1.149E-01 2.960E-02 0.000E+00 1.609E-03 0.000E+00 0.000E+00 
\end{verbatim}
}
\newpage
Au+Au \@ $E_{\rm CM}=62.5A~$GeV:
{\tiny
\begin{verbatim}
! m_t-m0 , 1/mt dN/dmt(pi+ pi- pi0 K+ K- P aP L+S0 a(L+S0) Xi- aXi- Om aOm) 
1.250E-02 4.423E+03 4.519E+03 4.724E+03 1.839E+02 1.457E+02 4.487E+01 1.100E+01 2.390E+01 6.213E+00 1.692E+00 6.679E-01 3.774E-01 2.384E-01 
3.750E-02 3.939E+03 3.992E+03 4.206E+03 1.735E+02 1.386E+02 4.427E+01 1.142E+01 2.265E+01 6.017E+00 1.645E+00 5.875E-01 3.132E-01 2.177E-01 
6.250E-02 3.401E+03 3.447E+03 3.608E+03 1.584E+02 1.291E+02 4.285E+01 1.177E+01 2.084E+01 5.868E+00 1.348E+00 4.494E-01 3.014E-01 1.953E-01 
8.750E-02 2.918E+03 2.941E+03 3.088E+03 1.449E+02 1.169E+02 4.077E+01 1.116E+01 1.951E+01 5.551E+00 1.264E+00 4.712E-01 2.591E-01 1.474E-01 
1.125E-01 2.491E+03 2.522E+03 2.635E+03 1.311E+02 1.073E+02 3.746E+01 1.120E+01 1.733E+01 5.305E+00 1.131E+00 4.131E-01 2.250E-01 1.406E-01 
1.375E-01 2.122E+03 2.155E+03 2.254E+03 1.171E+02 9.519E+01 3.619E+01 1.028E+01 1.567E+01 4.916E+00 8.609E-01 3.743E-01 1.826E-01 1.317E-01 
1.625E-01 1.812E+03 1.850E+03 1.920E+03 1.041E+02 8.525E+01 3.298E+01 1.038E+01 1.404E+01 4.548E+00 8.379E-01 3.482E-01 1.413E-01 1.345E-01 
1.875E-01 1.565E+03 1.585E+03 1.654E+03 9.360E+01 7.618E+01 3.108E+01 9.442E+00 1.285E+01 4.416E+00 7.516E-01 2.951E-01 1.394E-01 9.447E-02 
2.125E-01 1.345E+03 1.359E+03 1.418E+03 8.237E+01 6.855E+01 2.887E+01 8.830E+00 1.123E+01 3.977E+00 7.584E-01 3.121E-01 1.554E-01 7.768E-02 
2.375E-01 1.150E+03 1.168E+03 1.217E+03 7.324E+01 5.983E+01 2.734E+01 7.973E+00 1.084E+01 3.548E+00 5.280E-01 2.505E-01 1.073E-01 5.476E-02 
2.625E-01 9.984E+02 1.013E+03 1.047E+03 6.395E+01 5.383E+01 2.494E+01 7.602E+00 9.672E+00 3.317E+00 5.409E-01 2.360E-01 9.945E-02 3.459E-02 
2.875E-01 8.633E+02 8.695E+02 9.042E+02 5.801E+01 4.881E+01 2.334E+01 7.003E+00 8.710E+00 3.022E+00 4.280E-01 1.905E-01 7.897E-02 5.763E-02 
3.125E-01 7.454E+02 7.526E+02 7.805E+02 5.041E+01 4.306E+01 2.103E+01 6.629E+00 7.820E+00 2.958E+00 3.906E-01 1.825E-01 8.009E-02 3.794E-02 
3.375E-01 6.408E+02 6.507E+02 6.763E+02 4.469E+01 3.803E+01 1.951E+01 5.869E+00 7.179E+00 2.530E+00 3.569E-01 1.569E-01 4.579E-02 2.914E-02 
3.625E-01 5.586E+02 5.627E+02 5.886E+02 3.925E+01 3.426E+01 1.803E+01 5.474E+00 6.342E+00 2.264E+00 3.166E-01 1.396E-01 5.962E-02 3.700E-02 
3.875E-01 4.829E+02 4.911E+02 5.066E+02 3.441E+01 2.972E+01 1.623E+01 5.086E+00 5.664E+00 2.090E+00 2.555E-01 9.335E-02 3.249E-02 3.046E-02 
4.125E-01 4.219E+02 4.275E+02 4.432E+02 3.106E+01 2.699E+01 1.490E+01 4.385E+00 5.232E+00 1.908E+00 2.348E-01 7.989E-02 4.213E-02 3.010E-02 
4.375E-01 3.668E+02 3.690E+02 3.798E+02 2.709E+01 2.363E+01 1.360E+01 4.220E+00 4.833E+00 1.675E+00 2.052E-01 9.546E-02 4.362E-02 1.784E-02 
4.625E-01 3.178E+02 3.251E+02 3.345E+02 2.324E+01 2.103E+01 1.226E+01 3.524E+00 4.208E+00 1.556E+00 2.000E-01 9.176E-02 4.703E-02 2.547E-02 
4.875E-01 2.756E+02 2.814E+02 2.906E+02 2.123E+01 1.858E+01 1.146E+01 3.395E+00 3.852E+00 1.310E+00 1.462E-01 8.817E-02 3.486E-02 1.549E-02 
5.125E-01 2.416E+02 2.464E+02 2.540E+02 1.896E+01 1.641E+01 1.085E+01 2.915E+00 3.380E+00 1.236E+00 1.419E-01 6.179E-02 3.063E-02 1.532E-02 
5.375E-01 2.122E+02 2.138E+02 2.202E+02 1.653E+01 1.483E+01 9.584E+00 2.775E+00 3.056E+00 1.141E+00 1.467E-01 5.418E-02 3.218E-02 1.514E-02 
5.625E-01 1.850E+02 1.885E+02 1.959E+02 1.455E+01 1.318E+01 8.498E+00 2.475E+00 2.711E+00 1.064E+00 1.292E-01 4.901E-02 1.310E-02 9.359E-03 
5.875E-01 1.630E+02 1.633E+02 1.692E+02 1.320E+01 1.152E+01 7.868E+00 2.297E+00 2.487E+00 9.061E-01 9.453E-02 3.957E-02 1.296E-02 1.296E-02 
6.125E-01 1.419E+02 1.436E+02 1.497E+02 1.162E+01 1.071E+01 7.253E+00 2.080E+00 2.294E+00 7.163E-01 8.896E-02 4.123E-02 1.465E-02 1.098E-02 
6.375E-01 1.239E+02 1.279E+02 1.311E+02 1.000E+01 9.071E+00 6.366E+00 1.938E+00 1.932E+00 7.538E-01 6.212E-02 4.284E-02 1.630E-02 1.087E-02 
6.625E-01 1.092E+02 1.128E+02 1.155E+02 8.763E+00 8.245E+00 5.869E+00 1.484E+00 1.839E+00 7.080E-01 4.653E-02 2.749E-02 8.958E-03 0.000E+00 
6.875E-01 9.724E+01 9.791E+01 1.007E+02 8.184E+00 7.345E+00 5.190E+00 1.338E+00 1.707E+00 6.031E-01 5.848E-02 3.342E-02 1.773E-02 1.418E-02 
7.125E-01 8.562E+01 8.683E+01 8.966E+01 7.034E+00 6.566E+00 4.817E+00 1.363E+00 1.500E+00 5.056E-01 4.126E-02 3.300E-02 7.016E-03 1.052E-02 
7.375E-01 7.502E+01 7.646E+01 7.866E+01 6.341E+00 5.953E+00 4.476E+00 1.203E+00 1.361E+00 4.581E-01 5.094E-02 1.834E-02 8.679E-03 1.736E-03 
7.625E-01 6.615E+01 6.745E+01 7.047E+01 5.692E+00 5.329E+00 4.144E+00 1.021E+00 1.160E+00 4.298E-01 5.033E-02 1.812E-02 6.872E-03 1.718E-03 
7.875E-01 5.802E+01 5.952E+01 6.085E+01 5.104E+00 4.687E+00 3.556E+00 9.356E-01 1.096E+00 3.956E-01 3.978E-02 2.586E-02 3.401E-03 0.000E+00 
8.125E-01 5.223E+01 5.193E+01 5.422E+01 4.507E+00 4.101E+00 3.096E+00 8.721E-01 1.054E+00 3.557E-01 2.949E-02 1.966E-02 1.683E-03 1.683E-03 
8.375E-01 4.559E+01 4.674E+01 4.756E+01 4.080E+00 3.757E+00 3.072E+00 7.491E-01 9.120E-01 3.426E-01 2.332E-02 1.554E-02 0.000E+00 1.667E-03 
8.625E-01 4.132E+01 4.193E+01 4.316E+01 3.580E+00 3.413E+00 2.855E+00 6.643E-01 7.272E-01 2.601E-01 2.881E-02 1.344E-02 8.251E-03 1.650E-03 
8.875E-01 3.634E+01 3.647E+01 3.856E+01 3.312E+00 2.979E+00 2.481E+00 5.361E-01 7.098E-01 2.526E-01 1.709E-02 1.139E-02 6.536E-03 3.268E-03 
9.125E-01 3.218E+01 3.291E+01 3.379E+01 2.944E+00 2.762E+00 2.253E+00 5.289E-01 7.175E-01 2.165E-01 1.690E-02 1.314E-02 8.091E-03 3.236E-03 
9.375E-01 2.894E+01 2.843E+01 3.002E+01 2.501E+00 2.381E+00 2.014E+00 5.263E-01 4.847E-01 1.854E-01 1.857E-02 9.284E-03 6.411E-03 0.000E+00 
9.625E-01 2.544E+01 2.579E+01 2.612E+01 2.268E+00 2.157E+00 1.809E+00 4.335E-01 5.876E-01 2.274E-01 1.836E-02 3.673E-03 1.588E-03 1.588E-03 
9.875E-01 2.262E+01 2.300E+01 2.416E+01 2.114E+00 1.920E+00 1.714E+00 3.866E-01 4.931E-01 1.531E-01 2.180E-02 5.449E-03 1.573E-03 1.573E-03 
1.012E+00 2.015E+01 2.079E+01 2.108E+01 1.952E+00 1.749E+00 1.334E+00 3.388E-01 4.166E-01 1.513E-01 1.258E-02 3.594E-03 0.000E+00 0.000E+00 
1.038E+00 1.818E+01 1.854E+01 1.862E+01 1.720E+00 1.649E+00 1.363E+00 3.218E-01 4.312E-01 1.418E-01 5.333E-03 5.333E-03 0.000E+00 0.000E+00 
1.062E+00 1.632E+01 1.678E+01 1.725E+01 1.569E+00 1.419E+00 1.213E+00 2.801E-01 4.070E-01 8.833E-02 8.796E-03 0.000E+00 1.529E-03 1.529E-03 
1.087E+00 1.458E+01 1.441E+01 1.564E+01 1.346E+00 1.235E+00 1.094E+00 2.932E-01 3.360E-01 1.139E-01 2.089E-02 0.000E+00 1.516E-03 1.516E-03 
1.113E+00 1.303E+01 1.320E+01 1.332E+01 1.216E+00 1.169E+00 9.586E-01 2.182E-01 2.909E-01 1.089E-01 6.892E-03 0.000E+00 1.502E-03 0.000E+00 
1.137E+00 1.196E+01 1.174E+01 1.225E+01 1.084E+00 1.020E+00 8.766E-01 2.035E-01 2.673E-01 6.868E-02 3.411E-03 3.411E-03 2.977E-03 0.000E+00 
1.163E+00 1.003E+01 1.084E+01 1.104E+01 1.045E+00 8.711E-01 8.024E-01 1.951E-01 2.588E-01 7.343E-02 6.753E-03 0.000E+00 4.427E-03 1.476E-03 
1.188E+00 8.870E+00 9.466E+00 9.674E+00 8.407E-01 7.910E-01 7.812E-01 1.692E-01 1.906E-01 5.085E-02 3.343E-03 1.671E-03 1.463E-03 1.463E-03 
1.213E+00 8.101E+00 8.027E+00 8.634E+00 8.039E-01 6.936E-01 6.982E-01 1.459E-01 2.299E-01 5.030E-02 3.309E-03 1.655E-03 0.000E+00 0.000E+00 
1.238E+00 7.465E+00 7.486E+00 7.568E+00 7.222E-01 6.111E-01 6.229E-01 1.480E-01 1.724E-01 6.932E-02 1.147E-02 3.277E-03 1.437E-03 0.000E+00 
1.262E+00 6.734E+00 6.818E+00 6.728E+00 6.310E-01 6.429E-01 5.322E-01 1.083E-01 1.635E-01 5.276E-02 4.868E-03 1.623E-03 1.425E-03 0.000E+00 
1.288E+00 5.959E+00 6.370E+00 6.064E+00 5.165E-01 5.188E-01 4.961E-01 1.090E-01 1.618E-01 3.655E-02 6.428E-03 3.214E-03 2.826E-03 0.000E+00 
1.312E+00 5.499E+00 5.398E+00 5.675E+00 5.348E-01 4.561E-01 4.479E-01 8.549E-02 1.309E-01 3.789E-02 1.114E-02 1.592E-03 0.000E+00 1.401E-03 
\end{verbatim}
}
\newpage
Au+Au \@ $E_{\rm CM}=130A~$GeV:
{\tiny
\begin{verbatim}
! m_t-m0 , 1/mt dN/dmt(pi+ pi- pi0 K+ K- P aP L+S0 a(L+S0) Xi- aXi- Om aOm) 
1.250E-02 4.842E+03 4.889E+03 5.116E+03 2.045E+02 1.674E+02 3.670E+01 1.455E+01 1.975E+01 7.886E+00 1.601E+00 8.402E-01 4.664E-01 2.690E-01 
3.750E-02 4.389E+03 4.423E+03 4.644E+03 1.975E+02 1.652E+02 3.613E+01 1.554E+01 1.868E+01 8.251E+00 1.451E+00 7.675E-01 3.453E-01 2.042E-01 
6.250E-02 3.821E+03 3.847E+03 4.022E+03 1.840E+02 1.532E+02 3.498E+01 1.526E+01 1.734E+01 7.879E+00 1.424E+00 6.827E-01 3.192E-01 2.076E-01 
8.750E-02 3.291E+03 3.332E+03 3.474E+03 1.657E+02 1.405E+02 3.277E+01 1.484E+01 1.623E+01 7.868E+00 1.094E+00 6.560E-01 3.401E-01 1.955E-01 
1.125E-01 2.841E+03 2.881E+03 2.984E+03 1.506E+02 1.280E+02 3.119E+01 1.436E+01 1.514E+01 7.502E+00 9.582E-01 6.131E-01 2.760E-01 1.683E-01 
1.375E-01 2.443E+03 2.462E+03 2.575E+03 1.355E+02 1.169E+02 2.922E+01 1.412E+01 1.371E+01 6.921E+00 9.557E-01 5.661E-01 2.362E-01 1.342E-01 
1.625E-01 2.120E+03 2.145E+03 2.236E+03 1.223E+02 1.045E+02 2.740E+01 1.361E+01 1.253E+01 6.333E+00 8.293E-01 5.014E-01 1.908E-01 1.308E-01 
1.875E-01 1.831E+03 1.846E+03 1.914E+03 1.091E+02 9.509E+01 2.613E+01 1.272E+01 1.140E+01 6.315E+00 7.397E-01 4.416E-01 2.342E-01 1.312E-01 
2.125E-01 1.579E+03 1.588E+03 1.660E+03 9.803E+01 8.512E+01 2.503E+01 1.226E+01 1.075E+01 5.632E+00 6.263E-01 3.891E-01 1.663E-01 1.151E-01 
2.375E-01 1.368E+03 1.381E+03 1.434E+03 8.663E+01 7.735E+01 2.232E+01 1.142E+01 9.618E+00 5.185E+00 6.346E-01 3.251E-01 1.343E-01 9.034E-02 
2.625E-01 1.184E+03 1.193E+03 1.241E+03 7.765E+01 6.882E+01 2.157E+01 1.096E+01 8.933E+00 4.967E+00 5.239E-01 3.484E-01 1.031E-01 8.041E-02 
2.875E-01 1.032E+03 1.040E+03 1.075E+03 6.837E+01 6.208E+01 1.977E+01 1.028E+01 8.217E+00 4.528E+00 5.385E-01 3.226E-01 9.972E-02 4.596E-02 
3.125E-01 8.954E+02 9.064E+02 9.355E+02 6.064E+01 5.591E+01 1.858E+01 9.004E+00 7.324E+00 4.243E+00 4.127E-01 2.351E-01 6.974E-02 5.770E-02 
3.375E-01 7.772E+02 7.829E+02 8.134E+02 5.390E+01 4.869E+01 1.707E+01 8.464E+00 6.604E+00 3.812E+00 3.523E-01 2.759E-01 8.508E-02 5.363E-02 
3.625E-01 6.797E+02 6.843E+02 7.046E+02 4.836E+01 4.331E+01 1.566E+01 7.904E+00 5.862E+00 3.557E+00 3.567E-01 2.014E-01 8.004E-02 4.451E-02 
3.875E-01 5.914E+02 5.967E+02 6.141E+02 4.301E+01 3.862E+01 1.428E+01 7.168E+00 5.551E+00 3.184E+00 2.821E-01 1.913E-01 7.116E-02 3.557E-02 
4.125E-01 5.151E+02 5.225E+02 5.357E+02 3.827E+01 3.545E+01 1.350E+01 6.571E+00 4.826E+00 2.687E+00 2.380E-01 1.885E-01 6.249E-02 2.012E-02 
4.375E-01 4.521E+02 4.555E+02 4.693E+02 3.421E+01 3.153E+01 1.210E+01 6.274E+00 4.769E+00 2.686E+00 2.207E-01 1.022E-01 4.631E-02 3.319E-02 
4.625E-01 3.913E+02 3.993E+02 4.090E+02 3.090E+01 2.790E+01 1.145E+01 5.686E+00 4.041E+00 2.404E+00 1.992E-01 1.397E-01 2.670E-02 1.971E-02 
4.875E-01 3.451E+02 3.471E+02 3.604E+02 2.630E+01 2.525E+01 1.027E+01 4.860E+00 3.813E+00 2.247E+00 1.807E-01 1.174E-01 3.393E-02 2.602E-02 
5.125E-01 3.032E+02 3.063E+02 3.153E+02 2.397E+01 2.193E+01 9.604E+00 4.502E+00 3.465E+00 1.987E+00 1.604E-01 1.025E-01 2.982E-02 3.382E-02 
5.375E-01 2.674E+02 2.679E+02 2.774E+02 2.145E+01 2.018E+01 8.735E+00 4.111E+00 3.198E+00 1.817E+00 1.340E-01 7.911E-02 2.395E-02 1.116E-02 
5.625E-01 2.377E+02 2.376E+02 2.409E+02 1.871E+01 1.795E+01 8.022E+00 3.635E+00 2.731E+00 1.673E+00 1.236E-01 7.372E-02 3.279E-02 9.474E-03 
5.875E-01 2.073E+02 2.101E+02 2.139E+02 1.672E+01 1.640E+01 7.603E+00 3.413E+00 2.516E+00 1.503E+00 1.155E-01 7.489E-02 2.522E-02 1.407E-02 
6.125E-01 1.817E+02 1.844E+02 1.899E+02 1.466E+01 1.410E+01 6.771E+00 3.088E+00 2.266E+00 1.382E+00 1.267E-01 6.758E-02 2.138E-02 1.704E-02 
6.375E-01 1.599E+02 1.599E+02 1.669E+02 1.357E+01 1.315E+01 6.036E+00 2.682E+00 2.210E+00 1.133E+00 9.382E-02 6.880E-02 8.813E-03 7.672E-03 
6.625E-01 1.404E+02 1.423E+02 1.469E+02 1.183E+01 1.146E+01 5.601E+00 2.577E+00 1.952E+00 1.057E+00 6.999E-02 5.146E-02 1.744E-02 4.560E-03 
6.875E-01 1.242E+02 1.254E+02 1.315E+02 1.080E+01 1.017E+01 5.021E+00 2.126E+00 1.657E+00 9.164E-01 8.945E-02 5.692E-02 2.588E-02 9.036E-03 
7.125E-01 1.103E+02 1.115E+02 1.161E+02 9.532E+00 9.161E+00 4.674E+00 2.050E+00 1.583E+00 9.395E-01 7.830E-02 4.016E-02 1.366E-02 8.953E-03 
7.375E-01 9.905E+01 9.821E+01 1.026E+02 8.680E+00 8.119E+00 4.283E+00 1.822E+00 1.346E+00 7.621E-01 5.752E-02 3.967E-02 1.521E-02 4.436E-03 
7.625E-01 8.613E+01 8.786E+01 9.138E+01 7.393E+00 7.422E+00 3.883E+00 1.748E+00 1.222E+00 7.368E-01 4.311E-02 2.939E-02 1.338E-02 8.792E-03 
7.875E-01 7.688E+01 7.796E+01 8.045E+01 6.944E+00 6.798E+00 3.522E+00 1.390E+00 1.123E+00 6.437E-01 5.034E-02 3.098E-02 6.621E-03 5.809E-03 
8.125E-01 6.851E+01 6.970E+01 7.157E+01 6.263E+00 5.933E+00 3.212E+00 1.351E+00 1.074E+00 6.290E-01 4.401E-02 1.913E-03 8.192E-03 2.879E-03 
8.375E-01 6.031E+01 6.143E+01 6.382E+01 5.411E+00 5.381E+00 2.822E+00 1.229E+00 9.211E-01 5.168E-01 5.295E-02 2.080E-02 3.244E-03 4.280E-03 
8.625E-01 5.490E+01 5.507E+01 5.657E+01 5.123E+00 4.865E+00 2.602E+00 1.121E+00 8.354E-01 4.794E-01 4.487E-02 2.056E-02 9.637E-03 4.243E-03 
8.875E-01 4.840E+01 4.853E+01 5.078E+01 4.370E+00 4.261E+00 2.556E+00 1.099E+00 7.782E-01 4.755E-01 3.697E-02 2.033E-02 0.000E+00 2.805E-03 
9.125E-01 4.284E+01 4.269E+01 4.505E+01 3.812E+00 4.014E+00 2.105E+00 9.525E-01 7.004E-01 4.616E-01 2.924E-02 1.462E-02 3.150E-03 1.390E-03 
9.375E-01 3.858E+01 3.825E+01 4.002E+01 3.595E+00 3.467E+00 1.980E+00 7.836E-01 6.641E-01 3.727E-01 2.349E-02 2.530E-02 0.000E+00 2.757E-03 
9.625E-01 3.459E+01 3.467E+01 3.577E+01 3.189E+00 3.130E+00 1.797E+00 7.218E-01 5.543E-01 3.584E-01 3.396E-02 1.251E-02 4.636E-03 1.367E-03 
9.875E-01 3.092E+01 3.112E+01 3.226E+01 2.965E+00 2.792E+00 1.681E+00 6.913E-01 5.767E-01 3.561E-01 2.122E-02 1.061E-02 7.653E-03 4.067E-03 
1.012E+00 2.774E+01 2.830E+01 2.861E+01 2.724E+00 2.664E+00 1.442E+00 5.301E-01 4.820E-01 2.372E-01 2.973E-02 8.745E-03 3.033E-03 1.344E-03 
1.038E+00 2.449E+01 2.449E+01 2.598E+01 2.371E+00 2.294E+00 1.298E+00 5.049E-01 4.404E-01 2.174E-01 1.903E-02 1.211E-02 6.010E-03 0.000E+00 
1.062E+00 2.186E+01 2.242E+01 2.281E+01 2.145E+00 2.037E+00 1.270E+00 4.884E-01 4.186E-01 2.112E-01 1.541E-02 6.849E-03 4.466E-03 0.000E+00 
1.087E+00 2.000E+01 2.008E+01 2.070E+01 1.871E+00 1.853E+00 1.089E+00 4.482E-01 3.750E-01 1.847E-01 2.033E-02 1.017E-02 4.426E-03 2.624E-03 
1.113E+00 1.769E+01 1.803E+01 1.827E+01 1.645E+00 1.716E+00 1.080E+00 4.129E-01 3.635E-01 1.553E-01 1.174E-02 3.354E-03 1.462E-03 0.000E+00 
1.137E+00 1.588E+01 1.625E+01 1.633E+01 1.632E+00 1.552E+00 8.552E-01 3.589E-01 3.197E-01 1.427E-01 1.328E-02 4.980E-03 0.000E+00 0.000E+00 
1.163E+00 1.416E+01 1.442E+01 1.510E+01 1.457E+00 1.364E+00 7.907E-01 2.946E-01 2.537E-01 1.197E-01 9.859E-03 0.000E+00 2.872E-03 0.000E+00 
1.188E+00 1.253E+01 1.262E+01 1.340E+01 1.240E+00 1.278E+00 7.757E-01 2.662E-01 2.598E-01 1.325E-01 3.253E-03 4.880E-03 0.000E+00 1.271E-03 
1.213E+00 1.145E+01 1.149E+01 1.230E+01 1.181E+00 1.145E+00 7.420E-01 2.423E-01 2.308E-01 1.119E-01 6.442E-03 1.611E-03 0.000E+00 1.261E-03 
1.238E+00 1.036E+01 1.039E+01 1.086E+01 1.032E+00 9.522E-01 6.699E-01 2.470E-01 1.816E-01 1.003E-01 7.974E-03 0.000E+00 1.399E-03 1.251E-03 
1.262E+00 9.284E+00 9.394E+00 9.650E+00 9.803E-01 8.853E-01 5.365E-01 1.979E-01 2.002E-01 1.335E-01 6.317E-03 4.738E-03 0.000E+00 1.242E-03 
1.288E+00 8.159E+00 8.150E+00 9.064E+00 7.952E-01 8.158E-01 5.396E-01 2.085E-01 1.524E-01 8.130E-02 4.693E-03 1.564E-03 0.000E+00 0.000E+00 
1.312E+00 7.538E+00 7.732E+00 8.302E+00 7.797E-01 7.639E-01 4.685E-01 1.682E-01 1.408E-01 7.376E-02 3.099E-03 1.549E-03 1.364E-03 0.000E+00 
\end{verbatim}
}
\newpage
Au+Au \@ $E_{\rm CM}=200A~$GeV:
{\tiny
\begin{verbatim}
! m_t-m0 , 1/mt dN/dmt(pi+ pi- pi0 K+ K- P aP L+S0 a(L+S0) Xi- aXi- Om aOm) 
1.250E-02 5.014E+03 5.075E+03 5.311E+03 2.100E+02 1.726E+02 2.996E+01 1.546E+01 1.695E+01 8.409E+00 1.446E+00 8.370E-01 4.543E-01 2.881E-01 
3.750E-02 4.617E+03 4.635E+03 4.872E+03 2.045E+02 1.736E+02 3.067E+01 1.645E+01 1.635E+01 8.958E+00 1.349E+00 8.087E-01 4.148E-01 2.513E-01 
6.250E-02 4.031E+03 4.078E+03 4.290E+03 1.908E+02 1.643E+02 2.990E+01 1.631E+01 1.593E+01 8.940E+00 1.246E+00 8.160E-01 3.689E-01 2.133E-01 
8.750E-02 3.526E+03 3.550E+03 3.722E+03 1.796E+02 1.537E+02 2.918E+01 1.656E+01 1.530E+01 8.575E+00 1.131E+00 6.535E-01 3.317E-01 2.092E-01 
1.125E-01 3.057E+03 3.081E+03 3.231E+03 1.628E+02 1.407E+02 2.724E+01 1.612E+01 1.394E+01 8.330E+00 1.121E+00 6.966E-01 3.150E-01 1.889E-01 
1.375E-01 2.655E+03 2.662E+03 2.797E+03 1.490E+02 1.284E+02 2.652E+01 1.591E+01 1.288E+01 7.698E+00 1.054E+00 6.072E-01 2.222E-01 1.941E-01 
1.625E-01 2.299E+03 2.303E+03 2.420E+03 1.321E+02 1.173E+02 2.508E+01 1.539E+01 1.208E+01 7.693E+00 8.720E-01 5.530E-01 2.286E-01 1.387E-01 
1.875E-01 2.005E+03 2.016E+03 2.089E+03 1.203E+02 1.081E+02 2.295E+01 1.442E+01 1.083E+01 6.909E+00 7.999E-01 4.978E-01 2.372E-01 1.162E-01 
2.125E-01 1.738E+03 1.750E+03 1.821E+03 1.078E+02 9.636E+01 2.185E+01 1.350E+01 1.018E+01 6.444E+00 6.793E-01 4.415E-01 1.789E-01 1.115E-01 
2.375E-01 1.509E+03 1.525E+03 1.570E+03 9.575E+01 8.578E+01 2.039E+01 1.325E+01 9.551E+00 6.021E+00 5.987E-01 4.344E-01 1.494E-01 9.663E-02 
2.625E-01 1.313E+03 1.321E+03 1.366E+03 8.724E+01 7.906E+01 1.957E+01 1.225E+01 8.437E+00 5.874E+00 5.673E-01 3.563E-01 1.274E-01 7.517E-02 
2.875E-01 1.144E+03 1.156E+03 1.196E+03 7.679E+01 7.033E+01 1.836E+01 1.144E+01 8.092E+00 5.671E+00 4.452E-01 3.372E-01 1.236E-01 8.120E-02 
3.125E-01 9.950E+02 1.004E+03 1.042E+03 6.945E+01 6.255E+01 1.704E+01 1.057E+01 7.143E+00 4.957E+00 4.489E-01 3.028E-01 9.586E-02 6.534E-02 
3.375E-01 8.732E+02 8.758E+02 9.047E+02 6.193E+01 5.704E+01 1.639E+01 9.579E+00 6.639E+00 4.697E+00 4.552E-01 2.747E-01 1.119E-01 6.471E-02 
3.625E-01 7.654E+02 7.676E+02 7.912E+02 5.522E+01 5.119E+01 1.515E+01 9.348E+00 5.907E+00 4.296E+00 3.660E-01 2.191E-01 8.925E-02 4.930E-02 
3.875E-01 6.683E+02 6.692E+02 6.916E+02 4.972E+01 4.575E+01 1.401E+01 8.268E+00 5.702E+00 3.931E+00 3.276E-01 2.362E-01 8.397E-02 5.535E-02 
4.125E-01 5.820E+02 5.895E+02 6.079E+02 4.535E+01 4.095E+01 1.295E+01 7.798E+00 5.275E+00 3.688E+00 2.828E-01 1.927E-01 7.052E-02 5.484E-02 
4.375E-01 5.122E+02 5.170E+02 5.346E+02 3.839E+01 3.668E+01 1.186E+01 7.173E+00 4.898E+00 3.295E+00 2.664E-01 1.628E-01 5.329E-02 4.314E-02 
4.625E-01 4.487E+02 4.541E+02 4.668E+02 3.498E+01 3.325E+01 1.111E+01 6.693E+00 4.407E+00 3.024E+00 2.043E-01 1.727E-01 4.456E-02 2.533E-02 
4.875E-01 3.936E+02 3.971E+02 4.131E+02 3.124E+01 2.944E+01 1.061E+01 6.205E+00 3.972E+00 2.683E+00 1.943E-01 1.559E-01 3.604E-02 3.138E-02 
5.125E-01 3.465E+02 3.479E+02 3.636E+02 2.829E+01 2.777E+01 9.818E+00 5.335E+00 3.459E+00 2.419E+00 1.632E-01 1.301E-01 4.750E-02 2.954E-02 
5.375E-01 3.086E+02 3.094E+02 3.165E+02 2.573E+01 2.401E+01 8.588E+00 5.002E+00 3.313E+00 2.283E+00 1.540E-01 9.802E-02 4.696E-02 2.620E-02 
5.625E-01 2.717E+02 2.718E+02 2.833E+02 2.191E+01 2.168E+01 8.197E+00 4.633E+00 2.954E+00 2.045E+00 1.520E-01 9.671E-02 2.515E-02 1.527E-02 
5.875E-01 2.400E+02 2.411E+02 2.463E+02 1.953E+01 1.956E+01 7.632E+00 4.376E+00 2.624E+00 1.873E+00 1.023E-01 1.204E-01 4.210E-02 9.084E-03 
6.125E-01 2.110E+02 2.126E+02 2.181E+02 1.761E+01 1.732E+01 6.790E+00 3.996E+00 2.406E+00 1.656E+00 1.368E-01 8.972E-02 1.892E-02 1.951E-02 
6.375E-01 1.864E+02 1.881E+02 1.946E+02 1.622E+01 1.520E+01 6.287E+00 3.672E+00 2.360E+00 1.632E+00 1.085E-01 5.536E-02 1.498E-02 1.934E-02 
6.625E-01 1.651E+02 1.665E+02 1.726E+02 1.419E+01 1.409E+01 5.870E+00 3.266E+00 2.207E+00 1.395E+00 9.620E-02 4.154E-02 1.111E-02 7.376E-03 
6.875E-01 1.475E+02 1.490E+02 1.528E+02 1.305E+01 1.284E+01 5.452E+00 2.779E+00 1.848E+00 1.208E+00 8.420E-02 6.045E-02 1.099E-02 1.609E-02 
7.125E-01 1.312E+02 1.319E+02 1.361E+02 1.176E+01 1.180E+01 5.006E+00 2.769E+00 1.582E+00 1.175E+00 7.677E-02 3.625E-02 1.632E-02 7.252E-03 
7.375E-01 1.171E+02 1.154E+02 1.197E+02 1.049E+01 1.042E+01 4.420E+00 2.487E+00 1.509E+00 1.071E+00 5.266E-02 3.581E-02 1.256E-02 5.754E-03 
7.625E-01 1.027E+02 1.037E+02 1.078E+02 9.369E+00 9.311E+00 3.956E+00 2.125E+00 1.388E+00 9.620E-01 3.954E-02 4.786E-02 1.421E-02 9.986E-03 
7.875E-01 9.156E+01 9.261E+01 9.527E+01 8.232E+00 8.411E+00 3.758E+00 2.020E+00 1.338E+00 8.881E-01 6.580E-02 2.673E-02 8.789E-03 8.489E-03 
8.125E-01 8.043E+01 8.277E+01 8.548E+01 7.591E+00 7.591E+00 3.352E+00 1.951E+00 1.177E+00 7.555E-01 3.861E-02 4.268E-02 1.044E-02 5.614E-03 
8.375E-01 7.315E+01 7.416E+01 7.705E+01 6.861E+00 6.806E+00 3.095E+00 1.658E+00 1.131E+00 7.923E-01 3.816E-02 3.816E-02 1.206E-02 1.392E-03 
8.625E-01 6.395E+01 6.570E+01 6.783E+01 6.212E+00 6.272E+00 2.704E+00 1.443E+00 8.981E-01 6.643E-01 3.574E-02 3.772E-02 6.823E-03 5.524E-03 
8.875E-01 5.857E+01 5.890E+01 6.055E+01 5.639E+00 5.627E+00 2.473E+00 1.305E+00 9.646E-01 5.826E-01 3.141E-02 2.748E-02 1.351E-02 6.850E-03 
9.125E-01 5.207E+01 5.248E+01 5.386E+01 5.189E+00 5.038E+00 2.285E+00 1.093E+00 8.547E-01 5.136E-01 3.494E-02 2.523E-02 0.000E+00 0.000E+00 
9.375E-01 4.662E+01 4.612E+01 4.814E+01 4.606E+00 4.542E+00 2.151E+00 1.173E+00 8.000E-01 5.390E-01 2.495E-02 9.597E-03 6.627E-03 1.349E-03 
9.625E-01 4.150E+01 4.169E+01 4.341E+01 4.019E+00 4.046E+00 1.970E+00 9.554E-01 6.968E-01 4.618E-01 1.898E-02 1.329E-02 4.923E-03 4.015E-03 
9.875E-01 3.693E+01 3.733E+01 3.961E+01 3.546E+00 3.642E+00 1.697E+00 9.992E-01 6.330E-01 4.152E-01 2.629E-02 1.878E-02 4.877E-03 1.328E-03 
1.012E+00 3.304E+01 3.414E+01 3.518E+01 3.343E+00 3.386E+00 1.649E+00 8.113E-01 5.748E-01 4.103E-01 1.672E-02 1.486E-02 8.053E-03 1.318E-03 
1.038E+00 2.968E+01 2.975E+01 3.194E+01 2.987E+00 3.012E+00 1.475E+00 7.288E-01 4.979E-01 3.674E-01 2.573E-02 1.286E-02 3.191E-03 3.924E-03 
1.062E+00 2.683E+01 2.743E+01 2.808E+01 2.678E+00 2.761E+00 1.303E+00 6.419E-01 4.545E-01 3.195E-01 1.818E-02 5.455E-03 3.162E-03 3.894E-03 
1.087E+00 2.455E+01 2.452E+01 2.563E+01 2.441E+00 2.488E+00 1.287E+00 5.614E-01 4.120E-01 2.590E-01 1.260E-02 1.260E-02 3.133E-03 3.865E-03 
1.113E+00 2.154E+01 2.178E+01 2.276E+01 2.285E+00 2.468E+00 1.168E+00 5.756E-01 3.337E-01 2.600E-01 1.959E-02 1.425E-02 1.553E-03 0.000E+00 
1.137E+00 1.970E+01 1.992E+01 2.088E+01 1.958E+00 2.078E+00 1.012E+00 4.583E-01 3.684E-01 2.571E-01 7.051E-03 1.763E-03 0.000E+00 1.269E-03 
1.163E+00 1.765E+01 1.812E+01 1.849E+01 1.730E+00 1.845E+00 9.653E-01 4.899E-01 3.871E-01 1.992E-01 1.222E-02 1.222E-02 4.576E-03 0.000E+00 
1.188E+00 1.572E+01 1.627E+01 1.654E+01 1.566E+00 1.679E+00 8.930E-01 4.048E-01 3.285E-01 1.971E-01 1.209E-02 8.638E-03 3.024E-03 2.502E-03 
1.213E+00 1.434E+01 1.478E+01 1.531E+01 1.507E+00 1.611E+00 7.318E-01 3.498E-01 2.525E-01 1.300E-01 1.368E-02 1.539E-02 1.499E-03 3.726E-03 
1.238E+00 1.275E+01 1.343E+01 1.340E+01 1.366E+00 1.468E+00 7.731E-01 3.537E-01 2.756E-01 1.488E-01 8.469E-03 6.775E-03 0.000E+00 0.000E+00 
1.262E+00 1.127E+01 1.218E+01 1.240E+01 1.366E+00 1.137E+00 6.562E-01 3.124E-01 1.945E-01 1.745E-01 1.174E-02 6.709E-03 1.473E-03 1.225E-03 
1.288E+00 1.047E+01 1.072E+01 1.112E+01 1.048E+00 1.099E+00 5.945E-01 2.603E-01 1.871E-01 1.367E-01 8.306E-03 1.661E-03 0.000E+00 0.000E+00 
1.312E+00 9.627E+00 9.654E+00 1.017E+01 1.089E+00 1.127E+00 5.437E-01 2.613E-01 1.656E-01 1.193E-01 6.582E-03 6.582E-03 2.897E-03 3.622E-03 
\end{verbatim}
}
\newpage
\subsection*{Rapidity spectra}

For each beam energy there is a table containing rapidity spectra for all particle species.

Au+Au \@ $E_{\rm lab}=2A~$GeV:
{\tiny
\begin{verbatim}
! y, dN/dy (pi+ pi- pi0 K+ K- P aP L+S0 a(L+S0) Xi- aXi- Om aOm) 
-3.125E+00 4.000E-04 4.000E-04 0.000E+00 0.000E+00 0.000E+00 0.000E+00 0.000E+00 0.000E+00 0.000E+00 0.000E+00 0.000E+00 0.000E+00 0.000E+00 
-2.875E+00 4.000E-04 8.000E-04 1.200E-03 0.000E+00 0.000E+00 0.000E+00 0.000E+00 0.000E+00 0.000E+00 0.000E+00 0.000E+00 0.000E+00 0.000E+00 
-2.625E+00 8.800E-03 1.360E-02 1.000E-02 0.000E+00 0.000E+00 0.000E+00 0.000E+00 0.000E+00 0.000E+00 0.000E+00 0.000E+00 0.000E+00 0.000E+00 
-2.375E+00 4.640E-02 6.680E-02 5.320E-02 0.000E+00 0.000E+00 0.000E+00 0.000E+00 0.000E+00 0.000E+00 0.000E+00 0.000E+00 0.000E+00 0.000E+00 
-2.125E+00 1.692E-01 2.436E-01 1.984E-01 4.000E-04 0.000E+00 0.000E+00 0.000E+00 0.000E+00 0.000E+00 0.000E+00 0.000E+00 0.000E+00 0.000E+00 
-1.875E+00 5.528E-01 7.908E-01 6.536E-01 0.000E+00 0.000E+00 8.000E-04 0.000E+00 0.000E+00 0.000E+00 0.000E+00 0.000E+00 0.000E+00 0.000E+00 
-1.625E+00 1.263E+00 1.941E+00 1.650E+00 6.000E-03 0.000E+00 6.040E-02 0.000E+00 4.000E-04 0.000E+00 0.000E+00 0.000E+00 0.000E+00 0.000E+00 
-1.375E+00 2.559E+00 3.878E+00 3.390E+00 3.920E-02 1.200E-03 1.229E+00 0.000E+00 3.200E-03 0.000E+00 0.000E+00 0.000E+00 0.000E+00 0.000E+00 
-1.125E+00 4.521E+00 6.836E+00 5.796E+00 1.296E-01 2.000E-03 1.396E+01 0.000E+00 3.720E-02 0.000E+00 0.000E+00 0.000E+00 0.000E+00 0.000E+00 
-8.750E-01 6.688E+00 1.025E+01 8.812E+00 3.228E-01 8.800E-03 4.767E+01 0.000E+00 2.444E-01 0.000E+00 0.000E+00 0.000E+00 0.000E+00 0.000E+00 
-6.250E-01 8.902E+00 1.376E+01 1.153E+01 6.084E-01 1.360E-02 7.263E+01 0.000E+00 7.048E-01 0.000E+00 0.000E+00 0.000E+00 0.000E+00 0.000E+00 
-3.750E-01 1.106E+01 1.691E+01 1.410E+01 7.996E-01 1.920E-02 9.606E+01 0.000E+00 1.274E+00 0.000E+00 4.000E-04 0.000E+00 0.000E+00 0.000E+00 
-1.250E-01 1.204E+01 1.874E+01 1.545E+01 9.032E-01 2.680E-02 1.081E+02 0.000E+00 1.654E+00 0.000E+00 4.000E-04 0.000E+00 0.000E+00 0.000E+00 
1.250E-01 1.203E+01 1.874E+01 1.539E+01 9.588E-01 2.600E-02 1.083E+02 0.000E+00 1.585E+00 0.000E+00 8.000E-04 0.000E+00 0.000E+00 0.000E+00 
3.750E-01 1.116E+01 1.701E+01 1.411E+01 8.156E-01 1.800E-02 9.600E+01 0.000E+00 1.256E+00 0.000E+00 8.000E-04 0.000E+00 0.000E+00 0.000E+00 
6.250E-01 8.837E+00 1.403E+01 1.162E+01 5.940E-01 1.720E-02 7.223E+01 0.000E+00 7.216E-01 0.000E+00 4.000E-04 0.000E+00 0.000E+00 0.000E+00 
8.750E-01 6.620E+00 1.032E+01 8.682E+00 3.392E-01 8.000E-03 4.795E+01 0.000E+00 2.292E-01 0.000E+00 0.000E+00 0.000E+00 0.000E+00 0.000E+00 
1.125E+00 4.416E+00 6.866E+00 5.735E+00 1.412E-01 2.400E-03 1.388E+01 0.000E+00 3.400E-02 0.000E+00 0.000E+00 0.000E+00 0.000E+00 0.000E+00 
1.375E+00 2.609E+00 4.016E+00 3.346E+00 3.320E-02 1.200E-03 1.213E+00 0.000E+00 3.200E-03 0.000E+00 0.000E+00 0.000E+00 0.000E+00 0.000E+00 
1.625E+00 1.258E+00 1.896E+00 1.626E+00 6.800E-03 4.000E-04 5.680E-02 0.000E+00 0.000E+00 0.000E+00 0.000E+00 0.000E+00 0.000E+00 0.000E+00 
1.875E+00 5.088E-01 7.932E-01 6.276E-01 0.000E+00 0.000E+00 2.000E-03 0.000E+00 0.000E+00 0.000E+00 0.000E+00 0.000E+00 0.000E+00 0.000E+00 
2.125E+00 1.760E-01 2.208E-01 2.088E-01 4.000E-04 0.000E+00 0.000E+00 0.000E+00 0.000E+00 0.000E+00 0.000E+00 0.000E+00 0.000E+00 0.000E+00 
2.375E+00 3.320E-02 6.200E-02 4.560E-02 0.000E+00 0.000E+00 0.000E+00 0.000E+00 0.000E+00 0.000E+00 0.000E+00 0.000E+00 0.000E+00 0.000E+00 
2.625E+00 9.600E-03 1.200E-02 1.360E-02 0.000E+00 0.000E+00 0.000E+00 0.000E+00 0.000E+00 0.000E+00 0.000E+00 0.000E+00 0.000E+00 0.000E+00 
2.875E+00 1.200E-03 1.200E-03 2.000E-03 0.000E+00 0.000E+00 0.000E+00 0.000E+00 0.000E+00 0.000E+00 0.000E+00 0.000E+00 0.000E+00 0.000E+00 
\end{verbatim}
}

Au+Au \@ $E_{\rm lab}=4A~$GeV:
{\tiny
\begin{verbatim}
! y, dN/dy (pi+ pi- pi0 K+ K- P aP L+S0 a(L+S0) Xi- aXi- Om aOm) 
-3.125E+00 3.600E-03 4.000E-03 4.000E-03 0.000E+00 0.000E+00 0.000E+00 0.000E+00 0.000E+00 0.000E+00 0.000E+00 0.000E+00 0.000E+00 0.000E+00 
-2.875E+00 2.240E-02 2.480E-02 2.400E-02 0.000E+00 0.000E+00 0.000E+00 0.000E+00 0.000E+00 0.000E+00 0.000E+00 0.000E+00 0.000E+00 0.000E+00 
-2.625E+00 8.400E-02 1.152E-01 1.184E-01 0.000E+00 0.000E+00 0.000E+00 0.000E+00 0.000E+00 0.000E+00 0.000E+00 0.000E+00 0.000E+00 0.000E+00 
-2.375E+00 3.380E-01 4.312E-01 4.168E-01 0.000E+00 0.000E+00 0.000E+00 0.000E+00 0.000E+00 0.000E+00 0.000E+00 0.000E+00 0.000E+00 0.000E+00 
-2.125E+00 9.788E-01 1.237E+00 1.162E+00 2.000E-03 0.000E+00 2.400E-03 0.000E+00 0.000E+00 0.000E+00 0.000E+00 0.000E+00 0.000E+00 0.000E+00 
-1.875E+00 2.271E+00 2.961E+00 2.696E+00 1.800E-02 3.200E-03 7.840E-02 0.000E+00 0.000E+00 0.000E+00 0.000E+00 0.000E+00 0.000E+00 0.000E+00 
-1.625E+00 4.478E+00 5.793E+00 5.352E+00 1.088E-01 8.800E-03 1.258E+00 0.000E+00 6.000E-03 0.000E+00 0.000E+00 0.000E+00 0.000E+00 0.000E+00 
-1.375E+00 7.777E+00 1.013E+01 9.405E+00 4.048E-01 2.800E-02 1.312E+01 0.000E+00 1.052E-01 0.000E+00 4.000E-04 0.000E+00 0.000E+00 0.000E+00 
-1.125E+00 1.217E+01 1.603E+01 1.452E+01 9.976E-01 7.480E-02 3.625E+01 0.000E+00 6.032E-01 0.000E+00 4.000E-04 0.000E+00 0.000E+00 0.000E+00 
-8.750E-01 1.698E+01 2.221E+01 2.031E+01 2.052E+00 1.532E-01 4.852E+01 4.000E-04 1.858E+00 0.000E+00 1.600E-03 0.000E+00 0.000E+00 0.000E+00 
-6.250E-01 2.158E+01 2.812E+01 2.538E+01 3.082E+00 2.604E-01 6.790E+01 4.000E-04 3.709E+00 0.000E+00 8.400E-03 0.000E+00 0.000E+00 0.000E+00 
-3.750E-01 2.534E+01 3.321E+01 2.976E+01 3.917E+00 3.096E-01 8.319E+01 4.000E-04 5.390E+00 0.000E+00 1.320E-02 0.000E+00 0.000E+00 0.000E+00 
-1.250E-01 2.733E+01 3.596E+01 3.210E+01 4.382E+00 3.824E-01 9.035E+01 0.000E+00 6.434E+00 0.000E+00 1.840E-02 0.000E+00 0.000E+00 0.000E+00 
1.250E-01 2.734E+01 3.616E+01 3.241E+01 4.368E+00 3.640E-01 9.031E+01 0.000E+00 6.454E+00 0.000E+00 1.960E-02 0.000E+00 4.000E-04 0.000E+00 
3.750E-01 2.527E+01 3.319E+01 2.960E+01 3.979E+00 3.116E-01 8.273E+01 4.000E-04 5.402E+00 0.000E+00 1.720E-02 0.000E+00 0.000E+00 0.000E+00 
6.250E-01 2.168E+01 2.839E+01 2.544E+01 3.102E+00 2.368E-01 6.807E+01 0.000E+00 3.772E+00 0.000E+00 1.160E-02 0.000E+00 0.000E+00 0.000E+00 
8.750E-01 1.692E+01 2.224E+01 2.018E+01 2.076E+00 1.512E-01 4.903E+01 0.000E+00 1.881E+00 4.000E-04 2.800E-03 0.000E+00 0.000E+00 0.000E+00 
1.125E+00 1.211E+01 1.588E+01 1.442E+01 1.017E+00 7.800E-02 3.638E+01 0.000E+00 6.096E-01 0.000E+00 0.000E+00 0.000E+00 0.000E+00 0.000E+00 
1.375E+00 7.856E+00 1.010E+01 9.337E+00 4.188E-01 2.080E-02 1.286E+01 0.000E+00 1.012E-01 0.000E+00 0.000E+00 0.000E+00 0.000E+00 0.000E+00 
1.625E+00 4.502E+00 5.825E+00 5.320E+00 1.136E-01 1.200E-02 1.239E+00 0.000E+00 7.600E-03 0.000E+00 0.000E+00 0.000E+00 0.000E+00 0.000E+00 
1.875E+00 2.276E+00 2.900E+00 2.729E+00 2.440E-02 4.000E-04 6.720E-02 0.000E+00 0.000E+00 0.000E+00 0.000E+00 0.000E+00 0.000E+00 0.000E+00 
2.125E+00 9.584E-01 1.238E+00 1.115E+00 2.000E-03 0.000E+00 3.600E-03 0.000E+00 0.000E+00 0.000E+00 0.000E+00 0.000E+00 0.000E+00 0.000E+00 
2.375E+00 3.412E-01 4.408E-01 4.084E-01 0.000E+00 0.000E+00 0.000E+00 0.000E+00 0.000E+00 0.000E+00 0.000E+00 0.000E+00 0.000E+00 0.000E+00 
2.625E+00 9.720E-02 1.048E-01 1.176E-01 0.000E+00 0.000E+00 0.000E+00 0.000E+00 0.000E+00 0.000E+00 0.000E+00 0.000E+00 0.000E+00 0.000E+00 
2.875E+00 2.360E-02 2.240E-02 2.320E-02 0.000E+00 0.000E+00 0.000E+00 0.000E+00 0.000E+00 0.000E+00 0.000E+00 0.000E+00 0.000E+00 0.000E+00 
3.125E+00 4.000E-03 6.400E-03 2.000E-03 0.000E+00 0.000E+00 0.000E+00 0.000E+00 0.000E+00 0.000E+00 0.000E+00 0.000E+00 0.000E+00 0.000E+00 
\end{verbatim}
}
\newpage
Au+Au \@ $E_{\rm lab}=6A~$GeV:
{\tiny
\begin{verbatim}
! y, dN/dy (pi+ pi- pi0 K+ K- P aP L+S0 a(L+S0) Xi- aXi- Om aOm) 
-3.375E+00 3.600E-03 2.800E-03 1.600E-03 0.000E+00 0.000E+00 0.000E+00 0.000E+00 0.000E+00 0.000E+00 0.000E+00 0.000E+00 0.000E+00 0.000E+00 
-3.125E+00 1.480E-02 1.720E-02 1.800E-02 0.000E+00 0.000E+00 0.000E+00 0.000E+00 0.000E+00 0.000E+00 0.000E+00 0.000E+00 0.000E+00 0.000E+00 
-2.875E+00 6.560E-02 9.480E-02 8.320E-02 0.000E+00 0.000E+00 0.000E+00 0.000E+00 0.000E+00 0.000E+00 0.000E+00 0.000E+00 0.000E+00 0.000E+00 
-2.625E+00 2.944E-01 3.468E-01 3.180E-01 0.000E+00 0.000E+00 0.000E+00 0.000E+00 0.000E+00 0.000E+00 0.000E+00 0.000E+00 0.000E+00 0.000E+00 
-2.375E+00 8.484E-01 1.033E+00 9.804E-01 1.600E-03 0.000E+00 8.000E-04 0.000E+00 0.000E+00 0.000E+00 0.000E+00 0.000E+00 0.000E+00 0.000E+00 
-2.125E+00 2.144E+00 2.623E+00 2.501E+00 8.400E-03 8.000E-04 2.440E-02 0.000E+00 0.000E+00 0.000E+00 0.000E+00 0.000E+00 0.000E+00 0.000E+00 
-1.875E+00 4.484E+00 5.408E+00 5.233E+00 8.000E-02 6.400E-03 5.324E-01 0.000E+00 3.600E-03 0.000E+00 0.000E+00 0.000E+00 0.000E+00 0.000E+00 
-1.625E+00 8.108E+00 9.942E+00 9.618E+00 3.112E-01 3.560E-02 6.160E+00 0.000E+00 5.160E-02 0.000E+00 0.000E+00 0.000E+00 0.000E+00 0.000E+00 
-1.375E+00 1.334E+01 1.627E+01 1.556E+01 9.188E-01 1.292E-01 2.752E+01 4.000E-04 4.188E-01 0.000E+00 4.000E-04 0.000E+00 0.000E+00 0.000E+00 
-1.125E+00 1.957E+01 2.392E+01 2.269E+01 2.118E+00 2.596E-01 3.515E+01 0.000E+00 1.508E+00 0.000E+00 2.400E-03 0.000E+00 0.000E+00 0.000E+00 
-8.750E-01 2.628E+01 3.222E+01 3.011E+01 3.699E+00 4.600E-01 4.871E+01 4.000E-04 3.482E+00 0.000E+00 1.360E-02 0.000E+00 0.000E+00 0.000E+00 
-6.250E-01 3.213E+01 3.937E+01 3.651E+01 5.331E+00 6.924E-01 6.370E+01 8.000E-04 6.048E+00 0.000E+00 2.800E-02 0.000E+00 4.000E-04 0.000E+00 
-3.750E-01 3.667E+01 4.534E+01 4.160E+01 6.370E+00 8.224E-01 7.517E+01 8.000E-04 8.123E+00 8.000E-04 4.680E-02 0.000E+00 1.200E-03 0.000E+00 
-1.250E-01 3.913E+01 4.853E+01 4.450E+01 7.003E+00 9.504E-01 8.117E+01 4.000E-04 9.568E+00 0.000E+00 5.320E-02 0.000E+00 4.000E-03 0.000E+00 
1.250E-01 3.927E+01 4.856E+01 4.447E+01 6.883E+00 9.432E-01 8.111E+01 0.000E+00 9.463E+00 0.000E+00 5.640E-02 0.000E+00 2.000E-03 0.000E+00 
3.750E-01 3.657E+01 4.520E+01 4.185E+01 6.320E+00 8.424E-01 7.509E+01 4.000E-04 8.266E+00 4.000E-04 5.200E-02 0.000E+00 4.000E-04 0.000E+00 
6.250E-01 3.203E+01 3.960E+01 3.650E+01 5.338E+00 6.748E-01 6.354E+01 8.000E-04 5.930E+00 0.000E+00 2.840E-02 0.000E+00 0.000E+00 0.000E+00 
8.750E-01 2.609E+01 3.218E+01 2.987E+01 3.656E+00 4.728E-01 4.851E+01 4.000E-04 3.467E+00 0.000E+00 1.360E-02 0.000E+00 0.000E+00 0.000E+00 
1.125E+00 1.940E+01 2.396E+01 2.272E+01 2.122E+00 2.548E-01 3.507E+01 4.000E-04 1.465E+00 0.000E+00 3.200E-03 0.000E+00 0.000E+00 0.000E+00 
1.375E+00 1.339E+01 1.616E+01 1.552E+01 9.504E-01 1.096E-01 2.753E+01 0.000E+00 4.252E-01 0.000E+00 0.000E+00 0.000E+00 0.000E+00 0.000E+00 
1.625E+00 8.137E+00 9.955E+00 9.501E+00 3.140E-01 4.040E-02 6.107E+00 0.000E+00 4.600E-02 0.000E+00 0.000E+00 0.000E+00 0.000E+00 0.000E+00 
1.875E+00 4.468E+00 5.386E+00 5.076E+00 8.760E-02 1.120E-02 5.208E-01 0.000E+00 1.600E-03 0.000E+00 0.000E+00 0.000E+00 0.000E+00 0.000E+00 
2.125E+00 2.142E+00 2.645E+00 2.431E+00 1.160E-02 4.000E-04 2.600E-02 0.000E+00 0.000E+00 0.000E+00 0.000E+00 0.000E+00 0.000E+00 0.000E+00 
2.375E+00 8.716E-01 1.056E+00 9.708E-01 2.000E-03 0.000E+00 0.000E+00 0.000E+00 0.000E+00 0.000E+00 0.000E+00 0.000E+00 0.000E+00 0.000E+00 
2.625E+00 2.932E-01 3.232E-01 3.104E-01 0.000E+00 0.000E+00 0.000E+00 0.000E+00 0.000E+00 0.000E+00 0.000E+00 0.000E+00 0.000E+00 0.000E+00 
2.875E+00 7.240E-02 9.000E-02 8.240E-02 0.000E+00 0.000E+00 0.000E+00 0.000E+00 0.000E+00 0.000E+00 0.000E+00 0.000E+00 0.000E+00 0.000E+00 
3.125E+00 1.200E-02 2.320E-02 1.640E-02 0.000E+00 0.000E+00 0.000E+00 0.000E+00 0.000E+00 0.000E+00 0.000E+00 0.000E+00 0.000E+00 0.000E+00 
3.375E+00 1.200E-03 3.200E-03 2.400E-03 0.000E+00 0.000E+00 0.000E+00 0.000E+00 0.000E+00 0.000E+00 0.000E+00 0.000E+00 0.000E+00 0.000E+00 
3.625E+00 8.000E-04 0.000E+00 0.000E+00 0.000E+00 0.000E+00 0.000E+00 0.000E+00 0.000E+00 0.000E+00 0.000E+00 0.000E+00 0.000E+00 0.000E+00 

\end{verbatim}
}
\newpage
Au+Au \@ $E_{\rm lab}=8A~$GeV:
{\tiny
\begin{verbatim}
! y, dN/dy (pi+ pi- pi0 K+ K- P aP L+S0 a(L+S0) Xi- aXi- Om aOm) 
-3.625E+00 2.000E-03 1.200E-03 1.200E-03 0.000E+00 0.000E+00 0.000E+00 0.000E+00 0.000E+00 0.000E+00 0.000E+00 0.000E+00 0.000E+00 0.000E+00 
-3.375E+00 6.800E-03 9.200E-03 6.000E-03 0.000E+00 0.000E+00 0.000E+00 0.000E+00 0.000E+00 0.000E+00 0.000E+00 0.000E+00 0.000E+00 0.000E+00 
-3.125E+00 4.040E-02 4.320E-02 4.400E-02 0.000E+00 0.000E+00 0.000E+00 0.000E+00 0.000E+00 0.000E+00 0.000E+00 0.000E+00 0.000E+00 0.000E+00 
-2.875E+00 1.484E-01 1.960E-01 1.852E-01 0.000E+00 0.000E+00 0.000E+00 0.000E+00 0.000E+00 0.000E+00 0.000E+00 0.000E+00 0.000E+00 0.000E+00 
-2.625E+00 5.828E-01 6.664E-01 6.280E-01 0.000E+00 0.000E+00 0.000E+00 0.000E+00 0.000E+00 0.000E+00 0.000E+00 0.000E+00 0.000E+00 0.000E+00 
-2.375E+00 1.520E+00 1.806E+00 1.750E+00 4.000E-03 8.000E-04 7.600E-03 0.000E+00 0.000E+00 0.000E+00 0.000E+00 0.000E+00 0.000E+00 0.000E+00 
-2.125E+00 3.452E+00 4.152E+00 4.110E+00 3.000E-02 6.400E-03 1.384E-01 0.000E+00 4.000E-04 0.000E+00 0.000E+00 0.000E+00 0.000E+00 0.000E+00 
-1.875E+00 6.898E+00 8.047E+00 7.808E+00 1.808E-01 2.280E-02 1.738E+00 0.000E+00 1.480E-02 0.000E+00 0.000E+00 0.000E+00 0.000E+00 0.000E+00 
-1.625E+00 1.193E+01 1.407E+01 1.357E+01 6.064E-01 9.680E-02 1.581E+01 0.000E+00 1.752E-01 0.000E+00 0.000E+00 0.000E+00 0.000E+00 0.000E+00 
-1.375E+00 1.841E+01 2.179E+01 2.127E+01 1.574E+00 2.592E-01 2.990E+01 4.000E-04 8.348E-01 0.000E+00 1.600E-03 0.000E+00 0.000E+00 0.000E+00 
-1.125E+00 2.595E+01 3.112E+01 2.969E+01 3.091E+00 5.076E-01 3.514E+01 1.600E-03 2.397E+00 4.000E-04 5.200E-03 0.000E+00 0.000E+00 0.000E+00 
-8.750E-01 3.388E+01 4.014E+01 3.814E+01 5.036E+00 8.212E-01 4.815E+01 4.000E-04 4.753E+00 4.000E-04 2.440E-02 0.000E+00 0.000E+00 4.000E-04 
-6.250E-01 4.043E+01 4.833E+01 4.519E+01 6.745E+00 1.151E+00 6.006E+01 8.000E-04 7.516E+00 4.000E-04 4.960E-02 4.000E-04 1.600E-03 0.000E+00 
-3.750E-01 4.562E+01 5.435E+01 5.103E+01 8.135E+00 1.405E+00 6.983E+01 2.000E-03 9.788E+00 1.600E-03 7.360E-02 4.000E-04 4.000E-04 0.000E+00 
-1.250E-01 4.858E+01 5.807E+01 5.408E+01 8.817E+00 1.551E+00 7.517E+01 8.000E-04 1.121E+01 0.000E+00 9.520E-02 0.000E+00 4.000E-03 0.000E+00 
1.250E-01 4.861E+01 5.808E+01 5.407E+01 8.845E+00 1.564E+00 7.488E+01 2.000E-03 1.090E+01 4.000E-04 9.680E-02 0.000E+00 2.000E-03 0.000E+00 
3.750E-01 4.573E+01 5.484E+01 5.120E+01 8.120E+00 1.420E+00 6.964E+01 2.000E-03 9.808E+00 1.600E-03 6.920E-02 0.000E+00 1.600E-03 0.000E+00 
6.250E-01 4.037E+01 4.850E+01 4.542E+01 6.721E+00 1.118E+00 6.015E+01 8.000E-04 7.498E+00 4.000E-04 4.240E-02 0.000E+00 2.000E-03 0.000E+00 
8.750E-01 3.373E+01 4.022E+01 3.820E+01 5.118E+00 7.888E-01 4.795E+01 1.200E-03 4.790E+00 0.000E+00 2.320E-02 0.000E+00 0.000E+00 0.000E+00 
1.125E+00 2.615E+01 3.100E+01 3.002E+01 3.126E+00 5.328E-01 3.505E+01 1.200E-03 2.328E+00 4.000E-04 1.040E-02 0.000E+00 0.000E+00 0.000E+00 
1.375E+00 1.848E+01 2.178E+01 2.112E+01 1.517E+00 2.400E-01 2.983E+01 8.000E-04 8.332E-01 0.000E+00 8.000E-04 0.000E+00 0.000E+00 0.000E+00 
1.625E+00 1.191E+01 1.405E+01 1.366E+01 6.168E-01 9.560E-02 1.569E+01 4.000E-04 1.500E-01 0.000E+00 0.000E+00 0.000E+00 0.000E+00 0.000E+00 
1.875E+00 6.750E+00 8.142E+00 7.863E+00 1.792E-01 2.360E-02 1.774E+00 0.000E+00 1.360E-02 0.000E+00 0.000E+00 0.000E+00 0.000E+00 0.000E+00 
2.125E+00 3.519E+00 4.141E+00 4.038E+00 3.480E-02 3.600E-03 1.336E-01 0.000E+00 8.000E-04 0.000E+00 0.000E+00 0.000E+00 0.000E+00 0.000E+00 
2.375E+00 1.526E+00 1.844E+00 1.763E+00 3.200E-03 1.200E-03 4.800E-03 0.000E+00 0.000E+00 0.000E+00 0.000E+00 0.000E+00 0.000E+00 0.000E+00 
2.625E+00 5.644E-01 6.696E-01 6.676E-01 4.000E-04 0.000E+00 0.000E+00 0.000E+00 0.000E+00 0.000E+00 0.000E+00 0.000E+00 0.000E+00 0.000E+00 
2.875E+00 1.724E-01 1.860E-01 1.884E-01 0.000E+00 0.000E+00 0.000E+00 0.000E+00 0.000E+00 0.000E+00 0.000E+00 0.000E+00 0.000E+00 0.000E+00 
3.125E+00 4.640E-02 4.440E-02 4.920E-02 0.000E+00 0.000E+00 0.000E+00 0.000E+00 0.000E+00 0.000E+00 0.000E+00 0.000E+00 0.000E+00 0.000E+00 
3.375E+00 6.800E-03 5.200E-03 1.000E-02 0.000E+00 0.000E+00 0.000E+00 0.000E+00 0.000E+00 0.000E+00 0.000E+00 0.000E+00 0.000E+00 0.000E+00 
3.625E+00 4.000E-04 1.200E-03 4.000E-04 0.000E+00 0.000E+00 0.000E+00 0.000E+00 0.000E+00 0.000E+00 0.000E+00 0.000E+00 0.000E+00 0.000E+00 
\end{verbatim}
}
\newpage
Au+Au \@ $E_{\rm lab}=11A~$GeV:
{\tiny
\begin{verbatim}
 ! y, dN/dy (pi+ pi- pi0 K+ K- P aP L+S0 a(L+S0) Xi- aXi- Om aOm) 
-3.875E+00 8.000E-04 0.000E+00 4.000E-04 0.000E+00 0.000E+00 0.000E+00 0.000E+00 0.000E+00 0.000E+00 0.000E+00 0.000E+00 0.000E+00 0.000E+00 
-3.625E+00 3.600E-03 3.600E-03 3.600E-03 0.000E+00 0.000E+00 0.000E+00 0.000E+00 0.000E+00 0.000E+00 0.000E+00 0.000E+00 0.000E+00 0.000E+00 
-3.375E+00 2.160E-02 2.720E-02 2.920E-02 0.000E+00 0.000E+00 0.000E+00 0.000E+00 0.000E+00 0.000E+00 0.000E+00 0.000E+00 0.000E+00 0.000E+00 
-3.125E+00 9.680E-02 1.092E-01 1.100E-01 0.000E+00 0.000E+00 0.000E+00 0.000E+00 0.000E+00 0.000E+00 0.000E+00 0.000E+00 0.000E+00 0.000E+00 
-2.875E+00 3.824E-01 4.260E-01 4.068E-01 4.000E-04 0.000E+00 0.000E+00 0.000E+00 0.000E+00 0.000E+00 0.000E+00 0.000E+00 0.000E+00 0.000E+00 
-2.625E+00 1.135E+00 1.292E+00 1.278E+00 1.200E-03 4.000E-04 4.000E-04 0.000E+00 0.000E+00 0.000E+00 0.000E+00 0.000E+00 0.000E+00 0.000E+00 
-2.375E+00 2.757E+00 3.205E+00 3.160E+00 1.160E-02 4.000E-03 3.920E-02 0.000E+00 0.000E+00 0.000E+00 0.000E+00 0.000E+00 0.000E+00 0.000E+00 
-2.125E+00 5.675E+00 6.633E+00 6.538E+00 9.360E-02 2.080E-02 6.868E-01 0.000E+00 5.200E-03 0.000E+00 0.000E+00 0.000E+00 0.000E+00 0.000E+00 
-1.875E+00 1.036E+01 1.198E+01 1.182E+01 3.776E-01 6.400E-02 7.225E+00 0.000E+00 7.040E-02 0.000E+00 0.000E+00 0.000E+00 0.000E+00 0.000E+00 
-1.625E+00 1.704E+01 1.946E+01 1.951E+01 1.140E+00 2.188E-01 2.602E+01 8.000E-04 4.720E-01 0.000E+00 4.000E-04 0.000E+00 0.000E+00 0.000E+00 
-1.375E+00 2.517E+01 2.917E+01 2.852E+01 2.514E+00 5.116E-01 2.797E+01 2.400E-03 1.626E+00 4.000E-04 5.600E-03 0.000E+00 0.000E+00 0.000E+00 
-1.125E+00 3.420E+01 3.958E+01 3.831E+01 4.444E+00 8.984E-01 3.651E+01 4.800E-03 3.542E+00 1.200E-03 1.520E-02 4.000E-04 4.000E-04 0.000E+00 
-8.750E-01 4.299E+01 4.957E+01 4.774E+01 6.684E+00 1.372E+00 4.682E+01 3.200E-03 6.238E+00 2.400E-03 4.720E-02 0.000E+00 2.000E-03 0.000E+00 
-6.250E-01 5.015E+01 5.844E+01 5.550E+01 8.398E+00 1.902E+00 5.605E+01 4.000E-03 8.964E+00 2.000E-03 7.400E-02 4.000E-04 2.400E-03 0.000E+00 
-3.750E-01 5.577E+01 6.484E+01 6.153E+01 9.888E+00 2.233E+00 6.352E+01 5.600E-03 1.108E+01 1.600E-03 1.092E-01 8.000E-04 3.600E-03 0.000E+00 
-1.250E-01 5.876E+01 6.880E+01 6.455E+01 1.041E+01 2.439E+00 6.777E+01 4.800E-03 1.230E+01 2.800E-03 1.332E-01 4.000E-04 9.200E-03 4.000E-04 
1.250E-01 5.900E+01 6.843E+01 6.516E+01 1.044E+01 2.337E+00 6.791E+01 6.800E-03 1.233E+01 4.400E-03 1.336E-01 0.000E+00 4.800E-03 0.000E+00 
3.750E-01 5.575E+01 6.484E+01 6.144E+01 9.898E+00 2.154E+00 6.347E+01 5.600E-03 1.109E+01 3.200E-03 1.160E-01 0.000E+00 6.000E-03 0.000E+00 
6.250E-01 5.006E+01 5.825E+01 5.567E+01 8.446E+00 1.825E+00 5.641E+01 6.800E-03 8.882E+00 2.000E-03 7.640E-02 0.000E+00 3.600E-03 0.000E+00 
8.750E-01 4.267E+01 4.945E+01 4.762E+01 6.610E+00 1.374E+00 4.691E+01 3.200E-03 6.145E+00 1.600E-03 4.600E-02 8.000E-04 4.000E-04 0.000E+00 
1.125E+00 3.425E+01 3.938E+01 3.838E+01 4.470E+00 8.840E-01 3.645E+01 3.200E-03 3.538E+00 4.000E-04 1.320E-02 0.000E+00 8.000E-04 0.000E+00 
1.375E+00 2.502E+01 2.889E+01 2.868E+01 2.482E+00 4.856E-01 2.795E+01 2.000E-03 1.549E+00 4.000E-04 4.000E-03 0.000E+00 0.000E+00 0.000E+00 
1.625E+00 1.704E+01 1.957E+01 1.932E+01 1.146E+00 2.068E-01 2.604E+01 0.000E+00 4.796E-01 0.000E+00 8.000E-04 0.000E+00 0.000E+00 0.000E+00 
1.875E+00 1.041E+01 1.201E+01 1.188E+01 4.132E-01 7.320E-02 7.198E+00 0.000E+00 6.440E-02 0.000E+00 0.000E+00 0.000E+00 0.000E+00 0.000E+00 
2.125E+00 5.597E+00 6.676E+00 6.487E+00 8.800E-02 2.040E-02 6.528E-01 0.000E+00 5.600E-03 0.000E+00 0.000E+00 0.000E+00 0.000E+00 0.000E+00 
2.375E+00 2.824E+00 3.211E+00 3.140E+00 1.800E-02 1.200E-03 4.080E-02 0.000E+00 4.000E-04 0.000E+00 0.000E+00 0.000E+00 0.000E+00 0.000E+00 
2.625E+00 1.114E+00 1.338E+00 1.232E+00 1.600E-03 0.000E+00 4.000E-04 0.000E+00 0.000E+00 0.000E+00 0.000E+00 0.000E+00 0.000E+00 0.000E+00 
2.875E+00 3.624E-01 4.192E-01 4.124E-01 0.000E+00 0.000E+00 0.000E+00 0.000E+00 0.000E+00 0.000E+00 0.000E+00 0.000E+00 0.000E+00 0.000E+00 
3.125E+00 9.280E-02 1.184E-01 1.064E-01 0.000E+00 0.000E+00 0.000E+00 0.000E+00 0.000E+00 0.000E+00 0.000E+00 0.000E+00 0.000E+00 0.000E+00 
3.375E+00 2.440E-02 2.240E-02 2.280E-02 0.000E+00 0.000E+00 0.000E+00 0.000E+00 0.000E+00 0.000E+00 0.000E+00 0.000E+00 0.000E+00 0.000E+00 
3.625E+00 1.600E-03 4.000E-03 5.600E-03 0.000E+00 0.000E+00 0.000E+00 0.000E+00 0.000E+00 0.000E+00 0.000E+00 0.000E+00 0.000E+00 0.000E+00 
3.875E+00 8.000E-04 4.000E-04 0.000E+00 0.000E+00 0.000E+00 0.000E+00 0.000E+00 0.000E+00 0.000E+00 0.000E+00 0.000E+00 0.000E+00 0.000E+00 
\end{verbatim}
}
\newpage
Pb+Pb \@ $E_{\rm lab}=20A~$GeV:
{\tiny
\begin{verbatim}
! y, dN/dy (pi+ pi- pi0 K+ K- P aP L+S0 a(L+S0) Xi- aXi- Om aOm) 
-4.125E+00 1.600E-03 8.000E-04 1.200E-03 0.000E+00 0.000E+00 0.000E+00 0.000E+00 0.000E+00 0.000E+00 0.000E+00 0.000E+00 0.000E+00 0.000E+00 
-3.875E+00 4.800E-03 8.400E-03 6.000E-03 0.000E+00 0.000E+00 0.000E+00 0.000E+00 0.000E+00 0.000E+00 0.000E+00 0.000E+00 0.000E+00 0.000E+00 
-3.625E+00 2.560E-02 3.640E-02 3.960E-02 0.000E+00 0.000E+00 0.000E+00 0.000E+00 0.000E+00 0.000E+00 0.000E+00 0.000E+00 0.000E+00 0.000E+00 
-3.375E+00 1.296E-01 1.632E-01 1.532E-01 0.000E+00 0.000E+00 0.000E+00 0.000E+00 0.000E+00 0.000E+00 0.000E+00 0.000E+00 0.000E+00 0.000E+00 
-3.125E+00 5.176E-01 5.484E-01 5.620E-01 1.200E-03 4.000E-04 0.000E+00 0.000E+00 0.000E+00 0.000E+00 0.000E+00 0.000E+00 0.000E+00 0.000E+00 
-2.875E+00 1.433E+00 1.681E+00 1.630E+00 3.200E-03 2.400E-03 1.200E-03 0.000E+00 0.000E+00 0.000E+00 0.000E+00 0.000E+00 0.000E+00 0.000E+00 
-2.625E+00 3.471E+00 4.008E+00 3.974E+00 1.880E-02 3.200E-03 5.440E-02 0.000E+00 4.000E-04 0.000E+00 0.000E+00 0.000E+00 0.000E+00 0.000E+00 
-2.375E+00 7.156E+00 8.171E+00 7.968E+00 1.164E-01 3.000E-02 8.328E-01 0.000E+00 7.600E-03 0.000E+00 4.000E-04 0.000E+00 0.000E+00 0.000E+00 
-2.125E+00 1.284E+01 1.435E+01 1.444E+01 4.892E-01 1.100E-01 9.043E+00 8.000E-04 9.400E-02 0.000E+00 0.000E+00 0.000E+00 0.000E+00 0.000E+00 
-1.875E+00 2.077E+01 2.363E+01 2.356E+01 1.487E+00 3.484E-01 2.566E+01 2.000E-03 5.972E-01 4.000E-04 1.200E-03 0.000E+00 0.000E+00 0.000E+00 
-1.625E+00 3.108E+01 3.475E+01 3.472E+01 3.087E+00 7.892E-01 2.496E+01 7.600E-03 1.844E+00 3.600E-03 1.000E-02 0.000E+00 0.000E+00 0.000E+00 
-1.375E+00 4.203E+01 4.730E+01 4.681E+01 5.401E+00 1.478E+00 3.080E+01 1.240E-02 3.693E+00 6.400E-03 2.960E-02 4.000E-04 4.000E-04 0.000E+00 
-1.125E+00 5.349E+01 6.013E+01 5.904E+01 7.962E+00 2.253E+00 3.816E+01 2.640E-02 6.172E+00 1.520E-02 6.120E-02 4.000E-04 1.600E-03 0.000E+00 
-8.750E-01 6.395E+01 7.158E+01 7.010E+01 1.041E+01 3.079E+00 4.551E+01 3.080E-02 9.000E+00 2.040E-02 1.108E-01 8.000E-04 4.800E-03 0.000E+00 
-6.250E-01 7.315E+01 8.219E+01 7.948E+01 1.214E+01 3.889E+00 5.181E+01 4.320E-02 1.152E+01 1.960E-02 1.536E-01 2.400E-03 1.120E-02 4.000E-04 
-3.750E-01 8.024E+01 8.996E+01 8.705E+01 1.360E+01 4.421E+00 5.681E+01 4.400E-02 1.353E+01 2.640E-02 2.020E-01 2.000E-03 1.680E-02 0.000E+00 
-1.250E-01 8.371E+01 9.395E+01 9.134E+01 1.436E+01 4.864E+00 5.967E+01 5.520E-02 1.468E+01 2.680E-02 2.376E-01 1.600E-03 2.080E-02 1.200E-03 
1.250E-01 8.339E+01 9.435E+01 9.100E+01 1.446E+01 4.910E+00 5.943E+01 4.520E-02 1.476E+01 2.480E-02 2.400E-01 2.800E-03 2.000E-02 4.000E-04 
3.750E-01 7.968E+01 9.006E+01 8.682E+01 1.369E+01 4.400E+00 5.673E+01 4.240E-02 1.351E+01 2.800E-02 2.192E-01 2.400E-03 1.760E-02 1.200E-03 
6.250E-01 7.282E+01 8.199E+01 7.920E+01 1.222E+01 3.801E+00 5.189E+01 3.360E-02 1.157E+01 1.680E-02 1.596E-01 2.800E-03 1.000E-02 0.000E+00 
8.750E-01 6.353E+01 7.180E+01 6.956E+01 1.032E+01 3.035E+00 4.547E+01 2.720E-02 8.965E+00 1.560E-02 1.012E-01 2.400E-03 7.600E-03 0.000E+00 
1.125E+00 5.309E+01 5.981E+01 5.887E+01 7.934E+00 2.218E+00 3.807E+01 2.320E-02 6.136E+00 1.080E-02 5.480E-02 2.000E-03 4.400E-03 0.000E+00 
1.375E+00 4.174E+01 4.735E+01 4.677E+01 5.528E+00 1.379E+00 3.098E+01 2.200E-02 3.723E+00 6.000E-03 2.200E-02 8.000E-04 0.000E+00 0.000E+00 
1.625E+00 3.066E+01 3.461E+01 3.424E+01 3.117E+00 7.664E-01 2.495E+01 1.120E-02 1.720E+00 2.800E-03 6.800E-03 0.000E+00 4.000E-04 0.000E+00 
1.875E+00 2.076E+01 2.350E+01 2.322E+01 1.412E+00 3.308E-01 2.555E+01 2.000E-03 5.748E-01 4.000E-04 1.200E-03 0.000E+00 0.000E+00 0.000E+00 
2.125E+00 1.271E+01 1.432E+01 1.424E+01 5.300E-01 1.108E-01 9.105E+00 4.000E-04 1.056E-01 0.000E+00 0.000E+00 0.000E+00 0.000E+00 0.000E+00 
2.375E+00 7.052E+00 8.065E+00 8.006E+00 1.336E-01 2.600E-02 8.420E-01 0.000E+00 8.000E-03 0.000E+00 0.000E+00 0.000E+00 0.000E+00 0.000E+00 
2.625E+00 3.441E+00 3.970E+00 3.916E+00 1.680E-02 8.000E-03 4.800E-02 0.000E+00 4.000E-04 0.000E+00 0.000E+00 0.000E+00 0.000E+00 0.000E+00 
2.875E+00 1.459E+00 1.696E+00 1.661E+00 5.200E-03 4.000E-04 1.200E-03 0.000E+00 0.000E+00 0.000E+00 0.000E+00 0.000E+00 0.000E+00 0.000E+00 
3.125E+00 4.996E-01 5.676E-01 5.520E-01 0.000E+00 0.000E+00 0.000E+00 0.000E+00 0.000E+00 0.000E+00 0.000E+00 0.000E+00 0.000E+00 0.000E+00 
3.375E+00 1.260E-01 1.576E-01 1.464E-01 0.000E+00 0.000E+00 0.000E+00 0.000E+00 0.000E+00 0.000E+00 0.000E+00 0.000E+00 0.000E+00 0.000E+00 
3.625E+00 3.040E-02 2.920E-02 3.160E-02 0.000E+00 0.000E+00 0.000E+00 0.000E+00 0.000E+00 0.000E+00 0.000E+00 0.000E+00 0.000E+00 0.000E+00 
3.875E+00 5.200E-03 3.600E-03 6.000E-03 0.000E+00 0.000E+00 0.000E+00 0.000E+00 0.000E+00 0.000E+00 0.000E+00 0.000E+00 0.000E+00 0.000E+00 
4.125E+00 8.000E-04 4.000E-04 8.000E-04 0.000E+00 0.000E+00 0.000E+00 0.000E+00 0.000E+00 0.000E+00 0.000E+00 0.000E+00 0.000E+00 0.000E+00 
\end{verbatim}
}
\newpage
Pb+Pb \@ $E_{\rm lab}=30A~$GeV:
{\tiny
\begin{verbatim}
! y, dN/dy (pi+ pi- pi0 K+ K- P aP L+S0 a(L+S0) Xi- aXi- Om aOm) 
-4.375E+00 4.000E-04 0.000E+00 0.000E+00 0.000E+00 0.000E+00 0.000E+00 0.000E+00 0.000E+00 0.000E+00 0.000E+00 0.000E+00 0.000E+00 0.000E+00 
-4.125E+00 4.000E-03 2.800E-03 1.600E-03 0.000E+00 0.000E+00 0.000E+00 0.000E+00 0.000E+00 0.000E+00 0.000E+00 0.000E+00 0.000E+00 0.000E+00 
-3.875E+00 1.920E-02 2.760E-02 2.240E-02 0.000E+00 0.000E+00 0.000E+00 0.000E+00 0.000E+00 0.000E+00 0.000E+00 0.000E+00 0.000E+00 0.000E+00 
-3.625E+00 1.032E-01 1.124E-01 1.124E-01 0.000E+00 0.000E+00 0.000E+00 0.000E+00 0.000E+00 0.000E+00 0.000E+00 0.000E+00 0.000E+00 0.000E+00 
-3.375E+00 3.728E-01 4.316E-01 4.080E-01 0.000E+00 0.000E+00 0.000E+00 0.000E+00 0.000E+00 0.000E+00 0.000E+00 0.000E+00 0.000E+00 0.000E+00 
-3.125E+00 1.168E+00 1.313E+00 1.323E+00 1.600E-03 4.000E-04 4.000E-04 0.000E+00 0.000E+00 0.000E+00 0.000E+00 0.000E+00 0.000E+00 0.000E+00 
-2.875E+00 2.911E+00 3.347E+00 3.312E+00 1.360E-02 2.000E-03 3.080E-02 0.000E+00 0.000E+00 0.000E+00 0.000E+00 0.000E+00 0.000E+00 0.000E+00 
-2.625E+00 6.186E+00 6.984E+00 6.891E+00 7.600E-02 2.400E-02 4.932E-01 0.000E+00 4.800E-03 0.000E+00 0.000E+00 0.000E+00 0.000E+00 0.000E+00 
-2.375E+00 1.151E+01 1.281E+01 1.284E+01 3.772E-01 9.960E-02 5.541E+00 4.000E-04 6.640E-02 0.000E+00 4.000E-04 0.000E+00 0.000E+00 0.000E+00 
-2.125E+00 1.904E+01 2.149E+01 2.129E+01 1.182E+00 2.900E-01 2.294E+01 3.200E-03 4.104E-01 4.000E-04 1.200E-03 0.000E+00 0.000E+00 0.000E+00 
-1.875E+00 2.904E+01 3.233E+01 3.253E+01 2.720E+00 7.128E-01 2.294E+01 1.480E-02 1.386E+00 4.400E-03 6.400E-03 0.000E+00 0.000E+00 0.000E+00 
-1.625E+00 4.073E+01 4.534E+01 4.524E+01 4.902E+00 1.392E+00 2.611E+01 2.720E-02 3.002E+00 1.200E-02 1.840E-02 8.000E-04 4.000E-04 0.000E+00 
-1.375E+00 5.338E+01 5.907E+01 5.876E+01 7.551E+00 2.287E+00 3.151E+01 4.800E-02 5.167E+00 2.000E-02 4.080E-02 1.200E-03 4.000E-03 4.000E-04 
-1.125E+00 6.576E+01 7.274E+01 7.176E+01 1.007E+01 3.315E+00 3.715E+01 7.000E-02 7.712E+00 3.840E-02 9.360E-02 2.000E-03 5.200E-03 8.000E-04 
-8.750E-01 7.766E+01 8.567E+01 8.422E+01 1.234E+01 4.500E+00 4.254E+01 7.720E-02 1.013E+01 5.240E-02 1.624E-01 3.600E-03 1.440E-02 4.000E-04 
-6.250E-01 8.798E+01 9.703E+01 9.494E+01 1.426E+01 5.503E+00 4.733E+01 1.088E-01 1.233E+01 5.280E-02 2.156E-01 4.800E-03 2.400E-02 2.000E-03 
-3.750E-01 9.570E+01 1.057E+02 1.034E+02 1.549E+01 6.188E+00 5.101E+01 1.208E-01 1.407E+01 8.440E-02 2.832E-01 6.400E-03 2.440E-02 3.200E-03 
-1.250E-01 9.976E+01 1.101E+02 1.078E+02 1.634E+01 6.752E+00 5.287E+01 1.396E-01 1.505E+01 7.320E-02 3.120E-01 1.200E-02 3.880E-02 1.600E-03 
1.250E-01 9.997E+01 1.101E+02 1.079E+02 1.645E+01 6.737E+00 5.277E+01 1.400E-01 1.515E+01 8.680E-02 2.792E-01 1.000E-02 3.600E-02 2.800E-03 
3.750E-01 9.529E+01 1.053E+02 1.035E+02 1.558E+01 6.294E+00 5.074E+01 1.192E-01 1.408E+01 6.400E-02 2.540E-01 7.200E-03 2.880E-02 2.400E-03 
6.250E-01 8.759E+01 9.633E+01 9.449E+01 1.422E+01 5.518E+00 4.718E+01 9.920E-02 1.244E+01 6.360E-02 2.136E-01 6.400E-03 2.440E-02 2.800E-03 
8.750E-01 7.708E+01 8.519E+01 8.408E+01 1.239E+01 4.444E+00 4.262E+01 7.560E-02 1.003E+01 5.160E-02 1.608E-01 1.600E-03 1.160E-02 2.400E-03 
1.125E+00 6.576E+01 7.240E+01 7.156E+01 1.012E+01 3.303E+00 3.720E+01 6.360E-02 7.565E+00 3.000E-02 9.240E-02 3.200E-03 6.000E-03 0.000E+00 
1.375E+00 5.333E+01 5.897E+01 5.851E+01 7.490E+00 2.333E+00 3.153E+01 4.200E-02 5.230E+00 2.000E-02 4.560E-02 1.200E-03 1.600E-03 4.000E-04 
1.625E+00 4.066E+01 4.530E+01 4.503E+01 5.001E+00 1.414E+00 2.617E+01 3.240E-02 2.928E+00 1.080E-02 1.920E-02 8.000E-04 0.000E+00 0.000E+00 
1.875E+00 2.913E+01 3.229E+01 3.219E+01 2.747E+00 6.820E-01 2.284E+01 1.200E-02 1.368E+00 4.000E-03 6.000E-03 0.000E+00 0.000E+00 0.000E+00 
2.125E+00 1.895E+01 2.123E+01 2.121E+01 1.236E+00 2.916E-01 2.326E+01 2.000E-03 4.244E-01 4.000E-04 1.600E-03 0.000E+00 0.000E+00 0.000E+00 
2.375E+00 1.131E+01 1.273E+01 1.271E+01 3.908E-01 8.760E-02 5.538E+00 0.000E+00 6.160E-02 0.000E+00 0.000E+00 0.000E+00 0.000E+00 0.000E+00 
2.625E+00 6.151E+00 6.862E+00 6.917E+00 8.880E-02 2.240E-02 4.912E-01 0.000E+00 5.200E-03 0.000E+00 0.000E+00 0.000E+00 0.000E+00 0.000E+00 
2.875E+00 2.916E+00 3.310E+00 3.293E+00 1.200E-02 1.600E-03 2.880E-02 0.000E+00 0.000E+00 0.000E+00 0.000E+00 0.000E+00 0.000E+00 0.000E+00 
3.125E+00 1.188E+00 1.360E+00 1.296E+00 1.200E-03 0.000E+00 0.000E+00 0.000E+00 0.000E+00 0.000E+00 0.000E+00 0.000E+00 0.000E+00 0.000E+00 
3.375E+00 3.768E-01 4.420E-01 4.316E-01 0.000E+00 0.000E+00 0.000E+00 0.000E+00 0.000E+00 0.000E+00 0.000E+00 0.000E+00 0.000E+00 0.000E+00 
3.625E+00 9.920E-02 1.000E-01 1.200E-01 0.000E+00 0.000E+00 0.000E+00 0.000E+00 0.000E+00 0.000E+00 0.000E+00 0.000E+00 0.000E+00 0.000E+00 
3.875E+00 1.720E-02 2.040E-02 2.200E-02 0.000E+00 0.000E+00 0.000E+00 0.000E+00 0.000E+00 0.000E+00 0.000E+00 0.000E+00 0.000E+00 0.000E+00 
4.125E+00 4.800E-03 3.200E-03 4.400E-03 0.000E+00 0.000E+00 0.000E+00 0.000E+00 0.000E+00 0.000E+00 0.000E+00 0.000E+00 0.000E+00 0.000E+00 
4.375E+00 4.000E-04 4.000E-04 0.000E+00 0.000E+00 0.000E+00 0.000E+00 0.000E+00 0.000E+00 0.000E+00 0.000E+00 0.000E+00 0.000E+00 0.000E+00 
\end{verbatim}
}
\newpage
Pb+Pb \@ $E_{\rm lab}=40A~$GeV:
{\tiny
\begin{verbatim}
! y, dN/dy (pi+ pi- pi0 K+ K- P aP L+S0 a(L+S0) Xi- aXi- Om aOm) 
-4.375E+00 1.200E-03 1.200E-03 4.000E-04 0.000E+00 0.000E+00 0.000E+00 0.000E+00 0.000E+00 0.000E+00 0.000E+00 0.000E+00 0.000E+00 0.000E+00 
-4.125E+00 7.600E-03 1.200E-02 8.400E-03 0.000E+00 0.000E+00 0.000E+00 0.000E+00 0.000E+00 0.000E+00 0.000E+00 0.000E+00 0.000E+00 0.000E+00 
-3.875E+00 4.800E-02 5.360E-02 4.720E-02 0.000E+00 0.000E+00 0.000E+00 0.000E+00 0.000E+00 0.000E+00 0.000E+00 0.000E+00 0.000E+00 0.000E+00 
-3.625E+00 2.160E-01 2.384E-01 2.616E-01 0.000E+00 0.000E+00 0.000E+00 0.000E+00 0.000E+00 0.000E+00 0.000E+00 0.000E+00 0.000E+00 0.000E+00 
-3.375E+00 7.392E-01 8.532E-01 8.232E-01 4.000E-04 0.000E+00 0.000E+00 0.000E+00 0.000E+00 0.000E+00 0.000E+00 0.000E+00 0.000E+00 0.000E+00 
-3.125E+00 1.988E+00 2.327E+00 2.262E+00 4.000E-03 4.000E-03 4.000E-03 0.000E+00 0.000E+00 0.000E+00 0.000E+00 0.000E+00 0.000E+00 0.000E+00 
-2.875E+00 4.401E+00 4.978E+00 5.017E+00 4.120E-02 1.400E-02 1.436E-01 0.000E+00 0.000E+00 0.000E+00 0.000E+00 0.000E+00 0.000E+00 0.000E+00 
-2.625E+00 8.710E+00 9.751E+00 9.773E+00 2.108E-01 5.440E-02 1.833E+00 0.000E+00 2.040E-02 0.000E+00 0.000E+00 0.000E+00 0.000E+00 0.000E+00 
-2.375E+00 1.537E+01 1.715E+01 1.696E+01 7.004E-01 1.916E-01 1.566E+01 2.000E-03 1.752E-01 0.000E+00 4.000E-04 0.000E+00 0.000E+00 0.000E+00 
-2.125E+00 2.438E+01 2.698E+01 2.727E+01 1.928E+00 5.356E-01 2.408E+01 1.080E-02 8.296E-01 2.000E-03 4.000E-03 0.000E+00 0.000E+00 0.000E+00 
-1.875E+00 3.562E+01 3.943E+01 3.931E+01 3.918E+00 1.122E+00 2.206E+01 2.920E-02 2.084E+00 7.600E-03 1.080E-02 4.000E-04 0.000E+00 0.000E+00 
-1.625E+00 4.833E+01 5.314E+01 5.332E+01 6.368E+00 2.036E+00 2.659E+01 5.520E-02 3.808E+00 3.040E-02 3.280E-02 8.000E-04 1.600E-03 0.000E+00 
-1.375E+00 6.163E+01 6.760E+01 6.747E+01 9.088E+00 3.076E+00 3.169E+01 8.400E-02 6.042E+00 4.240E-02 6.400E-02 2.400E-03 5.200E-03 4.000E-04 
-1.125E+00 7.504E+01 8.180E+01 8.117E+01 1.148E+01 4.343E+00 3.611E+01 1.048E-01 8.303E+00 6.280E-02 1.304E-01 2.800E-03 1.320E-02 0.000E+00 
-8.750E-01 8.801E+01 9.562E+01 9.488E+01 1.371E+01 5.638E+00 4.048E+01 1.464E-01 1.077E+01 8.440E-02 1.840E-01 1.120E-02 2.160E-02 1.200E-03 
-6.250E-01 9.905E+01 1.078E+02 1.064E+02 1.572E+01 6.797E+00 4.420E+01 1.756E-01 1.269E+01 1.140E-01 2.392E-01 9.600E-03 2.920E-02 2.000E-03 
-3.750E-01 1.072E+02 1.175E+02 1.159E+02 1.714E+01 7.756E+00 4.674E+01 2.304E-01 1.430E+01 1.216E-01 3.184E-01 1.480E-02 4.040E-02 2.800E-03 
-1.250E-01 1.125E+02 1.223E+02 1.212E+02 1.776E+01 8.145E+00 4.823E+01 2.504E-01 1.502E+01 1.380E-01 3.216E-01 1.520E-02 4.760E-02 3.200E-03 
1.250E-01 1.124E+02 1.216E+02 1.214E+02 1.768E+01 8.378E+00 4.832E+01 2.548E-01 1.502E+01 1.508E-01 3.284E-01 1.680E-02 4.320E-02 4.800E-03 
3.750E-01 1.080E+02 1.170E+02 1.159E+02 1.703E+01 7.727E+00 4.704E+01 2.392E-01 1.418E+01 1.316E-01 2.912E-01 1.520E-02 4.280E-02 5.200E-03 
6.250E-01 9.897E+01 1.077E+02 1.066E+02 1.578E+01 6.868E+00 4.424E+01 1.944E-01 1.269E+01 1.092E-01 2.576E-01 1.160E-02 3.640E-02 2.800E-03 
8.750E-01 8.782E+01 9.582E+01 9.496E+01 1.381E+01 5.606E+00 4.053E+01 1.428E-01 1.070E+01 9.120E-02 1.836E-01 1.040E-02 1.760E-02 2.000E-03 
1.125E+00 7.500E+01 8.217E+01 8.168E+01 1.155E+01 4.380E+00 3.633E+01 1.148E-01 8.433E+00 6.040E-02 1.336E-01 4.400E-03 1.280E-02 2.000E-03 
1.375E+00 6.177E+01 6.782E+01 6.685E+01 9.162E+00 3.115E+00 3.184E+01 7.840E-02 6.056E+00 3.480E-02 7.360E-02 2.000E-03 4.400E-03 0.000E+00 
1.625E+00 4.842E+01 5.339E+01 5.274E+01 6.393E+00 1.972E+00 2.679E+01 5.640E-02 3.860E+00 2.040E-02 2.800E-02 4.000E-04 4.000E-04 4.000E-04 
1.875E+00 3.550E+01 3.936E+01 3.943E+01 3.862E+00 1.127E+00 2.211E+01 3.480E-02 2.050E+00 7.200E-03 1.000E-02 0.000E+00 8.000E-04 0.000E+00 
2.125E+00 2.425E+01 2.681E+01 2.696E+01 1.930E+00 5.036E-01 2.397E+01 1.160E-02 8.524E-01 2.400E-03 6.400E-03 4.000E-04 0.000E+00 0.000E+00 
2.375E+00 1.521E+01 1.683E+01 1.710E+01 7.416E-01 2.000E-01 1.552E+01 1.600E-03 1.808E-01 4.000E-04 4.000E-04 0.000E+00 0.000E+00 0.000E+00 
2.625E+00 8.677E+00 9.822E+00 9.863E+00 1.988E-01 4.920E-02 1.837E+00 0.000E+00 1.680E-02 0.000E+00 0.000E+00 0.000E+00 0.000E+00 0.000E+00 
2.875E+00 4.463E+00 5.081E+00 5.016E+00 3.880E-02 1.480E-02 1.524E-01 0.000E+00 4.000E-04 0.000E+00 0.000E+00 0.000E+00 0.000E+00 0.000E+00 
3.125E+00 2.024E+00 2.281E+00 2.182E+00 4.400E-03 1.200E-03 2.800E-03 0.000E+00 0.000E+00 0.000E+00 0.000E+00 0.000E+00 0.000E+00 0.000E+00 
3.375E+00 7.192E-01 8.428E-01 8.484E-01 0.000E+00 0.000E+00 4.000E-04 0.000E+00 0.000E+00 0.000E+00 0.000E+00 0.000E+00 0.000E+00 0.000E+00 
3.625E+00 2.100E-01 2.480E-01 2.336E-01 0.000E+00 0.000E+00 0.000E+00 0.000E+00 0.000E+00 0.000E+00 0.000E+00 0.000E+00 0.000E+00 0.000E+00 
3.875E+00 4.680E-02 5.280E-02 5.000E-02 0.000E+00 0.000E+00 0.000E+00 0.000E+00 0.000E+00 0.000E+00 0.000E+00 0.000E+00 0.000E+00 0.000E+00 
4.125E+00 6.800E-03 8.800E-03 1.120E-02 0.000E+00 0.000E+00 0.000E+00 0.000E+00 0.000E+00 0.000E+00 0.000E+00 0.000E+00 0.000E+00 0.000E+00 
\end{verbatim}
}

\newpage
Pb+Pb \@ $E_{\rm lab}=80A~$GeV:
{\tiny
\begin{verbatim}
 ! y, dN/dy (pi+ pi- pi0 K+ K- P aP L+S0 a(L+S0) Xi- aXi- Om aOm) 
-4.875E+00 4.418E-04 8.836E-04 4.418E-04 0.000E+00 0.000E+00 0.000E+00 0.000E+00 0.000E+00 0.000E+00 0.000E+00 0.000E+00 0.000E+00 0.000E+00 
-4.625E+00 2.209E-03 3.093E-03 3.093E-03 0.000E+00 0.000E+00 0.000E+00 0.000E+00 0.000E+00 0.000E+00 0.000E+00 0.000E+00 0.000E+00 0.000E+00 
-4.375E+00 1.325E-02 1.502E-02 1.723E-02 0.000E+00 0.000E+00 0.000E+00 0.000E+00 0.000E+00 0.000E+00 0.000E+00 0.000E+00 0.000E+00 0.000E+00 
-4.125E+00 8.792E-02 9.410E-02 9.057E-02 0.000E+00 0.000E+00 0.000E+00 0.000E+00 0.000E+00 0.000E+00 0.000E+00 0.000E+00 0.000E+00 0.000E+00 
-3.875E+00 3.384E-01 3.645E-01 3.698E-01 0.000E+00 0.000E+00 0.000E+00 0.000E+00 0.000E+00 0.000E+00 0.000E+00 0.000E+00 0.000E+00 0.000E+00 
-3.625E+00 1.033E+00 1.204E+00 1.230E+00 1.325E-03 0.000E+00 4.418E-04 0.000E+00 0.000E+00 0.000E+00 0.000E+00 0.000E+00 0.000E+00 0.000E+00 
-3.375E+00 2.616E+00 2.977E+00 2.928E+00 1.016E-02 1.325E-03 2.121E-02 0.000E+00 0.000E+00 0.000E+00 0.000E+00 0.000E+00 0.000E+00 0.000E+00 
-3.125E+00 5.496E+00 6.269E+00 6.182E+00 6.715E-02 2.209E-02 3.556E-01 4.418E-04 2.209E-03 0.000E+00 0.000E+00 0.000E+00 0.000E+00 0.000E+00 
-2.875E+00 1.050E+01 1.146E+01 1.173E+01 3.353E-01 8.924E-02 4.487E+00 0.000E+00 4.285E-02 4.418E-04 0.000E+00 0.000E+00 0.000E+00 0.000E+00 
-2.625E+00 1.784E+01 1.945E+01 1.973E+01 1.059E+00 2.898E-01 2.154E+01 5.302E-03 3.437E-01 8.836E-04 8.836E-04 0.000E+00 0.000E+00 0.000E+00 
-2.375E+00 2.763E+01 3.033E+01 3.041E+01 2.536E+00 7.639E-01 2.035E+01 2.253E-02 1.149E+00 7.069E-03 7.069E-03 0.000E+00 0.000E+00 0.000E+00 
-2.125E+00 3.988E+01 4.351E+01 4.380E+01 4.718E+00 1.489E+00 2.021E+01 6.008E-02 2.351E+00 2.386E-02 2.032E-02 1.325E-03 1.767E-03 0.000E+00 
-1.875E+00 5.375E+01 5.827E+01 5.844E+01 7.327E+00 2.579E+00 2.385E+01 1.144E-01 3.987E+00 5.522E-02 5.390E-02 3.976E-03 2.209E-03 4.418E-04 
-1.625E+00 6.854E+01 7.406E+01 7.398E+01 1.004E+01 3.951E+00 2.756E+01 1.750E-01 5.970E+00 9.499E-02 9.940E-02 7.510E-03 5.743E-03 2.651E-03 
-1.375E+00 8.432E+01 9.027E+01 9.078E+01 1.249E+01 5.584E+00 3.078E+01 2.244E-01 7.924E+00 1.396E-01 1.476E-01 1.370E-02 1.414E-02 5.743E-03 
-1.125E+00 1.002E+02 1.068E+02 1.067E+02 1.501E+01 7.256E+00 3.347E+01 3.256E-01 9.825E+00 2.001E-01 2.037E-01 2.032E-02 2.474E-02 5.302E-03 
-8.750E-01 1.152E+02 1.229E+02 1.235E+02 1.705E+01 8.951E+00 3.603E+01 4.581E-01 1.149E+01 2.620E-01 2.801E-01 2.474E-02 4.241E-02 7.952E-03 
-6.250E-01 1.283E+02 1.370E+02 1.378E+02 1.893E+01 1.046E+01 3.807E+01 5.827E-01 1.277E+01 3.437E-01 3.353E-01 2.916E-02 5.302E-02 1.016E-02 
-3.750E-01 1.393E+02 1.472E+02 1.490E+02 2.026E+01 1.146E+01 3.946E+01 7.175E-01 1.385E+01 3.861E-01 3.897E-01 3.799E-02 6.362E-02 1.590E-02 
-1.250E-01 1.442E+02 1.531E+02 1.555E+02 2.092E+01 1.196E+01 3.961E+01 7.515E-01 1.414E+01 4.546E-01 4.042E-01 4.065E-02 6.804E-02 1.900E-02 
1.250E-01 1.443E+02 1.529E+02 1.552E+02 2.071E+01 1.196E+01 3.970E+01 7.325E-01 1.423E+01 4.246E-01 3.958E-01 3.799E-02 5.920E-02 1.811E-02 
3.750E-01 1.385E+02 1.478E+02 1.492E+02 2.027E+01 1.152E+01 3.951E+01 7.020E-01 1.369E+01 4.118E-01 3.605E-01 3.755E-02 5.302E-02 1.546E-02 
6.250E-01 1.284E+02 1.369E+02 1.383E+02 1.884E+01 1.030E+01 3.798E+01 5.695E-01 1.290E+01 3.450E-01 3.300E-01 3.269E-02 5.257E-02 1.016E-02 
8.750E-01 1.153E+02 1.234E+02 1.235E+02 1.702E+01 8.994E+00 3.602E+01 4.427E-01 1.154E+01 2.739E-01 2.748E-01 2.518E-02 3.667E-02 8.836E-03 
1.125E+00 1.002E+02 1.071E+02 1.074E+02 1.489E+01 7.145E+00 3.365E+01 3.336E-01 9.808E+00 1.895E-01 2.054E-01 1.370E-02 3.137E-02 8.394E-03 
1.375E+00 8.431E+01 9.066E+01 9.071E+01 1.255E+01 5.506E+00 3.077E+01 2.169E-01 7.912E+00 1.334E-01 1.515E-01 1.060E-02 1.502E-02 3.534E-03 
1.625E+00 6.871E+01 7.341E+01 7.431E+01 9.976E+00 3.923E+00 2.768E+01 1.462E-01 5.944E+00 9.145E-02 9.322E-02 4.418E-03 7.510E-03 2.651E-03 
1.875E+00 5.387E+01 5.803E+01 5.834E+01 7.281E+00 2.609E+00 2.401E+01 1.078E-01 4.006E+00 5.434E-02 5.081E-02 1.325E-03 2.209E-03 4.418E-04 
2.125E+00 3.992E+01 4.328E+01 4.378E+01 4.703E+00 1.451E+00 2.047E+01 6.759E-02 2.315E+00 2.783E-02 1.767E-02 4.418E-04 0.000E+00 0.000E+00 
2.375E+00 2.767E+01 3.023E+01 3.072E+01 2.453E+00 7.400E-01 2.029E+01 2.121E-02 1.117E+00 5.743E-03 5.743E-03 0.000E+00 4.418E-04 0.000E+00 
2.625E+00 1.753E+01 1.956E+01 1.969E+01 1.082E+00 2.827E-01 2.153E+01 3.534E-03 3.207E-01 1.325E-03 0.000E+00 0.000E+00 0.000E+00 0.000E+00 
2.875E+00 1.038E+01 1.153E+01 1.157E+01 3.194E-01 7.864E-02 4.355E+00 0.000E+00 5.036E-02 0.000E+00 0.000E+00 0.000E+00 0.000E+00 0.000E+00 
3.125E+00 5.512E+00 6.219E+00 6.235E+00 6.273E-02 1.458E-02 3.910E-01 0.000E+00 4.860E-03 0.000E+00 0.000E+00 0.000E+00 0.000E+00 0.000E+00 
3.375E+00 2.630E+00 2.945E+00 2.864E+00 1.193E-02 3.976E-03 1.723E-02 0.000E+00 0.000E+00 0.000E+00 0.000E+00 0.000E+00 0.000E+00 0.000E+00 
3.625E+00 1.046E+00 1.229E+00 1.149E+00 4.418E-04 4.418E-04 0.000E+00 0.000E+00 0.000E+00 0.000E+00 0.000E+00 0.000E+00 0.000E+00 0.000E+00 
3.875E+00 3.327E-01 3.481E-01 3.835E-01 0.000E+00 0.000E+00 0.000E+00 0.000E+00 0.000E+00 0.000E+00 0.000E+00 0.000E+00 0.000E+00 0.000E+00 
4.125E+00 9.057E-02 1.012E-01 8.703E-02 0.000E+00 0.000E+00 0.000E+00 0.000E+00 0.000E+00 0.000E+00 0.000E+00 0.000E+00 0.000E+00 0.000E+00 
4.375E+00 1.149E-02 1.856E-02 2.165E-02 0.000E+00 0.000E+00 0.000E+00 0.000E+00 0.000E+00 0.000E+00 0.000E+00 0.000E+00 0.000E+00 0.000E+00 
4.625E+00 2.209E-03 3.976E-03 4.418E-03 0.000E+00 0.000E+00 0.000E+00 0.000E+00 0.000E+00 0.000E+00 0.000E+00 0.000E+00 0.000E+00 0.000E+00 
\end{verbatim}
}
\newpage
Pb+Pb \@ $E_{\rm lab}=160A~$GeV:
{\tiny
\begin{verbatim}
 ! y, dN/dy (pi+ pi- pi0 K+ K- P aP L+S0 a(L+S0) Xi- aXi- Om aOm) 
-4.875E+00 3.739E-03 3.323E-03 5.816E-03 0.000E+00 0.000E+00 0.000E+00 0.000E+00 0.000E+00 0.000E+00 0.000E+00 0.000E+00 0.000E+00 0.000E+00 
-4.625E+00 2.991E-02 3.614E-02 4.154E-02 0.000E+00 0.000E+00 0.000E+00 0.000E+00 0.000E+00 0.000E+00 0.000E+00 0.000E+00 0.000E+00 0.000E+00 
-4.375E+00 1.462E-01 1.516E-01 1.591E-01 0.000E+00 0.000E+00 0.000E+00 0.000E+00 0.000E+00 0.000E+00 0.000E+00 0.000E+00 0.000E+00 0.000E+00 
-4.125E+00 4.856E-01 5.770E-01 5.957E-01 4.154E-04 0.000E+00 0.000E+00 0.000E+00 0.000E+00 0.000E+00 0.000E+00 0.000E+00 0.000E+00 0.000E+00 
-3.875E+00 1.474E+00 1.650E+00 1.653E+00 8.308E-04 1.246E-03 4.154E-04 0.000E+00 0.000E+00 0.000E+00 0.000E+00 0.000E+00 0.000E+00 0.000E+00 
-3.625E+00 3.293E+00 3.861E+00 3.779E+00 1.537E-02 5.400E-03 6.148E-02 0.000E+00 4.154E-04 0.000E+00 0.000E+00 0.000E+00 0.000E+00 0.000E+00 
-3.375E+00 6.680E+00 7.545E+00 7.571E+00 1.242E-01 3.448E-02 8.055E-01 0.000E+00 7.477E-03 0.000E+00 0.000E+00 0.000E+00 0.000E+00 0.000E+00 
-3.125E+00 1.226E+01 1.355E+01 1.362E+01 4.981E-01 1.421E-01 9.744E+00 2.077E-03 9.222E-02 4.154E-04 0.000E+00 0.000E+00 0.000E+00 0.000E+00 
-2.875E+00 2.028E+01 2.210E+01 2.256E+01 1.476E+00 4.146E-01 2.317E+01 1.703E-02 5.550E-01 6.647E-03 3.323E-03 0.000E+00 0.000E+00 0.000E+00 
-2.625E+00 3.125E+01 3.369E+01 3.406E+01 3.061E+00 9.704E-01 1.732E+01 4.611E-02 1.352E+00 1.288E-02 1.163E-02 0.000E+00 0.000E+00 0.000E+00 
-2.375E+00 4.407E+01 4.755E+01 4.835E+01 5.461E+00 1.859E+00 1.940E+01 1.151E-01 2.613E+00 4.196E-02 3.406E-02 2.077E-03 8.308E-04 0.000E+00 
-2.125E+00 5.934E+01 6.329E+01 6.441E+01 8.079E+00 3.263E+00 2.241E+01 1.836E-01 4.203E+00 8.059E-02 5.317E-02 5.400E-03 2.908E-03 0.000E+00 
-1.875E+00 7.581E+01 8.053E+01 8.164E+01 1.071E+01 4.884E+00 2.518E+01 3.012E-01 5.992E+00 1.537E-01 9.554E-02 1.371E-02 9.970E-03 4.985E-03 
-1.625E+00 9.293E+01 9.837E+01 9.952E+01 1.338E+01 6.599E+00 2.732E+01 4.133E-01 7.629E+00 2.443E-01 1.616E-01 1.579E-02 2.119E-02 6.231E-03 
-1.375E+00 1.110E+02 1.176E+02 1.183E+02 1.607E+01 8.701E+00 2.939E+01 6.015E-01 9.166E+00 3.419E-01 2.247E-01 2.991E-02 2.866E-02 1.163E-02 
-1.125E+00 1.288E+02 1.357E+02 1.380E+02 1.822E+01 1.060E+01 3.081E+01 8.076E-01 1.064E+01 4.549E-01 2.883E-01 3.365E-02 5.068E-02 1.703E-02 
-8.750E-01 1.455E+02 1.529E+02 1.561E+02 2.054E+01 1.256E+01 3.202E+01 1.034E+00 1.163E+01 6.061E-01 3.514E-01 5.400E-02 5.525E-02 2.368E-02 
-6.250E-01 1.604E+02 1.676E+02 1.720E+02 2.201E+01 1.429E+01 3.283E+01 1.317E+00 1.249E+01 7.203E-01 4.034E-01 6.356E-02 7.560E-02 1.994E-02 
-3.750E-01 1.706E+02 1.788E+02 1.836E+02 2.324E+01 1.532E+01 3.303E+01 1.472E+00 1.296E+01 8.217E-01 4.187E-01 6.896E-02 8.267E-02 2.783E-02 
-1.250E-01 1.770E+02 1.843E+02 1.897E+02 2.371E+01 1.582E+01 3.319E+01 1.571E+00 1.296E+01 8.732E-01 4.582E-01 7.104E-02 8.973E-02 3.033E-02 
1.250E-01 1.762E+02 1.843E+02 1.891E+02 2.389E+01 1.583E+01 3.312E+01 1.561E+00 1.307E+01 8.686E-01 4.503E-01 6.397E-02 9.845E-02 2.908E-02 
3.750E-01 1.711E+02 1.785E+02 1.829E+02 2.291E+01 1.527E+01 3.323E+01 1.468E+00 1.285E+01 8.184E-01 4.378E-01 7.851E-02 8.848E-02 2.783E-02 
6.250E-01 1.606E+02 1.677E+02 1.718E+02 2.207E+01 1.421E+01 3.285E+01 1.252E+00 1.247E+01 7.386E-01 4.162E-01 6.605E-02 7.644E-02 2.409E-02 
8.750E-01 1.460E+02 1.529E+02 1.563E+02 2.040E+01 1.250E+01 3.237E+01 1.068E+00 1.166E+01 5.845E-01 3.369E-01 5.774E-02 6.314E-02 2.243E-02 
1.125E+00 1.290E+02 1.355E+02 1.376E+02 1.818E+01 1.074E+01 3.096E+01 8.271E-01 1.050E+01 4.661E-01 2.908E-01 3.863E-02 4.611E-02 1.662E-02 
1.375E+00 1.113E+02 1.169E+02 1.190E+02 1.596E+01 8.584E+00 2.915E+01 5.670E-01 9.240E+00 3.440E-01 2.218E-01 2.866E-02 3.365E-02 1.205E-02 
1.625E+00 9.270E+01 9.896E+01 9.970E+01 1.345E+01 6.740E+00 2.749E+01 4.154E-01 7.659E+00 2.372E-01 1.583E-01 1.454E-02 2.077E-02 6.647E-03 
1.875E+00 7.558E+01 8.039E+01 8.093E+01 1.081E+01 4.838E+00 2.506E+01 2.858E-01 6.006E+00 1.603E-01 1.101E-01 7.062E-03 8.308E-03 2.492E-03 
2.125E+00 5.944E+01 6.357E+01 6.391E+01 8.116E+00 3.235E+00 2.240E+01 2.011E-01 4.216E+00 1.034E-01 6.356E-02 2.908E-03 4.985E-03 1.246E-03 
2.375E+00 4.451E+01 4.760E+01 4.830E+01 5.458E+00 1.910E+00 1.934E+01 9.762E-02 2.638E+00 4.570E-02 2.492E-02 3.323E-03 8.308E-04 4.154E-04 
2.625E+00 3.127E+01 3.371E+01 3.389E+01 3.117E+00 1.013E+00 1.728E+01 5.110E-02 1.390E+00 1.371E-02 8.308E-03 0.000E+00 8.308E-04 0.000E+00 
2.875E+00 2.032E+01 2.204E+01 2.250E+01 1.406E+00 4.063E-01 2.310E+01 1.412E-02 5.180E-01 3.323E-03 2.492E-03 0.000E+00 0.000E+00 0.000E+00 
3.125E+00 1.228E+01 1.349E+01 1.366E+01 4.997E-01 1.425E-01 9.746E+00 2.908E-03 9.388E-02 1.246E-03 0.000E+00 0.000E+00 0.000E+00 0.000E+00 
3.375E+00 6.714E+00 7.452E+00 7.464E+00 1.346E-01 2.783E-02 8.275E-01 4.154E-04 6.231E-03 0.000E+00 0.000E+00 0.000E+00 0.000E+00 0.000E+00 
3.625E+00 3.306E+00 3.768E+00 3.720E+00 2.617E-02 5.400E-03 5.733E-02 0.000E+00 0.000E+00 0.000E+00 0.000E+00 0.000E+00 0.000E+00 0.000E+00 
3.875E+00 1.424E+00 1.644E+00 1.579E+00 8.308E-04 0.000E+00 4.154E-04 0.000E+00 0.000E+00 0.000E+00 0.000E+00 0.000E+00 0.000E+00 0.000E+00 
4.125E+00 4.948E-01 5.770E-01 5.749E-01 0.000E+00 0.000E+00 0.000E+00 0.000E+00 0.000E+00 0.000E+00 0.000E+00 0.000E+00 0.000E+00 0.000E+00 
4.375E+00 1.421E-01 1.462E-01 1.346E-01 0.000E+00 0.000E+00 0.000E+00 0.000E+00 0.000E+00 0.000E+00 0.000E+00 0.000E+00 0.000E+00 0.000E+00 
4.625E+00 2.451E-02 3.614E-02 3.282E-02 0.000E+00 0.000E+00 0.000E+00 0.000E+00 0.000E+00 0.000E+00 0.000E+00 0.000E+00 0.000E+00 0.000E+00 
4.875E+00 2.077E-03 5.400E-03 6.231E-03 0.000E+00 0.000E+00 0.000E+00 0.000E+00 0.000E+00 0.000E+00 0.000E+00 0.000E+00 0.000E+00 0.000E+00 
5.125E+00 4.154E-04 4.154E-04 4.154E-04 0.000E+00 0.000E+00 0.000E+00 0.000E+00 0.000E+00 0.000E+00 0.000E+00 0.000E+00 0.000E+00 0.000E+00 
\end{verbatim}
}
\newpage
Au+Au \@ $E_{\rm CM}=56A~$GeV:
{\tiny
\begin{verbatim}
! y, dN/dy (pi+ pi- pi0 K+ K- P aP L+S0 a(L+S0) Xi- aXi- Om aOm) 
-6.375E+00 4.229E-04 0.000E+00 0.000E+00 0.000E+00 0.000E+00 0.000E+00 0.000E+00 0.000E+00 0.000E+00 0.000E+00 0.000E+00 0.000E+00 0.000E+00 
-6.125E+00 1.692E-03 8.458E-04 1.692E-03 0.000E+00 0.000E+00 0.000E+00 0.000E+00 0.000E+00 0.000E+00 0.000E+00 0.000E+00 0.000E+00 0.000E+00 
-5.875E+00 1.184E-02 1.522E-02 1.057E-02 0.000E+00 0.000E+00 0.000E+00 0.000E+00 0.000E+00 0.000E+00 0.000E+00 0.000E+00 0.000E+00 0.000E+00 
-5.625E+00 6.343E-02 8.373E-02 6.470E-02 0.000E+00 0.000E+00 0.000E+00 0.000E+00 0.000E+00 0.000E+00 0.000E+00 0.000E+00 0.000E+00 0.000E+00 
-5.375E+00 2.584E-01 2.871E-01 2.854E-01 4.229E-04 0.000E+00 0.000E+00 0.000E+00 0.000E+00 0.000E+00 0.000E+00 0.000E+00 0.000E+00 0.000E+00 
-5.125E+00 8.263E-01 8.965E-01 8.969E-01 0.000E+00 4.229E-04 0.000E+00 0.000E+00 0.000E+00 0.000E+00 0.000E+00 0.000E+00 0.000E+00 0.000E+00 
-4.875E+00 2.003E+00 2.217E+00 2.169E+00 6.343E-03 3.383E-03 6.343E-03 0.000E+00 0.000E+00 0.000E+00 0.000E+00 0.000E+00 0.000E+00 0.000E+00 
-4.625E+00 4.132E+00 4.638E+00 4.675E+00 4.948E-02 1.692E-02 1.789E-01 0.000E+00 8.458E-04 0.000E+00 0.000E+00 0.000E+00 0.000E+00 0.000E+00 
-4.375E+00 8.190E+00 8.829E+00 8.981E+00 2.081E-01 7.189E-02 3.643E+00 8.458E-04 2.622E-02 0.000E+00 0.000E+00 0.000E+00 0.000E+00 0.000E+00 
-4.125E+00 1.429E+01 1.530E+01 1.562E+01 7.007E-01 2.580E-01 1.989E+01 9.726E-03 2.199E-01 1.269E-03 1.269E-03 0.000E+00 0.000E+00 0.000E+00 
-3.875E+00 2.275E+01 2.435E+01 2.496E+01 1.785E+00 7.011E-01 1.460E+01 4.821E-02 7.899E-01 1.522E-02 8.035E-03 4.229E-04 4.229E-04 0.000E+00 
-3.625E+00 3.448E+01 3.633E+01 3.746E+01 3.510E+00 1.528E+00 1.352E+01 1.362E-01 1.591E+00 4.144E-02 1.988E-02 3.383E-03 1.269E-03 0.000E+00 
-3.375E+00 4.886E+01 5.132E+01 5.211E+01 6.000E+00 2.816E+00 1.600E+01 2.795E-01 2.648E+00 1.142E-01 3.806E-02 6.343E-03 2.537E-03 8.458E-04 
-3.125E+00 6.513E+01 6.794E+01 6.978E+01 8.565E+00 4.467E+00 1.831E+01 4.888E-01 4.130E+00 2.195E-01 8.542E-02 1.776E-02 3.383E-03 4.652E-03 
-2.875E+00 8.370E+01 8.732E+01 8.901E+01 1.149E+01 6.550E+00 2.012E+01 6.842E-01 5.512E+00 3.827E-01 1.298E-01 2.199E-02 1.607E-02 4.652E-03 
-2.625E+00 1.040E+02 1.075E+02 1.102E+02 1.426E+01 8.901E+00 2.125E+01 9.777E-01 6.877E+00 5.442E-01 1.971E-01 4.440E-02 3.256E-02 1.480E-02 
-2.375E+00 1.256E+02 1.299E+02 1.332E+02 1.691E+01 1.141E+01 2.202E+01 1.324E+00 7.746E+00 7.675E-01 2.541E-01 6.005E-02 4.271E-02 2.537E-02 
-2.125E+00 1.482E+02 1.527E+02 1.564E+02 1.976E+01 1.394E+01 2.229E+01 1.830E+00 8.589E+00 1.066E+00 3.100E-01 8.035E-02 6.470E-02 3.679E-02 
-1.875E+00 1.701E+02 1.744E+02 1.793E+02 2.236E+01 1.637E+01 2.247E+01 2.441E+00 9.151E+00 1.387E+00 3.480E-01 1.057E-01 8.288E-02 4.144E-02 
-1.625E+00 1.905E+02 1.939E+02 2.012E+02 2.467E+01 1.864E+01 2.223E+01 3.016E+00 9.385E+00 1.703E+00 4.055E-01 1.383E-01 8.880E-02 5.751E-02 
-1.375E+00 2.084E+02 2.130E+02 2.204E+02 2.673E+01 2.097E+01 2.205E+01 3.719E+00 9.495E+00 2.008E+00 4.334E-01 1.544E-01 9.895E-02 5.709E-02 
-1.125E+00 2.234E+02 2.273E+02 2.363E+02 2.841E+01 2.297E+01 2.176E+01 4.308E+00 9.544E+00 2.269E+00 5.032E-01 1.683E-01 1.142E-01 5.540E-02 
-8.750E-01 2.352E+02 2.394E+02 2.486E+02 2.985E+01 2.436E+01 2.133E+01 4.726E+00 9.538E+00 2.522E+00 5.379E-01 1.772E-01 1.193E-01 7.316E-02 
-6.250E-01 2.437E+02 2.483E+02 2.575E+02 3.066E+01 2.531E+01 2.106E+01 4.978E+00 9.360E+00 2.635E+00 5.244E-01 2.072E-01 1.235E-01 8.204E-02 
-3.750E-01 2.489E+02 2.525E+02 2.627E+02 3.102E+01 2.581E+01 2.055E+01 5.212E+00 9.268E+00 2.698E+00 5.582E-01 2.220E-01 1.315E-01 7.020E-02 
-1.250E-01 2.516E+02 2.555E+02 2.656E+02 3.154E+01 2.606E+01 2.024E+01 5.318E+00 9.126E+00 2.756E+00 5.633E-01 2.220E-01 1.421E-01 7.866E-02 
1.250E-01 2.511E+02 2.560E+02 2.661E+02 3.164E+01 2.622E+01 2.048E+01 5.312E+00 9.150E+00 2.814E+00 5.747E-01 2.271E-01 1.362E-01 8.627E-02 
3.750E-01 2.486E+02 2.527E+02 2.633E+02 3.123E+01 2.577E+01 2.073E+01 5.242E+00 9.329E+00 2.733E+00 5.527E-01 2.119E-01 1.391E-01 7.696E-02 
6.250E-01 2.440E+02 2.482E+02 2.578E+02 3.048E+01 2.533E+01 2.103E+01 5.004E+00 9.471E+00 2.629E+00 5.552E-01 2.047E-01 1.163E-01 8.119E-02 
8.750E-01 2.357E+02 2.394E+02 2.493E+02 2.978E+01 2.437E+01 2.155E+01 4.624E+00 9.496E+00 2.529E+00 5.430E-01 2.068E-01 1.087E-01 7.992E-02 
1.125E+00 2.240E+02 2.282E+02 2.369E+02 2.845E+01 2.277E+01 2.189E+01 4.304E+00 9.656E+00 2.234E+00 5.146E-01 1.653E-01 1.104E-01 6.470E-02 
1.375E+00 2.084E+02 2.131E+02 2.201E+02 2.667E+01 2.087E+01 2.190E+01 3.690E+00 9.643E+00 2.007E+00 4.618E-01 1.535E-01 1.146E-01 6.555E-02 
1.625E+00 1.907E+02 1.953E+02 2.015E+02 2.454E+01 1.882E+01 2.230E+01 3.023E+00 9.418E+00 1.679E+00 4.144E-01 1.307E-01 9.642E-02 5.582E-02 
1.875E+00 1.699E+02 1.744E+02 1.795E+02 2.230E+01 1.643E+01 2.236E+01 2.460E+00 9.128E+00 1.360E+00 3.759E-01 1.138E-01 7.823E-02 3.890E-02 
2.125E+00 1.484E+02 1.524E+02 1.567E+02 1.968E+01 1.390E+01 2.215E+01 1.851E+00 8.755E+00 1.084E+00 3.083E-01 7.823E-02 5.413E-02 3.425E-02 
2.375E+00 1.259E+02 1.297E+02 1.333E+02 1.688E+01 1.123E+01 2.209E+01 1.341E+00 7.831E+00 7.671E-01 2.673E-01 5.751E-02 4.948E-02 2.114E-02 
2.625E+00 1.046E+02 1.080E+02 1.102E+02 1.427E+01 8.825E+00 2.141E+01 9.663E-01 6.798E+00 5.442E-01 1.933E-01 4.398E-02 2.960E-02 1.522E-02 
2.875E+00 8.398E+01 8.737E+01 8.941E+01 1.134E+01 6.541E+00 2.016E+01 6.495E-01 5.604E+00 3.628E-01 1.252E-01 2.368E-02 2.030E-02 6.766E-03 
3.125E+00 6.511E+01 6.810E+01 6.991E+01 8.657E+00 4.463E+00 1.833E+01 4.740E-01 4.126E+00 2.216E-01 7.950E-02 1.565E-02 1.057E-02 3.383E-03 
3.375E+00 4.833E+01 5.102E+01 5.217E+01 5.967E+00 2.857E+00 1.593E+01 2.766E-01 2.676E+00 1.133E-01 3.594E-02 5.075E-03 5.497E-03 8.458E-04 
3.625E+00 3.448E+01 3.629E+01 3.732E+01 3.564E+00 1.519E+00 1.354E+01 1.400E-01 1.591E+00 5.032E-02 2.368E-02 4.229E-04 1.269E-03 0.000E+00 
3.875E+00 2.307E+01 2.445E+01 2.507E+01 1.824E+00 7.083E-01 1.455E+01 4.694E-02 7.840E-01 1.311E-02 7.189E-03 8.458E-04 0.000E+00 0.000E+00 
4.125E+00 1.417E+01 1.535E+01 1.564E+01 7.223E-01 2.491E-01 1.986E+01 7.612E-03 2.081E-01 2.114E-03 8.458E-04 0.000E+00 0.000E+00 0.000E+00 
4.375E+00 8.117E+00 8.761E+00 8.995E+00 2.157E-01 7.443E-02 3.557E+00 1.269E-03 2.326E-02 0.000E+00 0.000E+00 0.000E+00 0.000E+00 0.000E+00 
4.625E+00 4.189E+00 4.715E+00 4.644E+00 4.229E-02 1.565E-02 1.831E-01 0.000E+00 1.269E-03 0.000E+00 0.000E+00 0.000E+00 0.000E+00 0.000E+00 
4.875E+00 1.952E+00 2.195E+00 2.226E+00 7.189E-03 1.692E-03 5.920E-03 0.000E+00 0.000E+00 0.000E+00 0.000E+00 0.000E+00 0.000E+00 0.000E+00 
5.125E+00 8.204E-01 9.003E-01 8.821E-01 1.692E-03 0.000E+00 0.000E+00 0.000E+00 0.000E+00 0.000E+00 0.000E+00 0.000E+00 0.000E+00 0.000E+00 
5.375E+00 2.444E-01 2.871E-01 2.922E-01 0.000E+00 0.000E+00 0.000E+00 0.000E+00 0.000E+00 0.000E+00 0.000E+00 0.000E+00 0.000E+00 0.000E+00 
5.625E+00 6.005E-02 7.104E-02 7.104E-02 0.000E+00 0.000E+00 0.000E+00 0.000E+00 0.000E+00 0.000E+00 0.000E+00 0.000E+00 0.000E+00 0.000E+00 
5.875E+00 9.303E-03 1.226E-02 1.311E-02 0.000E+00 0.000E+00 0.000E+00 0.000E+00 0.000E+00 0.000E+00 0.000E+00 0.000E+00 0.000E+00 0.000E+00 
6.125E+00 1.692E-03 8.458E-04 2.960E-03 0.000E+00 0.000E+00 0.000E+00 0.000E+00 0.000E+00 0.000E+00 0.000E+00 0.000E+00 0.000E+00 0.000E+00 
6.375E+00 4.229E-04 0.000E+00 0.000E+00 0.000E+00 0.000E+00 0.000E+00 0.000E+00 0.000E+00 0.000E+00 0.000E+00 0.000E+00 0.000E+00 0.000E+00 
\end{verbatim}
}
\newpage
Au+Au \@ $E_{\rm CM}=62.5A~$GeV:
{\tiny
\begin{verbatim}
! y, dN/dy (pi+ pi- pi0 K+ K- P aP L+S0 a(L+S0) Xi- aXi- Om aOm) 
-6.375E+00 4.182E-04 2.091E-03 8.365E-04 0.000E+00 0.000E+00 0.000E+00 0.000E+00 0.000E+00 0.000E+00 0.000E+00 0.000E+00 0.000E+00 0.000E+00 
-6.125E+00 3.346E-03 5.437E-03 1.255E-03 0.000E+00 0.000E+00 0.000E+00 0.000E+00 0.000E+00 0.000E+00 0.000E+00 0.000E+00 0.000E+00 0.000E+00 
-5.875E+00 2.426E-02 2.509E-02 2.551E-02 0.000E+00 0.000E+00 0.000E+00 0.000E+00 0.000E+00 0.000E+00 0.000E+00 0.000E+00 0.000E+00 0.000E+00 
-5.625E+00 1.221E-01 1.297E-01 1.217E-01 0.000E+00 0.000E+00 0.000E+00 0.000E+00 0.000E+00 0.000E+00 0.000E+00 0.000E+00 0.000E+00 0.000E+00 
-5.375E+00 4.320E-01 4.810E-01 4.655E-01 0.000E+00 0.000E+00 0.000E+00 0.000E+00 0.000E+00 0.000E+00 0.000E+00 0.000E+00 0.000E+00 0.000E+00 
-5.125E+00 1.175E+00 1.324E+00 1.273E+00 1.673E-03 0.000E+00 0.000E+00 0.000E+00 0.000E+00 0.000E+00 0.000E+00 0.000E+00 0.000E+00 0.000E+00 
-4.875E+00 2.763E+00 2.998E+00 3.078E+00 1.380E-02 8.365E-03 3.722E-02 0.000E+00 0.000E+00 0.000E+00 0.000E+00 0.000E+00 0.000E+00 0.000E+00 
-4.625E+00 5.501E+00 6.118E+00 6.145E+00 7.737E-02 3.890E-02 5.379E-01 0.000E+00 3.764E-03 0.000E+00 0.000E+00 0.000E+00 0.000E+00 0.000E+00 
-4.375E+00 1.037E+01 1.111E+01 1.139E+01 3.559E-01 1.309E-01 1.042E+01 2.091E-03 5.730E-02 4.182E-04 0.000E+00 0.000E+00 0.000E+00 0.000E+00 
-4.125E+00 1.745E+01 1.855E+01 1.908E+01 1.074E+00 4.391E-01 1.971E+01 2.008E-02 4.157E-01 4.601E-03 2.091E-03 0.000E+00 0.000E+00 0.000E+00 
-3.875E+00 2.750E+01 2.894E+01 2.966E+01 2.386E+00 9.950E-01 1.232E+01 7.194E-02 1.090E+00 1.840E-02 1.715E-02 0.000E+00 4.182E-04 0.000E+00 
-3.625E+00 4.011E+01 4.217E+01 4.355E+01 4.525E+00 1.999E+00 1.409E+01 1.899E-01 1.975E+00 7.319E-02 2.719E-02 1.673E-03 8.365E-04 0.000E+00 
-3.375E+00 5.518E+01 5.787E+01 5.940E+01 7.118E+00 3.532E+00 1.666E+01 3.601E-01 3.205E+00 1.627E-01 6.608E-02 1.380E-02 3.346E-03 2.091E-03 
-3.125E+00 7.313E+01 7.627E+01 7.771E+01 9.860E+00 5.463E+00 1.898E+01 6.069E-01 4.653E+00 2.840E-01 9.452E-02 2.384E-02 1.004E-02 7.946E-03 
-2.875E+00 9.288E+01 9.642E+01 9.870E+01 1.274E+01 7.747E+00 2.050E+01 8.754E-01 6.100E+00 4.550E-01 1.510E-01 2.384E-02 2.133E-02 1.129E-02 
-2.625E+00 1.141E+02 1.179E+02 1.208E+02 1.545E+01 1.019E+01 2.138E+01 1.188E+00 7.266E+00 6.445E-01 2.200E-01 4.517E-02 5.019E-02 1.798E-02 
-2.375E+00 1.368E+02 1.408E+02 1.443E+02 1.836E+01 1.274E+01 2.180E+01 1.650E+00 8.173E+00 9.084E-01 2.781E-01 6.859E-02 5.437E-02 2.677E-02 
-2.125E+00 1.591E+02 1.634E+02 1.683E+02 2.107E+01 1.535E+01 2.218E+01 2.176E+00 8.876E+00 1.239E+00 3.442E-01 9.243E-02 8.072E-02 3.639E-02 
-1.875E+00 1.812E+02 1.858E+02 1.913E+02 2.372E+01 1.810E+01 2.219E+01 2.796E+00 9.228E+00 1.549E+00 4.145E-01 1.184E-01 8.867E-02 6.148E-02 
-1.625E+00 2.018E+02 2.059E+02 2.123E+02 2.599E+01 2.037E+01 2.203E+01 3.415E+00 9.377E+00 1.936E+00 4.529E-01 1.485E-01 8.950E-02 5.855E-02 
-1.375E+00 2.190E+02 2.231E+02 2.312E+02 2.792E+01 2.230E+01 2.181E+01 4.046E+00 9.578E+00 2.223E+00 4.847E-01 1.585E-01 1.184E-01 7.152E-02 
-1.125E+00 2.335E+02 2.372E+02 2.470E+02 2.967E+01 2.416E+01 2.147E+01 4.574E+00 9.504E+00 2.507E+00 4.960E-01 2.058E-01 1.150E-01 6.817E-02 
-8.750E-01 2.442E+02 2.488E+02 2.583E+02 3.125E+01 2.545E+01 2.106E+01 5.089E+00 9.345E+00 2.754E+00 5.232E-01 2.095E-01 1.397E-01 6.734E-02 
-6.250E-01 2.529E+02 2.563E+02 2.671E+02 3.190E+01 2.655E+01 2.052E+01 5.504E+00 9.296E+00 2.854E+00 5.705E-01 2.171E-01 1.393E-01 7.361E-02 
-3.750E-01 2.575E+02 2.611E+02 2.718E+02 3.231E+01 2.695E+01 2.011E+01 5.588E+00 9.175E+00 2.927E+00 5.755E-01 2.120E-01 1.418E-01 8.323E-02 
-1.250E-01 2.598E+02 2.634E+02 2.743E+02 3.256E+01 2.739E+01 2.001E+01 5.746E+00 9.082E+00 2.965E+00 5.634E-01 2.292E-01 1.313E-01 8.825E-02 
1.250E-01 2.596E+02 2.635E+02 2.739E+02 3.248E+01 2.735E+01 2.025E+01 5.847E+00 9.010E+00 2.973E+00 5.956E-01 2.518E-01 1.476E-01 8.574E-02 
3.750E-01 2.571E+02 2.608E+02 2.718E+02 3.218E+01 2.704E+01 2.010E+01 5.650E+00 9.230E+00 2.915E+00 5.680E-01 2.388E-01 1.397E-01 8.992E-02 
6.250E-01 2.524E+02 2.568E+02 2.660E+02 3.190E+01 2.633E+01 2.059E+01 5.435E+00 9.350E+00 2.856E+00 5.487E-01 2.095E-01 1.422E-01 7.570E-02 
8.750E-01 2.444E+02 2.487E+02 2.580E+02 3.079E+01 2.560E+01 2.105E+01 5.109E+00 9.532E+00 2.712E+00 5.521E-01 2.200E-01 1.267E-01 7.570E-02 
1.125E+00 2.338E+02 2.370E+02 2.467E+02 2.974E+01 2.419E+01 2.124E+01 4.637E+00 9.564E+00 2.537E+00 5.136E-01 1.970E-01 1.192E-01 7.654E-02 
1.375E+00 2.190E+02 2.234E+02 2.316E+02 2.794E+01 2.247E+01 2.172E+01 4.150E+00 9.534E+00 2.237E+00 4.860E-01 1.690E-01 1.058E-01 7.110E-02 
1.625E+00 2.012E+02 2.053E+02 2.131E+02 2.587E+01 2.019E+01 2.205E+01 3.391E+00 9.405E+00 1.923E+00 4.333E-01 1.522E-01 1.012E-01 6.106E-02 
1.875E+00 1.813E+02 1.853E+02 1.915E+02 2.381E+01 1.776E+01 2.220E+01 2.809E+00 9.330E+00 1.578E+00 3.973E-01 1.280E-01 9.076E-02 5.102E-02 
2.125E+00 1.592E+02 1.638E+02 1.687E+02 2.100E+01 1.535E+01 2.233E+01 2.193E+00 8.905E+00 1.227E+00 3.363E-01 9.118E-02 7.779E-02 4.517E-02 
2.375E+00 1.362E+02 1.405E+02 1.442E+02 1.834E+01 1.271E+01 2.213E+01 1.629E+00 8.191E+00 8.946E-01 2.798E-01 6.859E-02 6.023E-02 3.053E-02 
2.625E+00 1.142E+02 1.179E+02 1.212E+02 1.546E+01 1.013E+01 2.156E+01 1.178E+00 7.307E+00 6.278E-01 2.317E-01 4.935E-02 3.680E-02 1.798E-02 
2.875E+00 9.255E+01 9.623E+01 9.851E+01 1.270E+01 7.621E+00 2.047E+01 8.331E-01 6.143E+00 4.722E-01 1.573E-01 3.430E-02 2.091E-02 1.631E-02 
3.125E+00 7.299E+01 7.621E+01 7.789E+01 9.842E+00 5.498E+00 1.889E+01 5.563E-01 4.721E+00 2.999E-01 1.041E-01 1.464E-02 1.464E-02 4.601E-03 
3.375E+00 5.527E+01 5.821E+01 5.952E+01 6.977E+00 3.542E+00 1.652E+01 3.572E-01 3.145E+00 1.610E-01 5.061E-02 9.201E-03 2.509E-03 1.255E-03 
3.625E+00 4.016E+01 4.230E+01 4.353E+01 4.535E+00 2.037E+00 1.406E+01 2.074E-01 1.899E+00 7.110E-02 2.593E-02 3.346E-03 1.673E-03 8.365E-04 
3.875E+00 2.754E+01 2.911E+01 2.991E+01 2.446E+00 1.039E+00 1.224E+01 7.068E-02 1.075E+00 2.384E-02 1.506E-02 8.365E-04 4.182E-04 0.000E+00 
4.125E+00 1.762E+01 1.873E+01 1.910E+01 1.125E+00 4.040E-01 1.949E+01 2.008E-02 4.086E-01 4.601E-03 2.509E-03 0.000E+00 0.000E+00 0.000E+00 
4.375E+00 1.032E+01 1.126E+01 1.133E+01 3.643E-01 1.393E-01 1.038E+01 2.509E-03 6.608E-02 1.255E-03 4.182E-04 0.000E+00 0.000E+00 0.000E+00 
4.625E+00 5.637E+00 6.140E+00 6.184E+00 9.368E-02 3.513E-02 5.115E-01 4.182E-04 4.182E-03 0.000E+00 0.000E+00 0.000E+00 0.000E+00 0.000E+00 
4.875E+00 2.744E+00 3.038E+00 3.031E+00 1.547E-02 4.182E-03 3.053E-02 0.000E+00 4.182E-04 0.000E+00 0.000E+00 0.000E+00 0.000E+00 0.000E+00 
5.125E+00 1.141E+00 1.339E+00 1.303E+00 1.255E-03 8.365E-04 4.182E-04 0.000E+00 0.000E+00 0.000E+00 0.000E+00 0.000E+00 0.000E+00 0.000E+00 
5.375E+00 4.253E-01 4.714E-01 4.797E-01 0.000E+00 0.000E+00 0.000E+00 0.000E+00 0.000E+00 0.000E+00 0.000E+00 0.000E+00 0.000E+00 0.000E+00 
5.625E+00 1.175E-01 1.380E-01 1.368E-01 0.000E+00 0.000E+00 0.000E+00 0.000E+00 0.000E+00 0.000E+00 0.000E+00 0.000E+00 0.000E+00 0.000E+00 
5.875E+00 2.677E-02 2.844E-02 2.802E-02 0.000E+00 0.000E+00 0.000E+00 0.000E+00 0.000E+00 0.000E+00 0.000E+00 0.000E+00 0.000E+00 0.000E+00 
6.125E+00 1.255E-03 3.764E-03 4.601E-03 0.000E+00 0.000E+00 0.000E+00 0.000E+00 0.000E+00 0.000E+00 0.000E+00 0.000E+00 0.000E+00 0.000E+00 
6.375E+00 4.182E-04 4.182E-04 0.000E+00 0.000E+00 0.000E+00 0.000E+00 0.000E+00 0.000E+00 0.000E+00 0.000E+00 0.000E+00 0.000E+00 0.000E+00 
\end{verbatim}
}
\newpage
Au+Au \@ $E_{\rm CM}=130A~$GeV:
{\tiny
\begin{verbatim}
 ! y, dN/dy (pi+ pi- pi0 K+ K- P aP L+S0 a(L+S0) Xi- aXi- Om aOm) 
-6.875E+00 2.035E-03 1.628E-03 2.850E-03 0.000E+00 0.000E+00 0.000E+00 0.000E+00 0.000E+00 0.000E+00 0.000E+00 0.000E+00 0.000E+00 0.000E+00 
-6.625E+00 1.221E-02 1.628E-02 1.058E-02 0.000E+00 0.000E+00 0.000E+00 0.000E+00 0.000E+00 0.000E+00 0.000E+00 0.000E+00 0.000E+00 0.000E+00 
-6.375E+00 7.531E-02 7.572E-02 8.142E-02 0.000E+00 0.000E+00 0.000E+00 0.000E+00 0.000E+00 0.000E+00 0.000E+00 0.000E+00 0.000E+00 0.000E+00 
-6.125E+00 2.854E-01 3.293E-01 3.289E-01 0.000E+00 0.000E+00 0.000E+00 0.000E+00 0.000E+00 0.000E+00 0.000E+00 0.000E+00 0.000E+00 0.000E+00 
-5.875E+00 7.979E-01 9.017E-01 8.935E-01 4.071E-04 0.000E+00 0.000E+00 0.000E+00 0.000E+00 0.000E+00 0.000E+00 0.000E+00 0.000E+00 0.000E+00 
-5.625E+00 1.960E+00 2.133E+00 2.183E+00 8.142E-03 1.628E-03 9.363E-03 0.000E+00 0.000E+00 0.000E+00 0.000E+00 0.000E+00 0.000E+00 0.000E+00 
-5.375E+00 4.275E+00 4.708E+00 4.736E+00 5.862E-02 1.832E-02 2.467E-01 0.000E+00 1.628E-03 0.000E+00 0.000E+00 0.000E+00 0.000E+00 0.000E+00 
-5.125E+00 8.332E+00 9.017E+00 9.163E+00 2.276E-01 8.712E-02 8.618E+00 8.142E-04 2.483E-02 4.071E-04 0.000E+00 0.000E+00 0.000E+00 0.000E+00 
-4.875E+00 1.516E+01 1.609E+01 1.646E+01 7.352E-01 3.073E-01 1.878E+01 1.588E-02 2.499E-01 1.628E-03 2.850E-03 4.071E-04 0.000E+00 0.000E+00 
-4.625E+00 2.480E+01 2.593E+01 2.689E+01 1.823E+00 7.975E-01 8.909E+00 6.513E-02 7.193E-01 1.628E-02 8.549E-03 4.071E-04 0.000E+00 0.000E+00 
-4.375E+00 3.791E+01 3.921E+01 4.061E+01 3.564E+00 1.856E+00 1.155E+01 1.921E-01 1.470E+00 5.129E-02 2.565E-02 2.442E-03 1.221E-03 0.000E+00 
-4.125E+00 5.359E+01 5.582E+01 5.717E+01 6.103E+00 3.438E+00 1.469E+01 4.189E-01 2.632E+00 1.437E-01 4.763E-02 7.327E-03 2.850E-03 1.628E-03 
-3.875E+00 7.187E+01 7.445E+01 7.660E+01 9.098E+00 5.613E+00 1.750E+01 6.953E-01 4.173E+00 3.045E-01 9.566E-02 1.791E-02 8.142E-03 4.478E-03 
-3.625E+00 9.189E+01 9.542E+01 9.776E+01 1.234E+01 8.166E+00 1.986E+01 1.033E+00 5.546E+00 5.117E-01 1.571E-01 3.867E-02 2.280E-02 1.181E-02 
-3.375E+00 1.148E+02 1.182E+02 1.216E+02 1.553E+01 1.104E+01 2.130E+01 1.395E+00 6.873E+00 7.336E-01 2.483E-01 6.066E-02 3.257E-02 1.954E-02 
-3.125E+00 1.386E+02 1.432E+02 1.465E+02 1.870E+01 1.380E+01 2.204E+01 1.943E+00 7.844E+00 1.010E+00 2.972E-01 8.264E-02 6.554E-02 3.053E-02 
-2.875E+00 1.638E+02 1.673E+02 1.729E+02 2.202E+01 1.682E+01 2.203E+01 2.515E+00 8.701E+00 1.386E+00 3.749E-01 9.607E-02 8.060E-02 4.641E-02 
-2.625E+00 1.882E+02 1.919E+02 1.985E+02 2.501E+01 1.992E+01 2.189E+01 3.122E+00 9.100E+00 1.747E+00 4.315E-01 1.400E-01 1.062E-01 5.048E-02 
-2.375E+00 2.106E+02 2.148E+02 2.220E+02 2.754E+01 2.226E+01 2.142E+01 4.012E+00 9.306E+00 2.198E+00 4.934E-01 1.661E-01 1.079E-01 6.920E-02 
-2.125E+00 2.317E+02 2.353E+02 2.432E+02 3.001E+01 2.483E+01 2.076E+01 4.727E+00 9.384E+00 2.588E+00 5.406E-01 2.080E-01 1.445E-01 8.060E-02 
-1.875E+00 2.497E+02 2.527E+02 2.627E+02 3.224E+01 2.715E+01 2.019E+01 5.453E+00 9.225E+00 3.039E+00 5.357E-01 2.329E-01 1.449E-01 8.264E-02 
-1.625E+00 2.653E+02 2.685E+02 2.787E+02 3.397E+01 2.932E+01 1.965E+01 6.137E+00 9.193E+00 3.317E+00 5.471E-01 2.467E-01 1.518E-01 9.281E-02 
-1.375E+00 2.785E+02 2.820E+02 2.921E+02 3.587E+01 3.104E+01 1.919E+01 6.807E+00 9.089E+00 3.698E+00 5.744E-01 2.789E-01 1.575E-01 9.851E-02 
-1.125E+00 2.898E+02 2.926E+02 3.034E+02 3.711E+01 3.227E+01 1.877E+01 7.248E+00 8.942E+00 3.890E+00 6.004E-01 2.915E-01 1.596E-01 1.176E-01 
-8.750E-01 2.975E+02 3.002E+02 3.121E+02 3.811E+01 3.342E+01 1.823E+01 7.610E+00 8.618E+00 4.151E+00 6.082E-01 3.204E-01 1.718E-01 1.189E-01 
-6.250E-01 3.029E+02 3.065E+02 3.175E+02 3.873E+01 3.411E+01 1.778E+01 8.006E+00 8.560E+00 4.275E+00 6.041E-01 3.314E-01 1.677E-01 1.286E-01 
-3.750E-01 3.069E+02 3.091E+02 3.213E+02 3.888E+01 3.429E+01 1.775E+01 8.131E+00 8.462E+00 4.364E+00 6.049E-01 3.428E-01 1.539E-01 1.233E-01 
-1.250E-01 3.077E+02 3.113E+02 3.231E+02 3.884E+01 3.448E+01 1.736E+01 8.212E+00 8.258E+00 4.397E+00 5.834E-01 3.187E-01 1.750E-01 1.225E-01 
1.250E-01 3.078E+02 3.109E+02 3.230E+02 3.903E+01 3.473E+01 1.749E+01 8.170E+00 8.372E+00 4.390E+00 5.821E-01 3.472E-01 1.848E-01 1.274E-01 
3.750E-01 3.065E+02 3.091E+02 3.213E+02 3.900E+01 3.454E+01 1.767E+01 8.095E+00 8.389E+00 4.390E+00 5.813E-01 3.546E-01 1.767E-01 1.225E-01 
6.250E-01 3.030E+02 3.061E+02 3.177E+02 3.898E+01 3.419E+01 1.793E+01 8.140E+00 8.542E+00 4.257E+00 6.025E-01 3.342E-01 1.702E-01 1.347E-01 
8.750E-01 2.977E+02 3.010E+02 3.117E+02 3.811E+01 3.333E+01 1.833E+01 7.800E+00 8.793E+00 4.136E+00 6.131E-01 3.196E-01 1.681E-01 1.197E-01 
1.125E+00 2.885E+02 2.929E+02 3.032E+02 3.712E+01 3.233E+01 1.858E+01 7.262E+00 8.885E+00 3.820E+00 5.785E-01 3.078E-01 1.559E-01 1.132E-01 
1.375E+00 2.789E+02 2.816E+02 2.921E+02 3.582E+01 3.085E+01 1.930E+01 6.700E+00 8.996E+00 3.695E+00 5.899E-01 2.793E-01 1.543E-01 1.099E-01 
1.625E+00 2.654E+02 2.687E+02 2.787E+02 3.430E+01 2.925E+01 1.978E+01 6.215E+00 9.111E+00 3.360E+00 5.443E-01 2.463E-01 1.498E-01 8.874E-02 
1.875E+00 2.492E+02 2.532E+02 2.621E+02 3.224E+01 2.710E+01 2.013E+01 5.453E+00 9.374E+00 2.977E+00 5.463E-01 2.394E-01 1.413E-01 1.005E-01 
2.125E+00 2.316E+02 2.344E+02 2.427E+02 2.980E+01 2.483E+01 2.082E+01 4.672E+00 9.453E+00 2.535E+00 5.023E-01 2.027E-01 1.266E-01 7.246E-02 
2.375E+00 2.107E+02 2.142E+02 2.219E+02 2.773E+01 2.210E+01 2.133E+01 3.946E+00 9.445E+00 2.200E+00 4.686E-01 1.502E-01 1.099E-01 6.473E-02 
2.625E+00 1.877E+02 1.913E+02 1.979E+02 2.507E+01 1.944E+01 2.200E+01 3.189E+00 9.154E+00 1.755E+00 4.071E-01 1.441E-01 9.933E-02 5.658E-02 
2.875E+00 1.636E+02 1.672E+02 1.724E+02 2.188E+01 1.676E+01 2.227E+01 2.493E+00 8.584E+00 1.339E+00 3.542E-01 1.181E-01 8.427E-02 5.251E-02 
3.125E+00 1.390E+02 1.428E+02 1.466E+02 1.876E+01 1.383E+01 2.210E+01 1.964E+00 8.044E+00 1.003E+00 3.277E-01 7.979E-02 5.984E-02 3.216E-02 
3.375E+00 1.144E+02 1.182E+02 1.216E+02 1.565E+01 1.087E+01 2.147E+01 1.461E+00 6.923E+00 7.486E-01 2.402E-01 6.228E-02 4.030E-02 1.995E-02 
3.625E+00 9.184E+01 9.501E+01 9.773E+01 1.230E+01 8.229E+00 1.988E+01 1.008E+00 5.628E+00 4.938E-01 1.388E-01 4.437E-02 1.954E-02 1.018E-02 
3.875E+00 7.136E+01 7.427E+01 7.637E+01 9.128E+00 5.552E+00 1.779E+01 6.884E-01 4.197E+00 2.976E-01 9.770E-02 1.913E-02 1.262E-02 3.257E-03 
4.125E+00 5.343E+01 5.563E+01 5.725E+01 6.186E+00 3.388E+00 1.482E+01 4.270E-01 2.642E+00 1.498E-01 4.641E-02 8.549E-03 1.221E-03 1.628E-03 
4.375E+00 3.747E+01 3.960E+01 4.068E+01 3.605E+00 1.837E+00 1.141E+01 1.864E-01 1.438E+00 5.251E-02 2.280E-02 2.442E-03 1.628E-03 0.000E+00 
4.625E+00 2.469E+01 2.597E+01 2.678E+01 1.785E+00 8.146E-01 9.024E+00 5.740E-02 7.238E-01 8.956E-03 7.735E-03 0.000E+00 0.000E+00 0.000E+00 
4.875E+00 1.501E+01 1.597E+01 1.641E+01 7.360E-01 2.817E-01 1.859E+01 1.750E-02 2.499E-01 2.442E-03 1.221E-03 0.000E+00 0.000E+00 0.000E+00 
5.125E+00 8.436E+00 8.925E+00 9.184E+00 2.426E-01 8.142E-02 8.609E+00 1.221E-03 3.053E-02 0.000E+00 0.000E+00 0.000E+00 0.000E+00 0.000E+00 
5.375E+00 4.425E+00 4.673E+00 4.825E+00 5.781E-02 1.343E-02 2.153E-01 0.000E+00 1.628E-03 0.000E+00 0.000E+00 0.000E+00 0.000E+00 0.000E+00 
5.625E+00 1.983E+00 2.231E+00 2.186E+00 7.735E-03 3.664E-03 1.303E-02 0.000E+00 0.000E+00 0.000E+00 0.000E+00 0.000E+00 0.000E+00 0.000E+00 
5.875E+00 8.357E-01 9.066E-01 9.143E-01 1.221E-03 0.000E+00 0.000E+00 0.000E+00 0.000E+00 0.000E+00 0.000E+00 0.000E+00 0.000E+00 0.000E+00 
6.125E+00 2.813E-01 3.155E-01 3.045E-01 0.000E+00 0.000E+00 0.000E+00 0.000E+00 0.000E+00 0.000E+00 0.000E+00 0.000E+00 0.000E+00 0.000E+00 
6.375E+00 8.101E-02 7.897E-02 8.671E-02 0.000E+00 0.000E+00 0.000E+00 0.000E+00 0.000E+00 0.000E+00 0.000E+00 0.000E+00 0.000E+00 0.000E+00 
6.625E+00 1.669E-02 1.873E-02 1.750E-02 0.000E+00 0.000E+00 0.000E+00 0.000E+00 0.000E+00 0.000E+00 0.000E+00 0.000E+00 0.000E+00 0.000E+00 
6.875E+00 2.035E-03 1.221E-03 3.257E-03 0.000E+00 0.000E+00 0.000E+00 0.000E+00 0.000E+00 0.000E+00 0.000E+00 0.000E+00 0.000E+00 0.000E+00 
7.125E+00 4.071E-04 0.000E+00 0.000E+00 0.000E+00 0.000E+00 0.000E+00 0.000E+00 0.000E+00 0.000E+00 0.000E+00 0.000E+00 0.000E+00 0.000E+00 
\end{verbatim}
}
\newpage
Au+Au \@ $E_{\rm CM}=200A~$GeV:
{\tiny
\begin{verbatim}
! y, dN/dy (pi+ pi- pi0 K+ K- P aP L+S0 a(L+S0) Xi- aXi- Om aOm) 
-7.375E+00 8.631E-04 8.631E-04 1.726E-03 0.000E+00 0.000E+00 0.000E+00 0.000E+00 0.000E+00 0.000E+00 0.000E+00 0.000E+00 0.000E+00 0.000E+00 
-7.125E+00 5.179E-03 8.631E-03 7.768E-03 0.000E+00 0.000E+00 0.000E+00 0.000E+00 0.000E+00 0.000E+00 0.000E+00 0.000E+00 0.000E+00 0.000E+00 
-6.875E+00 3.496E-02 4.445E-02 4.618E-02 0.000E+00 0.000E+00 0.000E+00 0.000E+00 0.000E+00 0.000E+00 0.000E+00 0.000E+00 0.000E+00 0.000E+00 
-6.625E+00 1.571E-01 1.756E-01 1.808E-01 0.000E+00 0.000E+00 0.000E+00 0.000E+00 0.000E+00 0.000E+00 0.000E+00 0.000E+00 0.000E+00 0.000E+00 
-6.375E+00 4.842E-01 5.256E-01 5.554E-01 4.315E-04 4.315E-04 0.000E+00 0.000E+00 0.000E+00 0.000E+00 0.000E+00 0.000E+00 0.000E+00 0.000E+00 
-6.125E+00 1.251E+00 1.412E+00 1.442E+00 1.295E-03 8.631E-04 2.158E-03 0.000E+00 0.000E+00 0.000E+00 0.000E+00 0.000E+00 0.000E+00 0.000E+00 
-5.875E+00 2.884E+00 3.159E+00 3.233E+00 2.503E-02 9.926E-03 5.308E-02 0.000E+00 0.000E+00 0.000E+00 0.000E+00 0.000E+00 0.000E+00 0.000E+00 
-5.625E+00 5.950E+00 6.375E+00 6.551E+00 1.213E-01 4.272E-02 4.030E+00 0.000E+00 6.905E-03 0.000E+00 0.000E+00 0.000E+00 0.000E+00 0.000E+00 
-5.375E+00 1.134E+01 1.204E+01 1.240E+01 4.182E-01 1.627E-01 1.909E+01 3.452E-03 1.174E-01 0.000E+00 8.631E-04 0.000E+00 0.000E+00 0.000E+00 
-5.125E+00 1.985E+01 2.085E+01 2.137E+01 1.121E+00 4.933E-01 8.266E+00 3.452E-02 4.531E-01 6.905E-03 7.768E-03 0.000E+00 0.000E+00 0.000E+00 
-4.875E+00 3.213E+01 3.357E+01 3.457E+01 2.589E+00 1.279E+00 8.386E+00 1.364E-01 9.226E-01 3.323E-02 1.424E-02 4.315E-04 0.000E+00 0.000E+00 
-4.625E+00 4.796E+01 4.979E+01 5.122E+01 4.817E+00 2.685E+00 1.189E+01 3.301E-01 1.950E+00 1.096E-01 3.107E-02 4.315E-03 1.726E-03 1.726E-03 
-4.375E+00 6.660E+01 6.923E+01 7.119E+01 7.744E+00 4.788E+00 1.615E+01 6.896E-01 3.466E+00 2.512E-01 5.912E-02 1.554E-02 6.042E-03 2.158E-03 
-4.125E+00 8.748E+01 9.054E+01 9.353E+01 1.116E+01 7.412E+00 1.944E+01 1.088E+00 5.146E+00 4.613E-01 1.342E-01 3.582E-02 1.985E-02 1.036E-02 
-3.875E+00 1.103E+02 1.140E+02 1.165E+02 1.455E+01 1.056E+01 2.149E+01 1.507E+00 6.559E+00 7.060E-01 2.326E-01 5.394E-02 2.546E-02 1.424E-02 
-3.625E+00 1.353E+02 1.386E+02 1.429E+02 1.848E+01 1.376E+01 2.269E+01 2.011E+00 7.737E+00 9.762E-01 3.219E-01 8.545E-02 5.955E-02 3.064E-02 
-3.375E+00 1.606E+02 1.638E+02 1.689E+02 2.181E+01 1.705E+01 2.306E+01 2.553E+00 8.545E+00 1.356E+00 3.780E-01 1.023E-01 7.811E-02 4.186E-02 
-3.125E+00 1.856E+02 1.894E+02 1.955E+02 2.535E+01 2.025E+01 2.277E+01 3.260E+00 8.996E+00 1.715E+00 4.471E-01 1.364E-01 9.839E-02 5.696E-02 
-2.875E+00 2.100E+02 2.140E+02 2.213E+02 2.847E+01 2.312E+01 2.199E+01 3.979E+00 9.402E+00 2.164E+00 4.881E-01 1.687E-01 1.286E-01 6.344E-02 
-2.625E+00 2.325E+02 2.361E+02 2.449E+02 3.117E+01 2.598E+01 2.130E+01 4.840E+00 9.463E+00 2.616E+00 5.437E-01 2.020E-01 1.359E-01 8.458E-02 
-2.375E+00 2.527E+02 2.565E+02 2.652E+02 3.337E+01 2.840E+01 2.054E+01 5.623E+00 9.492E+00 3.070E+00 5.614E-01 2.309E-01 1.618E-01 9.408E-02 
-2.125E+00 2.706E+02 2.743E+02 2.846E+02 3.521E+01 3.069E+01 1.984E+01 6.256E+00 9.319E+00 3.458E+00 5.519E-01 2.619E-01 1.623E-01 1.066E-01 
-1.875E+00 2.865E+02 2.895E+02 3.004E+02 3.717E+01 3.285E+01 1.896E+01 7.054E+00 9.021E+00 3.846E+00 5.865E-01 3.120E-01 1.843E-01 1.174E-01 
-1.625E+00 3.014E+02 3.042E+02 3.151E+02 3.901E+01 3.442E+01 1.842E+01 7.739E+00 8.810E+00 4.223E+00 5.744E-01 3.198E-01 1.804E-01 1.329E-01 
-1.375E+00 3.135E+02 3.164E+02 3.290E+02 4.066E+01 3.643E+01 1.800E+01 8.136E+00 8.777E+00 4.633E+00 5.916E-01 3.586E-01 1.942E-01 1.394E-01 
-1.125E+00 3.245E+02 3.258E+02 3.389E+02 4.223E+01 3.786E+01 1.771E+01 8.681E+00 8.678E+00 4.871E+00 6.193E-01 3.500E-01 1.899E-01 1.420E-01 
-8.750E-01 3.314E+02 3.343E+02 3.471E+02 4.303E+01 3.884E+01 1.739E+01 9.126E+00 8.637E+00 5.081E+00 6.478E-01 3.690E-01 1.769E-01 1.450E-01 
-6.250E-01 3.363E+02 3.396E+02 3.527E+02 4.360E+01 3.923E+01 1.712E+01 9.419E+00 8.419E+00 5.286E+00 6.145E-01 3.845E-01 1.942E-01 1.709E-01 
-3.750E-01 3.397E+02 3.423E+02 3.563E+02 4.350E+01 3.941E+01 1.661E+01 9.442E+00 8.303E+00 5.262E+00 6.266E-01 3.918E-01 2.115E-01 1.558E-01 
-1.250E-01 3.410E+02 3.431E+02 3.578E+02 4.357E+01 3.942E+01 1.665E+01 9.560E+00 8.181E+00 5.292E+00 5.770E-01 3.780E-01 1.856E-01 1.623E-01 
1.250E-01 3.415E+02 3.438E+02 3.572E+02 4.348E+01 3.948E+01 1.648E+01 9.633E+00 8.150E+00 5.250E+00 6.175E-01 4.022E-01 1.933E-01 1.640E-01 
3.750E-01 3.394E+02 3.425E+02 3.559E+02 4.368E+01 3.940E+01 1.674E+01 9.699E+00 8.356E+00 5.196E+00 5.942E-01 3.875E-01 1.825E-01 1.614E-01 
6.250E-01 3.367E+02 3.392E+02 3.532E+02 4.329E+01 3.917E+01 1.686E+01 9.411E+00 8.436E+00 5.222E+00 6.193E-01 3.802E-01 2.084E-01 1.497E-01 
8.750E-01 3.321E+02 3.347E+02 3.472E+02 4.301E+01 3.887E+01 1.738E+01 9.085E+00 8.633E+00 4.986E+00 6.119E-01 3.957E-01 2.046E-01 1.472E-01 
1.125E+00 3.238E+02 3.263E+02 3.387E+02 4.197E+01 3.772E+01 1.783E+01 8.776E+00 8.813E+00 4.930E+00 6.020E-01 3.526E-01 2.080E-01 1.450E-01 
1.375E+00 3.130E+02 3.166E+02 3.276E+02 4.053E+01 3.638E+01 1.801E+01 8.335E+00 8.878E+00 4.605E+00 6.059E-01 3.487E-01 1.851E-01 1.282E-01 
1.625E+00 3.010E+02 3.039E+02 3.152E+02 3.891E+01 3.454E+01 1.840E+01 7.629E+00 8.890E+00 4.292E+00 5.973E-01 3.047E-01 1.873E-01 1.299E-01 
1.875E+00 2.865E+02 2.894E+02 2.996E+02 3.725E+01 3.250E+01 1.899E+01 7.049E+00 9.193E+00 3.909E+00 5.753E-01 2.827E-01 1.705E-01 1.169E-01 
2.125E+00 2.705E+02 2.734E+02 2.844E+02 3.559E+01 3.047E+01 1.978E+01 6.277E+00 9.284E+00 3.518E+00 5.765E-01 2.611E-01 1.661E-01 8.545E-02 
2.375E+00 2.524E+02 2.558E+02 2.651E+02 3.329E+01 2.849E+01 2.056E+01 5.642E+00 9.469E+00 3.127E+00 5.796E-01 2.227E-01 1.415E-01 9.667E-02 
2.625E+00 2.325E+02 2.363E+02 2.446E+02 3.093E+01 2.595E+01 2.154E+01 4.829E+00 9.362E+00 2.669E+00 5.412E-01 2.002E-01 1.437E-01 8.199E-02 
2.875E+00 2.100E+02 2.137E+02 2.215E+02 2.823E+01 2.332E+01 2.218E+01 3.949E+00 9.328E+00 2.156E+00 4.872E-01 1.653E-01 1.213E-01 7.897E-02 
3.125E+00 1.859E+02 1.893E+02 1.954E+02 2.543E+01 2.023E+01 2.279E+01 3.266E+00 9.004E+00 1.727E+00 4.484E-01 1.325E-01 9.451E-02 4.272E-02 
3.375E+00 1.604E+02 1.637E+02 1.687E+02 2.179E+01 1.704E+01 2.306E+01 2.582E+00 8.450E+00 1.339E+00 4.009E-01 1.087E-01 7.509E-02 4.445E-02 
3.625E+00 1.349E+02 1.387E+02 1.429E+02 1.821E+01 1.380E+01 2.283E+01 2.029E+00 7.646E+00 1.012E+00 3.055E-01 7.638E-02 4.488E-02 2.805E-02 
3.875E+00 1.102E+02 1.137E+02 1.169E+02 1.445E+01 1.056E+01 2.161E+01 1.512E+00 6.531E+00 6.616E-01 2.179E-01 5.912E-02 2.719E-02 2.028E-02 
4.125E+00 8.743E+01 9.022E+01 9.325E+01 1.104E+01 7.529E+00 1.952E+01 1.062E+00 5.046E+00 4.471E-01 1.381E-01 2.891E-02 1.467E-02 6.905E-03 
4.375E+00 6.675E+01 6.925E+01 7.123E+01 7.786E+00 4.734E+00 1.623E+01 6.711E-01 3.428E+00 2.529E-01 7.466E-02 1.597E-02 6.905E-03 3.884E-03 
4.625E+00 4.797E+01 4.988E+01 5.149E+01 4.753E+00 2.641E+00 1.198E+01 3.362E-01 1.946E+00 1.057E-01 2.676E-02 6.042E-03 2.589E-03 1.295E-03 
4.875E+00 3.212E+01 3.354E+01 3.439E+01 2.541E+00 1.297E+00 8.487E+00 1.450E-01 9.775E-01 3.280E-02 1.726E-02 8.631E-04 4.315E-04 4.315E-04 
5.125E+00 1.974E+01 2.075E+01 2.138E+01 1.128E+00 4.833E-01 8.299E+00 3.452E-02 4.363E-01 5.179E-03 7.336E-03 4.315E-04 0.000E+00 0.000E+00 
5.375E+00 1.138E+01 1.197E+01 1.248E+01 4.311E-01 1.472E-01 1.905E+01 3.021E-03 1.226E-01 1.295E-03 0.000E+00 0.000E+00 0.000E+00 0.000E+00 
5.625E+00 6.023E+00 6.519E+00 6.596E+00 1.282E-01 4.488E-02 4.047E+00 4.315E-04 9.494E-03 0.000E+00 0.000E+00 0.000E+00 0.000E+00 0.000E+00 
5.875E+00 2.935E+00 3.235E+00 3.220E+00 2.546E-02 6.473E-03 5.265E-02 0.000E+00 0.000E+00 0.000E+00 0.000E+00 0.000E+00 0.000E+00 0.000E+00 
6.125E+00 1.256E+00 1.410E+00 1.411E+00 3.021E-03 2.158E-03 1.295E-03 0.000E+00 0.000E+00 0.000E+00 0.000E+00 0.000E+00 0.000E+00 0.000E+00 
6.375E+00 4.838E-01 5.619E-01 5.481E-01 0.000E+00 0.000E+00 0.000E+00 0.000E+00 0.000E+00 0.000E+00 0.000E+00 0.000E+00 0.000E+00 0.000E+00 
6.625E+00 1.381E-01 1.886E-01 1.778E-01 0.000E+00 0.000E+00 0.000E+00 0.000E+00 0.000E+00 0.000E+00 0.000E+00 0.000E+00 0.000E+00 0.000E+00 
6.875E+00 4.143E-02 4.186E-02 3.452E-02 0.000E+00 0.000E+00 0.000E+00 0.000E+00 0.000E+00 0.000E+00 0.000E+00 0.000E+00 0.000E+00 0.000E+00 
7.125E+00 8.631E-03 6.473E-03 7.768E-03 0.000E+00 0.000E+00 0.000E+00 0.000E+00 0.000E+00 0.000E+00 0.000E+00 0.000E+00 0.000E+00 0.000E+00 
7.375E+00 8.631E-04 2.589E-03 1.295E-03 0.000E+00 0.000E+00 0.000E+00 0.000E+00 0.000E+00 0.000E+00 0.000E+00 0.000E+00 0.000E+00 0.000E+00 
\end{verbatim}
}

\end{appendix}

\end{document}